\documentclass[1p,number]{elsarticle}

\usepackage{amsmath}
\usepackage{amsfonts}
\usepackage{amssymb}
\usepackage{graphicx}
\usepackage{color}
\usepackage{ulem}

\begin{document}

\begin{frontmatter}

%% Title, authors and addresses

%% use the tnoteref command within \title for footnotes;
%% use the tnotetext command for the associated footnote;
%% use the fnref command within \author or \address for footnotes;
%% use the fntext command for the associated footnote;
%% use the corref command within \author for corresponding author footnotes;
%% use the cortext command for the associated footnote;
%% use the ead command for the email address,
%% and the form \ead[url] for the home page:
%%
\title{Fractal superconductivity near localization threshold}

\author[Landau,MIPT]{M. V. Feigel'man}
\ead{feigel@landau.ac.ru}
\address[Landau]{L. D. Landau Institute for Theoretical Physics, Kosygin st. 2, 
Moscow 119334, Russia}
\address[MIPT]{Moscow Institute of Physics and Technology, Moscow 141700, Russia}

\author[Rutgers,Orsay,Landau]{L. B. Ioffe}
\address[Rutgers]{Department of Physics and Astronomy, Rutgers
University, Piscataway, NJ 08854, USA} 
\address[Orsay]{CNRS and Universit\'e Paris-Sud, UMR 8626, LPTMS, Orsay Cedex, F-91405
FRANCE} 

\author[ICTP,Landau]{V. E. Kravtsov}
\address[ICTP]{Abdus Salam International Center for Theoretical Physics, Trieste, 
Italy}

\author[Murcia]{E. Cuevas}
\address[Murcia]{Departamento de F\'{\i}sica, Universidad de Murcia, E-30071 Murcia, Spain}

\begin{abstract}
We develop a semi-quantitative theory of electron pairing and resulting superconductivity 
in bulk "poor conductors" in which Fermi energy $E_F$ is located in the region of localized
states not so far from the Anderson mobility edge $E_c$. 
We assume attractive interaction between electrons near the Fermi surface.
We review the existing theories and experimental data and argue that a large class of 
disordered films is described by this model.  

Our theoretical analysis is based on analytical treatment of pairing correlations, 
described in the basis of the exact single-particle eigenstates of the 3D Anderson model, 
which we combine with numerical data on eigenfunction correlations. Fractal nature of
critical wavefunction's correlations is shown to be crucial for the
physics of these systems. 

We identify three distinct phases: 'critical'
superconductive state formed at $E_F=E_c$, superconducting state with a
strong pseudogap, realized due to pairing of weakly localized electrons
and insulating state realized at $E_F$ still deeper inside localized band. 
The 'critical' superconducting phase is characterized by 
the enhancement of the transition temperature with respect to BCS result,
by the inhomogeneous spatial distribution of superconductive order parameter
and local density of states. 
The major new feature of the pseudo-gaped state is the presence of two independent 
energy scales: superconducting gap $\Delta$, that is due to many-body
correlations and a new "pseudogap" energy scale $\Delta_P$ which characterizes
typical binding energy of localized electron pairs and
leads to the insulating behavior of the resistivity as a function of temperature above
superconductive $T_c$. Two gap nature of the pseudogapped superconductor
is shown to lead to specific features seen in scanning tunneling
spectroscopy and point-contact Andreev spectroscopy. We predict that
pseudogaped superconducting state demonstrates
anomalous behavior of the optical spectral weight.
The insulating state is realized due to presence of local pairing gap but
without superconducting correlations; it is characterized by a hard
insulating gap in the density of single electrons and by purely activated 
low-temperature resistivity $\ln R(T) \sim 1/T$. 

Based on these results we propose a new "pseudospin" scenario of superconductor-insulator
transition and argue that it is realized in a particular class of disordered superconducting films.
We conclude by the discussion of the experimental predictions of the theory and the theoretical 
issues that remain unsolved. 
\end{abstract}

\end{frontmatter}

\tableofcontents

\section{Introduction\label{Intoduction}}

The purpose of this paper is to develop the semi-quantitative extension of
the BCS theory of superconductivity that describes strongly disordered
conductors which normal state is a weak Anderson insulator~\cite{AndersonLoc}
or a very poor metal. The paper focuses on the case of "uniformly
disordered" materials, which do not contain morphological structures such as
grains coupled by tunnel junctions. Below in this section we briefly review
the existing theoretical models of the superconductor-insulator transition
(SIT), compare them with the results of the experimental studies of the
uniformly disordered films and choose the appropriate model for the
superconductor-insulator transition in these materials. The conclusion of
this introductory part is that this quantum transition in uniformly
disordered films can be described by BCS pairing of electrons which
single-particle states are close to the mobility edge of Anderson
localization~\cite{AndersonLoc}. Because BCS pairing is most relevant for
electrons close to the Fermi surface, the transition occurs when Fermi-level 
$E_{F}$ is located in the region of localized single-electron states but
close to the mobility edge. In the vicinity of the transition localization
length is longer than typical distance between carriers while the relevant
single electron states have the statistical properties of "critical
wavefunctions". Section \ref{Model} formulates the model in more detail and
discusses the issue of \textit{wavefunction fractality}, which turns out to
be very important for the theory of superconductor-insulator transition
developed in this work.

The important difference between our approach and many other works on
superconductor--insulator transition is neglect of the effects of Coulomb
interaction but consistent treatment of moderately strong disorder. In this
respect our work is the extension of the approaches developed originally by
Ma and Lee~\cite{MaLee}, Kapitulnik and Kotliar~\cite{KapitulnikKotliar1986}%
, Bulaevskii and Sadovskii~\cite{BulaSad}, and more recently by Ghosal,
Trivedi and Randeria~\cite{Ghosal2001} that have considered competition of
superconducting pairing and Anderson localization without explicit account
for Coulomb interaction. We give detailed arguments which justify 
applicability of our approach to disordered films such as amorphous InO$_{x}$ and TiN on
both phenomenological and microscopic levels below in subsection \ref%
{Experimental results}. Briefly, one should distinguish two possible effects
of the Coulomb interaction: suppression of paring interaction between
individual fermions which occurs at short scales and enhancement of the
phase fluctuations of the order parameter at large scales. The first would
lead to a gap suppression in a direct contradiction with data while the
second would lead to the phenomenology similar to that of Josephson junction
arrays which display markedly different behavior. Our theory can be also
applied to the cold atoms in optical and magnetic lattices with a controlled
disorder. In such systems interaction is always attractive and effectively
short-range. Moreover, it is tunable by magnetic field due to the Feshbach
resonance~\cite{atoms} which would allow to test directly the detailed
predictions of the developed theory.

The important consequence of the wave function fractality is the formation
of the strongly bound electron pairs which survive deep in the insulating
regime. In this situation the pairing interaction reduces the mobility of
individual electons leading to "superconductivity-induced" insulator. We
discuss this behavior in section \ref{Insulating state}. Theory of Cooper
pairing of electrons populating critical fractal states is developed in
section \ref{Cooper instability}. Here we develop three different
approximations for the computation of the superconducting transition
temperature and other properties. Section \ref{Superconducting state} gives
the physical properties of the superconductor in this regime. We show that
three different approximations developed in section \ref{Cooper instability}
agree with each other and predict parametrically strong enhancement of $%
T_{c} $ with respect to its value given by the "Anderson theorem". Another
distinguishing feature of the emerging superconducting state is extremely
strong spatial inhomogeneity of superconducting order parameter and the
presence of a well-defined global $T_{c}$.

Section \ref{Superconductivity with a pseudogap} presents the theory of
superconductivity coexisting with a strong pseudogap. Here we show that
superconductivity survives when $E_{F}$ is located much deeper in the
localized band than it was previously expected. In this regime the
superconductivity develops against the background of the pseudogap and,
thus, is characterized by a number of unconventional properties. We present
the specific results for the tunneling density of states and tunneling
conductance, Andreev point-contact conductance, and spectral weight of
high-frequency conductivity in this pseudogap superconductive state.
Finally, section \ref{Summary of results} reviews the main results and
discusses a number of open problems. The Appendices present technical
details of virial expansion that was used as the one of methods for the
determination of $T_{c}$.

\subsection{Theoretical models\label{Theoretical models}}

Quantum phase transitions from superconducting to insulating state in
disordered conductors and artificial structures were studied intensively
since mid-1980s, for review see e.g. \cite%
{ReviewGoldman,Finkelstein1994,FazioZant}. A number of theoretical models
describing such transitions were proposed and studied but a fully coherent
theoretical picture of this phenomena has not been established. We believe
that such transitions might be driven by different mechanisms in different
materials and thus belong to different universality classes which should be
described by different models. Below in this section we briefly review the
alternative mechanisms and the theoretical methods employed for their
description (see also short review~\cite{Larkin1999}). To avoid confusion,
we note that the first and the second of the mechanisms described below will
not be studied in the main part of our paper. We describe them in some
detail mostly because we need these details in order to argue below in this
section that they are not relevant for the superconductor-insulator
transition in homogeneous amorphous films.

\subsubsection{Coulomb blockade versus Superconductivity in Josephson
junction arrays. \label{Coulomb blockade}}

Macroscopic conductor that looks homogeneous at a macroscopic scale might be
in fact composed of small grains or islands of \ a good superconducting
metal with transition temperature $T_{c0}$. These grains are coupled to each
other via low-transparency insulating tunnel barriers~\cite%
{ReviewGoldman,FazioZant}, characterized by dimensionless tunnel
conductances $G_{ij}=h/(2e)^{2}R_{ij}$. The transition is due to the
competition of the charging and Josephson energies. Exactly the same physics
is realized in the artificial Josephson junction arrays, the only difference
between the inhomogeneous films and artificial array is that the number of
neighbors and of plaquette areas in the former are random. The simplest
model describing this physics is given by effective Hamiltonian written in
terms of phases $\phi _{j}$ and charges $Q_{j}$ assigned to grains: 
\begin{equation}
H_{1}=\frac{(2e)^{2}}{2}\sum_{ij}C_{ij}^{-1}N_{i}N_{j}-\sum_{ij}E_{J}^{ij}%
\cos (\phi _{i}-\phi _{j})  \label{H1}
\end{equation}%
where $N_{j}=Q_{j}/2e$ is the number of Cooper pairs on the $j$-th grain, $%
C_{ij}$ is the matrix of mutual capacitances, and $E_{J}^{ij}$ is the
Josephson coupling energy, correspondingly. The model is often further
simplified by assuming that $E_{J}^{ij}=E_{J}$ is a non-zero constant only
for nearest-neighboring grains while\textbf{\ }the capacitance matrix is
diagonal, $C_{ij}=C_{0}\delta _{ij}$. This model neglects the effects of the
quasiparticles which might be (sometimes) justified for small grains at low
temperatures $T\ll T_{c0}$, due to exponentially small density of normal
electrons. The key parameter of the problem is the energy ratio $%
x=E_{J}/E_{C}$ where $E_{C}=(2e)^{2}/2C_{0}$ is the Coulomb charging energy
due to $2e$ charge transfer. As was shown in the paper~\cite{Efetov1980} the
ground-state of the Hamiltonian (\ref{H1}) is insulating at $x\ll 1$ and
superconducting at $x\gg 1$, thus a phase transition(s) takes place at $%
x\sim 1$. An essence of this phase transition is the Mott-Hubbard
localization of Cooper pairs, taking place when tunneling matrix element of
a pair ($E_{J}$) is much less than on-site repulsion $E_{C}$.

However, the model (\ref{H1}) is unlikely to describe correctly the physics
of superconductor-insulator transition in Josephson arrays or granular
materials, especially in its simplified version with diagonal capacitance
matrix. It has two important deficiencies.

First, realistic Josephson junction arrays and disordered films are poorly
described by the model of diagonal capacitance matrix $C_{ij}=C_{0}\delta
_{ij}$ because normally the charging effects are controlled by capacitances
of junctions $C\gg C_{0}$, not by the ground capacitances of the islands
(see Ref.~\cite{FazioZant}). It is in fact impossible to have a capacitance
matrix dominated by the ground capacitance in the arrays which dimensionless
normal state conductance is $G=h/[(2e)^{2}R_{T}]\gtrsim 1$ because in these
arrays the capacitance of the junctions cannot be small. The reason for this
is that apart from purely geometrical contribution $C_{\mathrm{geom}}=4\pi
S/d$, junction capacitance $C=C^{\mathrm{geom}}+C^{\mathrm{ind}}$ contains
additional induced term $C_{\mathrm{ind}}=\frac{3}{16}Ge^{2}/\Delta $, (this
expression is valid at $T\ll \Delta $). This induced contribution is due to
virtual electron transition across the gap~\cite{LarkinOvchinnikov1983,AES}.
As a result, the charging energy can not be made arbitrary large
(equivalently capacitance cannot be small): $E_{C}\leq 32\Delta /3G$.
Josephson energy of the symmetric junction at low temperatures is $%
E_{J}=G\Delta /2$, thus the condition $E_{C}\geq E_{J}$ cannot be realized
at large $G$. Furthermore, at temperatures above the parity effect threshold 
$T^{\ast }$, (see~\cite{FeigKor}) an additional contribution to the
screening of Coulomb interaction between Cooper pairs comes from
single-electron tunneling. Thus, in all cases the effect of capacitance
renormalization is that the ratio $x=E_{J}/E_{C}$ in granular arrays is
controlled by the dimensionless conductance $G$ in such a way that Coulomb
effects are always weak at $G\geq 1$. Ground capacitance larger than the
junction capacitance thus implies that the transition into the insulating
state would occur in the arrays characterized by very small $G\ll 1$. Such
behavior was never observed experimentally in Josephson arrays (see section %
\ref{Experimental results}).

More realistic model involves the capacitance matrix that is dominated by
the junctions capacitances with a small contribution from the ground
capacitance of each grain. In this case the arguments of preceding paragraph
show that the transition between the insulating and superconducting state
should occur at $G_{c}\sim 1$.\cite{Orr1986} Moreover, deep in the
insulating phase the electrostatic interaction between the charges in 2D
Josephson array becomes logarithmic in distance, similar to the one between
the vortices in the superconducting phase. Assumption of the full duality
between vortices in the superconducting state and charges in the insulating
state allows one to make a number of predictions.\cite%
{MFisher1990a,MFisher1990b} For instance, because the current of vortices
generates voltage while the current of pairs implies the electrical current
one expects that superconducting-insulator transition is characterized by
the universal value of $G_{c}=1$. \cite{MFisher1990b}

Unfortunately, despite a significant experimental effort the universal value
of the resistance was never experimentally confirmed for Josephson arrays
and for most disordered films. Moreover, in non-zero magnetic field the
Josephson arrays often show a large regime of the temperature-independent
resistance. The reason for this is likely to be due to the important
physical effects missed by the model (\ref{H1}), namely the presence of
random induced charge on the superconducting islands. As was shown in a
number of Josephson junction studies (see e.g. \cite{Zimmerman2008} ) the
induced charge on each island exhibits very slow random fluctuations and is
therefore inherently random variable. Most likely the time dependence of
these fluctuations can be neglected and the induced charge can be regarded
as a quenched random variable $q_{i}$ that should be added to the
Hamiltonian (\ref{H1}):

\begin{equation}
H_{1}=\frac{(2e)^{2}}{2}\sum_{ij}C_{ij}^{-1}(N_{i}-q_{i})(N_{j}-q_{j})-%
\sum_{ij}E_{J}^{ij}\cos (\phi _{i}-\phi _{j})  \label{H_JJ}
\end{equation}%
The properties of the model (\ref{H_JJ}) are not currently well understood;
in particular, it seems likely (see \cite{MullerIoffe}) but was not proven
that insulating state has many glassy features responsible for intermediate
'normal' phase. It is, however, clear that in this model the transition, or
a series of transitions occurs at $E_{J}/E_{C}\sim 1$, which corresponds to $%
G\sim 1$ in the normal state. Near the critical point the gapless
excitations correspond to collective modes build of electron pairs while the
electron spectrum remains fully gapped. In analogy with spin glasses, one
expects that effective frustration introduced by random charges and magnetic
field leads to a large density of low energy states. This might explain the
observed temperature independent resistance that varies at least by one order of
magnitude around $G\sim 1$ as a function of magnetic field.\cite{FazioZant}

As we show in section \ref{Experimental results} both the data and
theoretical expectations for models (\ref{H1},\ref{H_JJ}) differ markedly
from the behavior of the homogeneously disordered films.

A spectacular property of the superconductor-insulator transition in
granular materials is that a strong magnetic field applied to the system in
the insulating regime results in a dramatic increase of the conductance.
Qualitatively, the reason for this behavior in granular systems is that in
the absence of the field the single electron excitations are absent due to
superconducting gap while pairs are localized due to Coulomb energy and
random induced charges. Large field suppresses superconducting gap, which
allows transport by individual electrons that is characterized by a much
larger tunneling amplitude and lower (by a factor of four) effective
charging energy. This effect was reported by \cite{Gerber1997} where strong
magnetic field was applied to Al grains immersed into Ge insulating matrix
and giant negative magneto-resistance was observed. Similar behavior was
reported for homogeneous films of InO deep in the insulating regime\cite%
{Gantmakher1996}. This similarity indicates that the main reason for this
effect, which is that the pairing of the electrons survives deep in the
insulating state, also holds for homogeneous films of InO. The quantitative
theory of negative magnetoresistance in granular superconductors was
developed in~\cite{BeloborodovLarkinEfetov} for the case of relatively large
inter-grain conductances $G_{ij}\approx G\gg 1$, in which case the negative
magnetoresistance effect is small as $1/G$. Recent review of theoretical
results on normal and superconductive granular systems can be found in \cite%
{Beloborodov}.

\subsubsection{Coulomb suppression of $T_{c}$ in uniformly disordered thin
films. \label{Coulomb suppression}}

The scenario described above assumes that superconductivity remains intact
inside each grain. An alternative mechanism for the suppression of
superconductivity by Coulomb repulsion was developed by Finkelstein~\cite%
{Finkelstein1987,Finkelstein1994}, building upon earlier perturbative
calculations~\cite{CoulPert}. Finkelstein effect becomes important for very
thin strongly but homogeneously disordered films, as well as quasi-1D wires
made out of such films ~\cite{Oreg1999}. Contrary to the Coulomb blockade
scenario, the system is supposed to be "uniformly disordered", with no
superstructures such as grains coupled together by weak junctions. Somewhat
similar idea was proposed~\cite{AndersonMuttalibRamakrishnan1983,Coffey} for
three-dimensional materials near the localization threshold. The essence of
Finkelstein effect is that Coulomb repulsion between electrons gets enhanced
due to slow diffusion of electrons in highly disordered film, which results
in the \textit{negative} contribution to the effective Cooper attraction
amplitude at small energy transfer $\varepsilon $: 
\begin{equation}
\lambda (\varepsilon )=\lambda _{0}-\frac{1}{12\pi g}\ln \frac{1}{%
\varepsilon \tau _{\ast }}  \label{lambdaFin}
\end{equation}%
where $g=h/(2e)^{2}R_{\Box }$ is dimensionless film conductance, $\lambda
_{0}$ is the "bare" Cooper attraction constant defined at the scale of Debye
frequency $\omega _{D}$, and $\tau _{\ast }=\max {\tau ,\tau (b/l)^{2}}$,
where $\tau $ and $l=v_{F}\tau $ are the mean scattering time and mean free
path, and $b$ is the film thickness. The suppression of attraction constant
Eq.(\ref{lambdaFin}) leads immediately (we assume here $\omega _{D}\sim
1/\tau _{\ast }$) to the result obtained early on in the leading order of
the perturbation theory~\cite{CoulPert} 
\begin{equation}
\frac{\delta T_{c}}{T_{c}}=\frac{\delta \lambda }{\lambda ^{2}}=-\frac{1}{%
12\pi g}\ln^{3}\frac{1}{T_{c0}\tau _{\ast }}  \label{Tc1}
\end{equation}%
The leading terms in the perturbation theory for $T_{c}$ can be summed by
means of the renormalization group method developed in~\cite{Finkelstein1987}%
. In the leading order over $1/g\ll 1$ one gets 
\begin{equation}
\frac{T_{c}\tau _{\ast }}{\hbar }=\left[ \frac{\sqrt{8\pi g}-\ln (\hbar
/T_{c0}\tau _{\ast })}{\sqrt{8\pi g}+\ln (\hbar /T_{c0}\tau _{\ast })}\right]
^{\sqrt{2\pi g}},  \label{Tc-Fin}
\end{equation}%
According to Eq.~(\ref{Tc-Fin}), $T_{c}$ vanishes at the critical
conductance $g_{c\mathrm{F}}=\ln ^{2}(\hbar /T_{c0}\tau _{\ast })/(8\pi )$,
(which needs to be large enough for the theory to be self-consistent). At
lower (but still large compared to unity) conductances, the material never
becomes superconducting; it stays metallic at least down to very low
temperatures $T_{loc}\sim (\hbar /\tau _{\ast })\exp (-4\pi g)$ where weak
localization crosses over into the strong localization~\cite{Gershenson}.

This mechanism of superconductivity suppression, described by Eq.(\ref%
{Tc-Fin}), might be called "fermionic", as opposed to the "bosonic"
mechanism discussed in previous subsection~\cite{Larkin1999}. Within this
mechanism, superconductivity is destroyed at relatively large conductances $%
g_{cF}\geq 1$, thus a {\it direct} superconductor-insulator transition does
not seem to be a natural option.

The theory of the superconducting-insulator transition outlined above
neglects the mesoscopic fluctuations of the interaction constant, ${g(%
\mathbf{r)}}$. This assumption was questioned on phenomenological grounds by
Kowal and Ovadyahu \cite{Kowal1994}. The role of these fluctuations become
larger when superconductivity is strongly suppressed because as follows from
(\ref{lambdaFin}), in this regime even small mesoscopic fluctuations of $g(%
\mathbf{r})$ lead to a large spatial fluctuations of the effective coupling
constant $\lambda (\varepsilon ,\mathbf{r})$. In its turn, the fluctuations
of the effective coupling lead to the local spatial fluctuations of the
transition temperature $T_{c}(\mathbf{r})$, which becomes very strong, $%
\delta T_{c}/\bar{T_{c}}\geq 1$, for nearly-critical conductance $g\approx
g_{c\mathrm{F}}$, as shown in Ref.~\cite{Skvortsov2005}. Note that
mesoscopic fluctuations of $T_{c}$ remain small in the universal case of
short range disorder if Coulomb suppression of superconductivity is not
taken into account, even in the vicinity of the upper critical field $%
H_{c2}(0)$ at very low temperatures. \cite%
{SpivakZhou1995,GalitskiiLarkin2001}

These results demonstrate the inherent inhomogeneity of superconducting
state near the critical point at which it is destroyed by disorder. Thus, it
seems likely that the regime close to the superconductor-insulator
transition is described by the effective model of superconducting islands
(appeared due to spatial fluctuations of local attraction constant) coupled
by weak SNS junctions. The theoretically consistent description of this
physics is still lacking, the difficulty can be traced to the absence of
tunnel barriers separating fluctuation-induced superconducting islands from
the surrounding media. In the absence of these barriers charging effects
become non-local which makes the determination of the effective Coulomb
energy a nontrivial problem, furthermore the presence of a large normal part
implies dissipative (non-local in time) dynamics of the superconducting
phase. A toy model of this type was solved in~\cite%
{FeigelmanLarkinSkvortsov2001}; this model describes artificial
superconductive islands in a good contact with disordered thin film.

In conclusion, in the fermionic mechanism the Coulomb interaction is
enhanced by disorder leading to the suppression of superconductivity. The
state formed when the superconductivity is suppressed is likely to be a poor
metal characterized by a large resistivity and finite density of states at
the Fermi level.

\subsubsection{Localization versus Superconductivity. \label{Localization
versus superconductivity}}

The third alternative mechanism for the superconductor-insulator transition
is due to the localization of single electrons. In this scenario the effects
of Coulomb interaction are neglected, whereas local (in space) Cooper
attraction is treated within standard BCS approximation. Here we focus on
the case of bulk disordered materials or sufficiently thick films in which
localization remains a three dimensional effect. Abrikosov and Gor'kov~\cite%
{AG1959} and Anderson~\cite{Anderson1959} have shown that potential disorder
does not affect thermodynamic properties of usual s-wave superconductors.
More precisely, this statement (based upon the presence of time-reversal
symmetry and called "Anderson theorem") means that the parameter $T_{c}\tau
/\hbar $ does not appear in BCS theory as long as magnetic field and/or
supercurrent are absent. However, localization of single-electron
eigenstates at very strong disorder leads to appearance of an additional
energy scale $\delta _{L}=1/\nu _{0}L^{3}$, where $L$ is the single-electron
localization length and $\nu _{0}$ is the density of states (per single spin
projection). The meaning of $\delta _{L}$ is just average level spacing
inside typical volume where wavefunction is localized. One expects that at
large $\delta _{L}$ the superconducting pairing between electrons is
suppressed. Competition between superconductivity and Anderson localization
was studied originally in mid-80's~\cite{MaLee,BulaSad,KapitulnikKotliar1986}%
. Their major conclusion was that Anderson theorem is valid and
superconductivity survives provided that the condition 
\begin{equation}
T_{c}\gg \delta _{L}  \label{cond1}
\end{equation}%
is satisfied. The reasoning leading to Eq.(\ref{cond1}) is that for Cooper
instability to develop, characteristic energy spacing between hybridized
Cooper pairs (which are supposed to be localized in the same region of size $%
L$) should be smaller than typical energy scale $T_{c}$ corresponding to the
Cooper instability. On the contrary, no superconducting long-range pairing
seems possible when level spacing $\delta _{L}$ strongly exceeds $T_{c}$, in
spite of the presence of inter-electron attraction (we assume that Cooper
attraction survives when the single electron states are localized as long as 
$\delta _{L}$ is much smaller than Debye energy $\omega _{D}$).

We will show below that the analysis presented in~\cite%
{MaLee,BulaSad,KapitulnikKotliar1986} is not complete in two important
respects. First, the absence of long-range superconductive order does not
necessarily mean that Cooper pairing is totally negligible; we will show
that in the range $T_{c}\ll \delta _{L}\ll \omega _{D}$ Cooper correlations
leads to formation of the hard-gap insulator instead of usual variable-range
one. This gap is of the same origin as the "parity gap" studied by Matveev
and Larkin~\cite{Matveev1997} in the context of ultra-small grains of good
superconductive metal. Second, the notion of \textit{eigenfunction fractality%
} (that was not known when the theory\cite%
{MaLee,BulaSad,KapitulnikKotliar1986} was developed) has to be taken into
account and leads to important physical consequences. In the present paper
we are going to fill both these gaps; it will be shown that fractality of
electron eigenfunctions changes qualitative features of superconductive
state, and even modifies the condition $\delta _{L}\approx T_{c}$ for the
critical region where superconductivity is finally destroyed.

More recently the issue of competition between localization and
superconductivity was reconsidered in the important paper by Ghosal, Trivedi
and Randeria~\cite{Ghosal2001}. They considered two-dimensional lattice
model of superconductivity with moderately strong local attraction
(negative-U Hubbard model) and on-site disorder and studied it numerically
by two methods: \ by solving the self-consistent Bogolyubov-de Gennes
equations, and by solving BCS pairing equations in the basis of exact
single-electron eigenstates. They demonstrated that with increase of local
disorder superconducting state is transformed into the insulating one. The
latter possesses sharp gap in the density of states but does not show
coherence peaks. The energy gap was found to be non-monotonic as function of
the disorder strength. It was also shown that superconductivity is very
inhomogeneous in the crossover region, with disorder-generated "islands" of
large pairing amplitude. We will see below that qualitative features of the
results obtained in Ref.~\cite{Ghosal2001} are very robust and survive in a
continuum weak-coupling BCS model that we consider in the present paper (see
section \ref{Model}). The drawback of the treatment developed in Ref.~\cite%
{Ghosal2001} is that it does not allow the quantitative analysis of the
physical properties as function of main parameters of the problem (coupling
strength $\lambda \ll 1$ and proximity of the Fermi-energy to the
localization edge, $|E_{F}-E_{c}|\ll E_{F}$), due to limitations imposed by
purely numerical methods. A major drawback of most conventional numerical methods 
is their inability to study the regime characterized by dramatically different
energy scales, in particular $T_c \ll E_F$. Development of the method to study
this regime is the goal of the present paper.

The importance of the analytical treatment of the weak coupling regime
$T_c \ll E_F$ is demonstrated in particular, by the numerical work 
\cite{Shepelyansky2002} which has studied the 3D disordered Hubbard model with 
strong local attraction (4 times larger than bandwidth). In this regime the electrons 
are strongly bound to each other even in translationally invariant systems. One
expects that the mobility of the formed pairs is less than the mobility of the original 
electrons; this enhances the effect of the disorder. This expectation
is conformed by the data\cite{Shepelyansky2002}. We show that in the 
physically relevant regime of \textit{weak} attraction, the situation is opposite:
the superconductivity survives deep in the regime of localized states.

\subsection{Experimental results on S-I transitions. \label{Experimental
results}}

Phenomenologically one should distinguish at least three types of materials
that display superconductivity suppression with the increase of the
disorder: granular systems \cite{Gerber1997,Goldman2002}, nominally
homogeneous films that exhibit superconductor-metal-insulator transition and
homogeneous films that show direct superconductor-insulator transition.\cite%
{BaturinaReview2007} We shall discuss only the latter class in this paper,
the materials that exhibit it are thick (more than 20 nm) InO$_{x}$ films,
thinner (abound 4nm) TiN films and very thin (few atomic layers) Be
films. Recent work reports similar behavior also in disordered epitaxial
films of NbN with varying disorder~\cite{Pratap}. The goal of this section
is twofold: to discuss the data that allow to exclude Coulomb mechanisms
(both fermionic and bosonic) of the superconductor-insulator transition in
these films and to briefly summarize the data on these films that need
theoretical explanation. We begin with the first.

The most direct experimental evidence that allows to exclude the 'fermionic'
mechanism of superconductivity suppression discussed in section \ref{Coulomb
suppression} in homogeneously disordered films that exhibit direct
superconductor-insulator transition comes from the recent tunneling data. In
these experiments one observed that the suppression of the superconductivity
either by disorder or temperature is {\it not} accompanied by the
suppression of the gap, which remains intact or even increases reaching $%
2\Delta _{1}/T_{c}\in 6-9$. \cite{Sacepe2007} Instead, as the insulator is
approached, one observes the disappearance of the local coherence peaks
which, in addition, vary strongly from one point to another ( see also
Ref.~\cite{TiN-STM}) . This behavior
(and especially the temperature dependence of tunnelling conductance) is in
a striking contrast to what is expected for the fermionic mechanism. Less
direct evidence is provided by the data \cite{Shahar2004,Baturina2007a}
showing that superconductivity exists up to a very strong disorder
corresponding to $g\approx 1$, which is at least a factor of two smaller
than the one expected in the fermionic mechanism (\ref{Tc-Fin}).

As explained in section \ref{Coulomb suppression} the basis of the fermionic
mechanism is the idea that Coulomb repulsion is enhanced by disorder which
results in the effective suppression of the attractive interaction that
leads to superconductivity. The actual equations are derived in the
assumption that bare Coulomb repulsion is very strong but is reduced by
screening to the universal limit in which the effective Coulomb repulsion
constant is equal to unity. The dimensionless parameter characterizing the
strength of the Coulomb interaction and its screening is $2\sigma
/(T_{c}\kappa )$, where $\sigma =(e^{2}k_{F}/6\pi ^{2})\,(k_{F}l)$ is the
residual conductivity and $\kappa $ is the dielectric constant due to
electrons far from the Fermi energy ($|E-E_{F}|>\omega _{D}$). Coulomb
interaction is effectively strong provided that $2\sigma /(T_{c}\kappa )\sim
(\xi _{0}/a_{\mathrm{scr}})^{2}/\kappa \gg 1$, where $a_{\mathrm{scr}}$ is
the Thomas-Fermi screening length, $\xi _{0}$ is the coherence length in a
dirty superconductor. If instead $2\sigma /(T_{c}\kappa )\ll 1$ the
coefficient in front of the logarithm in (\ref{lambdaFin},\ref{Tc-Fin})
becomes small so that enhancement of Coulomb repulsion become important only
at exponentially low energy scales. In a very dirty metal (such InO$_{x}$%
film) with a short mean free path $k_{F}l\sim 0.3$ and low carrier density $%
e^{2}k_{F}\sim 5000K$ (see Ref.\cite{Shahar1992}) the ratio $\sigma
/T_{c}\sim 10$. Thus, the effects of Coulomb interaction in these films
become unimportant if dielectric constant $\kappa \gg 10$. The direct
measurements of the dielectric constant deep in the insulator regime give $%
\kappa \geq 30$ ~\cite{Zvi-Insulator}, one expects that it can be only
larger in the vicinity of superconductor-insulator transition, so $2\sigma
/(T_{c}\kappa )\ll 1$ in these films which makes the fermionic mechanism
irrelevant.

The microscopic origin of this large dielectric constant is likely to be due
to the low density of the carriers, $n_{e}\sim 10^{21}cm^{-3}$ in these
conductors, and the peculiar structure of their density of states in which
the Fermi level is located in a large dip.\cite{Anisimov2009}. In this
situation the density of the electrons distant from the Fermi level, $%
|E-E_{F}|>\omega _{D},$ is high, which results in a large screening of the
Coulomb interaction.

Now we turn to the possibility of the Coulomb driven transition similar to
the one of Josephson arrays (section \ref{Coulomb blockade}). This physics
is due to the long range nature of the Coulomb interaction; the estimates 
of the associated energy scales given below show that Coulomb interaction
is sufficiently well screened even in poor conductors so that the corresponding
energy is too small compared with all other energy scales; this 
rules out this possibility. These arguments are quite general and apply to 
other effects that originate from the long range part of the Coulomb interaction. 

We begin with  energy scales that  are known to  be relevant experimentally  for
this problem. Namely, the superconducting gap on the ordered side and activation
energy on the insulating side. These energies are large: the superconducting gap
is around $\Delta \sim  5K$ while the activation  gap is even larger  $T_{0}\sim
10-15K$ (see below). We now estimate the Coulomb interaction at the scale of the
superconducting coherence length, $\xi_0$. Because $T_c/E_F \lesssim 10^{-3}$ in
these materials,  this length,  even for  a very  poor conductor,  cannot be too
short: $\xi  _{0}\gtrsim 10nm$.  Large value  of the  dielectric constant in the
parent insulating compounds, $\kappa \geq 30$  , implies that in the absence  of
the mobile electrons with $|E-E_{F}|<\omega _{D}$, the effective charging energy
at  scales  $\xi  _{0}$ would  be  $E_{c0}=e^{2}/\kappa  \xi _{0}\lesssim 50\; K$.
Screening  by  conduction electrons  with  energies $T_c<|E-E_{F}|<\omega  _{D}$
decreases it further. To estimate this effect we note that at scales  less than
$\xi_{0}$ the properties of the electrons are similar to those at the mobility 
threshold. At the threshold, the dielectric constant grows with scale, $L$, 
according to the scaling law $(L/l)^x$ with $x\gtrsim 1$. This results in the 
additional factor $(l/\xi _{0})^{x}$ in the effective Coulomb energy at scales 
$\xi_0$ which reduces it to $E_{c}\lesssim 1K$. Thus, the effective Coulomb energy is
much smaller that all relevant energy scales and cannot be the driving force  of
the transition.  Note that  this estimate  becomes incorrect  in the presence of
thin insulating barriers  between the grains,  thus allowing for  Coulomb driven
transition in inhomogeneous materials. The absence of structural inhomogeneities
that  might  lead  to  such  barriers  in  films  of  InO$_{x}$  was  shown  in
\cite{Kowal1994}; later studies\cite{Baturina2008} also reported  the absence
of inhomogeneities in TiN films.

Another argument against Coulomb driven transition is provided by a
completely different phenomenology of the transition in the films and in
Josephson arrays: in the former one observes direct transition to the
insulating state characterized by a large gap and activation behavior of
resistivity, the transition can be driven either by the increase of disorder
or by magnetic field. In contrast, in Josephson arrays the transition driven
by the field is characterized by a large intermediate regime of temperature
independent resistivity.\cite{FazioZant,Serret2002}. Furthermore, there is
no reason to expect the disappearance of coherence peaks in some places and
not in others as one approaches the transition (reported in \cite{Sacepe2007}%
) in the array of superconducting grains.

Finally, Coulomb mechanism cannot explain the large value of the activation
energy in the insulating state in the vicinity of the transitions. One
expects that when charging energy becomes large enough to suppress transport
by Cooper pairs, single electrons should dominate transport which implies
activation energy equal to superconducting gap. Instead one often observes
large activation energies: $T_{0}\approx 15K$ in InO$_{x}$ films (see Fig.1
of \cite{Gantmakher1996}) and even larger in Be ones\cite{Wu05}.

We now summarize (see also the review \cite{BaturinaReview2007}) the
phenomenology of the direct superconductor--insulator transition in
homogeneous films as exhibited by three different systems: thick amorphous
InO$_{x}$ films \cite%
{Shahar1992,Kowal1994,Gantmakher1996,Shahar2004,Shahar2005,Steiner2005},
thin TiN films ~\cite{Baturina2004,Baturina2007a,Baturina2007} and
extra-thin (below 1 nm) Berillium films~\cite{Wu01,Wu05,Adams2001};
somewhat similar phenomena were observed recently in the patterned Bi film 
with honeycomb array of holes~\cite{Valles2009}.

\begin{enumerate}
\item {\ On insulating side of SIT, low-temperature resistivity curves show
simple activated behavior, $R(T)\propto \exp (T_{0}/T)$, which crosses over
into Mott~\cite{Kowal1994}, $R(T)\propto \exp (T_{M}/T)^{1/4}$, or
Efros-Shklovsky~\cite{Baturina2007a,Wu05}, $R(T)\propto \exp
(T_{ES}/T)^{1/2} $ variable-range hopping {\it at higher} temperatures.
This behavior is highly unusual: in hopping insulators where the activation
is frequently observed at high temperatures it crosses over to some
fractional (variable-range) behavior upon the temperature decrease. }

\item {\ At high magnetic fields the films on both sides of the SIT\ show
large negative magnetoresistance \cite%
{Gantmakher1996,Shahar2004,Shahar2005,Steiner2005,Adams2001,Baturina2007,Baturina2007a}%
. It is important that such behavior was observed even in films
characterized by a very high activation energy. }

\item {\ At low fields all insulating films close to SIT\ show positive
magneto-resistance at low fields \cite%
{Gantmakher1996,Baturina2007,Baturina2007a,Adams2001,Wu012}. }

\item {\ Ultra-low-temperature measurements on nearly-critical samples of
a-InO$_{x}$~\cite{Shahar2005} and TiN~\cite{Baturina2007a} revealed a very
sharp jump (by several orders of magnitude in current) in nonlinear $I(V)$
curves. }

\item {\ The resistance of Be \cite{Adams2001} and TiN films~\cite%
{Baturina2007} approach the quantum resistance $h/e^{2}$ at very strong
magnetic fields and low temperatures. Very recently the importance of Zeeman
pair-breaking for the properties in this regime was demonstrated in Ref.~%
\cite{Adams2009}. }

\item The properties of the quantum critical point that separates
superconductor and insulator are not fully established. {The samples
corresponding to the critical disorder which separates superconducting and
insulating behavior~ display insulating behavior of the $R(T)$\cite%
{Gantmakher1998,Baturina2007a,Wu05}; this suggests that the disorder driven
superconductor-insulator transition is not described by a self-dual theory
proposed in~\cite{MFisher1990a} or that the critical regime where this
behavior sets in is very narrow. Weakly superconducting samples can be
driven into insulating state by the application of magnetic field. Scaling
(or a lack of thereof) near this quantum critical point is subject of
controversy in both the value of the critical resistance and the scaling
exponents. Some works\cite{Baturina2007,Gantmakher1998,Sacepe2007} report
critical value of the resistance larger than }$R_{Q}=6.5k\Omega $ and
exponents that do not agree with the theoretical predictions\cite%
{MFisher1990b}. In contrast, recent paper\cite{Steiner2008} reports both the
critical value of the resistance and exponent in a perfect agreement with
the theoretical predictions based on dirty boson scenario\cite{MFisher1990b}.
\end{enumerate}

\begin{figure}[tph]
\includegraphics[width=12cm]{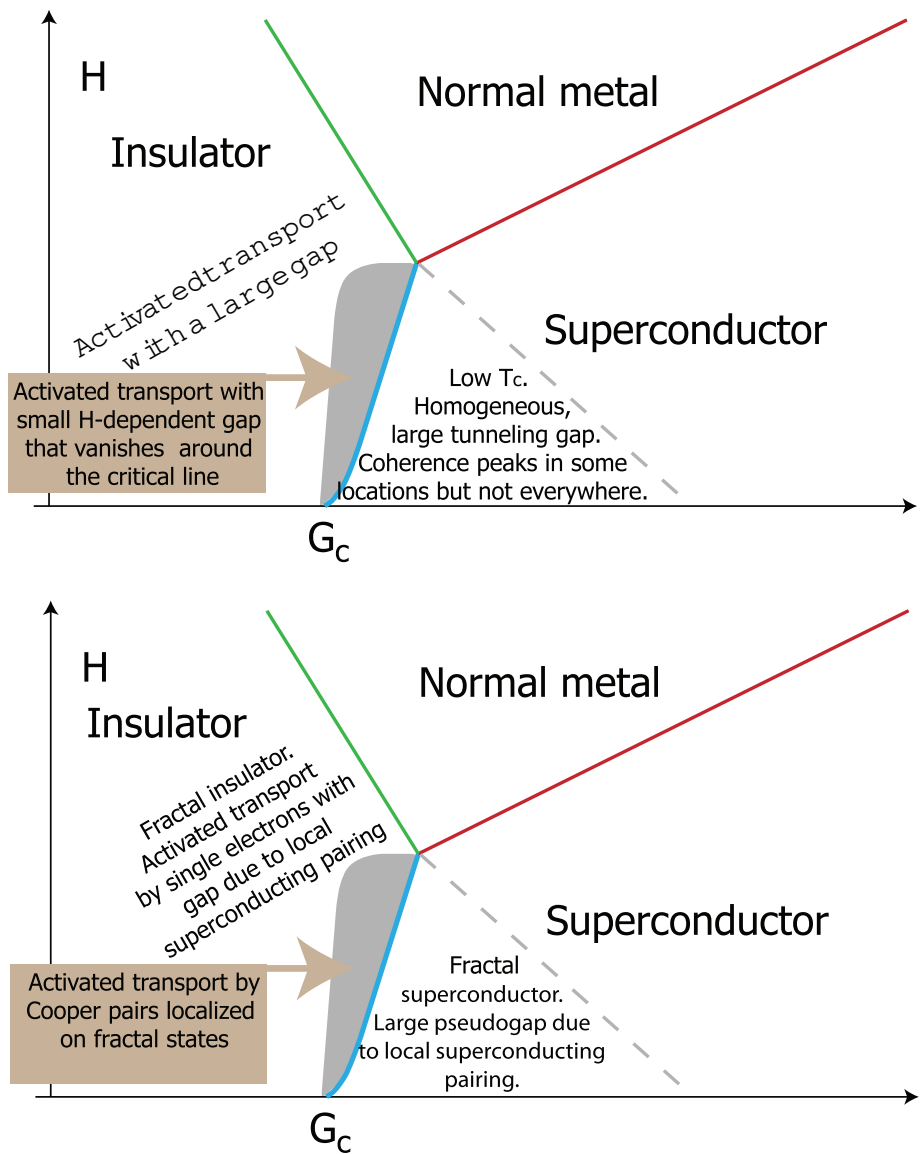}
\caption{(Color online) Sketch of the experimental phase diagram of
homogeneously disordered films (upper panel) at $T\rightarrow 0$ and its
interpretation in the theory of superconductivity developed in the electron
system close to mobility edge (lower panel).}
\label{PhaseDiagram}
\end{figure}

These results can be summarized by the low temperature phase diagram in $%
(H,G)$ plane sketched in Figure~\ref{PhaseDiagram}.

It is tempting to explain these observations by the scenario in which the
disorder destroys the global superconducting coherence while preserving the
local superconducting gap, $\Delta $, even in the insulating phase. Indeed,
in the absence of pair coherence the conductivity is due to the thermally
excited fermionic quasiparticles. The density of these excitations would be
additionally suppressed by local superconductive gap $\Delta $ that is
formed at low temperatures: $n_{1}(T)\sim \exp (-\Delta /T)$ . This would
explain the crossover to activation behavior at low $T$. High magnetic field
suppresses local $\Delta (H)$ thus leading to very large negative
magnetoresistance. Low-field positive magneto-resistance could be then
associated with frustration induced by magnetic field which eliminates the
last vestiges of superconducting coherence thereby shifting the array
further into insulating side (whereas not yet affecting local
superconducting gap).

This scenario would be naturally realized in the granular material where
superconductivity remains intact in each grain whereas global coherence
appears only due to Josephson coupling which competes with charging energy
(see section \ref{Coulomb blockade})~\cite{Dubi06}. However, this assumption
of hidden granularity of the SIT\ films is not plausible for many reasons
explained above.

The correct theory should be also able to explain the large value of the
activation energy on the insulating side of the transition. Assumption that
it is due to the local superconducting gap in weakly coupled grains is not
sufficient because this energy is higher than the gap in a less disordered
film of the same series that shows superconductivity. Indeed, maximal
observed values of $T_{0}$ were about $15K$ in work\cite{Gantmakher1996} and 
$11K$ in~work\cite{Kowal1994} which are significantly larger than maximal
superconducting gap $\Delta \approx 5K$ shown in Fig.1 of Ref.~\cite%
{Shahar1992} (note that samples studied in~\cite{Gantmakher1996} were from
the same source as in~\cite{Shahar1992,Kowal1994}). Furthermore, this
assumption would lead to the conclusion that the transition is due to
competition between Coulomb and Josephson energies which are both much
smaller than $T_{0}$ ; because $E_{J}=G\Delta /2$ this would imply that $%
G\ll 1$ in a direct contradiction with the data.

Large value of the activation energy, together with inefficiency of
electron-phonon coupling at low $T$ provides a natural explanation for a
large jump in $I(V)$ observed in the insulating state~\cite{Jumps}. The
detailed predictions of the theory \cite{Jumps} were recently verified
experimentally in~\cite{SacepeShahar09}. The important ingredient of this
theory is a hard gap for single electron excitations, similar to the one
directly observed in \cite{Sacepe2007} and inferred from the resistivity
data. Thus, the phenomena of $I(V)$ jumps does not impose additional
constraints on microscopic theory. Moreover, very similar jumps were observed
previously in the system which seems to have nothing to do with superconductivity
~\cite{Sanquer1996}.

Real challenge to the theory is presented by the magnetoresistance data from
Ref.\cite{Gantmakher1996}: straightforward interpretation of the negative
magnetoresistance data as being due to superconducting gap suppression in
individual grains leads to unphysically large values of critical field
needed to destroy the superconductivity in such grains. For instance, the
field of $8$ T was observed to produced only moderate ( $R(H=8\mathrm{T}%
)/R(0)\approx 0.5$ )\ negative magneto-resistance in a sample characterized
by $T_{0}\approx 15\,\mathrm{K}$ (as determined in the temperature range $%
1.3-5$ K). Interpreting this effect as being due to the suppression of the
pairing gap, we find $T_{0}-T_{0}(H=8\mathrm{T})\approx 0.7\,\mathrm{K}$,
which is about 5\% of $T_{0}$. Interpolating this dependence we find that
the field necessary to destroy completely the superconductivity in each
grain is huge: $H_{cg}^{\exp }\sim 50-80\,\mathrm{T}$. \footnote{%
Data of Ref.\cite{Steiner2005} show that 32 Tesla field is not sufficient to
fully suppress negative magnetoresistance.}

Such large values of the critical fields are impossible for realistic grain
sizes. Indeed, critical orbital magnetic field for a small (radius $R<\xi $,
where $\xi =\sqrt{\hbar D/\Delta }$ is the coherence length) superconducting
grain is~\cite{Larkin1965,BeloborodovLarkinEfetov} 
\begin{equation*}
H_{cg}^{est}\approx \frac{1000\,\mathrm{T}}{R\xi }
\end{equation*}%
where $R$ and $\xi $ are measured in nanometers. Using a typical diffusion
constant for a poor metal $D\approx 1\mathrm{cm}^{2}/\mathrm{s}$ and
allowing for a very high gap value $\Delta =10\,\mathrm{K}$, we find $\xi
=8.5\,\mathrm{nm}$. Together with the lowest bound for the grain radius $%
R=6\,\mathrm{nm}$ in which the distance between the levels does not exceed
the superconducting gap 
\begin{equation*}
\delta =(4\nu _{0}R^{3})^{-1}<\Delta
\end{equation*}%
it leads to $H_{cg}^{est}\lesssim 20\,\mathrm{Tesla}$\thinspace\ which is
still much smaller than $H_{cg}^{\exp }$ above. These estimates did not take
into account the spin effect of magnetic field that would further decrease
the value of $H_{cg}^{est}$.

We conclude that the observed activation gap cannot be explained as BCS gap
in small grains composing the material: it is too wide and too stable with
respect to magnetic field.

A number of works argued that mesoscopic fluctuations might lead to the
appearance of inhomogeneous superconductivity (self-induced granularity)\ in
the vicinity of the transition even in the absence of structural granularity~%
\cite{Kowal1994,BaturinaReview2007,Baturina2007a}. The direct computation
shows that these speculations are correct for the fermionic mechanism of the
superconductivity suppression in two dimensions ~\cite{Skvortsov2005}. We
expect that the self-induced granularity that appears due to this mechanism
does not lead to thin insulating barriers. It is therefore characterized by
a small value of the Coulomb interaction between the 'grains'. Thus, it can
be ruled out as a mechanism of direct superconductor-insulator transition in
homogeneous films by the energy scale arguments given above.

An important unresolved issue is the nature of carriers responsible for the
transport in the insulating state: are they Cooper pairs or single
electrons?\ One expects that the presence of the superconducting gap in the
insulating state implies that the transport is dominated by Cooper pairs in
the vicinity of the transition and this was indeed observed in ultrathin Bi
films\cite{Valles2e}. However, one expects that the transport is dominated
by single electrons further in the insulating state where activation
behavior was observed. Unfortunately there are no data to confirm this.

To summarize: experimental data on SIT in amorphous materials call for a new
mechanism of a gap formation, which is somehow related to the
superconductivity, but is different from the usual BCS gap formation. In the
vicinity of SIT this mechanism should lead to a "pseudo-gap" features in $%
R(T)$ behavior and tunneling data.

\bigskip

\subsection{Main features of the fractal pseudospin scenario. \label{Main
features}}

The fractal pseudospin mechanism (briefly presented in~\cite{FIKY2007})
should be viewed as alternative to both 'boson' and 'fermion' scenarios.
Here we argue that it is fully compatible with the data. The key elements of
this approach are in fact quite old: (i) Anderson's reformulation~\cite%
{Anderson1958} of the BCS theory in terms of "pseudo-spins", (ii) Matveev -
Larkin theory~\cite{Matveev1997} of parity gap in ultra-small
superconducting grains and (iii) fractal properties of single-electron
eigenfunctions with energies near the Anderson mobility edge~\cite%
{Chalker1990,KrMut1997,MirlinReview2000}.

Qualitatively, in this scenario the electrons near the mobility edge form
strongly coupled but localized Cooper pairs (notion first introduced in~\cite%
{Gantmakher1996,Gantmakher1998}, see also~\cite{Ghosal2001}) due to the
attraction of two electrons occupying the same localized orbital state.
These pairs are characterized by a large binding energy which is responsible
for the single electron gap $T_{0}$ observed in transport measurement in the
insulating state. At temperatures below $T_{0}$ the system can be described
as a collection of Anderson's $S=\frac{1}{2}$ pseudo-spins, whose $S_{j}^{z}$
components measure the Cooper pair occupation number and $S_{j}^{\pm }$
components correspond to pair creation/annihilation operators.
Superconductivity in this system is due to the tunneling of Cooper pairs
from one state to another. It is essential that it competes not with the
Coulomb repulsion but with the random energy of the pair on each orbital
state. In spin language it is described as a formation of non-zero averages $%
\langle S^{\pm }\rangle $ due to "off-diagonal" $%
S_{i}^{-}S_{j}^{+}+S_{i}^{+}S_{j}^{-}$ coupling in the effective Hamiltonian
which competes with random field in $z-$direction term $h_{i}S_{i}^{z}$.
Large values of the binding energy and off-diagonal interactions are due to
the properties of localized nearly-critical wavefunctions; the main features
of these wavefunctions are their strong correlations both in real space and
in energy space, and their sparsity in real space (see section \ref%
{Fractality and correlations}). The resulting phase diagram is shown in
Figure \ref{PhaseDiagram}.

The fractal pseudospin scenario has many common features with the bosonic
mechanism, but it is distinct from it in a few important respects:

\begin{itemize}
\item {pseudo-gap energy scale $T_{0}$ is independent from the collective
energy gap $\Delta $}

\item {fractal nature of individual eigenstates implies a large
"coordination number" $Z\gg 1$ of interacting pseudospins away from the
superconductor-insulator transition. Close to the transition $Z$ drops,
resulting in very inhomogeneous superconductive state and an abrupt decrease
of $T_c$.}

\item {distribution of superconducting order parameter in real space is
extremely inhomogeneous, thus usual notion of space-averaged order parameter 
$\overline{\Delta }$ is useless even qualitatively, and the "Anderson
theorem" is not applicable.}
\end{itemize}

In the main part of the paper we present theoretical arguments in support of
this new scenario. We restrict our discussion to the three-dimensional
problem which is appropriate for the electron wave function behavior in most
films. It is possible that the physics in the near vicinity of the
transition is dominated by large scales where the two dimensional nature of
the films become important, the details of the crossover to this critical
regime is beyond the developed theory. We will assume below that
localization effects are not very strong, allowing for the presence of
phonon-induced attraction between electrons. Clearly, the necessary
condition for that is $\delta _{L}\ll \omega _{D}$.

The theory that we develop starts with the single electron states of the
non-interacting problem, so it is not applicable to describe the physics in
high magnetic fields where these states change significantly. Thus,
interesting physics of the metallic state with resistance approaching $%
h/e^{2}$ is beyond the applicability limits of our theory.

\section{Model \label{Model}}

\subsection{BCS Hamiltonian for electrons in localized eigenstates. \label%
{BCS Hamiltonian}}

We consider simplest model of space-local BCS-type electron-electron
attraction, $V_{int}=g\delta (\mathbf{r})$. It is assumed, as usual, that
this attraction is present for electrons with energies $E$ in the relatively
narrow stripe $E\in E_{F}\pm \omega _{D}$ around Fermi energy. However, we
will see below that in contrast with the usual BCS theory, the parameter $%
\omega _{D}$ will not enter our final results. The Hamiltonian represented
in the basis of exact single-electron eigenstates $\psi _{j}(\mathbf{r})$
becomes%
\begin{equation}
H=\sum_{j\sigma }\xi _{j}c_{j\sigma }^{\dagger }c_{j\sigma }-\frac{\lambda }{%
\nu _{0}}\sum_{i,j,k,l}M_{ijkl}c_{i\uparrow }^{\dagger }c_{j\downarrow
}^{\dagger }c_{k\downarrow }c_{l\uparrow }\,,  \label{Ham1}
\end{equation}%
where 
\begin{equation}
M_{ijkl}=\int d\mathbf{r}\psi _{i}^{\ast }(\mathbf{r})\psi _{j}^{\ast }(%
\mathbf{r})\psi _{k}(\mathbf{r})\psi _{l}(\mathbf{r})\,,  \label{Melements1}
\end{equation}%
$\xi _{j}=E_{j}-E_{F}$ is the single-particle energy of the eigenstate $j$
counted from Fermi level, $c_{j\sigma }$ is the corresponding electron
annihilation operator for the spin projection $\sigma $, $\nu _{0}$ is the
density of states (per single spin projection) and $\lambda =g\nu _{0}\ll 1$
is dimensionless Cooper coupling constant.

Note that writing Hamiltonian in the form (\ref{Ham1}) we omitted the
Hatree-type terms which do not contribute directly to the Cooper
instability. Such terms are known to be negligible when single-electron
states are extended (see discussion in Ref.~\cite{MaLee}); the issue of
their importance for critical and, especially, localized states is more
delicate. Below in section \ref{Transition temperature} we present results
for the superconducting transition temperature obtained with and without
account of the Hatree-type terms. The comparison, shown in Fig.~\ref%
{TcVersusE}, demonstrates that these terms, while changing the quantitative
results somewhat, do not affect our main qualitative conclusions. Therefore,
in order to keep the arguments as simple as possible, we neglect Hatree
terms in the main part of the following text. \footnote{%
However, those terms are important for a quantitative description of
superconductivity in the region of localized single-particle states, as
shown in Ref.~\cite{Ghosal2001} where 2D problem in the limit of strong
disorder and strong attraction was studied numerically. In particular, they
might lead to additional inhomogeneous broadening of the coherence peaks
observed in tunneling experiments, see section \ref{Point contact tunneling}}

Unless specified, we will not consider magnetic field effects, thus
eigenfunctions $\psi _{j}(\mathbf{r})$ can be chosen real. In the following
we will use frequently a simplified Hamiltonian (\ref{Ham1}) where only
pair-wise terms $i=j$ and $k=l$ are taken into account: 
\begin{equation}
H_{2}=\sum_{j\sigma }\xi _{j}c_{j\sigma }^{\dagger }c_{j\sigma }-\frac{%
\lambda }{\nu _{0}}\sum_{jk}M_{jk}c_{j\uparrow }^{\dagger }c_{j\downarrow
}^{\dagger }c_{k\downarrow }c_{k\uparrow }\,,  \label{Ham2}
\end{equation}%
where 
\begin{equation}
M_{jk}=\int d\mathbf{r}\psi _{j}^{2}(\mathbf{r)}\psi _{k}^{2}(\mathbf{r})\,.
\label{Melements2}
\end{equation}%
Eq.(\ref{Ham2}) is the minimal Hamiltonian that includes hopping of pairs
necessary to establish a global superconducting order. It plays the same
role for our theory as the BCS Hamiltonian with $i=\mathbf{p}$, $j={-\mathbf{%
p}}$, $k=\mathbf{p^{\prime }}$, $l=-\mathbf{p^{\prime }}$ for usual theory
of superconductivity. We will discuss the accuracy of this approximation
below in Sec.~\ref{Ginzburg-Landau functional}

\subsubsection{Ultra-small metallic grain. \label{Ultra-small metallic grain}%
}

Here we rederive the known results for the model (\ref{Ham2}) applied to
ultra-small metal grains; this derivation will provide the starting point
for our solution of the model Hamiltonian (\ref{Ham1}) or (\ref{Ham2}) for
the electrons with Fermi level near mobility edge.

Pairing correlations in metallic grains of very small volume $V$, with level
spacing $\delta =(\nu _{0}V)^{-1}$ comparable to the bulk superconductive
gap $\Delta $ were considered in many papers, see review~\cite{UltraSmall}.
The issue which is most important for this work is the parity gap introduced
in~\cite{Matveev1997} to characterize pairing effects in ultra-small grains
with $\delta \ll \Delta $. The work \cite{Matveev1997} assumed the
simplified Hamiltonian (\ref{Ham2}) with identical matrix elements $%
M_{jk}=1/V$.

More complete treatment of a weak electron-electron interaction in small
metallic grains is given by~\cite{Kurland00} where it was argued that to the
leading order in the small parameter $\tau _{0}\delta $, where $\tau _{0}$
is the flight (or diffusion) time for electron motion inside grain, all
off-diagonal terms $M_{ijkl}$ can be neglected. In the relevant terms the
indices of $M_{ijkl}$\ should be pairwise equal. Because for small grains
the wavefunctions $\psi _{j}(\mathbf{r})$ are essentially random Gaussian
variables subject only to orthogonality and normalization conditions, the
matrix elements (\ref{Melements2}) appearing in the Hamiltonian (\ref{Ham2})
are given by 
\begin{equation}
M_{j\neq k}=\frac{1}{V}\,\quad M_{jj}\equiv M_{j}=\frac{3}{V}\,,  \label{Mg}
\end{equation}%
so that the full Hamiltonian can be expressed in terms of the total number
of electrons $\hat{n}$, the total spin $\hat{\mathbf{S}}$ and the operator $%
\hat{T}=\sum_{k}c_{k\downarrow }c_{k\uparrow }$ related to Cooper pairing
correlations: 
\begin{equation}
H_{uni}=\lambda \delta \left[ 2\hat{\mathbf{S}}^{2}-\frac{1}{2}\hat{n}^{2}%
\right] -\lambda \delta T^{\dagger }T  \label{kurland}
\end{equation}%
It is essential for the validity of (\ref{kurland}) that all matrix elements 
$M_{jk}$ with $i\neq j$ are equal to $1/V$, while diagonal terms are three
times larger, only in this case it is possible to represent Eq.(\ref{kurland}%
) in terms of the total density, spin and pairing operators.

Attractive interaction implies that $S=0$ in the ground state. Because $n$
is conserved for isolated grain the properties of this model in $n=0$, $S=0$
sector are equivalent to the properties of the simplified Matveev-Larkin
model (\ref{Ham2}) which takes into account only the last term in (\ref%
{kurland}). \ The first term in (\ref{kurland}) is important for the correct
evaluation of the coefficient of the interaction term with $j=k$ in Eq.(\ref%
{Ham2}) because only $1/3$ of it should be assigned to the interaction in
the Cooper channel, since other $2/3$ contribute to the "n" and "S" terms of
the Hamiltonian.

In the limit $\delta \gg \Delta $ one can use the perturbation theory with
respect to pairing Hamiltonian (\ref{kurland}). In the lowest order in $%
\lambda $, neglecting all terms except diagonal ones, one finds that the
energy of two identical grains with even number of electrons, $n=2k,$ and
zero spin is by $\Delta E=3\lambda \delta $ less than the energy of the same
two grains with $2k+1$ and $2k-1$ electrons and spin $1/2$. \footnote{%
For different grains containing different number of particles one needs to
take into account different chemical potentials in these grains but the
final conclusion remains unchanged.} It means~\cite{Matveev1997} that the
average ground-state energy of a grain with even number of electrons is
lower by "parity gap" $\Delta _{P}=\frac{3}{2}\lambda \delta $ than the
energy of the same grain with odd number of electrons. Note that Cooper
pairing contributes $1/3$ of this energy difference. This result is valid
only in the limit of a very small coupling constant $\lambda $, when all the
terms with $j\neq k$ in Eq.(\ref{Ham2}) can be neglected. In a more general
case these terms must be taken into account which leads~\cite{Matveev1997}
to the renormalization of the coefficient of the $T^{\dagger }T$ term in the
Hamiltonian (\ref{kurland}) which becomes 
\begin{equation}
\lambda _{R}=\lambda /(1-\lambda \ln (\omega _{D}/\delta )).
\label{lambdaR}
\end{equation}%
After this renormalization the coefficient of the Cooper pairing becomes
dominant. Introducing bulk energy gap $\Delta =\omega _{D}e^{-1/\lambda }$,
one finds~\cite{Matveev1997} parity gap which is valid for all $\Delta \ll
\delta $: 
\begin{equation}
\Delta _{P}=\frac{\delta }{2\ln \frac{\delta }{\Delta }}+\lambda  \label{ML}
\end{equation}%
where the second term is due to the first term in the Hamiltonian (\ref%
{kurland}) and is small compared to the main term. The result (\ref{ML})
shows that parity gap grows with the decrease in the grain size. Note that
parity gap (\ref{ML}) would not appear if one does note take into account
the double-diagonal terms in (\ref{Ham2}), which are totally irrelevant in
the usual BCS theory of bulk superconductivity. Below we will find somewhat
similar behavior in the case of bulk Anderson insulators.

\subsubsection{Vicinity of the mobility edge. \label{Vicinity of the
mobility edge}}

\begin{figure}[tbp]
\includegraphics[width=8cm]{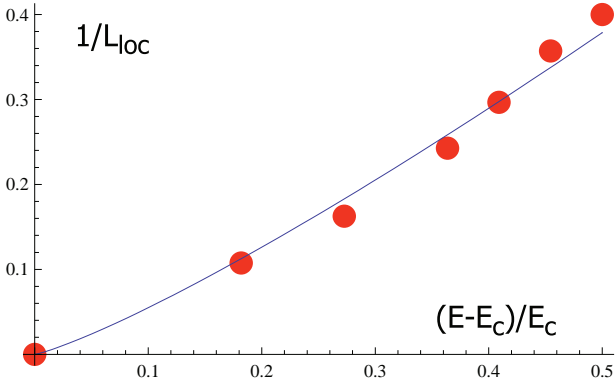}
\caption{Inverse localization length as function of proximity to the
mobility edge, obtained numerically for 3D Anderson model with Gaussian
disorder of width $W=4$. The values shown here were extracted from the
numerical computation of the inverse participation ratio in this model and
its conversion into the localization length by Eqs.(\protect\ref{delta_L_def},%
\protect\ref{d-L-IPR}). Full line is a fit to 
$1/L_{loc}=1.87\cdot (E/E_{c}-1)^{1.2}$.}
\label{nuFig}
\end{figure}

In the bulk Anderson insulator with the Fermi energy near the mobility edge,
the typical energy scale replacing $\delta $ is 
\begin{equation}
\delta _{L}=1/(\nu _{0}L_{loc}^{3}),  \label{delta_L_def}
\end{equation}%
where $L_{loc}$ is the localization length. It was argued in~\cite{MaLee}
that localization is irrelevant for superconductivity if $T_{c}\gg \delta
_{L}$. In the opposite limit $T_{c}\ll \delta _{L}$ pairing correlations
between electrons localized on different orbitals are irrelevant.
Localization length depends on the Fermi-energy (in the scaling region $%
L_{loc}\gg \ell $) as 
\begin{equation}
L_{loc}\approx \ell \left( \frac{E_{0}}{{E_{c}-E_{F}}}\right) ^{\nu }\,,
\label{Lloc}
\end{equation}%
where $E_{c}$ is the position of the mobility edge, $\nu $ is the
localization length exponent. Numerical data \cite{nu} show that in a very
narrow vicinity of the mobility edge ($({E_{c}-E_{F})/E}_{c}\lll 1$) of 3D
Anderson model the localization length obeys the scaling dependence (\ref%
{Lloc}) with $\nu \approx 1.57$. As we shall see below, the range of
energies relevant for the superconductor-insulator transition is relatively
wide, $({E_{F}-E_{c})/E}_{c}\lesssim 0.5$; in this broader range the
localization length follows the same scaling behavior (\ref{Lloc}) but with
as somewhat different exponent $\nu \approx 1.2$. We illustrate this by Fig.~%
\ref{nuFig} that shows the localization length obtained for 3D Anderson
model with Gaussian disorder (see section~\ref{Fractality and correlations}
and Eq.(\ref{G-d}) below). 
The parameter $\ell $ in Eq.(\ref{Lloc}) is the short-scale cutoff of the
order of the elastic scattering length. The associated energy scale 
\begin{equation}
E_{0}=1/(\nu _{0}\ell ^{3})
\end{equation}%
depends on the microscopic details of the model of disorder and can be small
compared to Fermi-energy $E_{F}$ (see next subsection for the discussion of
this issue).

We will assume that Fermi energy is not too close to the mobility edge so
that 
\begin{equation}
t\equiv \frac{E_{c}-E_{F}}{E_{0}}\gg \frac{T_{c}}{E_{0}}  \label{tau_bound1}
\end{equation}%
The condition (\ref{tau_bound1}) means that localization properties of
eigenstates $\psi _{j}(\mathbf{r})$ do not vary appreciably within the
energy stripe $E_{F}\pm T_{c}$ mainly responsible for development of
superconducting correlations. Violation of the condition (\ref{tau_bound1})
implies the absence of the particle-hole symmetry in the superconducting
state; we expect anomalous Hall effect in the superconducting state and near
transition in this regime. Using Eq.(\ref{Lloc}) we find 
\begin{equation}
\delta _{L}=E_{0}t^{3\nu }  \label{delta_tau}
\end{equation}%
Because $3\nu \approx 4\gg 1$, the condition (\ref{tau_bound1}) is
compatible with $t\ll 1$ in a wide range of parameters which include both
small and large ratios $\delta _{L}/T_{c}$. In the limiting case $\delta
_{L}\ll T_{c}$ the scale set by localization is much larger than the one set
by superconductivity, so all relevant statistical properties of matrix
elements (\ref{Melements2}) can be computed at $t=0$. We will refer to this
case as the critical regime. We will see in section \ref{Cooper instability}
that deep in the limit $\delta _{L}/T_{c}\rightarrow 0$, the transition
temperature approaches its limiting value, which we denote as $T_{c}^{0}$.
In contrast with the conclusions of Ref.~\cite{MaLee}, we will find that $%
T_{c}^{0}$ may differ substantially from the usual BCS value $T_{c}^{\mathrm{%
BCS}}=\omega _{D}e^{-1/\lambda }$ for the metal with the same value of the
Cooper attraction constant. Moreover, we find that the values of $T_{c}^{0}$
are typically \textit{larger} than $T_{c}^{\mathrm{BCS}}$ for weak couplings 
$\lambda \ll 1$. This unexpected result is related to the \textit{fractality}
of electron wavefunctions with energies close to the mobility edge.

\subsection{Fractality and correlations of the wave functions near the
mobility edge. \label{Fractality and correlations}}

The exact single-particle eigenfunctions $\psi _{j}(\mathbf{r})$ and
eigenvalues $E_{j}$ that enter the model Hamiltonian Eq.(\ref{Ham1})should
be found from the single-particle Hamiltonian with disorder. The
conventional models of disorder are the continuous model of free electrons
in a Gaussian random potential $U(\mathbf{r})$: 
\begin{equation}
H_{1a}=\frac{\mathbf{p}^{2}}{2m}+U(\mathbf{r}),  \label{cont-Ham}
\end{equation}%
or the tight-binding model with on-site energies $\varepsilon _{\mathbf{n}}$
being random variables with the probability distribution $\mathcal{P}%
(\{\varepsilon _{\mathbf{n}}\})=\prod_{\mathbf{n}}p(\varepsilon _{\mathbf{n}%
})$. The latter is known as the Anderson model, it is described by the
Hamiltonian 
\begin{equation}
H_{1b}=\sum_{\mathbf{n}}\varepsilon _{\mathbf{n}}\,a_{\mathbf{n}}^{\dagger
}a_{\mathbf{n}}-\sum_{\mathbf{n},\mathbf{m}=\mathbf{n}+\mathbf{a}}a_{\mathbf{%
n}}^{\dagger }a_{\mathbf{n}+\mathbf{a}}.  \label{AM}
\end{equation}%
The most common choice of the distribution function $p(x)$ are the box
distribution 
\begin{equation}
p(\varepsilon )=\left\{ 
\begin{matrix}
W^{-1},\;\;\mathrm{if}\;\;x<|W/2|\cr0,\;\;\mathrm{if}\;\;x>|W/2|.%
\end{matrix}%
\right.  \label{box}
\end{equation}%
or the Gaussian distribution 
\begin{equation}
p(\varepsilon )=\frac{1}{\sqrt{2\pi }\,W}\,\mathrm{exp}\left[ -\frac{%
\varepsilon ^{2}}{2W^{2}}\right] .  \label{G-d}
\end{equation}

Increasing the disorder parameter $W$ in the $3d$ Anderson model (\ref{AM}),(%
\ref{box}) at a fixed Fermi energy $E_{F}$ results in the Anderson
localization transition at the critical disorder $W=W_{c}$ (at $E_{F}=0$ the
critical value of disorder is $W_{c}=16.5$ for the box distribution, Eq.~\ref%
{box}). Alternatively, the localization transition occurs when $E_{F}$ is
increased at a fixed disorder $W<W_{c}$ beyond the mobility edge $E_{c}$; in
the following we will mainly use the Gaussian model, Eq.~(\ref{G-d}). The
same type of transition takes place in the continuous model Eq.(\ref%
{cont-Ham}). The changes in the statistics of wavefunctions resulting from
this transition do not merely reduce to their localization. Well before all
wavefunctions become localized they acquire a certain structure where a
wavefunction occupies not all available space but a certain \textit{fractal}
inside the \textit{correlation radius} $L_{\mathrm{corr}}$. The global
picture of an extended wavefunction resembles a "mosaic" made of such pieces
of fractal with the characteristics size $L_{\mathrm{corr}}$. This peculiar
phase (called the "multifractal metal" in Ref.\cite{CueKra}) appears in the
vicinity of the Anderson transition. It persists down to relatively weak
disorder as long as the decreasing correlation length $L_{\mathrm{corr}}$
exceeds a microscopic length $\ell $ which has a meaning of the minimal
length (a pixel) of the fractal structure. In Anderson model with the box
probability distribution the length $\ell \approx a\,W_{c}^{1/3}\geq 2.5a$,
where $a$ is the lattice constant; fractal effects disappear in this model
at $W<3\ll W_{c}$ only, see Ref.~\cite{CueKra}. For the continuous model
defined by Eq.(\ref{cont-Ham}) it is of the order of the elastic scattering
mean free path.

As one approaches the mobility edge or the critical value of disorder, the
correlation radius $L_{\mathrm{corr}}$ diverges so that the critical
wavefunctions are pure fractal (or, strictly speaking \textit{multifractal} 
\cite{CueKra}). On the localized side of the transition the wavefunctions
inside the \textit{localization radius} $L_{\mathrm{loc}}$ resemble the one
inside an element of the mosaic structure of the multifractal metal. This
"multifractal insulator" \cite{CueKra} exists in the vicinity of the
Anderson transition and becomes an ordinary insulator at strong disorder
when $L_{\mathrm{loc}}<\ell $.

\subsubsection{Wavefunction correlations at the mobility edge: algebra of
multi-fractal states. \label{Wavefunction correlations}}

We start by describing the multi-fractal correlations of wavefunctions
exactly at the mobility edge. To avoid confusion we note that for a finite $%
3d$ sample of the size $L\times L\times L$ the mobility edge is smeared out.
The critical multi-fractal states live in a spectral window around $E_{c}$
of the width $\delta E\propto L^{-1/\nu }$, where $\nu $ is the exponent of
the localization (correlation) length $L_{\mathrm{loc}}(L_{\mathrm{corr}%
})\propto |E-E_{c}|^{-\nu }$, such that the value of $L_{\mathrm{loc}}(L_{%
\mathrm{corr}})$ inside this window is larger than $L$. The number of
single-particle states in this window is proportional to $L^{3(1-\frac{1}{%
3\nu })}$. Because $\nu $ is definitely larger than $\frac{1}{3}$ (in fact,
Harris criterion tells that $\nu \geq 2/d=2/3)$ it tends to infinity as $%
L\rightarrow \infty $, .

There is a vast numerical and analytical evidence \cite{MirlinNewRep} that
the critical wavefunctions at the mobility edge obeys the multifractal
statistics. This can be seen, for instance, in the behavior of the moments
of the inverse participation ratio: 
\begin{equation}
P_{q}=\nu _{0}^{-1}\,\sum_{j}\int d^{d}\mathbf{r}\,|\psi _{j}(\mathbf{r}%
)|^{2q}\,\delta (E-E_{j}).  \label{mom-IPR}
\end{equation}%
The moments (\ref{mom-IPR}) describe the effective volume occupied by the
the wave function. At the mobility edge they scale with the size of the
sample 
\begin{equation}
\langle P_{q}\rangle \sim \ell ^{-(d-d_{q})(q-1)}L^{-d_{q}(q-1)}\propto
L^{-d_{q}(q-1)},  \label{def-multi}
\end{equation}%
where $d_{q}\leq 3$ is the corresponding fractal dimension. For the 3d
Anderson model of the orthogonal symmetry class (real Hamiltonian) we obtain
by numerical diagonalization the following results for the first two fractal
dimensions: \footnote{%
The fractal dimensions for bigger sample sizes have been studied recently by
Rodriguez, Vasquez and Roemer\cite{Roemer}. They have found $d_{2}=1.24\pm
0.07$, $d_{4}=0.63\pm 0.07$, $d_{2}^{\mathrm{typ}}=1.35\pm 0.07$, $d_{4}^{%
\mathrm{typ}}=1.02\pm 0.2$. They also point out on a large systematic error
for $d_{4}^{\mathrm{typ}}$ related with the finite-size effect. In view of
the fact that the critical $q_{c}\approx 2.1...2.2$ is close to 2, the
typical $d_{4}^{\mathrm{typ}}$ should be found from the condition Eq.(\ref%
{typ}). This gives an estimate $d_{4}^{\mathrm{typ}}=0.84\pm 0.04$.}

\begin{equation}
d_{2}\approx 1.29\pm 0.1,\,\,\,\,d_{4}\approx 0.72\pm 0.1.  \label{d-2-4}
\end{equation}%
The fact that the fractal dimensions $d_{q}$ depend on the order of the
moment $q$ implies the \textit{multiractality} of the wave functions. The
scaling arguments show that such behavior of $\langle P_{q}\rangle $ implies
the power-law correlations of wavefunction amplitudes at different space
points: 
\begin{equation}
C_{q}(0,\mathbf{r})=\langle |\psi _{j}(\mathbf{r})|^{2q}\,|\psi
_{j}(0)|^{2q}\rangle \sim L^{-2qd}\,(L/\ell )^{\beta _{q}}\,\left( \frac{L}{r%
}\right) ^{d-\alpha _{q}},  \label{corr-dif-r}
\end{equation}%
where $\ell <r<L$, and the exponents are equal to 
\begin{eqnarray}
\alpha _{q} &=&d_{2q}(2q-1)-2d_{q}(q-1),  \label{alpha} \\
\beta _{q} &=&2(q-1)(d-d_{q})
\label{alpha1}
\end{eqnarray}%
Note that the sign of $\alpha _{q}$ is positive provided that the moments $%
P_{q}$ are only \textit{moderately} fluctuating, so that the scaling
behavior of $\langle P_{q}^{2}\rangle $ and $\langle P_{q}\rangle ^{2}$ is
the same. This follows from the inequality 
\begin{equation*}
P_{q}^{2}=\left( \sum_{\mathbf{r}}|\psi (\mathbf{r})|^{2q}\right)
^{2}>P_{2q}=\sum_{\mathbf{r}}|\psi (\mathbf{r})|^{4q}
\end{equation*}%
and the definition of the fractal dimensions Eq.(\ref{def-multi}). However,
from Eq.(\ref{d-2-4}) is follows that 
\begin{equation}
\alpha _{2}=3d_{4}-2d_{2}\approx -0.43\pm 0.5.  \label{ineqality}
\end{equation}%
Although the error bars are rather large, it is likely that $\alpha _{2}$ is 
\textit{negative}.

This means that the second moment $P_{2}$ is strongly, not moderately
fluctuating, and the $L$-scaling of $\langle P_{2}^{2}\rangle $ is different
from that of $\langle P_{2}\rangle ^{2}$ in agreement with the early
conjecture\cite{Ioffe1985}. Indeed, our numerical simulations on the 3D
Anderson model of the orthogonal symmetry class show that 
\begin{equation*}
\langle P_{2}^{2}\rangle \propto L^{-2.16\pm 0.1},\;\;\;\;\;\;\langle
P_{2}\rangle ^{2}\propto L^{-2.58\pm 0.1}.
\end{equation*}%
This is consistent with the observation \cite{Mirlin2002} that the
distribution function of the second moment $\mathcal{P}_{2}(P_{2})\propto
P_{2}^{-p_{2}}$ has a power-law tail with $p_{2}\approx 2.6$ at relatively
large values of $P_{2}\gg \langle P_{2}\rangle \sim L^{-d_{2}}$ (which is
cut at $P_{2}=1$). As a result the average $\langle P_{2}^{2}\rangle $ is
dominated by the far tail where the one-parameter scaling $\mathcal{P}%
_{2}(P_{2})=f_{2}(P_{2}/\langle P_{2}\rangle )$ is no longer true. The
distribution function $\mathcal{P}_{4}(P_{4})$ of the fourth moment $P_{4}$
possesses even stronger tail with $\mathcal{P}_{4}\propto P_{4}^{-2.0}$ at $%
P_{4}\gg \langle P_{4}\rangle $. In this case the average $\langle
P_{4}\rangle $ is considerably contributed by the rare events in the far
tail of the distribution.

In the situation where the assumption of moderate fluctuations of $P_{q}$ no
longer holds and the rare events are important for averaging \cite%
{MirlinNewRep} the correct physical quantity is typical average $\langle
P_{q}\rangle _{\mathrm{typ}}=\mathrm{exp}[\langle \ln P_{q}\rangle ]$
instead of the usual one and the corresponding fractal dimensions should be
defined by 
\begin{equation}  \label{d-typ}
d_{q}^{\mathrm{typ}}\,(q-1)=-d\ln \langle P_{q}\rangle _{\mathrm{typ}}/d\ln
L.
\end{equation}%
Unlike the usual average, the typical average is determined by the body of
the corresponding distribution function $\mathcal{P}_{q}(P_{q})$ provided
that the tail exponent $p_{q}$ is larger than $1$. Computing the typical
averages $\langle P_{q}\rangle_{\mathrm{typ}}$ and using Eq.(\ref{d-typ}) we
obtained (see also a footnote $^4$ and Ref.~\cite{Roemer} for comparison)
for the orthogonal 3D Anderson model: 
\begin{eqnarray}
d_{2}^{\mathrm{typ}} &\approx &1.40\pm 0.1,\;\;\;\;d_{4}^{\mathrm{typ}%
}\approx 1.08\pm 0.1,  \label{alpha-typ} \\
\alpha _{2}^{\mathrm{typ}} &=&3d_{4}^{\mathrm{typ}}-2d_{2}^{\mathrm{typ}%
}\approx +0.44\pm 0.5.  \notag
\end{eqnarray}%
It is interesting to note \cite{MirlinNewRep} that if the tail exponent for
a given $q$-th moment lies in the region $1<p_{q}<2$ (i.e. the typical
fractal dimension is determined by the body of the distribution for the
corresponding moment $P_{q}$ but the averaged moment is dominated by the
rare events) the dependence of the typical fractal dimension on $q$ is
linear. \footnote{%
This follows from the fact that the spectrum of fractal dimensions given by
the Legendre transformation $f(\alpha )=-\tau_{q}+q\alpha,\;\;\;\alpha=\frac{%
d\tau_{q}}{dq},\;\;\;\tau_{q}=(q-1)d_{q}$ determines the scaling with the
system size $L$ of the number of sites $M\propto L^{f(\alpha)}$ where $|\psi(%
\mathbf{r}_{i})|^{2}\propto L^{-\alpha}$. As a typical event cannot occur at
a number of sites $M<1$ (which is only possible for a rare event), $f^{%
\mathrm{typ}}(\alpha)$ found from $d^{\mathrm{typ}}_{q}$ must be either
positive (if $\langle P_{q}\rangle=\langle P_{q}\rangle_{\mathrm{typ}}$) or
zero (if $\langle P_{q} \rangle$ is dominated by the rare events and is
different from $\langle P_{q}\rangle_{\mathrm{typ}}$). In the latter case
the Legendre transformation implies $[1-qd/dq]\{(q-1)d^{\mathrm{typ}%
}_{q}\}=0 $, leading to the fact that $(q-1)d^{\mathrm{typ}}_{q}$ is
proportional to $q $.}

\begin{equation}  \label{typ}
d_{q}^{typ}(q-1)=q\,\alpha_{+},\;\;\;\;q>q_{c},
\end{equation}
where $\alpha_{+}$ is a constant.

Substituting Eq.(\ref{typ}) into Eq.(\ref{alpha}) we immediately obtain: 
\begin{equation}
\alpha _{q}^{\mathrm{typ}}=(2q-1)d_{2q}^{\mathrm{typ}}-2(q-1)d_{q}^{\mathrm{%
typ}}=0,\;\;\;\;q>q_{c}.  \label{alpha-typ1}
\end{equation}%
The equality (\ref{alpha-typ1}) is valid if the condition $q>q_{c}$ is
satisfied for both terms entering (\ref{alpha}); in all other cases $\alpha
_{q}^{\mathrm{typ}}>0$. Given also that for $q>1$ $d_{2q}<d_{q}<d$, we
obtain: 
\begin{equation}
d>\alpha _{q}^{\mathrm{typ}}>0.  \label{pos-typ}
\end{equation}%
For 3D Anderson model of orthogonal symmetry class $q_{c}\approx 2.2\pm 0.1$
is very close to (but larger than) 2 (see Ref.\cite{Roemer}). This means
that in a typical sample the $r$-dependence of the correlation function Eq.(%
\ref{corr-dif-r}) is such that the integral over $r$ is dominated by large
distances. This will be important for the estimations of the matrix elements
done below.

Long expressions Eqs.(\ref{alpha},\ref{alpha1}) for the exponents $\alpha _{q}$ and $%
\beta _{q}$ reflect few simple rules: \newline
\textbf{Rule(i)}: for $r\sim L$ the two wavefunctions are statistically
independent and the result can be found using Eq.(\ref{def-multi}): 
\begin{equation*}
\langle |\psi _{j}(0)|^{2q}\rangle ^{2}\sim =[L^{-d}\langle P_{q}\rangle
]^{2}\sim L^{-2d_{q}(q-1)-2d}
\end{equation*}
\textbf{Rule (ii)}: for $r\sim \ell $ the result should be the same as for
the coincident points $r=0$, i.e. 
\begin{equation*}
\langle |\psi _{j}(\mathbf{0})|^{4q}\rangle =L^{-d}\langle P_{2q}\rangle
\sim L^{-d_{2q}(2q-1)-d}.
\end{equation*}

These rules can be generalized to the case of correlation of \textit{%
different} eigenfunctions 
\begin{equation*}
C_{q}(\omega ,\mathbf{r})=\langle |\psi _{i}(\mathbf{r})|^{2q}\,|\psi
_{j}(0)|^{2q}\rangle
\end{equation*}%
with the energy difference $E_{i}-E_{j}=\omega $. In this case the
eigenfunctions become statistically independent for $|\mathbf{r}|>L_{\omega
} $, where 
\begin{equation}
L_{\omega }^{d}=\frac{1}{\nu _{0}\omega }.  \label{Lom}
\end{equation}%
With this correction, applying the rules (i) and (ii) one can show that at $%
\omega >\delta $ the only change in Eq.(\ref{corr-dif-r}) is that the $|%
\mathbf{r}|$-dependent factor becomes $(L_{\omega }/|\mathbf{r}|)^{d-\alpha
_{q}}$ instead of $(L/|\mathbf{r}|)^{d-\alpha _{q}}$.

This allows us to estimate the \textit{local} averages of \textit{different
wave functions}. In these averages the power law correlations will be seen
in the energy space. Indeed, substituting $|\mathbf{r}|\sim \ell $ in $%
(L_{\omega }/|\mathbf{r}|)^{d-\alpha _{q}}$ one obtains: 
\begin{eqnarray}
C_{q}(\omega ,0) &=&\nu _{0}^{-2}\sum_{i,j}\langle |\psi _{i}(\mathbf{0}%
)|^{2q}\,|\psi _{j}(\mathbf{0})|^{2q}\,\delta (E-E_{i})\delta (E^{\prime
}-E_{j})\rangle  \notag  \label{corr-dif-E} \\
&\sim &L^{-2qd}\,(L/\ell )^{\beta _{q}}\,\left( \frac{E_{0}}{\omega }\right)
^{\gamma _{q}},\;\;\;\gamma _{q}=1-\frac{\alpha _{q}}{d}.
\end{eqnarray}%
where $(\nu _{0}L^{d})^{-1}=\delta <\omega <E_{0}=(\nu _{0}\ell ^{d})^{-1}$.
For $\omega <\delta $ the correlation function $C_{q}(\omega )$ saturates.

Here, essentially the same rules work: \newline
\textbf{(i)} at large energy separations $\omega \sim E_{0}$ the
wavefunctions are statistically independent and\newline
\textbf{(ii)} at small energy separations $\omega \sim \delta =(\nu
_{0}L^{d})^{-1}$ the result should be of the same order as for the one
single wavefunction ($\omega =0$).

\begin{figure}[tbp]
\includegraphics[width=8cm]{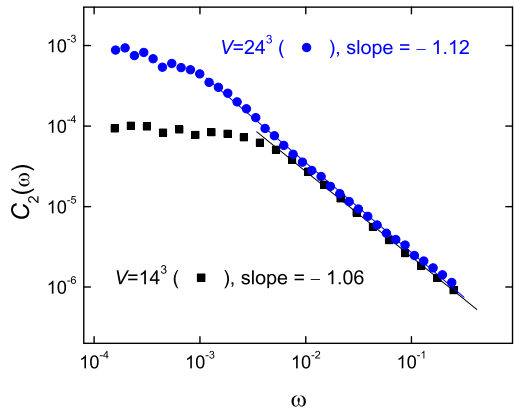}
\caption{(Color online) Scaling relationship between $\protect\gamma_{2}$
and $\protect\alpha_{2}$ for the 3D Anderson model of the orthogonal
symmetry class at two different system sizes L=14 and L=24. The theoretical
slope computed from Eqs.(\protect\ref{corr-dif-E}),(\protect\ref{d-2-4})
gives $\protect\gamma_{2}\approx 1.14\pm 0.03$.}
\label{FigC2}
\end{figure}

This immediately gives rise to the relationship between $\alpha_{q}$ and $%
\gamma_{q}$ given in Eq.(\ref{corr-dif-E}). In particular for $q=1$ one
finds: 
\begin{equation}  \label{Chalk}
\gamma_{1}\equiv\gamma=1-\frac{d_{2}}{d}.
\end{equation}
The scaling relationship Eq.(\ref{Chalk}) has been first suggested by
Chalker \cite{Chalker1990} and checked numerically in a number of works \cite%
{CueKra}. Fig.\ref{FigC2} gives an evidence that the scaling relationship
Eq.(\ref{corr-dif-E}) between $\gamma_{q}$ and $\alpha_{q}$ holds true for $%
q=2$ as well. An important point is that finite-size corrections are quite
large and should be taken into account for an accurate determination of the
exponents $\gamma_q$ (an example is presented in Figs.\ref{Cw2},\ref{E01}
below). 

So far we have discussed the correlation functions containing $|\psi_{i}(%
\mathbf{r})|^{2}$ and thus independent of the random eigenfunction \textit{%
phase}. The simplest \textit{phase-dependent} correlation function is
involved in the density-density correlation function: 
\begin{eqnarray}  \label{d-d}
K(\omega,\mathbf{r-r^{\prime }})&=&\nu_0^{-2}\sum_{i,j}\langle%
\delta(E-E_{i})\delta(E^{\prime }-E_{j})  \notag \\
&\times& \psi_{i}(\mathbf{r})\psi_{i}^{*}(\mathbf{r}^{\prime })\psi_{j}^{*}(%
\mathbf{r})\psi_{j}(\mathbf{r}^{\prime })\rangle
\end{eqnarray}

According to Ref.\cite{Chalker1990} the Fourier-transform of this
correlation function is equal to: 
\begin{equation}
K(\omega ,\mathbf{q})=\frac{1}{2\pi \nu _{0}}\,\frac{D(\omega ,q)q^{2}}{%
(D(\omega ,q)q^{2})^{2}+\omega ^{2}},  \label{d-d-ChAn}
\end{equation}%
which is a generalization of the correlation function in the diffusion
approximation for the case where the diffusion coefficient may depend on $q$
and $\omega $. The main assumption \cite{Chalker1990} here is that at the
critical point and for $q\ll \ell ^{-1}$: 
\begin{equation}
D(\omega ,q)=q^{d-2}\,F(qL_{\omega })=\left\{ 
\begin{matrix}
\begin{array}{cc}
L_{\omega }^{2-d}, & qL_{\omega }\ll 1 \\ 
q^{d_{2}}\,L_{\omega }^{2-d+d_{2}}, & qL_{\omega }\gg 1%
\end{array}%
\end{matrix}%
.\right.  \label{D-ChAn}
\end{equation}%
One can easily see that this assumption is equivalent to the statement that
at distances $\ell <|\mathbf{r-r^{\prime }}|<L_{\omega }$ the correlation
function $K(\omega ,\mathbf{r-r^{\prime }})$ has the same $\mathbf{%
r-r^{\prime }}$-dependence Eq.(\ref{corr-dif-r})(but with $L\rightarrow
L_{\omega }$ as explained above) as the correlation function of the local
DoS $C_{1}(\omega ,\mathbf{r-r^{\prime }})$ that we discussed above. This
requires non-trivial \textit{phase correlations} in the eigenfunctions which
allow to replace the \textit{phase-dependent} average by its \textit{%
phase-independent} counterpart: 
\begin{equation}
\langle \psi _{i}(\mathbf{r})\psi _{i}^{\ast }(\mathbf{r}^{\prime })\psi
_{j}^{\ast }(\mathbf{r})\psi _{j}(\mathbf{r}^{\prime })\rangle \Rightarrow
\langle |\psi _{i}(\mathbf{r})|^{2}|\psi _{j}(\mathbf{r}^{\prime})|^2\rangle
\label{replacement}
\end{equation}%
The replacement Eq.(\ref{replacement}) illustrates the general rule \newline
\textbf{Rule (iii)}: To estimate the averages that cannot be expressed in
terms of $|\psi |^{2}$ (\textit{phase-dependent averages}) one should find
by permutation of space and/or energy variables the corresponding \textit{%
phase-independent average} and apply rules (i) and (ii) in order to estimate
the latter.

One should note, however, that this rule does not apply for $|\mathbf{r}%
|>L_{\omega}$. At such distances the phase correlations are no longer
present, and the phase-dependent average vanishes exponentially while the $%
\mathbf{r}$-dependence of its phase-independent counterpart saturates.

\subsubsection{Scaling estimates for matrix elements: mobility edge. \label%
{Scaling estimates A}}

Now let us show how to use phenomenology of multifractal wavefunction
statistics to estimate the matrix elements of local interaction $M_{ijkl}$
given by Eq.(\ref{Melements1}). The simplest one $M_{i}\equiv M_{iiii}$, so
called \textit{super-diagonal} matrix element, is proportional to the
inverse participation ratio $P_{2}$: 
\begin{equation}  \label{i}
\langle M_{i}\rangle \approx 3 \ell^{-(d-d_{2})}\,L^{-d_{2}}.
\end{equation}
Here and below we define microscopic length-scale $\ell$ via its relation to
the upper energy cutoff $E_0 = 1/\nu_0\ell^3$. The factor 3 in the above
equation is of the same origin as in the statistics of eigenvectors of real
Gaussian random matrices. Applicability of such a relation to the case of
fractal wavefunctions was demonstrated in Ref.~\cite{MirlinFyodorov97} for
the case of Anderson transition on a Cayley tree. Qualitative picture,
emerging from their analysis, is that typical wavefunctions can be
represented as products $\psi_j(\mathbf{r}) = \chi_j(\mathbf{r})\Phi_\xi(%
\mathbf{r})$ where $\chi_j(\mathbf{r})$ describes fast-oscillating functions
specific for each eigenstate $j$, but insensitive to the vicinity of
Anderson transition, whereas $\Phi_\xi(\mathbf{r})$ has a meaning of smooth
envelope function, which is however sensitive to the proximity of energy $%
\xi $ to the mobility edge $E_c$. Combinatorial factor 3 in Eq.(\ref{i}) is
due to the fast-fluctuating ``Gaussian'' component $\chi_j(\mathbf{r})$.

\begin{figure}[tbp]
\includegraphics[width=8cm]{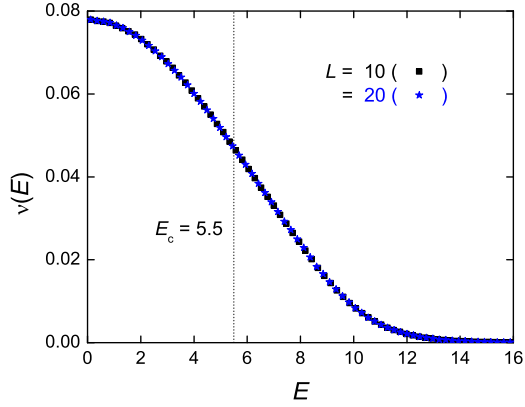}
\caption{Density of states for the 3D Anderson model with Gaussian disorder
with the width $W=4$ for the system size $L=10$ (black squares) and $L=20$
(blue stars).}
\label{FigDoS1}
\end{figure}

\begin{figure}[tbp]
\includegraphics[width=8cm]{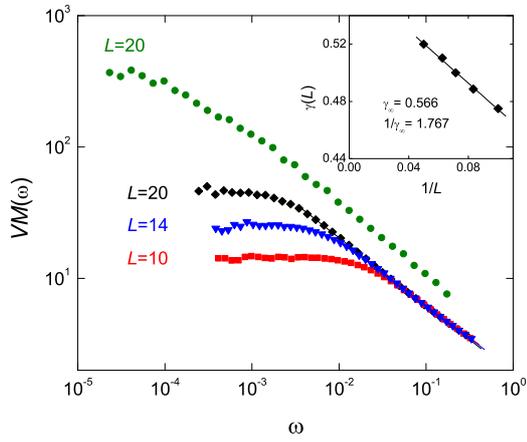}
\caption{(Color online) Correlation function $M(\protect\omega)$ for 3DAM
with Guassian disorder and lattice sizes $L=10,14,20$ at the mobility edge $%
E=5.5$ (red squares, blue triangles and black diamonds) and at the energy $%
E=8$ inside localized band (green dots). Inset shows $\protect\gamma$ values
for $L=10.12.14.16.20$.}
\label{Cw2}
\end{figure}

The \textit{diagonal} matrix element $M_{ij}\equiv M_{ijij}$ with $i\neq j$
can be easily found from Eq.(\ref{corr-dif-E}); 
\begin{equation}  \label{ij}
\langle M_{ij} \rangle \equiv \mathcal{V}^{-1}\, M(\omega) \approx \mathcal{V%
}^{-1}\, \left(\frac{E_{0}}{\omega}\right)^{\gamma},
\end{equation}
where $\mathcal{V}=L^{d}$ and $\omega = \xi_i-\xi_j$. The detailed numerical
analysis of this matrix element has been done in Ref.\cite{CueKra}. Here we
present in Fig.\ref{Cw2} data for $C_1(\omega) \equiv M(\omega)$ correlator
for 3D Anderson model with Gaussian disorder ($W=4$) at the mobility edge,
computed for three different system sizes. The plot of the density of states
for the same model is presented in Fig.~\ref{FigDoS1}. The values of $\gamma 
$ extracted for all three sizes are shown in the inset, together with
extrapolation to $L\to \infty$, which leads to 
\begin{equation}
\gamma \approx 0.57\pm 0.02  \label{gamma}
\end{equation}
Below we will use this value of $\gamma$ in our analysis. as well as the
value of the pre-factor 
\begin{equation}
E_0 = 2.08 \pm 0.25  \label{E00}
\end{equation}
which is also extracted via large-$L$ extrapolation, as shown in Fig.\ref%
{E01}. In addition, we present in Fig.\ref{Cw2} data for the same
correlation function at $E=8.0$ in the localized part of the spectrum (to be
discussed below in Sec.~\ref{Matrix elements} and used in Sec.~\ref%
{Superconductivity with a pseudogap}). Here we just note that the
logarithmic slope of the $M(\omega)$ function (rather deep inside localized
band) is close to its value $\gamma$ for the critical eigenstates.

\begin{figure}[tbp]
\includegraphics[width=8cm]{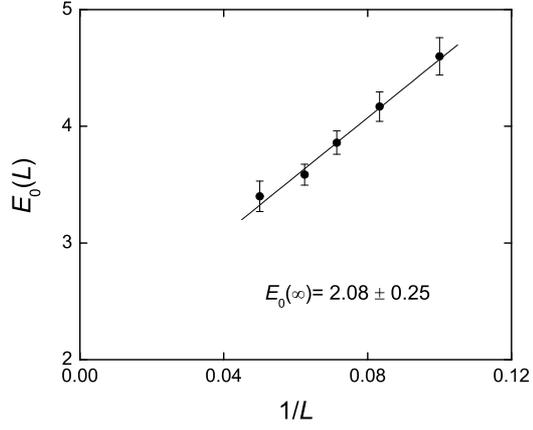}
\caption{Pre-factor $E_0$ for 3DAM with Guassian disorder width $W=4$ at
the mobility edge $E_c = 5.5$, for 
lattice sizes $L=10,12,14,16,20$}
\label{E01}
\end{figure}

\begin{figure}[tbp]
\includegraphics[width=8cm]{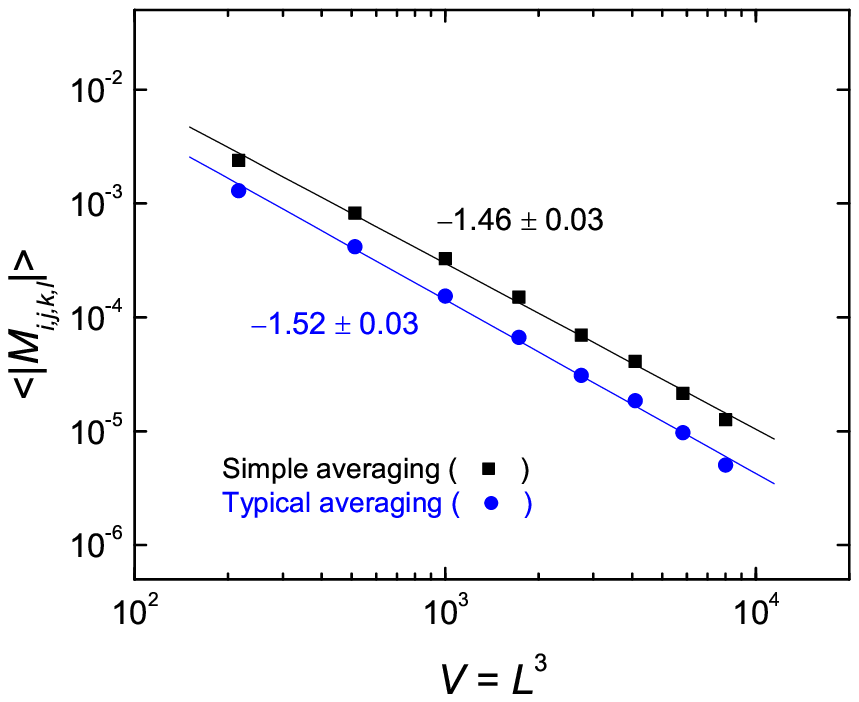}
\caption{(Color online) Scaling with the system size of the off-diagonal
matrix element. The theoretical exponent is $-3/2$ according to Eq.(\protect
\ref{off-square22}).}
\label{FigScalingOff1}
\end{figure}
A straightforward generalization of Eq.(\ref{ij}) is the matrix element 
\begin{equation}  \label{multi-ij}
\int d^{d}\mathbf{r}\,\langle |\psi_{i_{1}}(\mathbf{r})|^{2}...|\psi_{i_{n}}(%
\mathbf{r})|^{2}\rangle \sim \mathcal{V}^{-(n-1)}\,\left(\frac{E_{0}}{\omega}
\right)^{\gamma_{n}(n-1)},
\end{equation}
where $\gamma_{n}$ is defined in Eq.(\ref{corr-dif-E}), and all energy
differences $|\xi_{i_{m}}-\xi_{i_{m^{\prime }}}|$ are assumed to be of the
same order $\omega$.

The \textit{off-diagonal} matrix element $M_{ijkl}$ with all indices
different and all energy differences of the same order $\omega $ strongly
fluctuates and therefore has zero average value. However, its average square
modulus can be computed as following. We start by writing 
\begin{eqnarray}
&&\langle |M_{ijkl}|^{2}\rangle =\int d^{d}\mathbf{r}d^{d}\mathbf{r^{\prime }%
}  \label{off-square1} \\
&&\langle \psi _{i}^{\ast }(\mathbf{r})\psi _{i}(\mathbf{r^{\prime }})\psi
_{j}^{\ast }(\mathbf{r})\psi _{j}(\mathbf{r^{\prime }})\psi _{k}^{\ast }(%
\mathbf{r}^{\prime })\psi _{k}(\mathbf{r})\psi _{l}^{\ast }(\mathbf{%
r^{\prime }})\psi _{l}(\mathbf{r})\rangle  \notag
\end{eqnarray}%
Now we apply the rule $(iii)$ of the previous subsection and define the
phase-independent correlation function: 
\begin{equation}
\langle |\psi _{i}(\mathbf{r})|^{2}|\psi _{j}(\mathbf{r}^{\prime
})|^{2}|\psi _{k}(\mathbf{r})|^{2}|\psi _{l}(\mathbf{r^{\prime }}%
)|^{2}\rangle  \label{off-square-phase-ind}
\end{equation}%
Using the rules $(i),(ii)$, we estimate the phase-independent correlator as: 
\begin{equation}
\frac{1}{\mathcal{V}^{4}}\left( \frac{L_{\omega }}{|\mathbf{r-r^{\prime }}|}%
\right) ^{d-\alpha _{2}}\;\left( \frac{L_{\omega }}{\ell }\right)
^{2(d-d_{2})}  \label{off1}
\end{equation}%
In order to obtain an estimation for $\langle |M_{ijkl}|^{2}\rangle $ one
has to integrate Eq.(\ref{off1}) over $\mathbf{r}$ and $\mathbf{r-r^{\prime }%
}$. The first integration is trivial and results in the factor $\mathcal{V}$%
. To estimate the result of the second integration we note that according to
Eq.(\ref{pos-typ})in a typical sample the exponent $d-\alpha _{2}^{\mathrm{%
typ}}$ in the power law in Eq.(\ref{off1}) is smaller than $d$. This means
that the integral over $\mathbf{r-r^{\prime }}$ is dominated by large
distances $|\mathbf{r-r^{\prime }}|\sim L_{\omega }$. Finally we obtain: 
\begin{equation}
\langle |M_{ijkl}|^{2}\rangle _{\mathrm{typ}}\sim \frac{L_{\omega }^{d}}{%
\mathcal{V}^{3}}\,\left( \frac{E_{0}}{\omega }\right) ^{2\gamma }.
\label{off-square22}
\end{equation}

\begin{figure}[tbp]
\includegraphics[width=8cm]{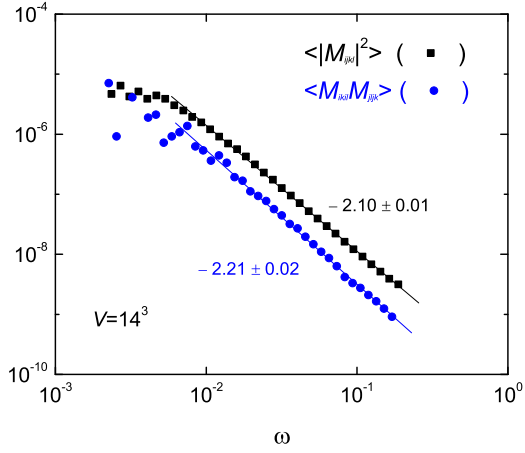}
\caption{(Color online) Scaling with energy difference $|E_{i}-E_{j}|\sim
|E_{i}-E_{k}|\sim |E_{i}-E_{l}|\sim...\sim \protect\omega$ of the
off-diagonal $\langle |M_{ijkl}|^{2} \rangle$ (blue dots) and the product of
semi-diagonal $\langle M_{iikl}M_{jjkl}\rangle$ (black squares) matrix
elements. The theoretical exponent found from Eqs.(\protect\ref{off-square22}%
),(\protect\ref{off-dia}) and (\protect\ref{d-2-4})is equal to $-2.13\pm 0.1$%
.}
\label{FigScalingOff2}
\end{figure}

The same rules applied to the correlation function $\langle
M_{ikil}M_{jljk}\rangle$ of \textit{semi-diagonal} matrix elements give
exactly the same answer: 
\begin{equation}  \label{off-dia}
\langle M_{ikil}M_{jljk}\rangle_{\mathrm{typ}}\sim\frac{L_{\omega}^{d}}{%
\mathcal{V}^{3}}\,\left( \frac{E_{0}}{\omega}\right)^{2\gamma}.
\end{equation}
\begin{figure}[tbp]
\includegraphics[width=8cm]{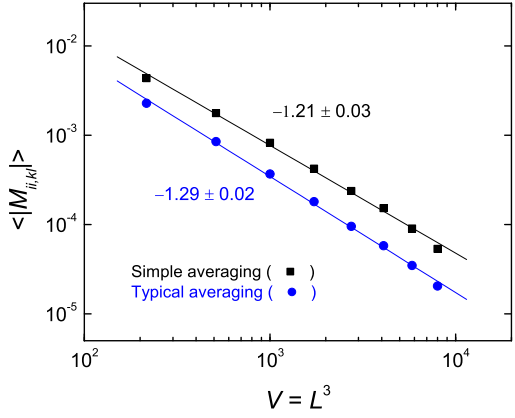}
\caption{(Color online) System size scaling of semi-diagonal matrix element
for the 3D AM. The theoretical exponent found from Eq.(\protect\ref{semi-dia}%
),(\protect\ref{d-2-4}) is equal to $-1.21\pm 0.02$ for the simple averaging
and $-1.24\pm 0.02$ for the typical averaging. }
\label{FigScalingOff3}
\end{figure}

\begin{figure}[tbp]
\includegraphics[width=8cm]{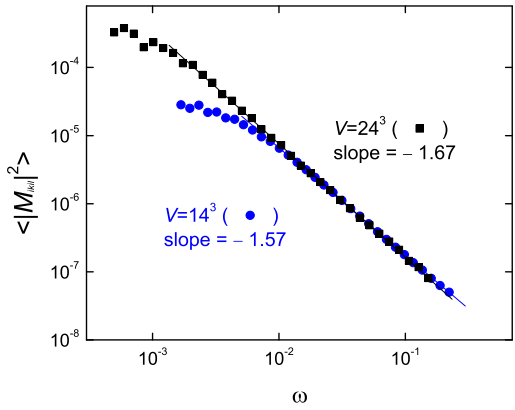}
\caption{(Color online) Energy difference scaling of semi-diagonal matrix
element for the 3D AM with two system sizes $L=14$ and $L=24$. The
theoretical exponent found from Eq.(\protect\ref{semi-dia}),(\protect\ref%
{d-2-4}) is equal to $-1.57\pm 0.03$. }
\label{FigScalingOff4}
\end{figure}
However, the square modulus of the semi-diagonal matrix elements is much
larger: 
\begin{equation}
\langle |M_{ikil}|^{2}\rangle _{\mathrm{typ}}\sim L_{\omega }^{d}\ell
^{-(d-d_{2})}\mathcal{V}^{-2-d_{2}/d}\,(E_{0}/\omega )^{\gamma }.
\label{semi-dia}
\end{equation}
The relevant phase-dependent correlation function and its phase-independent
counterpart according to rule (iii) are: 
\begin{eqnarray}
&&\langle |\psi _{i}(\mathbf{r})|^{2}\psi _{k}(\mathbf{r})\psi _{l}^{\ast }(%
\mathbf{r})|\psi _{i}(\mathbf{r^{\prime }})|^{2}\psi _{k}^{\ast }(\mathbf{r}%
^{\prime })\psi _{l}(\mathbf{r^{\prime }})\rangle \Rightarrow  \notag
\label{semi-off} \\
&&\langle |\psi _{i}(\mathbf{r})|^{2}|\psi _{i}(\mathbf{r^{\prime }}%
)|^{2}|\psi _{k}(\mathbf{r})|^{2}|\psi _{l}(\mathbf{r^{\prime }})|^{2}\rangle
\end{eqnarray}%
The latter for $\ell \ll |\mathbf{r-r^{\prime }}|\ll L_{\omega }$ can be
found in the following form: 
\begin{eqnarray}
&&\langle |\psi _{i}(\mathbf{r})|^{2}|\psi _{i}(\mathbf{r^{\prime }}%
)|^{2}|\psi _{k}(\mathbf{r})|^{2}|\psi _{l}(\mathbf{r^{\prime }}%
)|^{2}\rangle \sim  \notag  \label{semi-off-conj} \\
&\sim &\frac{1}{\mathcal{V}^{4}}\,\left( \frac{L_{\omega }}{|\mathbf{%
r-r^{\prime }}|}\right) ^{\nu _{1}}\left( \frac{L_{\omega }}{\ell }\right)
^{\nu _{2}}\,\left( \frac{L}{|\mathbf{r-r^{\prime }}|}\right) ^{\nu _{3}}.
\end{eqnarray}%
At $|\mathbf{r-r^{\prime }}|>L_{\omega }$ we assume that the $\mathbf{%
r-r^{\prime }}$ dependence in the first factor in r.h.s. of Eq.(\ref%
{semi-off-conj}) saturates.

Indeed, according to rule (i), at $|\mathbf{r-r^{\prime }}|\sim L\gg
L_{\omega}$ the following decoupling can be done: 
\begin{eqnarray}  \label{semi-off-large}
&&\mathcal{V}^{4}\langle |\psi_{i}(\mathbf{r})|^{2}|\psi_{i}(\mathbf{%
r^{\prime }})|^{2}|\psi_{k}(\mathbf{r})|^{2}|\psi_{l}(\mathbf{r^{\prime }}%
)|^{2} \rangle\approx  \notag \\
&& \mathcal{V}^{4}\langle |\psi_{i}(\mathbf{r})|^{2}|\psi_{k}(\mathbf{r}%
)|^{2}\rangle \langle|\psi_{i}(\mathbf{r^{\prime }})|^{2}|\psi_{l}(\mathbf{%
r^{\prime }})|^{2} \rangle\sim \left( \frac{E_{0}}{\omega}\right)^{2\gamma}.
\end{eqnarray}
This suggests that $\nu_{2}=2(d-d_{2})$.

At $|\mathbf{r-r^{\prime }}|\sim \ell$ according to rule (ii) all
wavefunctions are effectively in one space point: 
\begin{eqnarray}  \label{semi-off-small}
&& \mathcal{V}^{4}\langle |\psi_{i}(\mathbf{r})|^{2}|\psi_{i}(\mathbf{%
r^{\prime }})|^{2}|\psi_{k}(\mathbf{r})|^{2}|\psi_{l}(\mathbf{r^{\prime }}%
)|^{2} \rangle  \notag \\
&\approx & \mathcal{V}^{4}\langle |\psi_{i}(\mathbf{r})|^{4}|\psi_{k}(%
\mathbf{r})|^{2}|\psi_{l}(\mathbf{r})|^{2} \rangle  \notag \\
&\sim &\left( \frac{L}{\ell}\right)^{d-d_{2}}\,\left( \frac{L_{\omega}}{\ell}%
\right)^{-3d_{4}+d_{2}+2d}
\end{eqnarray}

Comparing Eqs.(\ref{semi-off-small}),(\ref{semi-off-large}) with Eq.(\ref%
{semi-off-conj}) we find: 
\begin{equation}  \label{nu-nu}
\nu_{1}=3(d_{2}-d_{4}),\;\;\;\;\nu_{2}=2(d-d_{2}),\;\;\;\;\;\nu_{3}=d-d_{2}.
\end{equation}

The $|\mathbf{r-r^{\prime }}|$ dependence in Eq.(\ref{semi-off-conj}) is the
power law $|\mathbf{r-r^{\prime }}|^{-(\nu_{1}+\nu_{3})}$ with the exponent $%
(d-\alpha_{2})$. As in the case of the off-diagonal matrix elements
considered above, for a typical realization of disorder this exponent is
smaller than $d$. Therefore, the integral of the phase-dependent correlation
function over $(\mathbf{r-r^{\prime }})$ which determines the matrix element 
$\langle|M_{ikil}|^{2} \rangle_{\mathrm{typ}}$ is dominated by large
distances $|\mathbf{r-r^{\prime }}|\sim L_{\omega}$ (for $|\mathbf{%
r-r^{\prime }}|>L_{\omega}$ the phase-dependent correlation function decays
exponentially). Thus plugging $|\mathbf{r-r^{\prime }}|\sim L_{\omega}$ in
Eq.(\ref{semi-off-conj}) and multiplying the result by $L_{\omega}^{d}$ and $%
\mathcal{V}$ (for integration over the remaining coordinate $\mathbf{r}$) we
arrive at the announced result Eq.(\ref{semi-dia}).

We conclude our analysis by noting that if instead of typical averaging the
full ensemble averaging is performed, the exponent $\alpha
_{2}=3d_{4}-2d_{2} $ may be (and in the 3D Anderson model of orthogonal
symmetry class is) negative, due to the contribution of rare untypical
realizations. In this case the principle contributions to all correlations
functions come from small distances $|\mathbf{r-r^{\prime }}|\sim \ell $.
This changes the estimates in Eqs.(\ref{off-square22}),(\ref{off-dia}),(\ref%
{semi-dia}): all those equations will acquire an extra factor 
\begin{equation}
(\omega /E_{0})^{\alpha _{2}/d}  \label{extra}
\end{equation}%
which explicitly depends on the fractal dimensionality $d_{4}$ and not only
on $d_{2}$. Note, however, that in the particular case of the 3D orthogonal
symmetry class, the exponent $|\alpha _{2}/d|\sim 0.1$ is extremely small,
so that an extra factor Eq.(\ref{extra}) is of order one for most of the
practical purposes.

The predictions made on the basis of rules (i)-(iii) of the algebra of
multifractal states are checked by numerical diagonalization of the 3d
Anderson model of the orthogonal symmetry class and summarized in figures
Fig.\ref{FigScalingOff1}-Fig.\ref{FigScalingOff4}. One can see a very
satisfactory agreement for exponents of various power laws which were found
numerically and derived theoretically using only one fractal dimension $%
d_{2} $.

Closing this subsection we conclude that the rules $(i)-(iii)$ and the
definition of multifractal dimensions Eq.(\ref{def-multi}) constitute the
full set of rules necessary to estimate any correlation function of critical
wavefunctions. This set of algebraic operations will be the main \textit{%
analytical} tool to deal with the strongly disordered case considered in
this paper.

\subsubsection{Scaling estimates for matrix elements: multifractal
insulator. \label{Scaling estimates B}}

Estimates for the matrix elements on the localized side of the Anderson
transition can be obtained from the corresponding formulae of the preceding
subsection provided that the localization radius is larger than the
characteristic length $\ell $ (multifractal insulator). The only
modification to be done in all the local averages $\langle ...\rangle $ is
to replace $L$ by $L_{\mathrm{loc}}$ and to add a factor $(L_{\mathrm{loc}%
}/L)^{nd}$ (where $n$ is the number of different eigenfunctions in the
average) that accounts for the probability for a point $\mathbf{r}$ to be
inside the localization radius of each of the wavefunctions. To apply these
simple rules for the matrix elements of different wavefunctions with the
energy separations $\omega $, one has also to make sure that all the
corresponding length scales $L_{\omega }=(\nu _{0}\omega )^{-1/d}$ are
smaller than the localization radius $L_{\mathrm{loc}}$. This sets up an
important condition which determines the frequency domain of the
multifractal correlations (\textit{multifractal frequency domain}): 
\begin{equation}
E_{0}>\omega >\delta _{L}=(\nu _{0}L_{\mathrm{loc}}^{d})^{-1}.
\label{delta-xi}
\end{equation}%
In particular, under this condition we have: 
\begin{equation}
\langle M_{i}\rangle =3\ell ^{-(d-d_{2})}\,L_{\mathrm{loc}}^{-d_{2}},
\label{Mi}
\end{equation}%
which is much larger than the corresponding critical value Eq.(\ref{i}). The
combinatorial factor 3 in Eq.(\ref{Mi}) arises because of the statistics of
phase of wave function as explained above, see note after Eq.(\ref{i}).

However, the estimate for the \textit{average} diagonal matrix element $%
M_{ij}$ does not change and is still described by Eq.(\ref{ij}). The reason
is the additional (compared to $M_{i}$) factor $(L_{\mathrm{loc}}/L)^{d}$
due to a small probability for \textit{two} wavefunctions being localized
close in space which exactly compensates for the replacement of $\mathcal{V}%
\rightarrow L_{\mathrm{loc}}^{d}$ in Eq.(\ref{ij}). One can easily check
that the estimates for the average square of the off-diagonal matrix element
Eq.(\ref{off-square22}), and the average of different semi-diagonal matrix
element (\ref{off-dia}) also remain unchanged under the condition Eq.(\ref%
{delta-xi}).

The average square of the semi-diagonal matrix element gets enhanced with
respect to the critical case Eq.(\ref{semi-dia}): 
\begin{equation}
\langle |M_{ikil}|^{2}\rangle _{\mathrm{typ}}\sim \frac{1}{\mathcal{V}^{2}}%
\ell ^{-(d-d_{2})}\,L_{\mathrm{loc}}^{-d_{2}}\,L_{\omega
}^{d}\;(E_{0}/\omega )^{\gamma }.  \label{semi-dia-ins}
\end{equation}%
This enhancement factor of $(L/L_{\mathrm{loc}})^{d_{2}}$ results in a
regular scaling of $\langle |M_{ikil}|^{2}\rangle \propto \mathcal{V}^{-2}$
with the total volume.

Note, however, that in all cases the typical value of a matrix element $%
|M^{typ}|$ of \textit{well overlapping states} is much greater than the
typical \textit{average} value $\langle |M| \rangle_{\mathrm{typ}}$, and it
can be obtained from the corresponding expression for $\langle M\rangle_{%
\mathrm{typ} }$ (or $\langle M^{2}\rangle_{\mathrm{typ} }$) by replacing the
total volume $\mathcal{V}$ by the localized volume $L_{\mathrm{loc}}^{d}$.
In particular, the typical value of the diagonal matrix element for well
overlapping states is: 
\begin{equation}
M_{ij}^{typ}\sim L_{\mathrm{loc}}^{-d}\,\left( \frac{E_{0}}{\omega}%
\right)^{\gamma}.
\end{equation}
This difference between the average value of a matrix element and the
typical value for well overlapping states is due to the fact that the most
of matrix elements in insulator are very small due to poor overlap of the
corresponding states.

One can easily check that at $\omega=\delta_{L}$ the typical absolute value
of matrix elements of well overlapping states $i,j,k,l$ does not depend on
the number of different states and has a order of magnitude of the inverse
participation ratio: $|M^{typ}|\sim \ell^{-d}\,(\ell/L_{\mathrm{loc}%
})^{d_{2}}$. For $\omega>\delta_{L}$ the matrix elements $M^{typ}$ of well
overlapping states get suppressed, and the suppression is stronger when the
number of different states in the matrix element increases.

%%%%%%%%%%%%%%%%%%%%%%%%%%%%%%%%%%%%%%%%%%%%%%%%%%%

\subsubsection{Scaling estimates for matrix elements: multifractal metal. 
\label{Scaling estimates C}}

On the metal side of the Anderson transition, the wave function is globally
not a fractal (or multifractal), as the moments $P_{q}$ are proportional to $%
L^{-d(q-1)}$. However, the correlations of different eigenfunctions (with
the energy difference $\omega $) show the same power-law $\omega $-behavior
as the critical eigenfunctions \cite{CueKra} provided that $\ell \ll
L_{\omega }\ll L_{\mathrm{corr}}$. The physical meaning of the correlation
length $L_{\mathrm{corr}}$ is a typical size of a fractal element of which
the entire eigenfunction support is built. Thus, locally a wavefunction in a
multifractal metal is identical to the one in a multifractal insulator
inside the localization radius. However, the global normalization $\sum_{%
\mathbf{r}}|\psi (\mathbf{r})|^{2}=1$ requires the reduction of $|\psi |^{2}$
by a factor of $L_{\mathrm{corr}}^{d}/\mathcal{V}$ compared to the case of
insulator. Thus, we can formulate the rule (valid provided that $\ell \ll
L_{\omega }\ll L_{\mathrm{corr}}$) for the estimation of the matrix elements
in the multifractal metal if their counterparts in the multifractal
insulator are known. One has (i) to multiply the result for insulator by a
factor $(L_{\mathrm{corr}}^{d}/\mathcal{V})^{q}$, where $q$ is the total
number of $|\psi |^{2}$ in the matrix element in order to take into account
the normalization and (ii) to eliminate the overlap probability factor $(L_{%
\mathrm{corr}}^{d}/\mathcal{V})^{n}$, where $n$ is the number of different
wavefunctions, which is no longer needed for the extended metal states. Thus
the overall factor to add is $(L_{\mathrm{corr}}^{d}/\mathcal{V})^{q-n}$.
This immediately leads to 
\begin{eqnarray}
&&\langle M_{i}\rangle \sim \frac{1}{\mathcal{V}}\,(L_{\mathrm{corr}}/\ell
)^{d-d_{2}},  \label{mm} \\
&&\langle |M_{ikil}|^{2}\rangle _{\mathrm{typ}}\sim \frac{1}{\mathcal{V}^{3}}%
\,(L_{\mathrm{corr}}/\ell )^{(d-d_{2})}\,L_{\omega }^{d}\,(E_{0}/\omega
)^{\gamma }.
\end{eqnarray}%
The averages $\langle M_{ij}\rangle $, $\langle |M_{ijkl}|^{2}\rangle $, and 
$\langle M_{ikil}M_{jljk}\rangle $ do not change and have the same order of
magnitude on the both side of the transition provided that the condition Eq.(%
\ref{delta-xi}) (with $L_{\mathrm{corr}}$ replacing $L_{\mathrm{loc}}$) is
respected.

\subsubsection{Matrix elements of the off-critical states beyond the
multifractal frequency domain. \label{Matrix elements}}

As we have seen in the previous subsections, in the multifractal metal and
insulator characterized by the large correlation/localization length $L_{%
\mathrm{loc}},L_{\mathrm{corr}}\gg \ell $, the wavefunction correlations are
very similar to those of the critical multifractal states at the Anderson
transition point. However, this correspondence is only valid if the energy
separation $\omega $ between the states lies in the multifractal frequency
domain, $\delta _{L}<\omega <E_{0}$, bounded by effective level spacing, $%
\delta _{L}$, and high frequency cutoff, $E_{0}$ . If one or several energy
separations are beyond the multifractal frequency domain, the
frequency-dependent factors in the estimates change. If all energy
separations are larger than $E_{0}$ the frequency-dependent factors decrease
very fast, so $E_{0}$ provides the high-frequency cut-off for all wave
function correlations.

More delicate is the case where all energy separations are smaller than $%
\delta _{L}$. The behavior of the matrix elements in this region is
different in the multifractal metal and in the insulator. In the
multifractal metal, the $\omega $-dependent factors simply saturate \cite%
{CueKra}. In contrast, in the multifractal insulator they acquire additional
powers of $\ln (\delta _{L}/\omega )$.

The logarithmic enhancement factor appearing in the insulator reflects the
Mott's physics of the resonance mixing of states; this effect is also
responsible for the logarithmic factors in the expression for the
low-frequency conductivity: $\sigma \sim \omega ^{2}\,\ln ^{d+1}(\delta
_{L}/\omega )$. The key element of this phenomenon is that localized states
with close energies cannot be considered independent even when the distance
between centers of localization is large compared to $L_{\mathrm{loc}}$. To
understand the origin of the correlations, we repeat the Dyson arguments for
the level statistics. Consider two typical localized states with small
energy separation $\delta E<\delta _{L}$ at distance $R>L_{\mathrm{loc}}$
from each other and vary the disorder potential. The states cease to become
orthogonal to each other, their typical overlap decreases as $t\sim \delta
_{L}\exp (-R/L_{\mathrm{loc}})$ with distance. The energy splitting of the
two states becomes $\omega =((\delta E)^{2}+t^{2})^{1/2}$. If $R\lesssim L_{%
\mathrm{loc}}\,\ln (\delta _{L}/\omega )$ the typical overlap between two
states at distance $R$ is larger than $\omega $ which implies that small
energy splittings are dominated by rare events when $t$ $\sim \delta E\sim
\omega $. This implies that in a typical situation the states at distance $R$
hybridize forming superpositions $\psi _{\pm }(\mathbf{r})=\cos \alpha _{\pm
}\psi _{1}(\mathbf{r})+\sin \alpha _{\pm }\psi _{2}(\mathbf{r})$ of the
parent states $\psi _{1,2}(\mathbf{r})$ with $\alpha _{\pm }\sim 1$. This
resonance hybridization makes even remote in space parent states mix with
each other, provided that the distance between their centers of localization
is less than the optimal one $R\lesssim R_{0}\sim L_{\mathrm{loc}}\,\ln
(\delta _{L}/\omega )\gg L_{\mathrm{loc}}$.

\begin{figure}[tbp]
\label{FigR-0}\includegraphics[width=8cm]{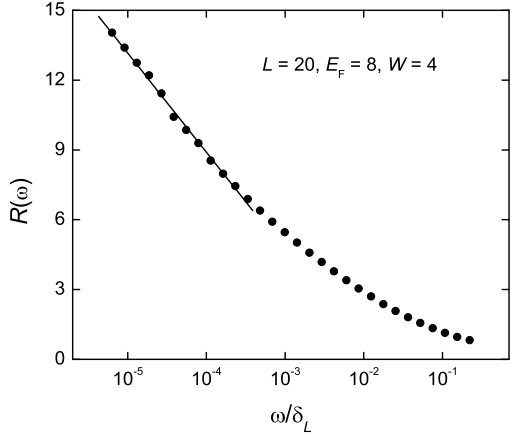}
\caption{Logarithmic dependence of $R_{0}$ on $\protect\omega $ for the 3D
Anderson model with $W=4$ and $E_{F}=8.0$ well in the localized region. At
this value of the energy the localization radius $L_{\mathrm{loc}}=(\protect%
\nu _{0}\protect\delta_{L})^{-1/3}\approx 3$ (this estimate follows from the
data for $\protect\nu_{0}$ shown in Fig.~\protect\ref{FigDoS1} and for $%
\protect\delta _{L}$ shown in Fig.~\protect\ref{TcDeltaPdeltaL}). The linear
in $\ln \protect\omega $ fit at small $\protect\omega $ corresponds to $R(%
\protect\omega )=A\ln \protect\delta _{L}/\protect\omega +B$ with $A=1.9\pm
0.05$, $B=-8.1\pm 0.4$. It is valid in the region where $R(\protect\omega )$
is larger than $2L_{\mathrm{loc}}$. }
\end{figure}

The definition of $R_{0}$ convenient for numerical study is given in terms
of the dipole-moment matrix element: 
\begin{equation}
R_{0}^{2}(\omega )=d\,\frac{\sum_{i\neq j}x_{ij}^{2}M_{ij}\delta (\omega
-\xi _{i}+\xi _{j})}{\sum_{i\neq j}M_{ij}\delta (\omega -\xi _{i}+\xi _{j})}
\label{R-X}
\end{equation}
where 
\begin{equation}
x_{ij}=2\int d^{d}\mathbf{r}\,\psi _{i}^{\ast }(\mathbf{r})\,x\,\psi _{j}(%
\mathbf{r}).  \label{X}
\end{equation}
The definition (\ref{R-X},\ref{X}) is useful in the range of relatively
well-localized states, at $\omega \ll \delta_L$. One can easily check that
for the extreme strong localization $|\psi _{1,2}(\mathbf{r})|^{2}=\delta (
\mathbf{r}-\mathbf{r}_{1,2})$ the matrix element $x_{ij}=x_{1}-x_{2}$
corresponding to $\psi _{i,j}=\psi _{\pm }$ is equal to the $x$-component of
the distance between centers of localization of the parent states. The
results of numerical computation of (\ref{R-X}), (\ref{X}) for the 3D
Anderson model are presented in Fig.\ref{FigR-0} for $E_{F}=8.0$. 
As expected, $R_{0}(\omega )$ is linear in $\ln (\omega )$ at very low
values of $\omega /\delta _{L}$. However, there is a broad transient regime
with essentially non-linear in $\ln \omega$ behavior for moderately small $%
\omega $. It is important to note that the values of $R(\omega )$ in this
regime are smaller or comparable than $2L_{\mathrm{loc}}$ (estimated from
the inverse participation ratio, with the use of data for the DoS value and
typical level spacing $\delta _{L}$ ) which is the minimum distance between
centers of localization where the Mott's physics of resonance mixing
strictly applies.

It was shown in Ref.\cite{CueKra} that the matrix element $M_{ij}$ is
proportional to $R_{0}^{d-1}(\omega )$. This result can be also justified by
the perturbation theory arguments similar to those used in the derivation $%
R_{0}\sim L_{\mathrm{loc}}\,\ln (\delta _{L}/\omega )$. Consider the parent
states that are strongly localized at sites $m,n$ and have a distance $R=|%
\mathbf{R}_{m}-\mathbf{R}_{n}|$ between centers of their localization. We
now vary the realization of disorder and treat the matrix element of the
corresponding change of the disorder potential $H_{mn}$ as a perturbation.
The amplitude of the eigenfunction $|\psi _{m}(\mathbf{R_{n}})|^{2}$ at a
(remote from its center of localization $\mathbf{R}_{m}$) site $\mathbf{R}%
_{n}$ becomes of the order of 
\begin{equation*}
|\psi _{m}(\mathbf{R_{n}})|^{2}\sim \frac{|H_{nm}|^{2}}{(E_{n}-E_{m})^{2}}%
\sim \left( \frac{\delta _{L}}{\omega }\right) ^{2}\,e^{-R/L_{\mathrm{loc}}}.
\end{equation*}%
The matrix element is given by 
\begin{equation*}
M_{ij}\propto \sum_{\mathrm{parent}\;\;\mathrm{states}}\sum_{r}|\psi _{m}(%
\mathbf{r})|^{2}|\psi _{n}(\mathbf{r})|^{2}\approx 2|
\psi _{m}(\mathbf{R}_{n})|^2\propto \left( \frac{\delta _{L}}{\omega }\right)
^{2}\int_{R_{0}}^{\infty }dR\,R^{d-1}\,e^{-R/L_{\mathrm{loc}}}.
\end{equation*}%
Here instead of summing over the parent states we integrate over the
distance between the centers of localization $R$ taking into account the
statistical repulsion of centers of localization at $R<R_{0}\sim L_{\mathrm{%
loc}}\,\ln (\delta _{L}/\omega )$. The final expression in terms of the
integral over $R$ is similar to the corresponding expression for the Mott's
frequency-dependent conductivity but differs from it by an extra $R^{2}$
because of the square of the dipole moment matrix element and an extra $%
\omega ^{2}$ in front of the integral. The estimate of the integral in the
limit $R_{0}\gg L_{\mathrm{loc}}$ finally gives $M_{ij}\propto R_{0}^{d-1}$
and $\sigma (\omega )\propto \omega ^{2}\,R_{0}^{d+1}$.

Summarizing the results of this analysis we conclude that the Eq.(\ref{ij})
for $M_{ij}$ in the multifractal insulator becomes 
\begin{equation}
\langle M_{ij}\rangle =\mathcal{V}^{-1}\,\left( \frac{E_{0}}{\delta _{L}}%
\right) ^{\gamma ^{\mathrm{eff}}}\,\left\{ 
\begin{array}{cc}
\ln ^{2}\left( \frac{\delta _{L}}{\omega }\right) , & \omega \ll \delta _{L}
\\ 
\left( \frac{\delta _{L}}{\omega }\right) ^{\gamma ^{\mathrm{eff}}}, & 
\omega \gg \delta _{L}%
\end{array}%
\right.  \label{MI-ij}
\end{equation}%
A simple analytic expression that smoothly interpolates between these two
asymptotes in Eq.(\ref{MI-ij}) can be written for the function $M(\omega )=%
\mathcal{V}\langle M_{ij}\rangle $ with $\omega =\xi _{i}-\xi _{j}$: 
\begin{equation}
M(\omega )=\frac{\left( \frac{E_{0}}{\delta _{L}}\right) ^{\gamma ^{\mathrm{%
eff}}}\,\ln ^{2}\left( \frac{\delta _{L}}{\omega }+c\right) }{\left( \frac{%
\omega }{\delta _{L}}\right) ^{\gamma ^{\mathrm{eff}}}\;\ln ^{2}\left( \frac{%
\delta _{L}}{\omega }+c\right) +1},  \label{interp}
\end{equation}%
where the constant $c>1$. One should also take into account small variations
of the effective fractal dimension $d_{2}^{\mathrm{eff}}$ as $E_{F}$ moves
away from the mobility edge or the disorder parameter $W$ moves away from
the critical value \cite{CueKra}. This results in the dependence of the
effective exponent: 
\begin{equation}
\gamma ^{\mathrm{eff}}=1-d_{2}/d+a\,(\delta _{L}/E_{0}),\;\;\;\;\;\;a>0.
\label{gamma-eff}
\end{equation}%
The energy scale $\delta _{L}$ can be expressed through the inverse
participation ratio $\langle M_{i}\rangle $ in the form which is convenient
for numerical simulations: 
\begin{equation}
\delta _{L}=E_{0}\,\left( \frac{\langle M_{i}\rangle }{3\nu _{0}\,E_{0}}%
\right) ^{\frac{3}{d_{2}}}  \label{d-L-IPR}
\end{equation}%
Finally, the upper fractality scale $E_{0}$ can be found from the condition $%
\mathcal{V}\,\langle M_{ij}\rangle \approx 1$ at $\omega =E_{0}$.

The enhancement, Eqs.(\ref{MI-ij},\ref{interp}), of the overlap of two
localized wavefunctions results from the fact that the eigenfunctions which
are anomalously close to each other in the energy space (closer than $\delta
_{L}$ for the localized wavefunctions) are automatically well overlapping,
as they are just the symmetric and anti-symmetric combinations of the one
and the same pair of localized wavefunctions with the optimal distance
between their centers of localization being $R_{\omega } \sim L_{\mathrm{loc}%
}\,\ln (\delta _{L}/\omega )$. The most important consequence of this
phenomenon is that in 3D case the correlation function $M(\omega )$
continues to grow at low frequencies as $\log ^{2}\frac{\delta _{L}}{\omega }
$, which makes it possible to establish the superconductive order in some
part of the domain $T_{c}\ll \delta _{L}$, as will be discussed below in
section \ref{Superconductivity with a pseudogap}.

\section{Insulating state.\label{Insulating state}}

In this section we discuss the physical properties of the insulating state
in the vicinity of superconductor-insulator transition and show that it is
characterized by the large single particle gap which is responsible for the
activation temperature dependence of conductivity, $\sigma (T)\propto \exp
(-T_{I}/T)$, observed~in many works\cite{Shahar1992,Kowal1994,Gantmakher1996}
at low temperatures on the insulating side of the transition. We start by
assuming that  Fermi energy $E_{F}$ is deep inside the region of
localized states, so that the interaction of electrons from different
localized orbitals is weak and leads merely to a small perturbation. Then
the Hamiltonian (\ref{Ham2}) can be further simplified to 
\begin{equation}
H_{3}=\sum_{j\sigma }\xi _{j}c_{j\sigma }^{\dagger }c_{j\sigma
}-g\sum_{j}M_{j}c_{j\uparrow }^{\dagger }c_{j\downarrow }^{\dagger
}c_{j\downarrow }c_{j\uparrow }\,,  \label{Ham3}
\end{equation}%
where the scaling estimate for the typical value of matrix elements $%
M_{j}=\int dr\psi _{j}^{4}(r)$ is given by Eq.(\ref{Mi}) above. We will
refer to the last term in (\ref{Ham3}) as to the \textit{local pairing
coupling}; formally it looks like the \textquotedblleft
negative-U\textquotedblright\ local attraction considered in~\cite%
{Ghosal2001}.

The operator product in the last term in Eq.(\ref{Ham3}) is identical to the
occupation number product $n_{j\uparrow } n_{j\downarrow}$ which is equal to
1 if both available electrons states are populated, and to 0 otherwise. Thus
the only role of the interaction term in (\ref{Ham3}) is to shift down
energies of all double-occupied orbitals. Note that one does not encounter
such a term in usual theory of disordered superconductors, since $M_{i}$
vanishes in the thermodynamic limit for delocalized electronic states
(compare Eqs.(\ref{Mi}),(\ref{mm})).

Let us order all eigenstates $\psi _{j}$ according to the increase of
eigenvalues $\xi _{j}$. Then the last filled eigenstate $\psi _{m}$ of the
Fermi-sea (at $T=0$) for \textit{even} total number of electrons is defined
by inequality 
\begin{equation}
2\xi _{m}-gM_{m}<0<2\xi _{m+1}-gM_{m+1}  \label{Fermisea}
\end{equation}%
(we count all single-particle energies from the Fermi-energy). In a
macroscopic system, the energy interval in (\ref{Fermisea}) vanishes as the
inverse volume, $1/\mathcal{V}$. Within the "even subspace" of the whole
Hilbert space (i.e. each orbital is either empty of double-occupied), the
local pairing can be fully accounted for by the redefinition $\xi
_{j}\rightarrow \tilde{\xi _{j}}=\xi _{j}-\frac{g}{2}M_{j}$. However,
single-occupied orbitals are not involved in this interaction.

The increase of thermodynamic potential $\Omega $ due to addition of \textit{%
odd} electron to the ground-state is 
\begin{eqnarray}
\delta \Omega _{\mathrm{oe}} &=&\xi _{m+1}=\xi _{m+1}-\tilde{\xi}_{m+1}+%
\tilde{\xi}_{m+1}= \\
&&\frac{g}{2}M_{m+1}+O(\mathcal{V}^{-1})  \notag  \label{deltaOmega1}
\end{eqnarray}%
Using the Eq.(\ref{Mi}) we estimate typical value of $\delta \Omega _{%
\mathrm{oe}}$: 
\begin{equation}
\delta \Omega _{\mathrm{oe}}^{typ}=\frac{3}{2}g\ell ^{-3}(L_{loc}/\ell
)^{-d_{2}}.  \label{deltaOmega2}
\end{equation}%
where $L_{loc}$ is the localization length for the states with $E=E_{F}$.

Consider now two single-electron excitations on top of a fully paired
Fermi-sea defined by Eq.(\ref{Fermisea}), which are produced by transferring
of one electron from the $m$-th state to the $m+1$-th one. The energy of the
two single-particle excitations (which results from the depairing) is 
\begin{equation}
2\Delta _{P}^{(m)}=\xi _{m+1}-\xi _{m}+gM_{m}=\frac{g}{2}(M_{m}+M_{m+1})+O(%
\mathcal{V}^{-1})  \label{exit1}
\end{equation}%
thus the typical value of the pairing energy $\Delta _{P}$ is also given by
Eq.(\ref{deltaOmega2}): 
\begin{equation}
\Delta _{P}=\frac{3}{2}g\ell ^{-3}(L_{loc}/\ell )^{-d_{2}}=\frac{3\lambda }{2%
}E_{0}\left( \frac{E_{c}-E_{F}}{E_{0}}\right) ^{\nu d_{2}}  \label{DeltaP}
\end{equation}%
In the right-hand side of (\ref{DeltaP}) we reintroduced the dimensionless
coupling $\lambda =g\nu _{0}$ and the parameter $E_{0}=1/(\nu _{0}\ell ^{3})$
which determines the high-energy cutoff for fractal correlations and we also
employed Eq.(\ref{Lloc}). The average single-particle density of states $\nu
(\varepsilon )$ is determined by the probability distribution $\mathcal{P}%
(M) $ of inverse participation ratios $M_{j}$. Namely, the probability to
find an excitation with the energy $\varepsilon $ coincides (in the limit $%
\mathcal{V}\rightarrow \infty $ ) with the probability to find a value of
inverse participation ratio $M\leq \frac{2\varepsilon }{g}$. Therefore 
\begin{equation}
\nu (\varepsilon )=\nu _{0}\int_{0}^{2\varepsilon /g}\mathcal{P}(M)dM
\label{DoS}
\end{equation}%
We have generated the distribution function $\mathcal{P}(M)$ numerically,
using the three dimensional Anderson model with the Gaussian distribution of
local energies, Eq.(\ref{G-d}) with $W=4$. The mobility edge in such a model
is located at $|E|=5.5$.

\begin{figure}[tbp]
\includegraphics[width=8cm]{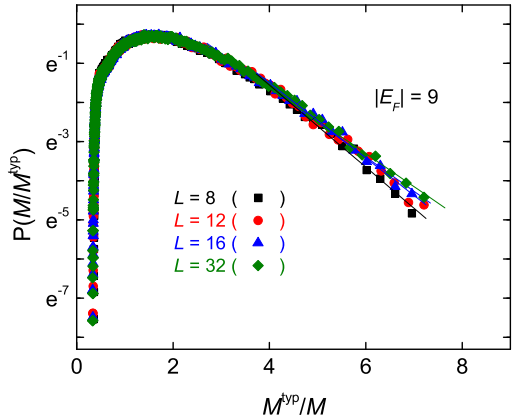}
\caption{(Color online) Distribution of the inverse participation ratios $%
P_{2}$ for the 3d Anderson model at the Fermi energy $E_{F}=9.0$ on the
insulating side of SIT (the mobility edge $E_{c}=5.5$) for different system
sizes.}
\label{FigP22}
\end{figure}
Numerical data for the distribution function $\mathcal{P}(M)$ are shown in
Fig.\ref{FigP22} for several sizes $L$ and the Fermi energy $E=9.0$ in the
localized part of the spectrum. These data demonstrate a sharp drop of $%
\mathcal{P}(M)$ at the values of $M$ much smaller than the typical value $M^{%
\mathrm{typ}}$, as well as a considerable size-dependence of the slope. Fig.%
\ref{FigP22} shows that the low $M$ tail can be well approximated by the
exponential dependence 
\begin{equation}
\mathcal{P}(M)\propto e^{-c(M^{\mathrm{typ}}/M)}\quad \mathrm{for}\,\,M\ll
M^{\mathrm{typ}}  \label{PM0}
\end{equation}%
where the coefficient $c$ depends on the energy $E_{F}$. We used the
finite-size scaling together with an extrapolation to large $L$ limit in
order to get the values of $c$ appropriate for a macroscopic system. The
results obtained for energies $E_{F}=8.0$ and $9.0$ are shown in Fig.\ref%
{FigP22a}. 
\begin{figure}[tbp]
\includegraphics[width=8cm]{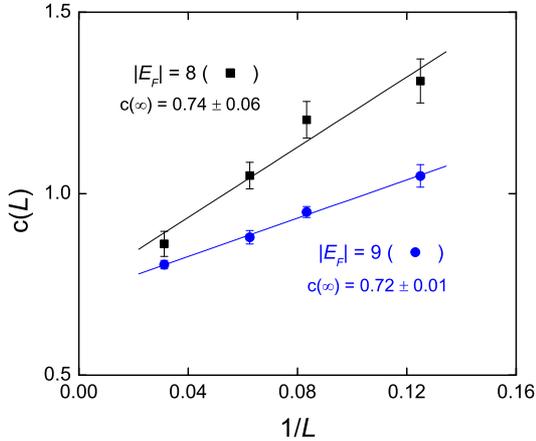}
\caption{(Color online) The exponent $c$ of the small-IPR tail of the
distribution function $P(M/M^{typ})$ at the energies $E_{F}=8.0$, and $9.0$
for different sizes $L$, together with extrapolated value $c_{\infty }$ for
each energy.}
\label{FigP22a}
\end{figure}
The data shown in Fig.\ref{FigP22a} demonstrate that the extrapolated (at $%
L\rightarrow \infty $) value of the coefficient $c$ saturates in the
interval 
\begin{equation}
c\approx 0.73\pm 0.05  \label{c}
\end{equation}%
for the Fermi energies deep enough in the localized band. This is in
agreement with the one-parameter character of the distribution $\mathcal{P}
(M)\equiv \mathcal{P}(M/M^{\mathrm{typ}})$ in the $L\rightarrow \infty $
limit.

\begin{figure}[tbp]
\includegraphics[width=8cm]{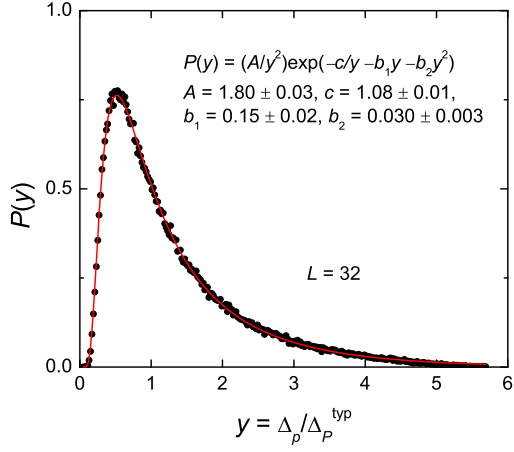}
\caption{(Color online) Distribution function for the normalized local gaps $%
y=\Delta _{P}/\Delta _{P}^{\mathrm{typ}}$, at the Fermi energy $E_{F}=8.0$
on the insulating side of SIT (the mobility edge $E_{c}=5.5$).}
\label{PDelta}
\end{figure}

In Fig. \ref{PDelta} we also present the linear scale data for the
distribution function $P(y)$ of local gaps $\Delta _{P}$,(at the
Fermi-energy $E_{F}=8.0$), normalized to the typical value $\Delta _{P}^{
\mathrm{typ}}$. Note that the data both for large and for small $y$ can be
fitted quite accurately by an analytical expression shown on the plot which
contains exponential factors and $1/y^{2}$ dependence relevant for the
intermediate $1<y<6$. This modification leads to a somewhat lager
coefficient $c$ in the exponential dependence $e^{-c/y}$.

\begin{figure}[tbp]
\includegraphics[width=8cm]{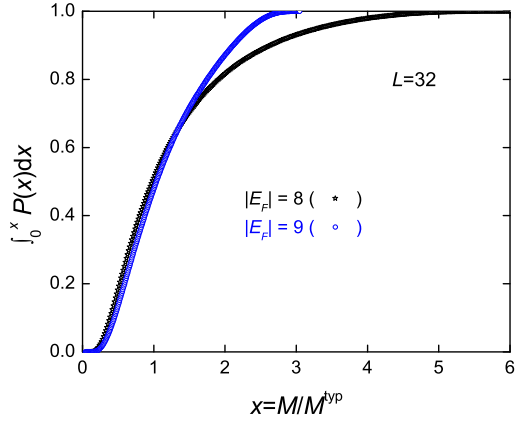}
\caption{(Color online) Gap in the one-particle density of states due to
local paring of electrons on the localized states at the energies $E_F=8.0$
(black stars) and $E_F=9.0$ (blue circles). Scaling with $x = M_{i}/M_{typ}$
for relatively small $x$ is demonstrated. }
\label{FigDoS}
\end{figure}

We use the numerical data for $\mathcal{P}(M)$ to obtain, according to Eq.(%
\ref{DoS}), the average density of single-electron states. The results are
shown (for the two Fermi-level positions $E_{F}=8.0$ and $E_{F}=9.0$) in Fig.%
\ref{FigDoS}. Both plots for DoS $\nu(\varepsilon)$ nearly coincide while
expressed as functions of the reduced variables $\varepsilon/\Delta_P$ in
the most interesting region $x \leq 1.5 $. At the lowest energies DoS decays
exponentially, $\nu_{<<}(\varepsilon) \approx \nu_0
e^{-c\Delta_P/\varepsilon}$. The similar results for the low energy DoS were
obtained in ~\cite{Ghosal2001} (see Fig.16b of Ref.~\cite{Ghosal2001}).

Practically, the shape shown in Fig.\ref{FigDoS} implies the existence of a
nearly hard gap 
\begin{equation}
\Delta _{1}=c_{1}\Delta _{P}=\frac{3c_{1}}{2}\lambda E_{0}\left( \frac{%
E_{c}-E_{F}}{E_{0}}\right) ^{\nu d_{2}}\quad c_{1}\approx 0.2-0.3
\label{hard}
\end{equation}

The average density of states could be measured directly by the tunneling
conductance via a large-area tunnel junction. The problem with such
measurement in an insulator is that electrons should be evacuated somewhere
after tunneling, otherwise the strong Coulomb blockade would make the
measurement impossible. A possible method to avoid the Coulomb blockade
might be to study tunneling conductance through a relatively thin film with
both its surfaces weakly coupled to metal contacts.

The local tunneling conductance measured by STM is expected to show a
threshold behavior with $\Delta _{P}^{(m)}$ corresponding to the state $m$
localized near the observation point in real space and with $\tilde{\xi}_{m}$
close to the Fermi energy. The local gap $\Delta _{P}^{(m)}$ fluctuates from
point to point and is distributed according to Fig.\ref{FigDoS}. We are not
aware of such measurements in the insulating state, the data on the
superconducting side of SIT are given in Ref.~\cite{Sacepe2007}, for
temperatures both above and below $T_{c}$. We present detailed results for
the tunneling conductance as a function of temperature in section~\ref%
{Tunneling conductance}.

Above the transition temperature the data \cite{Sacepe2007} show large
suppression of the density of states without any coherence peak that appear
at the gap edges below $T_{c}$. The absence of coherence peaks at the gap
threshold shows the qualitative difference between the local pairing gap due
to simple binding of two localized electrons and the BCS gap appearing due
to the \textit{many-body correlations} within the energy range $\sim \Delta
_{\mathrm{BCS}}$ around the Fermi-level that one expects\ in the small grain
of superconductor. In the case of fractal localized states \textit{all}
filled (double-occupied) levels are shifted down in energy, thus the total
number of states near the Fermi-energy (in a stripe of several $\Delta _{P}$
width) is \textit{not} conserved and there is no reason for appearance of a
peak above the gap. In that sense, the local pairing (due to attraction!)
plays the role similar to that of the Coulomb repulsion in suppressing the
tunneling conductance.

The above results were obtained neglecting all matrix elements of
interaction except from \textit{super-diagonal} ones, $M_{j}$. We now
discuss the validity of this approximation. Indeed, in the problem of
ultra-small grains treated by Matveev and Larkin~\cite{Matveev1997}, the 
\textit{diagonal} matrix elements $M_{ij}$ were taken into account while
calculating the parity gap (\ref{ML}), via the renormalization (\ref{lambdaR}%
) of the local pairing energy. This renormalization was necessary (even for $%
\lambda \ll 1$) due to the ultraviolet (UV) divergency in the Cooper loop
diagrams. Pairing on the fractal eigenstates is of different nature: the
frequency-dependence (\ref{ij}) of matrix elements $M_{ij}$ eliminates the
UV divergence, thus the virtual $i\rightarrow j$ transitions can be
neglected and the high-energy cutoff $\omega _{D}$ is not necessary as long
as $\delta _{L}\gg T_{c}^{0}$. Here $T_{c}^{0}$ is the superconducting
transition temperature for $E_{F}$ at the mobility edge $E_{c}$ given
(within modified mean-field approximation developed below in section \ref%
{Cooper instability}) by Eq.(\ref{Tc11}). Using of Eqs.(\ref{d-L-IPR},\ref%
{hard},\ref{Tc11}) one can eliminate the model parameters $E_{0}$ and $%
\lambda $ and express the local gap $\Delta _{1}$ via the observable
quantities $T_{c}^{0}$ and $\delta _{L}$: 
\begin{equation}
\Delta _{1}=\frac{3c_{1}}{2C^{\gamma }}\frac{\delta _{L}}{\left( {\delta _{L}%
}/{T_{c}^{0}}\right) ^{\gamma }}\approx 0.2\frac{\delta _{L}}{\left( {\delta
_{L}}/{T_{c}^{0}}\right) ^{\gamma }}  \label{D01}
\end{equation}%
where for the 3D Anderson transition $\gamma =0.57$ and $C\equiv
C(0.57)\approx 3.1$. The relation (\ref{D01}) shows that in the insulating
region $\delta _{L}\gg T_{c}^{0}$ due to the nonzero fractal exponent $%
\gamma $ the local gap value $\Delta _{1}$ grows with $\delta _{L}$ much
weaker than $\delta _{L}$ itself. In the limit $\gamma \rightarrow 0$ Eq.(%
\ref{D01}) transforms into the analog of the Matveev-Larkin relation (\ref%
{ML}), after a sub-leading term in the denominator of (\ref{D01}) is taken
into account via a substitution $\gamma \,(\delta _{L}/T_{c}^{0})^{\gamma
}\rightarrow (\delta _{L}/T_{c}^{0})^{\gamma }-1$.

Now we turn to the discussion of the intrinsic low-temperature conductivity
of the insulator with localized pairs. The binding of electrons into local
pairs diminishes the single-particle DoS and thus suppresses the
variable-range hopping conductivity. A classical example of such an effect
is the "Coulomb gap" due to Efros and Shklovsky: the soft gap $\nu (E)\sim
E^{2}$ in the average DoS leads to a transformation of the Mott law $\sigma
(T)\propto e^{-(T_{M}/T)^{1/4}}$ into the Efros-Shklovsky law $\sigma
(T)\propto e^{-(T_{ES}/T)^{1/2}}$. In our case the low-energy states are
exponentially rare, and their account leads to the logarithmic in
temperature corrections to the activation energy determined by the hard gap (%
\ref{hard}): 
\begin{equation}
\sigma (T)\propto \mathrm{exp}\left[ -\frac{T_{0}}{T\,\ln \left(
T_{0}/T\right) }\right] .  \label{sigma}
\end{equation}%
where 
\begin{equation}
T_{0}=\frac{c}{3}\Delta _{P}\approx 0.25\Delta _{P}  \label{T01}
\end{equation}%
Note that the nearly activated behavior given by Eqs.(\ref{sigma}) and (\ref%
{T01}) can hardly be distinguished from the purely activate one with the
activation gap $\Delta _{1}$ given by (\ref{hard}) or (\ref{D01}) in the
limited temperature range available in most experiments.

\begin{figure}[th]
\includegraphics[width=2in]{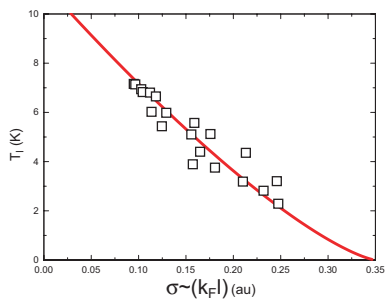}
\caption{Experimental values of the gap from Ref.\protect\cite{Shahar1992}, $%
T_{I}$ (boxes) and a fit to the equation (\protect\ref{ti}) with $\protect%
\nu _{\mathrm{av}}=1$, $d_{2}=1.3$.}
\label{DataFit}
\end{figure}

We thus associate the spectral gap $\Delta _{1}$ with the measured~\cite%
{Shahar1992,Kowal1994} activation energy $T_{I}$. The external parameter $%
(E_{c}-E_{F})$ representing the disorder strength in Eq.~(\ref{hard}) can be
replaced with an experimentally more accessible parameter $(1-\sigma /\sigma
_{c})\propto (E_{c}-E_{F})/E_{0}$. Here $\sigma $ is the high temperature
conductivity and $\sigma _{c}$ is the value of the conductivity where the
parity gap $\Delta _{P}$ first develops. We obtain 
\begin{equation}
T_{I}=A(1-\sigma /\sigma _{c})^{\nu d_{2}},\quad A\approx 0.5\lambda E_{0}
\label{ti}
\end{equation}%
where $A$ is conductivity-independent. This equation predicts a moderate
increase of $T_{I}$ with disorder strength in agreement with the
experimental data \cite{Shahar1992}, see Fig. \ref{DataFit}. 
The only fitting parameter used in Fig. \ref{DataFit} was the value of the
constant $A=0.5\lambda E_{0}\approx 10K$. Assuming the applicability of the
BCS theory for (less disordered) superconductive InO$_{x}$ samples~\cite%
{Shahar1992}, and using~\cite{ZviDan} the estimates $\omega _{D}\approx 500K$
for Debye frequency, we find $\lambda \approx 0.2$ and $E_{0}\approx 100K$.

Applicability of the scaling formulas to the above analysis is not obvious.
Indeed, the value of $1-\sigma /\sigma _{c}\approx 2/3$ for the most
insulating sample on the plot of Fig.\ref{DataFit}. We have demonstrated,
however, that the  correlation function $M(\omega)$ (see Fig~\ref{Cw2}) 
which is closely related to the inverse participation ratio, 
is approximately described by the
critical scaling even deeply inside the insulating phase. Then assuming that
Eq.(\ref{Lloc}) also holds approximately in a relatively wide interval of $%
1-\sigma /\sigma _{c}$ (although, perhaps, with some "average" exponent $\nu
_{\mathrm{av}}$ instead of $\nu $) we obtain an extrapolation of Eq.(\ref{ti}%
) for an entire interval of $\sigma /\sigma _{c}$ relevant for the
experiment. Note that the classical experiments on Si:P system have shown
that unless the special care was taken to shrink the interval of $1-\sigma
/\sigma _{c}$ to be much smaller than 1, the observed scaling of
localization radius was characterized by the exponent $\nu _{\mathrm{av}}<
1.5$, see Fig.~\ref{nuFig}. That is why it is not surprising that reasonable fit to the
experimental data shown in Fig. \ref{DataFit} corresponds to the choice $\nu
_{\mathrm{av}}=1$.

There are a few reasons why the estimated value of $E_{0}\approx 100K$
turned out to be low in comparison with the Fermi energy $E_{F}\sim 0.3eV$
\thinspace\ for amorphous InO$_{x}$ samples with the electron density in the
range of $10^{21}cm^{-3}$. This value implies that ratio $E_{0}/E_{F}\sim
0.03$ is one order of magnitude lower than our estimate (\ref{E00}) obtained
for the 3D Anderson model with Gaussian disorder (which, however, should not
be expected to be quantitatively correct for a-InO$_{x}$). First, this
estimate of $E_{0}$ uses BCS formula to related the value of the critical
temperature in the less disordered samples $T_{c}\approx 3K$ \cite%
{Shahar1992} \ to the value of the coupling constant in all materials of
these series. The BCS formula is however qualitatively wrong for the samples
close to mobility edge, which are in the regime discussed in the section \ref%
{Cooper instability}. In this regime the critical temperature is given by (%
\ref{Tc11}) which leads to a much lower value of the interaction constant $%
\lambda \approx 0.02$. The estimate for the upper energy cutoff $E_{0}$
becomes $1000K$ which is in agreement with (\ref{E00}). This estimate
neglects the effect of thermal fluctuations discussed in section \ref%
{Ginzburg parameters}, the effect of these fluctuations is to reduce $T_{c}$
so the actual value of the interaction constant corresponding to the sample
close to the mobility edge with $T_{c}\approx 3K$ might be slightly larger $%
\lambda \approx 0.03-0.05$ corresponding to $E_{0}\approx 400-600K$. These
values of $E_{0}/E_{F}\sim 0.1-0.3$ are roughly what one expects from the
analysis of the three dimensional Anderson model. Finally, we note that
appearance of low energy scales in the insulating samples of InO$_{x}$ was
conjectured in the early paper \cite{Zvi1986} for completely different
reasons. All these arguments demonstrate that the values of $\lambda $ and $%
E_{0}$ that we obtain from the fit of the experimental data are roughly what
one should expect in these samples.

The reasonable fit to the data was made possible by a small value of the
exponent in Eq.(\ref{ti}) which is substantially less than $\nu d\approx 3$
due to eigenfunction fractality. Thus, the data \cite{Shahar1992} provide
the indirect evidence for the eigenfunctions fractality in the insulating
samples of InO$_{x}$.

\section{Cooper instability near the mobility edge: the formalism. \label%
{Cooper instability}}

In this section we develop approximation schemes to treat the Cooper
instability and superconducting order formation in the regime when the
Anderson theorem~\cite{AG1959,Anderson1959} is not valid due to very strong
disorder. Namely, in section \ref{Modified mean-field} we give two versions
of the modified mean-field approximations, MFA, for determination of the
superconductive transition temperature $T_{c}$; in section \ref%
{Ginzburg-Landau functional} we derive an analog of the Ginzburg-Landau
functional which is necessary to estimate the role of fluctuations beyond
the modified MFA, and in section \ref{Virial expansion} we give an
alternative method of $T_{c}$ determination: the virial expansion. The
nature of approximations involved in the modified MFA and the virial
expansion methods are very different, thus reasonable agreement between the
results obtained by these methods indicates the validity of both of them. We
show that in a wide region near the mobility edge the dependence of the
superconducting critical temperature on the interaction constant can be
found \textit{analytically} with the accuracy up to a pre-factor of the
order of one despite of the presence of strong thermal and mesoscopic
fluctuations. In a more disordered sample the critical temperature can be
determined using the semi-analytical approach of virial expansion applied to
the \textit{pseudo-spin Hamiltonian}. The results of this section allow to
suggest the pseudo-spin Hamiltonian that provides a unified description of
superconductivity in homogeneously disordered (\textit{non-granular})
systems including the BCS regime and the region close to the
superconductor-insulator transition.

\subsection{Modified mean-field approximation. \label{Modified mean-field}}

The goal of this subsection is to develop two versions of a modified
mean-field approximations that can be used to determine the critical
temperature of a fractal superconductor. We start from the standard
Abrikosov-Gor'kov-Anderson~\cite{AG1959,Anderson1959} mean-field equation
for $T_{c}$ of a disordered superconductor: 
\begin{equation}
\Delta (\mathbf{r})=\int d^{d}\mathbf{r^{\prime }}\,K(\mathbf{r},\mathbf{%
r^{\prime }})\,\Delta (\mathbf{r^{\prime }}).  \label{MF}
\end{equation}%
The kernel $K(\mathbf{r},\mathbf{r^{\prime }})$ is defined by 
\begin{equation}
K(\mathbf{r},\mathbf{r^{\prime }})=\frac{g}{2}\,\sum_{ij}\eta _{ij}\,\psi
_{i}(\mathbf{r})\psi _{j}(\mathbf{r})\psi _{j}(\mathbf{r}^{\prime })\psi
_{i}(\mathbf{r}^{\prime }),  \label{K}
\end{equation}%
where $\psi _{j}(\mathbf{r})$ are exact single-electron wavefunctions and 
\begin{equation}
\eta _{ij}=\frac{\tanh (\xi _{i}/2T)+\tanh (\xi _{j}/2T)}{\xi _{i}+\xi _{j}}%
,\;\;\;\;\eta _{i}\equiv \eta _{ii}.  \label{eta}
\end{equation}

Within the standard mean-field approximation, one neglects the spatial
variations of $\Delta(\mathbf{r})$. Then integrating over $\mathbf{r}$ in
Eq.(\ref{MF}) and using the orthogonality and normalization of different
single-particle wavefunctions one can eliminate all $\psi_{i}(\mathbf{r})$
out of Eq.(\ref{MF}). Note that it is only possible if the wave functions
are real, $\psi_{i}^{*}(\mathbf{r})=\psi_{i}(\mathbf{r})$, i.e. the
time-reversal symmetry is preserved. This leads immediately to the
well-known equation for $T_c$ in terms of DoS function $\nu(\xi)$: 
\begin{equation}
1 = \frac{g}{2}\sum_i \eta_i \equiv \frac{\lambda}{2\nu_0}\int d\xi
\nu(\xi)\eta(\xi)  \label{MF0}
\end{equation}
where $\eta_i \equiv \eta_{ii} = \xi_i^{-1}\tanh(\xi_i/2T)$.

However, the approximation of a constant $\Delta $ cannot be used under
strong disorder conditions near the mobility edge. Physically it is due to
strong mesoscopic fluctuations of the local DoS function $\nu (\xi ,\mathbf{r%
})$, see \cite{BulaSad}. Below we propose a modification of the MFA scheme
which makes it possible to account for the major part of mesoscopic DoS
fluctuations.

To construct the modified MFA, we note that the appearance of the solution
to the Eq.(\ref{MF}) is equivalent to the divergence of the series $\mathrm{%
Tr}(1-K)^{-1}=\sum_{n=0}^{\infty }\mathrm{Tr}K^{n}$. By d'Alambert criterion
the latter is equivalent to the condition 
\begin{equation}
\lim_{n\rightarrow \infty }\frac{\mathrm{Tr}K^{n+1}}{\mathrm{Tr}K^{n}}=1.
\label{Dal}
\end{equation}%
Explicitly, the trace $\mathrm{Tr}K^{n}$ can be written as 
\begin{eqnarray}
\mathrm{Tr}K^{n} &=&(g/2)^{n}\sum_{i_{1}j_{1}...i_{n}j_{n}}\eta
_{i_{1}j_{1}}...\eta _{i_{n}j_{n}} \\
&&\times
M_{j_{n}i_{n}i_{1}j_{1}}%
\,M_{j_{1}i_{1}i_{2}j_{2}}...M_{j_{n-1}i_{n-1}i_{n}j_{n}},  \notag
\label{tr-tr}
\end{eqnarray}%
where $M_{ijkl}$ is defined in Eq.(\ref{Melements1}).

Now we make a crucial approximation: we neglect all the \textit{off
-diagonal matrix elements} in Eq.(\ref{Melements1}) with more than two
different indices. The \textit{diagonal} matrix elements $%
M_{ij}=M_{iijj}=M_{ijij}=M_{ijji}$ with only two different indices will be
retained. By so doing, we neglect thermal fluctuations of the
superconductive order parameter; we also neglect a part of mesoscopic
fluctuations, whereas the most important type of the them (i.e. fluctuations
of the local density of states) will be taken into account. Thus our
approach has a meaning of a modified mean-field approximation. The neglected
off-diagonal terms determine the strength of fluctuations on top of the
modified MFA solution. We derive the corresponding Ginzburg-Landau
functional and estimate the role of such fluctuations in subsection \ref%
{Ginzburg-Landau functional} below.

There are two ways to impose the \textquotedblleft diagonal
constraints\textquotedblright\ the for matrix elements $M_{ijkl}$: (i) to
set $i_{l}=j_{l}$ thus imposing $n$ constrains in Eq.(\ref{tr-tr}), or (ii)
to set $j_{l}=j_{l^{\prime }}$ or $j_{l}=i_{l^{\prime }}$ with $l\neq
l^{\prime }$. In this case one imposes $2n-2$ constrains, so that only 2
summations will remain. As any summation gives a macroscopically large
number of terms, one can neglect all terms corresponding to (ii) in the
thermodynamic limit $N\rightarrow \infty $. Thus, upon neglect of
off-diagonal matrix elements one finds eventually 
\begin{equation}
\mathrm{Tr}K^{n}\approx (g/2)^{n}\,\sum_{i_{1}...i_{n}}\eta _{i_{1}}\eta
_{i_{2}}...\eta _{i_{n}}\;M_{i_{n}i_{1}}M_{i_{1}i_{2}}...M_{i_{n-1}i_{n}},
\label{diag-app}
\end{equation}%
where $M_{ik}$ are defined in Eq.(\ref{Melements2}). It is easy to show now
that the condition given by Eq.(\ref{Dal}), with $\mathrm{Tr}K^{n}$ defined
by Eq.(\ref{diag-app}), is equivalent to the solvability condition for the
equation 
\begin{equation}
\Delta _{i}=\frac{g}{2}\sum_{k}\Delta _{k}\,\eta _{k}\,M_{ki},
\label{tilde-fin}
\end{equation}%
which is the basis of the modified MFA we will be using below.

The set of new order parameters $\Delta _{i}$, entering Eq.(\ref{tilde-fin}%
), represent the superconducting "ordering field" acting onto a pair of
electrons occupying the $i$-th orbital $\psi _{i}(\mathbf{r})$. The idea of
the approximation is that, instead of using a constant (in real space) order
parameter $\Delta (\mathbf{r})=\Delta $, we assume a \textit{smooth}
dependence of $\Delta _{i}$ on the single-electron energies $\xi _{i}$.
Under such an assumption, together with the replacement of $M_{ik}$ matrix
elements by their averages (according to Eq.(\ref{ij})), the Eq.(\ref%
{tilde-fin}) can be transformed into the integral equation 
\begin{equation}
\Delta (\xi )=\frac{\lambda }{2}\int d\zeta \eta (\zeta )M(\xi -\zeta
)\Delta (\zeta )  \label{MMFA}
\end{equation}%
Eq.(\ref{MMFA}) is a natural generalization of the BCS mean field equation
which follows from it at $M(\omega )=1$. However, allowing for the energy
dependence of $M(\omega )$ may lead to drastic consequences. Indeed, a
simple scaling analysis of this equation with "critical" $M(\omega
)=(E_{0}/\omega )^{\gamma }$ given by Eq.(\ref{ij}) leads to the power-law
dependence of the critical temperature on the attraction interaction
constant $\lambda $: 
\begin{equation}
T_{c}\propto E_{0}\lambda ^{1/\gamma }\,,  \label{Tc0}
\end{equation}%
where $\gamma $ is defined in Eqs.(\ref{Chalk},\ref{gamma}).

We encounter an unexpectedly strong increase of $T_{c}$ in the small $%
\lambda $ limit with respect to the usual result, $T_{c}\propto \mathrm{exp}%
(-1/\lambda )$ for a conventional BCS superconductor (with the same $\lambda 
$). The price to pay for this increase is a very strong inhomogeneity in the
real space of the local pairing amplitude: 
\begin{equation}
\tilde{\Delta}(\mathbf{r})=\frac{g}{2}\sum_{k}\Delta _{k}\eta _{k}\psi
_{k}^{2}(\mathbf{r}).  \label{Deltar}
\end{equation}%
The corresponding analysis will be presented in subsection \ref{Pairing
amplitude}.

To demonstrate the validity of the solution (\ref{Deltar}), one needs to
plug it into Eq.(\ref{MF}) and to use "diagonal approximation" 
\begin{equation*}
\int d^{d}\mathbf{r}\,\psi_{l}^{2}\psi_{j}(\mathbf{r})\psi_{i}(\mathbf{r}%
)=M_{il}\,\delta_{ij}
\end{equation*}
together with Eq.(\ref{tilde-fin}).

The transformation from Eq.(\ref{tilde-fin}) to Eq.(\ref{MMFA}) is not
exact: we replaced the fluctuating matrix elements $M_{ik}$ by their
averages according to Eq.(\ref{ij}). Thus Eq.(\ref{MMFA}) contains an
additional (with respect to Eq.(\ref{tilde-fin}) mean-field-type
approximation. One can eliminate this additional approximation. For this
purpose, let us define the new matrix $\hat{Q}$ and vectors $\phi _{i}$: 
\begin{equation}
Q_{ik}=\frac{g}{2}\sqrt{\eta _{i}\eta _{k}}M_{ik}\qquad \phi _{i}=\Delta _{i}%
\sqrt{\eta _{i}}  \label{Qmat}
\end{equation}%
Then the solvability condition for Eq.(\ref{tilde-fin}) transforms into the
condition that the largest eigenvalue $k_{\mathrm{max}}$ of the symmetric
matrix $\hat{Q}$ becomes equal to unity. This condition is equivalent to the
instability onset with respect to formation of a superconducting order
parameter $\Delta _{i}$, as will be seen below in subsection \ref{Transition
temperature: coefficient a(T)}.

In the thermodynamic limit the condition $k_{\mathrm{max}}=1$ implies global
superconductivity only if the corresponding eigenvector of the matrix $\hat{Q%
}$ is \textit{extended}. Otherwise, $k_{\mathrm{max}}=1$ means formation of
local "islands" of new phase with uncorrelated local order parameters in
these islands, it does not  immediately result in any global order parameter
(for the discussion of relevant examples see ~\cite%
{IoffeFeigelman1985,DFI1990}). Eigenfunctions of the $\hat{Q}$ matrix may be
localized, if the effective "coordination number" (the bandwidth) $Z$ of
this matrix stays finite in thermodynamic limit. General estimate for the
coordination number is $Z\sim \nu _{0}T_{c}L_{loc}^{d}$ because the relevant
energy window populated by "active" single-particle states ( taking part in
formation of the many-body superconductive state) is of the order of $T_{c}$%
, and each eigenstate $|\,i>$ is coupled by the the matrix elements $M_{ij}$
to all neighbors $|\,j>$ in the localization volume $L_{loc}^{d}$. This
argument shows that  as long as the single electron eigenstates are
delocalized, $E_{F}<E_{c}$ and \textit{within diagonal approximation}, the
effective coordination number $Z=\infty $ and no localization of the $\hat{Q}
$ eigenstates is possible. However, this conclusion becomes invalid when the
off-diagonal matrix elements $M_{ijkl}$ are taken into account: the lowest
eigenstates of the full kernel $K(r,r^{\prime })$ get localized. The most
important physical consequence of this localization is that, as the
temperature is decreased, the superconductivity first appears in small, well
separated regions, similarly to the situation realized in superconductors
with inhomogeneous $T_{c}$ \cite{IoffeLarkin1981}. In the regime of
delocalized single electron states a further decrease of temperature results
in a more homogeneous superconductivity. As the Fermi energy is increased
past the mobility edge, single electron states get localized and the issue
of the $\hat{Q}$-eigenfunction localization and resulting inhomogeneity
become relevant even within the "diagonal approximation", we discuss this
regime in section \ref{Superconductivity with a pseudogap}.

\begin{figure}[tbp]
\includegraphics[width=8cm]{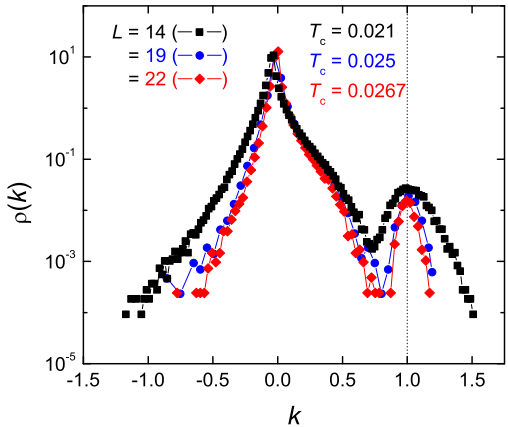}
\caption{(Color online) Density of states $\protect\rho(k)$ of the matrix $%
\hat{Q}$, with peak near $k=1$, for $L=14$ (black squares), $L=19$ (blue
dots) and $L=22$ (red diamonds). }
\label{FigDosK}
\end{figure}

Here we present, as an example, the numerically obtained spectrum of the $%
\hat{Q}$ matrix for a finite system at the mobility edge. Fig.~\ref{FigDosK}
shows the averaged (over 2000 realizations) density of states for the $\hat{Q%
}$ matrix generated for the 3D Anderson model with Gaussian on-site disorder
of the strength $W=4$, at the energy $E=5.5$ corresponding to the mobility
edge. Three system sizes, $L=14,19,22$ were analyzed. In a finite system the
peaks of the finite width are seen in the density of states $\rho (k)$ near $%
k=1$. One sees that with the system size increase, the peaks become more
narrow, so it is natural to assume that they evolve into a $\delta $%
-function peak in the $L\rightarrow \infty $ limit. The corresponding
temperature (chosen so that to pin the DoS peak position to $k=1$) is thus
associated with $T_{c}$.

In subsection \ref{Comparison of Tc values} we compare results for the
critical temperature $T_{c}$ obtained by the three methods: modified
analytical MFA equation (\ref{MMFA}), numerical generation of the spectrum
of the temperature-dependent matrix $\hat{Q}$, with the temperature $T=T_{c}$
adjusted so to locate the peak in $\rho (k)$ at $k=1$, and virial expansion
method, to be described in subsection \ref{Virial expansion}..

Our modified MFA method is similar in spirit to the usual BCS mean-field
approximation: it neglects both thermal and mesoscopic fluctuations of the
pairing field $\Delta (\mathbf{r})$ with respect to its "background
configuration". The essence of the modification we used is that our
background configuration is \textit{not uniform in real space}. Instead it
is given by the function $\tilde{\Delta}(\mathbf{r})$ defined in Eq.(\ref%
{Deltar}). In order to estimate the strength of fluctuations we neglected,
we will develop, in the next subsection \ref{Ginzburg-Landau functional}, a
Ginzburg-Landau description for the order parameter field configurations $%
\Delta (\mathbf{r})$ which are \textit{close to the background field} $%
\tilde{\Delta}(\mathbf{r})$.

\subsection{Ginzburg - Landau functional. \label{Ginzburg-Landau functional}}

Here we derive the free energy functional $F[\Psi (\mathbf{r})]$ defined in
terms of a \textit{smooth} envelope function 
\begin{equation}
\Psi (\mathbf{r})=\Delta (\mathbf{r})/\tilde{\Delta}(\mathbf{r})  \label{Psi}
\end{equation}%
and valid near the transition temperature $T_{c}$ for $\Psi (\mathbf{r})$
field which smoothly varies in space.

We start by writing the action in terms of the Grassmann fields $\varphi
_{n} $ 
\begin{eqnarray}
S &=&\sum_{n}\varphi _{n}^{\ast }(-i\omega _{n}+H_{1})\varphi
_{n}+\sum_{n}\varphi _{-n}^{\ast }(i\omega _{n}+H_{1})\varphi _{-n}
\label{ac} \\
&&-g\sum_{n_{1},n_{2},n_{3}}\varphi _{n_{1}+n_{3}}^{\ast }\varphi
_{n_{1}-n_{3}}^{\ast }\varphi _{-n_{1}+n_{2}}\varphi _{-n_{1}-n_{2}}.  \notag
\end{eqnarray}%
where $H_{1}$ is the single-particle part of the Hamiltonian, given e.g. by
Eq.(\ref{cont-Ham}) or Eq.(\ref{AM}); $g$ is the interaction constant and $n$
and $n_{1,2,3}$ are the Matsubara frequency summation indices. We decouple
the interaction term via the Hubbard-Stratonovich transformation involving
the auxiliary field $\Delta (\mathbf{r};\Omega _{n})$. Below we focus on the
static $n=0$ component of this field. In doing so we neglect quantum
fluctuations of the order parameter but take into account the thermal and
mesoscopic fluctuations.

We expand the Grassmann fields over the eigenstates $\psi _{i}(\mathbf{r})$
of the single-electron problem 
\begin{equation}
\varphi _{n}(\mathbf{r})=\sum_{j}\chi _{j}^{(n)}\psi _{j}(\mathbf{r})
\label{varphi}
\end{equation}%
and represent the thermodynamic potential $S[\Delta ,\Phi ]$ as 
\begin{equation}
S=\frac{1}{g}\sum_{r}|\Delta (\mathbf{r})|^{2}+\sum_{n}\sum_{i,j}\bar{\Phi}%
_{i}^{(n)}\,\left( 
\begin{array}{cc}
(-i\omega _{n}+\xi _{i})\delta _{ij} & \Delta _{ij} \\ 
-\Delta _{ij}^{\ast } & (-i\omega _{n}-\xi _{i})\delta _{ij}%
\end{array}%
\right) \Phi _{j}^{(n)},  \label{HS2}
\end{equation}%
where 
\begin{equation*}
\Phi _{i}^{(n)}=\left( 
\begin{array}{c}
\chi _{i}^{(n)} \\ 
\chi _{i}^{(-n)\ast }%
\end{array}%
\right) ,\;\;\;\;\;\bar{\Phi}_{i}^{(n)}=(\chi _{i}^{(n)\ast }\;\;\chi
_{i}^{(-n)}).
\end{equation*}%
and 
\begin{equation}
\Delta _{ij}=\sum_{\mathbf{r}}\Delta (\mathbf{r})\,\psi _{i}(\mathbf{r})\psi
_{j}(\mathbf{r})\equiv \sum_{k}D_{k}R_{kij}.  \label{Dab}
\end{equation}%
Here we defined 
\begin{equation}
R_{kij}=\int d\mathbf{r}\psi _{k}(\mathbf{r})\psi _{i}(\mathbf{r})\psi _{j}(%
\mathbf{r})\quad D_{k}=\int d\mathbf{r}\Delta (\mathbf{r})\psi _{k}(\mathbf{r%
})  \label{newnot}
\end{equation}

Note that the way we decoupled the interaction term is specific to the
superconductive correlations. A generic Hubbard-Stratonovich field should
contain also a component $V(\mathbf{r};\Omega _{n})$ coupled to the
combination $(\chi _{i}^{(n)})^{\ast }\chi _{i}^{(n)}$ which corresponds to
interaction in the particle-hole channels. We neglect such interactions in
this subsection and will study their effect in subsection \ref{Transition
temperature} in connection with the $S_{i}^{z}S_{j}^{z}$ terms in the 
\textit{pseudo-spin} Hamiltonian.

The matrix variables $\Delta _{ij}$, which appeared naturally in the second
term of the action \textit{are not mutually independent}. Indeed, the number
of independent components of the (discretized) field $\Delta (\mathbf{r})$
scales with system volume as $\propto \mathcal{V}$, whereas the number of
matrix elements $\Delta _{ij}$ scales as $\mathcal{V}^{2}$. The matrix
elements $\Delta _{ij}$ are mutually constrained due to the ortho-normality
conditions $\int d\mathbf{r}\psi _{i}(\mathbf{r})\psi _{j}(\mathbf{r}%
)=\delta _{ij}$.

Now one can complete the Hubbard-Stratonovich transformation by performing
the Gaussian integration over $\Phi _{j}^{(n)}$ to obtain the $\mathrm{Tr}%
\ln $ of the corresponding matrix in Eq.(\ref{HS2}). In addition, one can
represent the first term in Eq.(\ref{HS2}) in terms of the coefficients $%
D_{i}$ of expansion of $\Delta (\mathbf{r})$ over the full set of single
particle wave functions $\psi _{i}(\mathbf{r})$. Expanding the $\mathrm{Tr}%
\ln $ over $D_{i}$ up to fourth order, we obtain free energy functional in
the form 
\begin{equation}
F[D_{i}]=\frac{1}{g}\sum_{ij}D_{i}^{\ast }\mathcal{K}_{ij}D_{j}+\frac{1}{4}%
\,\sum_{ijkl}D_{i}D_{j}^{\ast }\,\mathcal{J}_{ijkl}\,D_{k}D_{l}^{\ast },
\label{F1}
\end{equation}%
where 
\begin{equation}
\mathcal{K}_{ij}=\delta _{ij}-\frac{g}{2}\sum_{\mu \nu }R_{i\mu \nu }\eta
_{\mu \nu }R_{j\mu \nu }.  \label{calK}
\end{equation}%
In Eqs.(\ref{F1}),(\ref{calK}) we denote 
\begin{equation}
\mathcal{J}_{ijkl}=\sum_{\nu _{1},\nu _{2},\nu _{3}\nu _{4}}R_{i\nu _{4}\nu
_{1}}R_{j\nu _{1}\nu _{2}}^{\ast }R_{k\nu _{2}\nu _{3}}R_{l\nu _{3}\nu
_{4}}^{\ast }\,\;\zeta _{\nu _{1}\nu _{2},\nu _{3}\nu _{4}}\,  \label{J}
\end{equation}%
where the function $\zeta (...)$ is defined by 
\begin{eqnarray}
&&\zeta _{\nu _{1}\nu _{2},\nu _{3}\nu _{4}}=\zeta _{\nu _{2}\nu _{1},\nu
_{4}\nu _{3}}=\zeta _{\nu _{3}\nu _{4},\nu _{1}\nu _{2}}=  \label{zeta} \\
&&\frac{1}{(\xi _{\nu _{1}}-\xi _{\nu _{3}})(\xi _{\nu _{4}}-\xi _{\nu _{2}})%
}\,\left[ \eta _{\nu _{1}\nu _{2}}+\eta _{\nu _{3}\nu _{4}}-\eta _{\nu
_{3}\nu _{2}}-\eta _{\nu _{1}\nu _{4}}\right] .  \notag
\end{eqnarray}

Below we will use Eqs.(\ref{F1})-(\ref{zeta}) to derive the effective
Ginzburg-Landau functional $F[\Psi (\mathbf{r})]$ in the form 
\begin{equation}
F_{GL}[\Psi (\mathbf{r})]=\nu _{0}T_{c}^{2}\int d\mathbf{r}\left( a(\mathbf{r%
})\Psi ^{2}(\mathbf{r})+\frac{b}{2}\Psi ^{4}(\mathbf{r})+C|\nabla \Psi (%
\mathbf{r})|^{2}\right)  \label{FGL}
\end{equation}%
The factor $\nu _{0}T_{c}^{2}$ in (\ref{FGL}) is introduced so that to keep $%
\Psi $ and $a(T)$ dimensionless. The physical properties of the
superconductor described by functional (\ref{FGL}) are controlled by four
parameters: $a=\langle a(\mathbf{r})\rangle $, $b$, $C$, and by the strength 
$W$ of fluctuations $\delta a(\mathbf{r})=a(\mathbf{r})-a$ defined as 
\begin{eqnarray}
W &=&\int d\mathbf{r}\,\langle \delta a(\mathbf{r})\delta a(\mathbf{%
r^{\prime }})\rangle =  \label{W} \\
&=&\mathcal{V}^{-1}\int d\mathbf{r}_{1}d\mathbf{r}_{2}d\mathbf{r}_{3}d%
\mathbf{r}_{4}\langle \delta K(\mathbf{r}_{1},\mathbf{r_{2}})\delta K(%
\mathbf{r}_{3},\mathbf{r}_{4})\rangle .  \notag
\end{eqnarray}%
where $\delta K(\mathbf{r,r^{\prime }})=K(\mathbf{r,r^{\prime }})-\langle K(%
\mathbf{r,r^{\prime }})\rangle $. We are able to describe the spatial
fluctuation by one parameter, $W$, because $\delta a(\mathbf{r})$
correlations are short-ranged compared to the typical scale of $\Psi (%
\mathbf{r})$ variations.

\subsubsection{Transition temperature: coefficient $a(T)$. \label{Transition
temperature: coefficient a(T)}}

We begin by evaluating the coefficient $a$ that vanishes at the transition
point, $a(T)=\tilde{a}(T-T_{c})$. For this calculation it is sufficient to
use a constant $\Psi (\mathbf{r})=\Psi $. Now let us transform the first
term of Eq.(\ref{F1}), using Eqs.(\ref{calK},\ref{Deltar},\ref{Dab},\ref%
{newnot}). We assume here that $\Delta _{j}$ obey the matrix equation (\ref%
{tilde-fin}), i.e. the local order parameter coincides with $\tilde{\Delta}(%
\mathbf{r})$ defined in Eq.(\ref{Deltar}). The result reads 
\begin{equation}
\frac{F_{2}}{\Psi ^{2}}=\frac{g}{4}\sum_{ij}\Delta _{i}\Delta _{j}\eta
_{i}\eta _{j}M_{ij}-\frac{g^{2}}{8}\sum_{ijk}\Delta _{i}\Delta _{j}\eta
_{i}\eta _{j}\eta _{kl}M_{iikl}M_{kljj}  \label{F21}
\end{equation}%
where we used the "fusion rule" following from the completeness of the set
of $\psi _{j}(\mathbf{r})$: 
\begin{equation}
\sum_{i}R_{kli}R_{pqi}=M_{klpq}.  \label{fus}
\end{equation}%
and an equivalent form of Eq.(\ref{tilde-fin}) written in terms of $D_{i}$
and $\Delta _{jk}$ variables: 
\begin{equation}
D_{i}=\frac{g}{2}\sum_{kl}\eta _{kl}\Delta _{kl}\,R_{ikl}.  \label{D-Dij}
\end{equation}

To determine the coefficient $a$ we neglect the off-diagonal terms $M_{iikl}$
with $k\neq l$ in Eq.(\ref{F21}) and thus reduce it to 
\begin{equation}
F_{2}^{(0)}=\frac{1}{2}\Psi ^{2}\sum_{ij}\left[ \left( \hat{Q}\right)
_{ij}-\left( \hat{Q^{2}}\right) _{ij}\right] \phi _{i}\phi _{j}  \label{F22}
\end{equation}%
where $\hat{Q}$ and $\phi _{i}$ are defined in Eq.(\ref{Qmat}). In the
continuum limit one writes $\xi _{i}\rightarrow \xi $, $M_{ij}\rightarrow 
\frac{1}{\mathcal{V}}M(\xi -\zeta )$, $\sum_{i}\rightarrow \mathcal{V}\nu
_{0}\int d\xi $ and thus the bilinear form (\ref{F22}) transforms into 
\begin{equation}
F_{2}^{(0)}=\frac{\nu _{0}\mathcal{V}}{2}\Psi ^{2}\int \int d\xi d\zeta
L(\xi ,\zeta )\phi (\xi )\phi (\zeta )\,  \label{F23}
\end{equation}%
where 
\begin{equation}
L(\xi ,\zeta )=Q(\xi ,\zeta )-\int d\xi _{1}Q(\xi ,\xi _{1})Q(\xi _{1},\zeta
)\,,  \label{L2}
\end{equation}%
with the kernel 
\begin{equation}
{Q}(\xi ,\zeta )=\frac{\lambda }{2}\sqrt{\eta (\xi )\eta (\zeta )}M(\xi
-\zeta )  \label{kernel}
\end{equation}%
The function $\phi (\xi )$ is, by construction, an eigenfunction of the
integral equation with the kernel $Q(\xi ,\zeta )$; its eigenvalue $k(T)$
approaches 1 at $T=T_{c}$. At this stage we need to specify the
normalization condition for the function $\phi (\xi )$: 
\begin{equation}
\int d\xi \Delta ^{2}(\xi )\eta (\xi )\equiv \int d\xi \phi ^{2}(\xi )=\frac{%
2T_{c}^{2}}{\gamma }  \label{norm1}
\end{equation}%
The form of the normalization condition Eq.(\ref{norm1}) was chosen to
enable a smooth crossover to the conventional (non-fractal) BCS case $\gamma
\rightarrow 0$, when $\phi _{\gamma =0}=T\sqrt{\eta (\xi )}$. To get the
equations valid in the crossover regime we need to introduce the high-energy
cutoff $\Omega _{D}$ so that all integrals over $d\xi $ go over the range $%
|\xi |<\Omega _{D}$. Then a straightforward modification of the
normalization condition 
\begin{equation}
\int_{-\Omega _{D}}^{\Omega _{D}}d\xi \Delta ^{2}(\xi )\eta (\xi )=\frac{%
2T_{c}^{2}\ln \frac{\Omega _{D}}{T_{c}}}{\gamma \ln \frac{\Omega _{D}}{T_{c}}%
+1}  \label{norm2}
\end{equation}%
(that reduces to (\ref{norm1}) when $\gamma \gg 1/\ln \frac{\Omega _{D}}{%
T_{c}}=\lambda $) allows one to recover the BCS limit at at a fixed $\Omega
_{D}$ and $\gamma \rightarrow 0$. We are mainly interested in the case of
intermediate to strong fractality, $\gamma \sim 0.6$, and weak interaction $%
\lambda \ll 1$, so we shall use the simplified condition (\ref{norm1})
appropriate in this limit.

Making use of Eq.(\ref{norm1}) we reduce the expression (\ref{F23}) to 
\begin{equation}
F_{2}^{(0)}=\nu _{0}\mathcal{V}T^{2}\Psi ^{2}\gamma ^{-1}[k(T)-k^{2}(T)]
\label{F24}
\end{equation}%
Since $1-k(T)\ll 1$ at $T\approx T_{c}$, it is sufficient to evaluate the
derivative $dk/dT|_{k=1}$. We find it by rewriting the kernel (\ref{kernel})
in a dimensionless form: 
\begin{equation*}
\bar{Q}(x,y)=\frac{\lambda _{T}}{2|x-y|^{\gamma }}\sqrt{\frac{\tanh (x)\tanh
(y)}{xy}}\,,
\end{equation*}%
where $\lambda _{T}=\lambda (E_{0}/2T)^{\gamma }$. Clearly, $d\ln k(T)/d\ln
(T)=-\gamma $, and, finally, the coefficient $a(T)$ is expressed only in
terms of the ratio of $T/T_{c}$: 
\begin{equation}
a(T)=\frac{T-T_{c}}{T_{c}}  \label{alphaT}
\end{equation}%
Note that the order parameter $\Psi $ introduced in Eq.(\ref{Psi}) is
dimensionless, whereas $\Delta $ has a dimension of energy.

\subsubsection{Quartic term: coefficient $b$}

Now we turn to the computation of the coefficient $b$ in the expansion (\ref%
{FGL}). Our starting points are Eqs.(\ref{F1}) and (\ref{J},\ref{zeta}). As
in the previous part we may use here $\Psi (\mathbf{r})=\mathrm{const}$,
thus $\Delta (\mathbf{r})$ is proportional to $\tilde{\Delta}(\mathbf{r})$
defined in Eq.(\ref{Deltar}). Therefore using Eqs.(\ref{Deltar}), (\ref%
{newnot}) we may substitute $D_{i}=\frac{g}{2}\sum_{a}\Delta _{a}\eta
_{a}R_{iaa}$ in the fourth order in $D_{i}$ term of Eq.(\ref{F1}) reducing
this term to the following form: 
\begin{eqnarray}
F_{4} &=&\frac{1}{4}\left( \frac{g}{2}\right)
^{4}\sum_{ijkl}\sum_{abcd}\sum_{\nu _{1}..\nu _{4}}\Delta _{a}\Delta
_{b}\Delta _{c}\Delta _{d}\,\eta _{a}\eta _{b}\eta _{c}\eta _{d}\,
\label{F40} \\
&&\times R_{i\nu _{4}\nu _{1}}R_{j\nu _{1}\nu _{2}}R_{k\nu _{2}\nu
_{3}}R_{l\nu _{3}\nu _{4}}\,R_{iaa}R_{jbb}R_{kcc}R_{ldd}\,\zeta _{\nu
_{1}\nu _{2},\nu _{3}\nu _{4}}  \notag
\end{eqnarray}%
Now we do summations over $i,j,k,l$ using the fusion rule (\ref{fus}), and
then neglect the off-diagonal matrix elements, i.e. we set everywhere $%
M_{\nu _{1}\nu _{2}aa}\rightarrow \delta _{\nu _{1}\nu _{2}}M_{\nu _{1}a}$.
Summations over $a,b,c,d$ can be done now with the use of modified MFA
equations (\ref{tilde-fin}). Finally, we proceed from the last remaining
summation over $\nu _{1}$ to integration, and find 
\begin{equation}
F_{4}=\frac{\nu _{0}\mathcal{V}}{32T^{2}}\Psi ^{4}\int \frac{d\xi }{2T}\Phi
\left( \frac{\xi }{2T}\right) \Delta ^{4}(\xi )  \label{F4}
\end{equation}%
where 
\begin{equation*}
\Phi (x)\equiv 16T^{3}\lim_{\forall \xi _{\alpha }\rightarrow 2Tx}\zeta
_{ij,kl}=\frac{1}{x^{2}}\left( \frac{\tanh (x)}{x}-\frac{1}{\cosh ^{2}(x)}%
\right)
\end{equation*}%
with the function $\zeta _{ijkl}$ defined in Eq.(\ref{zeta}). In the limit $%
\gamma \rightarrow 0$ we have $\Delta (\xi )=\mathrm{const}$ and the
integral in (\ref{F4}) reduces to 
\begin{equation*}
\frac{1}{32}\int dx\Phi (x)=\frac{7\zeta (3)}{16\pi ^{2}}=\frac{1}{2}b_{%
\mathrm{BCS}}\,,
\end{equation*}%
leading to the standard BCS result for the $b$ coefficient in Eq.(\ref{FGL}%
). In general case we find 
\begin{equation}
b=\frac{1}{4}\int dx\,\Phi (x)\,x^{2}\coth ^{2}(x)\,\bar{\phi}^{4}(x)
\label{beta}
\end{equation}%
where $\bar{\phi}(x)=T^{-1/2}\phi (2Tx)$ and $\phi (\xi )$ is the $k=1$
eigenfunction of the kernel (\ref{kernel}) subject to the normalization
condition (\ref{norm1}).

\subsubsection{Gradient term: coefficient $C$.\label{Gradient term}}

We now turn to calculation of the gradient term of the functional (\ref{FGL}%
). Note that previously employed "diagonal approximation" for the matrix
elements $M_{ijkl}$ is not sufficient for that purpose. In order to find the
coefficient $C$ we need to take into account the off-diagonal matrix
elements $M_{ijkl}$ which contain pairs of different levels with nearby
energies: $|\xi _{i}-\xi _{j}|\sim |\xi _{k}-\xi _{l}|\sim \omega (q)$, with 
$\omega (q\rightarrow 0)\rightarrow 0$. The reason for that is similar to
the one which applies to the standard BCS Hamiltonian with the reduced
interaction term $g\,\sum_{\mathbf{p,p^{\prime }}}c_{\uparrow ,\mathbf{p}%
}\,c_{\downarrow ,\mathbf{-p}}\,c_{\downarrow ,\mathbf{p^{\prime }}%
}^{\dagger }\,c_{\uparrow ,\mathbf{-p^{\prime }}}^{\dagger }$. One has to
allow for the non-zero momentum of a pair and thus go beyond the "diagonal
approximation" in $\mathbf{p}$ and $\mathbf{p^{\prime }}$ in order to be
able to compute the \textit{phase rigidity} $C$ which is related with the
supercurrent.

We first illustrate this statement using the standard theory of disordered
superconductors as an example, see e.g.~\cite{DeGennesBook}. In that case,
quadratic term of the free energy expansion (\ref{F1}) can be represented in
the coordinate space: 
\begin{equation}
F[\Delta (\mathbf{r})]=\int d\mathbf{r}d\mathbf{r^{\prime }}\left[ \frac{1}{g%
}\delta (\mathbf{r}-\mathbf{r^{\prime }})-K(\mathbf{r},\mathbf{r^{\prime }})%
\right] \Delta (\mathbf{r})\Delta (\mathbf{r^{\prime }})  \label{FGLs}
\end{equation}%
where 
\begin{equation}
K(\mathbf{r},\mathbf{r^{\prime }})=\nu _{0}T\sum_{\omega _{n}}\sum_{ij}\frac{%
\psi _{i}(\mathbf{r})\psi _{j}(\mathbf{r})\psi _{i}^{\ast }(\mathbf{%
r^{\prime }})\psi _{j}^{\ast }(\mathbf{r^{\prime }})}{(\xi _{i}-i\omega
_{n})(\xi _{j}+i\omega _{n})}  \label{Kold}
\end{equation}%
Here $\omega _{n}=\pi T(2n+1)$ and (\ref{Kold}) is the equivalent form of
Eqs.(\ref{K},\ref{eta}). As the standard procedure goes, one averages the
kernel (\ref{Kold}) over disorder in the semiclassical approximation and
obtains the Fourier-transformed kernel 
\begin{equation}
\overline{K}(q)=2\pi \nu _{0}T\sum_{\omega _{n}}\frac{1}{Dq^{2}+2|\omega
_{n}|}  \label{Kq}
\end{equation}%
where $D$ is the diffusion coefficient. Finally one obtains for the
coefficient $C$ in Eq.(\ref{FGL}) the well known result%
\begin{equation}
C^{(0)}=-\frac{1}{\nu _{0}}\left. \frac{dK(q)}{dq^{2}}\right\vert
_{q\rightarrow 0}=\frac{\pi }{8}\frac{D}{T_{c}}  \label{gamma0}
\end{equation}%
In order to generalize this derivation to the case of strong disorder in is
convenient to introduce two particle spectral function 
\begin{eqnarray}
g(q,\Omega ) &=&\frac{1}{\mathcal{V}\nu _{0}}\sum_{ij}\int d\mathbf{r}d%
\mathbf{r^{\prime }}\,e^{i\mathbf{q}(\mathbf{r}-\mathbf{r^{\prime }}%
)}\,\delta (\xi _{i}-\xi _{j}-\Omega )  \notag  \label{gq0} \\
&\times &\langle \psi _{i}(\mathbf{r})\psi _{j}(\mathbf{r})\psi _{i}^{\ast }(%
\mathbf{r^{\prime }})\psi _{j}^{\ast }(\mathbf{r^{\prime }})\rangle
\,\,\delta (E_{F}-\xi _{i}).
\end{eqnarray}%
The average kernel in Eq.(\ref{Kold}) can be expressed through this two
particle function: 
\begin{equation}
\overline{K}(q)=\nu _{0}T\sum_{\omega _{n}}\int \frac{d\xi d\xi ^{\prime
}g(q,\xi -\xi ^{\prime };E)}{(\xi -i\omega _{n})(\xi ^{\prime }+i\omega _{n})%
}  \label{Kq2}
\end{equation}%
In the diffusive limit the spectral function is 
\begin{equation}
g(q,\Omega )=\frac{1}{\pi }\frac{Dq^{2}}{\Omega ^{2}+(Dq^{2})^{2}}.
\label{gq}
\end{equation}%
and its substitution into Eq.(\ref{Kq2}) gives back Eq.(\ref{Kq}).

This derivation of the standard result demonstrates that the small $q$%
-dependence of the averaged kernel $\overline{K}(q)$ comes from matrix
elements of the operator $e^{iq(\mathbf{r}-\mathbf{r^{\prime }})}$ between
eigenstates $\psi _{i}(\mathbf{r}),\psi _{j}(\mathbf{r})$ with nearby
energies $|\xi _{i}-\xi _{j}|\sim Dq^{2}$. Notice that diagonal
approximation used in previous sections corresponds to $i=j$ in Eq.(\ref%
{Kold}). In this approximation one neglects the off-diagonal terms $M_{ijkl}$
completely and gets $g(q,\Omega )=\delta (\Omega )f(q)$, where 
\begin{equation*}
f(q)=\int d\mathbf{r}d\mathbf{r^{\prime }}\,\langle \psi _{i}^{2}(\mathbf{r}%
)\psi _{i}^{2}(\mathbf{r^{\prime }})\,e^{i\mathbf{q}(\mathbf{r}-\mathbf{%
r^{\prime }})}\rangle
\end{equation*}%
For delocalized wavefunctions which occupy all the available volume one can
write 
\begin{equation*}
\mathcal{V}^{2}\,\langle \psi _{i}^{2}(\mathbf{r})\psi _{i}^{2}(\mathbf{%
r^{\prime }})\rangle \approx \mathcal{V}^{2}\,\langle \psi _{i}^{2}(\mathbf{r%
})\rangle \,\langle \psi _{i}^{2}(\mathbf{r^{\prime }})\rangle =1+O(1/g),
\end{equation*}%
where $g\gg 1$ is the dimensionless conductance Then $f(q=0)=1+O(1/g)$ and $%
f(q\neq 0)=O(1/g)$. The jump in $f(q)$ at $q=0$ implies that in the diagonal
approximation the coefficient $C$ is infinite in a metal which is obviously
wrong. However, the value of $K$ at $q=0$ can be determined correctly in the
diagonal approximation because at $q=0$ the spectral function $g(q,\Omega
)\rightarrow \delta (\Omega )$ both in the diagonal and in the diffusive
approximation Eq.(\ref{gq}).

The correct derivation of the coefficient $C~$\ in case of a strong disorder
has to take into account a non-uniformity of the "background configuration"
of the superconducting order parameter given by Eq.(\ref{Deltar}). We follow
the same logics as in the derivation of the $a$ and $b$ coefficients in
previous sections. Note that the $q$-dependence comes from the second term
of the matrix (\ref{calK}) which leads to the second sum in Eq.(\ref{F21}).
We focus on this term and repeat the steps of the derivation which led to
Eq.(\ref{F21}) using now a weakly modulated $\Psi (\mathbf{r})=\Psi _{q}e^{i%
\mathbf{qr}}$ with small $q$. The second term of (\ref{F21}) becomes (in the
continuous form, after replacing $\sum_{j}\rightarrow \mathcal{V}\nu
_{0}d\xi $): 
\begin{eqnarray}
\int d\mathbf{q}\,F_{2}^{(2)}(q) &=&-\int d\mathbf{q}\,|\Psi _{q}|^{2}\,%
\frac{g^{2}}{8}\int d\xi d\omega d\omega _{1}d\omega _{2}\Delta (\xi
_{+}+\omega _{1})\,\Delta (\xi _{-}+\omega _{2})\;  \label{Fq1} \\
&&\times \eta (\xi _{+},\xi _{-})\,\eta (\xi _{+}+\omega _{1})\,\eta (\xi
_{-}+\omega _{2})\;M_{4}(q;\omega ,\omega _{1},\omega _{2})  \notag
\end{eqnarray}%
where $\xi _{\pm }=\xi \pm \omega /2$; the function $\Delta (\xi )=\phi (\xi
)/\sqrt{\eta (\xi )}$ obeys Eq.(\ref{MMFA}) and is normalized according to
Eq.(\ref{norm1}), and the function $M_{4}(q;\omega ,\omega _{1},\omega _{2})$
is defined by 
\begin{equation*}
M_{4}(q;\omega ,\omega _{1},\omega _{2})=\mathcal{V}^{-1}\left\langle
\sum_{ijkl}\delta (\xi _{ij}-\omega )\,\delta (\xi _{ki}-\omega
_{1})\,\delta (\xi _{lj}-\omega _{2})\,\delta (E-\xi
_{i})\,I_{ijkl}(q)\right\rangle ,
\end{equation*}%
where $\xi _{ij}=\xi _{i}-\xi _{j}$ 
\begin{equation}
I_{ijkl}(q)=\int d\mathbf{r}d\mathbf{r^{\prime }}\,e^{i\mathbf{q(r-r^{\prime
})}}\,\psi _{i}(\mathbf{r})\psi _{j}(\mathbf{r})\psi _{i}(\mathbf{r^{\prime }%
})\psi _{j}(\mathbf{r^{\prime }})\psi _{k}^{2}(\mathbf{r})\psi _{l}^{2}(%
\mathbf{r^{\prime }})  \label{Iijkl}
\end{equation}

The function $M_{4}(q;\omega ,\omega _{1},\omega _{2})$ replaces function $%
g(q,\omega )$ employed in the standard derivation of the $|\nabla \Psi |^{2}$
term for diffusive superconductors. The additional terms $\psi _{k}^{2}(%
\mathbf{r})$ and $\psi _{l}^{2}(\mathbf{r^{\prime }})$ (and the summation
over the corresponding states $k$ and $l$) arise due to the \textit{%
essentially $\mathbf{r}-dependent$} "background configuration" which $%
\mathbf{r}$-dependence is expressed through $\psi ^{2}(\mathbf{r})$ by Eq.(%
\ref{Deltar}).

The main source of fluctuations of $M_{4}$ are the spatial fluctuation of
the eigenfunctions whereas the spectral fluctuations are relatively weak.
Neglecting spectral fluctuations, one can switch from summation over indices 
$i,j,k,l$ to integration over the energy variables $\xi _{i}$, $\xi _{j}$, $%
\xi _{k}$, $\xi _{l}$ with the constant DoS and consider the average $%
I(q,\{\omega \})=\langle I_{ijkl}(q)\rangle $ as a function of all energy
differences $\omega $, $\omega _{1}$, $\omega _{2}$. Then the expression for
the coefficient $C$ in the Ginzburg-Landau functional Eq.(\ref{FGL}) takes
the form

\begin{eqnarray}
C &=&\frac{1}{8T_{c}^{2}}\lambda ^{2}\nu _{0}\int d\xi d\omega d\omega
_{1}d\omega _{2}\,\Delta (\xi +\omega /2+\omega _{1})\,\Delta (\xi -\omega
/2+\omega _{2})\,  \label{C} \\
&&\times \eta (\xi +\omega /2,\xi -\omega /2)\,\eta (\xi +\omega /2+\omega
_{1})\,\eta (\xi -\omega /2+\omega _{2})\;J(\{\omega \}),  \notag
\end{eqnarray}

where $J(\{ \omega\})= -\mathcal{V}^{3}\left.\frac{dI(q,\{\omega\})}{dq^{2}}%
\right|_{q=0}$.

The next step is to estimate $I(\{\omega \})$ using the rules formulated in
section \ref{Fractality and correlations}.\ We begin by considering the
critical wavefunction statistics. Applying the rule $\mathbf{(iii)}$ we find
the phase-independent counterpart to the combination of eigenfunctions in
Eq.(\ref{Iijkl}). It happens to coincide with the one for the square of the
off-diagonal matrix element $M_{ijkl}$ and is given by Eqs.(\ref%
{off-square-phase-ind}),(\ref{off1}). Then expanding Eq.(\ref{Iijkl}) up to $%
q^{2}$ and integrating over $(\mathbf{r}-\mathbf{r^{\prime }})$ up to $|%
\mathbf{r}-\mathbf{r^{\prime }}|\sim L_{\omega }$ we obtain an expression
similar to Eq.(\ref{off-square22}) but containing an extra factor $L_{\omega
}^{2}$ due to the expansion of $e^{i\mathbf{q(r-r^{\prime })}}$ up to $q^{2}$%
: 
\begin{equation}
J(\{\omega \})\sim \mathcal{V}^{3}\,L_{\omega }^{2}\,\langle
|M_{ijkl}|^{2}\rangle _{\mathrm{typ}}\sim L_{\omega }^{d+2}\,\left( \frac{%
E_{0}}{\omega }\right) ^{2\gamma }.  \label{deltaI}
\end{equation}%
In Eq.(\ref{deltaI}) we assumed all energy differences to be of the same
order $\omega _{1}\sim \omega _{2}\sim \omega $. The estimate Eq.(\ref%
{deltaI}) holds true on the insulator side of the Anderson transition as
long as $\omega \gg \delta _{L}$ (see the discussion in section \ref{Scaling
estimates C} and Eq.(\ref{delta-xi})).

Now we have to estimate the result of integration over the energy variables
in Eq.(\ref{C}). Using the asymptotic behavior of $\Delta (\omega )\propto
\omega ^{-\gamma }$ \ which follows from (\ref{MMFA}), (see section \ref%
{Modified mean-field}) and $\eta (\omega )\propto \omega ^{-1}$ and Eq.(\ref%
{deltaI}) power counting shows that for $T_{c}>\delta _{L}$ the integral is
dominated by $\omega \sim \omega _{1}\sim \omega _{2}\sim T_{c}$ and can be
estimated by 
\begin{equation}
C\sim \left( \frac{1}{\nu _{0}T_{c}}\right) ^{\frac{2}{d}}\,\lambda
^{2}\left( \frac{E_{0}}{T_{c}}\right) ^{2\gamma }\sim \left( \frac{1}{\nu
_{0}T_{c}}\right) ^{\frac{2}{d}},\;\;\;\;(T_{c}\gg \delta _{L}).  \label{CC}
\end{equation}%
Remarkably, the enhancement factor depending on the fractal exponent $\gamma 
$ is canceled by the coupling constant $\lambda $ due to Eq.(\ref{Tc0}), and
the result is not sensitive to fractality and is essentially the same as in
Refs.\cite{BulaSad,KapitulnikKotliar1986}. This cancellation occurs due to the
presence of the additional (with respect to the standard expression Eq.~\ref%
{gq0}) factors $\psi _{k}^{2}(\mathbf{r})$ and $\psi _{l}^{2}(\mathbf{r})$
in the integral $I_{ijkl}(q)$ defined in Eq.(\ref{Iijkl}); without these
factors, the final result for $C$ would contain extra small factor $\sim
(T_{c}/E_{0})^{\gamma }$.

Note that the coefficient $C$ is dominated by the off-diagonal matrix
elements only in metal or in very weak insulator. As one moves towards a
strong insulator where $T_{c}<\delta _{L}$, the main contribution to $C$
becomes the one that originates from the diagonal approximation and can be
roughly estimated as 
\begin{equation}
C\sim L_{\mathrm{loc}}^{2},\;\;\;\;(\delta _{L}>T_{c})  \label{C-ins}
\end{equation}%
in the diagonal approximation. The estimate (\ref{C-ins}) is based on the
simplest picture that wavefunctions localized at distances larger than
localization length $L_{\mathrm{loc}}$ do not overlap. In fact, as we
discuss in section \ref{Transition temperature}, this approximation is a bit
too crude as it misses an important logarithmic factor which increases $C$
values in the range $T_{c}\ll \delta _{L}$, see Eq.(\ref{Zeff}).

\subsubsection{Mesoscopic fluctuations: coefficient $W$. \label{Mesoscopic
fluctuations}}

Disorder always leads to spatial fluctuations of parameters which enter the
Ginzburg-Landau functional; the major effect is due to fluctuations of $a(T,%
\mathbf{r})$. Universal mesoscopic fluctuations (which provide a lower bound
for the strength of this effect) were studied in Ref.~\cite{BulaSad} for
usual disordered superconductors and more recently in~\cite{Skvortsov2005}
for 2D films with the strong Finkelstein effect. Here we follow the same
general approach but implement it in the Fock space instead of the
coordinate space.

We start from the Eq.(\ref{F21}) for the quadratic part of the free energy.
Previously we neglected off-diagonal matrix elements $M_{ijkk}$ entering the
second term of (\ref{F21}); now our goal is to estimate the strength of
mesoscopic fluctuations produced by these matrix elements. Thus we represent
the second term as 
\begin{equation}
F_{22}=-\frac{\Psi ^{2}}{2}\sum_{ij}\left( (\hat{Q}^{2})_{ij}+\Gamma
_{ij}\right) \phi _{i}\phi _{j}\,  \label{F2m}
\end{equation}%
where $\hat{Q}$ is defined in Eq.(\ref{Qmat}) and 
\begin{equation}
\Gamma _{ij}=\frac{g^{2}}{4}\sum_{k\neq l}M_{iikl}M_{kljj}\,\eta _{kl}(\eta
_{i}\eta _{j})^{\frac{1}{2}}  \label{Gamma}
\end{equation}%
The matrix $\Gamma _{ij}$ contains corrections from the (previously
neglected) off-diagonal matrix elements. Its average $\overline{\Gamma _{ij}}
$ contributes to the shift of the critical temperature which can be
estimated using scaling arguments developed in section \ref{Scaling
estimates A}. The result is that the relative shift due to the off-diagonal
matrix elements is $(\delta _{\mathrm{off}}T_{c})/T_{c}\sim 1$.

Here we are interested in the strength of \textit{mesoscopic fluctuations} $%
W $ defined in Eq.(\ref{W}) and thus calculate the free energy cumulant: 
\begin{eqnarray}
(\nu _{0}T^{2})^{2}\,\,W &=&\frac{1}{\Psi ^{4}\mathcal{V}}\langle \langle
F_{22}F_{22}\rangle \rangle = \\
&&\frac{1}{4\mathcal{V}}\sum_{i_{1}i_{2}i_{3}i_{4}}\phi _{i_{1}}\phi
_{i_{2}}\phi _{i_{3}}\phi _{i_{4}}\,\langle \langle \Gamma
_{i_{1},i_{2}}\Gamma _{i_{3},i_{4}}\rangle \rangle  \notag  \label{F22cu}
\end{eqnarray}%
Using Eq.(\ref{Gamma}) and switching to a continuum representation, we
obtain the irreducible correlation function $\langle \langle \Gamma \Gamma
\rangle \rangle $:%
\begin{gather*}
\langle \langle \Gamma _{i_{1},i_{2}}\Gamma _{i_{3},i_{4}}\rangle \rangle =
\\
\left( \frac{g^{2}}{4}\right) ^{2}\left( \nu _{0}\mathcal{V}\right) ^{4}%
\sqrt{\eta _{i_{1}}\eta _{i_{2}}\eta _{i_{3}}\eta _{i_{4}}}\int ...\int d\xi
_{m}d\xi _{n}d\xi _{k}d\xi _{l}\,\eta _{mn}\eta _{kl}\int ...\int d^{d}%
\mathbf{r}_{1}d^{d}\mathbf{r}_{2}d^{d}\mathbf{r}_{3}d^{d}\mathbf{r}_{4} \\
\langle \langle \psi _{i_{1}}^{2}(\mathbf{r}_{1})\psi _{i_{2}}^{2}(\mathbf{r}%
_{2})\psi _{i_{3}}^{2}(\mathbf{r}_{3})\psi _{i_{4}}^{2}(\mathbf{r}_{4})\psi
_{m,n}^{(2)}(\mathbf{r}_{1})\psi _{m,n}^{(2)}(\mathbf{r}_{2})\psi
_{k,l}^{(2)}(\mathbf{r}_{3})\psi _{k,l}^{(2)}(\mathbf{r}_{4})\rangle \rangle
,
\end{gather*}%
where the double brackets $\langle \langle ...\rangle \rangle $ denote the
cumulant average defined by (we use below indices $i_{1}...$ instead of $%
\psi _{i_{1}}^{2}...$ etc.): $\langle \langle
i_{1}i_{2}i_{3}i_{4}mnkl\rangle \rangle =\langle
i_{1}i_{2}i_{3}i_{4}mnkl\rangle -\langle i_{1}i_{2}mn\rangle \langle
i_{3}i_{4}kl\rangle $ and $\psi _{m,n}^{(2)}(\mathbf{r})\equiv \psi _{m}(%
\mathbf{r})\psi _{n}(\mathbf{r})$.

We now estimate the coefficient $W$ for the critical states near the
Anderson transition point. First of all we note that the decoupled average $%
\langle i_{1}i_{2}mn\rangle \langle i_{3}i_{4}kl\rangle $ depends only on
the two of the three independent differences in coordinates $R_{s,s^{\prime
}}=|\mathbf{r}_{s}-\mathbf{r}_{s^{\prime }}|$ ($s,s^{\prime }=1,...4$),
while the cumulant average depends on all the three of them and vanishes
when $R_{s,s^{\prime }}>L_{\omega }$ (as before we assume that all the
energy differences are of the same order $\omega $). In the region where all 
$R_{s,s^{\prime }}<L_{\omega }$ the decoupled average is smaller than the
cumulant one. Assuming all the differences of coordinates are of the same
order $R_{s,s^{\prime }}\sim R$ and applying the rules $\mathbf{(i)-(iii)}$
of section \ref{Wavefunction correlations} we obtain at $R<L_{\omega }$: 
\begin{gather*}
\left\langle \left\langle \psi _{i_{1}}^{2}(\mathbf{r}_{1})\psi _{i_{2}}^{2}(%
\mathbf{r}_{2})\psi _{i_{3}}^{2}(\mathbf{r}_{3})\psi _{i_{4}}^{2}(\mathbf{r}%
_{4})\psi _{m,n}^{(2)}(\mathbf{r}_{1})\psi _{m,n}^{(2)}(\mathbf{r}_{2})\psi
_{k,l}^{(2)}(\mathbf{r}_{3})\psi _{k,l}^{(2)}(\mathbf{r}_{4})\right\rangle
\right\rangle \sim \\
\left\langle \psi _{i_{1}}^{2}(\mathbf{r}_{1})\psi _{i_{2}}^{2}(\mathbf{r}%
_{2})\psi _{i_{3}}^{2}(\mathbf{r}_{3})\psi _{i_{4}}^{2}(\mathbf{r}_{4})\psi
_{m,n}^{(2)}(\mathbf{r}_{1})\psi _{m,n}^{(2)}(\mathbf{r}_{2})\psi
_{k,l}^{(2)}(\mathbf{r}_{3})\psi _{k,l}^{(2)}(\mathbf{r}_{4})\right\rangle
\sim \\
\left\langle \psi _{i_{1}}^{2}(\mathbf{r}_{1})\psi _{i_{2}}^{2}(\mathbf{r}%
_{2})\psi _{i_{3}}^{2}(\mathbf{r}_{3})\psi _{i_{4}}^{2}(\mathbf{r}_{4})\psi
_{m}^{2}(\mathbf{r}_{1})\psi _{n}^{2}(\mathbf{r}_{2})\psi _{k}^{2}(\mathbf{r}%
_{3})\psi _{l}^{2}(\mathbf{r}_{4})\right\rangle \sim \\
\sim \mathcal{V}^{-8}\left( \frac{L_{\omega }}{R}\right) ^{3d-2\alpha
_{2}-\alpha _{4}}\,\left( \frac{L_{\omega }}{\ell }\right) ^{4(d-d_{2})},
\end{gather*}%
where $\alpha _{2}=3d_{4}-2d_{2}$, $\alpha _{4}=7d_{8}-6d_{4}$ . Indeed, at $%
R\sim L_{\omega }$ eigenfunctions in different space points are
statistically independent and the corresponding averages can be decoupled.
The result is $\mathcal{V}^{-8}(L_{\omega }/\ell )^{4(d-d_{2})}$ in
agreement with Eq.(\ref{corr-dif-E}). At $R\sim \ell $ all eigenfunctions
can be considered in one space point and the averaging then gives $\mathcal{V%
}^{-8}(L_{\omega }/\ell )^{7(d-d_{8})}$. The last estimate coincides with
Eq.(\ref{def-multi}) at small energy separations $L_{\omega }\sim L$ where
one can consider all eigenfunctions to be identical, and it corresponds to
all eigenfunctions independently averaged at large energy separations when $%
L_{\omega }\sim \ell $. The decoupled average estimated in the same way
using Eqs.(\ref{off-square-phase-ind}),(\ref{off1}) is of the order of $%
\mathcal{V}^{-8}\,(L_{\omega }/R)^{2d-2\alpha _{2}}\,(L_{\omega }/\ell
)^{4(d-d_{2})}$ and thus is smaller at $R\ll L_{\omega }$ than the cumulant
average by the factor $(L_{\omega }/R)^{d-\alpha _{4}}$.

Now we estimate the result of the four spatial integrations: over difference
of coordinates $R_{12}=|\mathbf{r}_{1}-\mathbf{r}_{2}|$, $R_{13}=|\mathbf{r}%
_{1}-\mathbf{r}_{3}|$ and $R_{23}=|\mathbf{r}_{3}-\mathbf{r}_{2}|$ and one
free integration which results in the factor $\mathcal{V}$. At this point it
is important that the dependence of the cumulant average on $%
R_{12},R_{13},R_{23}$ is symmetric and such that the power of $R$ is
typically smaller than $3d$, as $\alpha _{q}^{\mathrm{typ}}>0$ (see Eq.(\ref%
{pos-typ})). This means that the main contribution to the integrals over $%
R_{12},R_{13},R_{23}$ comes from the region $R_{12}\sim R_{13}\sim
R_{23}\sim L_{\omega }$. This is a crucial circumstance that eliminates the
dependence on higher fractal dimensions.

Thus we arrive at 
\begin{gather*}
\int d\{\mathbf{r}\}\left\langle \left\langle \psi _{i_{1}}^{2}(\mathbf{r}%
_{1})\psi _{i_{2}}^{2}(\mathbf{r}_{2})\psi _{i_{3}}^{2}(\mathbf{r}_{3})\psi
_{i_{4}}^{2}(\mathbf{r}_{4})\psi _{m,n}^{(2)}(\mathbf{r}_{1})\psi
_{m,n}^{(2)}(\mathbf{r}_{2})\psi _{k,l}^{(2)}(\mathbf{r}_{3})\psi
_{k,l}^{(2)}(\mathbf{r}_{4})\right\rangle \right\rangle \\
\sim \mathcal{V}^{-7}\,L_{\omega }^{3d}\,\left( \frac{E_{0}}{\omega }\right)
^{4\gamma }.
\end{gather*}%
The remaining energy integration is estimated in the same way as in the
previous subsection. As before, the dominant contribution comes from the
energies within a strip of width $T_{c}$ near the Fermi energy. The final
result reads 
\begin{equation}
\langle \langle F_{22}F_{22}\rangle \rangle \sim \lambda ^{4}\,\nu _{0}%
\mathcal{V}\,T_{c}^{3}\,\left( \frac{E_{0}}{T_{c}}\right) ^{4\gamma }\sim
\nu _{0}\mathcal{V}\,T_{c}^{3}.  \label{fluct-fin}
\end{equation}%
Again, as in Eq.(\ref{CC}), the enhancement factor which depends on the
fractal exponent $\gamma $ cancels out exactly by the coupling constant $%
\lambda $, and the final result for the coefficient $W$ is 
\begin{equation}
W\sim \frac{1}{\nu _{0}T_{c}},\;\;\;\;(\delta _{L}\ll T_{c}).  \label{WW}
\end{equation}%
This result was obtained using the definition (\ref{W}) which makes sense if
the spatial scale $L_{a}$ of $a(\mathbf{r})$ fluctuations is small compared
to the scale of $\Psi (\mathbf{{r})}$ variation. We expect that the same
estimate (\ref{WW}) is valid if both length-scales are of the same order.

In the limit of strong insulator, one can repeat the above analysis to
arrive at 
\begin{equation}
W\sim L_{\mathrm{loc}}^{3},\;\;\;\;\;(\delta _{L}>T_{c}).  \label{W-ins}
\end{equation}%
This result is obtained in the diagonal approximation.

\subsubsection{Ginzburg parameters for thermal and mesoscopic fluctuations 
\label{Ginzburg parameters}}

Now we use the results given by Eqs.(\ref{alphaT},\ref{beta},\ref{CC},\ref%
{WW}) to estimate the relative width of the fluctuation region near the
thermal transition into a fractal superconductor state. First we estimate
the Ginzburg parameter $\mathrm{Gi}$ which determines the reduced
temperature range $|1-T/T_{c}|<\mathrm{Gi}$ where thermal fluctuations are
strong in a $3D$ system \cite{LarkinVarlamovBook}: 
\begin{equation}
\mathrm{Gi}\sim \frac{b^{2}}{C^{3}(\nu _{0}T_{c})^{2}}\sim 1,\;\;\;\;(\delta
_{L}\ll T_{c})  \label{Gi}
\end{equation}%
The relative width of \textquotedblleft smearing\textquotedblright\ of the
superconductive transition due to positional disorder is given by the
parameter $\mathrm{Gi_{d}}$ defined as follows (see e.g. Ref.\cite%
{IoffeLarkin1981}): 
\begin{equation}
\mathrm{Gi}_{d}\sim \frac{W^{2}}{C^{3}}\sim 1,\;\;\;\;(\delta _{L}\ll T_{c})
\label{Gid}
\end{equation}

The estimates (\ref{Gi}) and (\ref{Gid}) demonstrate that the modified
mean-field approximation developed in this section can be used (with
relative accuracy of the order of unity) in order to estimate $T_{c}$ of a
fractal superconductor. Based on this result we conclude that in the region
of extended and weakly localized single-particle states with $\delta _{L}\ll
T_{c}$ the thermal fluctuations of the order parameter phase and the
mesoscopic fluctuations of the local $T_{c}$ which were not taken into
account in the modified MFA can at most reduce the global $T_{c}$ by a
factor of the order of one compared to the modified MFA result, Eq.(\ref{Tc0}%
) but can hardly lead to a modification of the functional dependence of $%
T_{c}$ on the interaction constant $\lambda $. This is the most important
conclusion of this subsection.

Another conclusion concerns the \textit{role of the off-diagonal matrix
elements} in the region of extended and weakly localized single-particle
states. These matrix elements are completely neglected in modified MFA which
nevertheless gives accurate results. They are, however, necessary for the
correct account of the thermal fluctuations and the local $T_{c}$ mesoscopic
fluctuations. The off-diagonal matrix elements also determine the
electromagnetic response and are necessary for calculation of the critical
current, as both properties are related with the gradient term in the
Ginzburg-Landau functional Eq.(\ref{FGL}).

In the region $\delta _{L}\gg T_{c}$ the use of the simplified estimate $%
C\sim L_{\mathrm{loc}}^{2}$ would lead to the conclusion that the parameters 
$\mathrm{Gi}$ and $\mathrm{Gi_{d}}$ are large: 
\begin{equation*}
\mathrm{Gi}\sim \mathrm{Gi_{d}}\sim \frac{1}{(\nu _{0}T_{c}L_{\mathrm{loc}%
}^{3})^{2}}\sim \frac{1}{Z_{\mathrm{eff}}^{2}},\;\;\;\;(\delta _{L}>T_{c})
\end{equation*}%
where $Z_{\mathrm{eff}}\sim T_{c}/\delta _{L}$ is the effective coordination
number to be discussed later on in relation with the \textit{pseudo-spin
Hamiltonian}. In fact, more accurate estimate for $Z_{\mathrm{eff}}$ given
by Eq.(\ref{Zeff}) in section \ref{Transition temperature} below shows that
both $\mathrm{Gi}$ and $\mathrm{Gi_{d}}$ remain of the order of one in the
broad range of large $\delta _{L}/T_{c}$ ratios.

\subsection{Pseudo-spin Hamiltonian. \label{Pseudospin Hamiltonian}}

As we will see below, there is a sufficiently wide range of parameters on
the insulating side of the Anderson transition where the superconducting
transition temperature $T_{c}$ is of the order of its value $T_{c}^{(0)}$
right at the Anderson transition point while the paring gap $\Delta _{P}$
introduced in section \ref{Insulating state} is much larger than $T_{c}$.
This means that practically the entire region where $T_{c}$ gradually
decreases with increasing disorder falls into this \textit{pseudo-gap regime}%
. The modified MFA does not work in this regime because Ginzburg parameters $%
\mathrm{Gi}$ and $\mathrm{Gi_{d}}$ are larger than 1. However, the problem
can be significantly simplified by making use of the large value of the gap $%
\Delta _{P}$ between states with even and odd number of particles occupying
any orbital $\psi _{i}(\mathbf{r})$. Namely, as the creation of the \textit{%
odd state} with one particle on an orbital takes a large energy $2\Delta
_{P} $ to break the pair, the transitions between the even (having two or no
particles on any orbital) and odd states described by the \textit{%
off-diagonal} matrix elements can be neglected. Technically, it is
equivalent to neglecting all the off-diagonal matrix elements and
considering the sectors of Hilbert space with even and odd states as
completely decoupled. Restricting ourselves to the low-energy \textit{even}
sector one can rewrite \cite{Anderson1959} the Hamiltonian Eq.(\ref{Ham1})
in the form of a \textit{pseudo-spin Hamiltonian} with the spin operators $%
S_{i}^{\pm }$, $S_{i}^{z}$ acting in the Fock space of orbitals $|i\rangle
\equiv \psi _{i}(\mathbf{r})$ and the \textit{diagonal matrix elements} $%
M_{ij}$ playing a role of the coupling matrix. The most general form of this
Hamiltonian is: 
\begin{eqnarray}
H_{PS} &=&\sum_{j}\xi _{j}(2S_{j}^{z}+1)-\frac{g_{\perp }}{2}%
\sum_{ij}M_{ij}(S_{i}^{+}S_{j}^{-}+S_{i}^{-}S_{j}^{+})  \notag \\
&-&g_{\parallel }\sum_{ij}M_{ij}S_{i}^{z}S_{j}^{z}.  \label{HamSpin}
\end{eqnarray}%
where the set of operators 
\begin{equation}
S_{j}^{+}=c_{j,\uparrow }^{+}c_{j,\downarrow }^{+}\,\quad
S_{j}^{-}=(S_{j}^{+})^{\dagger }\quad S^{z}=\frac{1}{2}\left( \sum_{\sigma
}c_{j,\sigma }^{+}c_{j,\sigma }-1\right)
\end{equation}%
is equivalent to the set of spin-$\frac{1}{2}$ operators $\mathbf{S}=\frac{1%
}{2}\mathbf{\sigma }$. Here $\xi $ is random energy distributed with density 
$\nu $ in some interval around $0.$

The Hamiltonian Eq.(\ref{HamSpin}) is the basis of theory of the \textit{%
pseudogap superconductivity} we will develop in section \ref%
{Superconductivity with a pseudogap}. However, it is valid in \textit{any}
case where the off-diagonal matrix elements may be neglected for this or
another reason. It was originally suggested by Anderson \cite{Anderson1959}
for a BCS superconductor where the off-diagonal matrix elements are small in
the parameter $T_{c}/E_{F}$. It is parametrically justified in the
pseudo-gap region where the energy denominator associated with the even-odd
transitions is large since $\Delta _{P}\gg T_{c}$. However, the estimate of
the Ginzburg parameters $\mathrm{Gi}\sim \mathrm{Gi_{d}}\sim 1$ (see
discussion in section \ref{Transition temperature} and (\ref{Zeff})) shows
that it is also useful for a \textit{semi-quantitative} (up to a factor of
order one) determination of the transition temperature of the fractal
superconductor near the Anderson localization transition. We will use this
Hamiltonian in order to obtain the phase diagram as a function of $(T,E_{F})$
in the entire region spanning BCS and the pseudo-gap regime.

\subsection{Virial expansion method. \label{Virial expansion}}

In this subsection we develop a new approach based on the virial expansion
method applied to the pseudo-spin Hamiltonian Eq.(\ref{HamSpin}) an use it
to determine the superconductive transition temperature and thus the full
phase diagram of the disordered superconductor. The approximations implied
by this method are completely different from the ones of modified MFA. It
will turn out to be very useful in order to find the precise limit of
applicability of modified MFA and determine $T_{c}$ in the region of
localized single-particle states where the parameters $\mathrm{Gi}$ and $%
\mathrm{Gi_{d}}$ can be large.

Developing our scheme, we follow the approach of Larkin and Khmelnitsky~\cite%
{Virial} who first used the virial expansion method to determine the
temperature of magnetic phase transition in metallic alloys (see also~\cite%
{FeigTsvel1979}). The idea of this method is to express the free energy as a
series, where each term contains an \textit{exact} contribution from a fixed
number of local variables (e.g. spins for the problem of magnetic
impurities): 
\begin{gather}
F=\sum_{n=1}^{\infty }\mathcal{F}^{(n)}=\sum_{i}F_{i}+%
\sum_{i>j}(F_{ij}-F_{i}-F_{j})  \label{virial1} \\
+\sum_{i>j>k}(F_{ijk}-F_{ij}-F_{jk}-F_{ik}+F_{i}+F_{j}+F_{k})+\dots  \notag
\end{gather}%
Here $F_{i}=-T\ln \mathrm{Tr}e^{-H_{i}/T}$ is the free energy of a single $i$%
-th spin in a field, $F_{ij}=-T\ln \mathrm{Tr}e^{-(H_{i}+H_{j}+H_{ij})/T}$
is the exact free energy of two interacting spins, etc. The terms in the
brackets in the second sum and higher order sums cancel each other for large
space separations (e.g. $|\mathbf{r}_{i}-\mathbf{r}_{j}|$ for the second
order term) so that corresponding variables become essentially independent.

When the system is approaching a phase transition, all terms of the virial
expansion become relevant and an actual calculation of critical
singularities becomes impossible. However, the virial expansion method can
be used in order to find an approximate \textit{location} of the transition
point. Indeed, consider the virial expansion for some susceptibility 
\begin{equation}
\chi (T)=\sum_{n=1}^{\infty }\chi _{n}(T)  \label{vir-ser}
\end{equation}%
which must diverge at $T=T_{c}$. The value of $T_{c}$ can be determined from
the condition that $T_{c}$ corresponds to the \textit{limit of convergence }
of the series Eq.(\ref{vir-ser}): 
\begin{equation}
\lim_{n\rightarrow \infty }\frac{\chi _{n+1}(T_{c})}{\chi _{n}(T_{c})}=1.
\label{virial2}
\end{equation}%
In practice, an exact \textit{analytical} calculation of $\chi _{n}$ with
large $n$ is very cumbersome, such calculations being usually limited by the
first few terms, $n=1,2,3...$. In the following we will use instead of Eq.(%
\ref{virial2}) the approximate truncated criterion 
\begin{equation}
\chi _{2}(T_{c})=\chi _{3}(T_{c}).  \label{virial3}
\end{equation}%
This step constitutes the key approximation of the virial method of
calculation of $T_{c}$. It consists in an extrapolation (in general,
uncontrolled) into the thermodynamic limit of properties found with exact
treatment of few-spin systems.

We now apply this idea to the specific problem of the calculation of
superconducting transition temperature corresponding to the Hamiltonian Eq.(%
\ref{HamSpin}). The relevant susceptibility is defined with respect to the
"ordering" field $\Delta $ which enters the Hamiltonian via the source term 
\begin{equation}
V_{\Delta }=-\sum_{j}(\Delta S_{j}^{+}+\Delta ^{\ast }S_{j}^{-})
\label{Vdelta}
\end{equation}%
Transition into a superconducting state is signaled by the divergence of the
Cooper susceptibility 
\begin{equation}
\chi (T)=-\frac{\partial ^{2}F}{\partial \Delta \partial \Delta ^{\ast }}%
=\sum_{n=1}^{\infty }\chi _{n}(T)  \label{chi1}
\end{equation}%
Below we will describe calculation of the lowest-order virial expansion
terms $\chi _{n}(T)$ with $n=1,2,3$ defined in accordance with Eq.(\ref%
{virial1}) as: 
\begin{eqnarray}
\chi _{1} &=&\sum_{i}\chi _{i}^{(1)}  \label{chi2} \\
\chi _{2} &=&\sum_{n>m}(\chi _{nm}^{(2)}-\chi _{n}^{(1)}-\chi _{m}^{(1)}) 
\notag \\
\chi _{3} &=&\sum_{n>m>l}(\chi _{nml}^{(3)}-\chi _{nl}^{(2)}-\chi
_{ml}^{(2)}-\chi _{nm}^{(2)}+\chi _{n}^{(1)}+\chi _{m}^{(1)}+\chi _{l}^{(1)})
\notag
\end{eqnarray}%
where $\chi _{i}^{(1)},\chi _{ij}^{(2)},\chi _{ijl}^{(3)}...$ are the Cooper
susceptibilities of the system of 1, 2, 3 ... spins. These susceptibilities
can be expressed through the eigenvalues $\lambda _{\alpha }(\Delta )$ of
the $N$-spin Hamiltonian: 
\begin{equation}
\chi ^{(N)}=-\frac{1}{Z_{0}}\,\sum_{\alpha }e^{-\lambda _{\alpha
}^{(0)}/T}\,\gamma _{\alpha },  \label{chi-N}
\end{equation}%
where $Z_{0}=\sum_{\alpha }e^{-\lambda _{\alpha }^{(0)}/T}$, and $\gamma
_{\alpha }$ is the sensitivity of the $\alpha $-th eigenvalue of the N-spin
Hamiltonian to the $\Delta $ perturbation: 
\begin{equation}
\lambda _{\alpha }(\Delta )=\lambda _{\alpha }^{(0)}+|\Delta |^{2}\,\gamma
_{\alpha }+o(|\Delta |^{2}).  \label{eigen}
\end{equation}%
In order to compute the $N$-spin susceptibility $\chi _{nml...N}^{(N)}$ one
has to represent the $N$-spin Hamiltonian as a $2^{N}\times 2^{N}$ matrix.
The single-spin susceptibility can be found easily: 
\begin{equation}
\chi _{1}(T)=\sum_{i}\frac{1}{2\xi _{i}}\tanh \frac{\xi _{i}}{T}=\nu _{0}\ln
\left( \frac{4e^{\mathrm{\mathbf{C}}}E_{b}}{\pi T}\right)  \label{1spin}
\end{equation}%
where $E_{b}$ the the upper energy cutoff and $\mathrm{\mathbf{C}}=0.577...$
is the Euler constant.

The two-spin Hamiltonian is given by the 4 by 4 matrix 
\begin{eqnarray}
H^{(2)} &=&\left( 
\begin{array}{cccc}
\xi _{+}-\frac{1}{2}J_{12}^{\parallel } & \Delta & \Delta & 0 \\ 
\Delta ^{\ast } & -\xi _{-}+\frac{1}{2}J_{12}^{\parallel } & -(J_{12}^{\perp
})^{\ast } & \Delta \\ 
\Delta ^{\ast } & -J_{12}^{\perp } & \xi _{-}+\frac{1}{2}J_{12}^{\parallel }
& \Delta \\ 
0 & \Delta ^{\ast } & \Delta ^{\ast } & -\xi _{+}-\frac{1}{2}%
J_{12}^{\parallel }%
\end{array}%
\right)  \label{H-2} \\
&\equiv &H_{0}^{(2)}+V,  \notag
\end{eqnarray}%
where $\xi _{\pm }=\xi _{1}\pm \xi _{2}$, $J_{ij}^{\parallel }=\frac{%
g_{\parallel }}{2}\,M_{ij}$, $J_{ij}^{\perp }=\frac{g_{\perp }}{2}\,M_{ij}$
; and the perturbation term $V\propto \Delta $. Using the Hamiltonian (\ref%
{H-2}) we calculate $\lambda _{\alpha }^{(0)}$ and $\gamma _{\alpha }$ and
then use Eq.(\ref{chi-N}) to obtain second virial term $\chi _{2}(T)$. We
give here its simplest form corresponding to $J_{12}^{\parallel }=0$, the
full answer being given in the Appendix A: %\begin{eqnarray}
%\chi^{(2)}_{ij}=\frac{2J_{ij}}{E_{+}E_{-}}\,\tanh\left(
%\frac{E_{+}}{2T}\right)\,\tanh\left(
%\frac{E_{-}}{2T}\right) \, + \, \\ \nonumber
%\frac{1}{E_{+}}\,\tanh\left(
%\frac{E_{+}}{2T}\right)+\frac{1}{E_{-}}\,\tanh\left(
%\frac{E_{-}}{2T}\right),
%\label{2spin}
%\end{eqnarray}
{\small 
\begin{eqnarray}
\chi _{2} &=&\sum_{i>j}\left[ \frac{J_{ij}^{\perp }}{2E_{+}E_{-}}\,\tanh
\left( \frac{E_{+}}{T}\right) \,\tanh \left( \frac{E_{-}}{T}\right) +\frac{1%
}{2E_{+}}\,\tanh \left( \frac{E_{+}}{T}\right) \right.  \label{2-vir} \\
&&\left. +\frac{1}{2E_{-}}\,\tanh \left( \frac{E_{-}}{T}\right) -\frac{1}{%
2\xi _{i}}\,\tanh \left( \frac{\xi _{i}}{T}\right) -\frac{1}{2\xi _{j}}%
\,\tanh \left( \frac{\xi _{j}}{T}\right) \right]  \notag
\end{eqnarray}%
} where 
\begin{equation}
E_{\pm }=\frac{1}{2}(\xi _{i}+\xi _{j})\pm \frac{1}{2}\sqrt{(\xi _{i}-\xi
_{j})^{2}+|J_{ij}^{\perp }|^{2}}.  \label{E+-}
\end{equation}%
are the exact energies of the two-spin problem.

The three-spin susceptibility $\chi _{3}$ requires the solution of the cubic
equations which is too long to be written here. The corresponding derivation
is given in Appendix \ref{Virial expansion in pseudospin subspace} in a form
suitable for numerical calculations.

For the standard BCS problem with $M_{ij}=\frac{1}{\mathcal{V}}\rightarrow 0$%
, the renormalized energies $E_{\pm }$ coincide with the bare ones $\xi _{i}$
and $\xi _{j}$ , and the first term in Eq.(\ref{2-vir}) is the leading one.
In this case the summations over the two energy variables become independent
and one gets $\chi _{2}(T)=g^{2}\nu _{0}\ln ^{2}\left( \frac{4e^{\mathrm{%
\mathbf{C}}}E_{b}}{\pi T}\right) $. Applying the simplest truncated
criterion $\chi ^{(1)}(T_{c})=\chi ^{(2)}(T_{c})$, one finds the correct
result: $T_{c}=(4e^{\mathrm{\mathbf{C}}}/\pi )E_{b}e^{-1/\lambda }$. In the
same way one can show that \textit{any} truncated criterion $\chi
_{n}(T_{c})=\chi _{n+1}(T_{c})$ gives the same correct result for $T_{c}$ of
the BCS problem.

This is no longer true for the case where $M_{ij}$ has essential energy
dependence, as it has in the fractal case where $M_{ij}$ is given by Eq.(\ref%
{ij}). In this situation the double sum in $\chi _{2}$ and the corresponding
multiple sums in the higher-order susceptibilities $\chi _{n}$ \ do not
factorize. Moreover, all $\chi _{n}$ with $n\geq 2$ are dominated by the
low-energy region $\xi \sim T$ and do not contain any logarithmic
divergence, in contrast to $\chi _{1}$. This is why the simplest truncated
criterion $\chi _{1}(T)=\chi _{2}(T)$ does not make much sense: the
corresponding solution depends on the upper energy cutoff $E_{b}$, while for
any other truncated criterion $\chi _{n}(T)=\chi _{n+1}(T)$ with $n\geq 2$
the solution is independent of $E_{b}$. It is for this reason that we used
Eq.(\ref{virial3}) to find an approximate transition temperature. Note,
however, that the temperature $T_{1-2}$ found numerically for the 3D
Anderson model with critical disorder from the simplest truncated criterion $%
\chi _{1}(T_{1-2})=\chi _{2}(T_{1-2})$ gives values of $T_{1-2}$ which are
close to $T_{2-3}$ found from Eq.(\ref{virial3}) for a meaningful values of
the upper energy cutoff $E_{b}$.

So far we completely neglected the presence of the odd sector of the Hilbert
space. This is justified in the region of the pseudo-gap superconductivity
where the paring gap $\Delta _{P}\gg T_{c}$. However, even when being
decoupled from the even sector, the odd sector cannot be ignored completely
in a general case of $\Delta _{P}\sim T_{c}$. The reason for that can be
easily seen from Eq.(\ref{chi-N}) which is valid in a general case too.
Decoupling of the two sectors manifests itself in vanishing of all "partial"
susceptibilities $\gamma _{\alpha }$ corresponding to the states $\alpha $
of the odd sector. Thus only the eigenvalues $\lambda _{\alpha }^{(0)}$
corresponding to the even sector appear in the sum in Eq.(\ref{chi-N}). This
is not true for the partition function $Z_{0}$, where \textit{all} states
contribute. For $\Delta _{P}\gg T$ one can neglect $e^{-\lambda _{\mathrm{odd%
}}/T}$ compared to $e^{-\lambda _{\mathrm{even}}/T}$ and arrive at the
result given in Appendix \ref{Virial expansion in pseudospin subspace}.
However, in the BCS and the fractal region where $\Delta _{P}<T_{c}$ the
susceptibility will be substantially decreased by "statistical dilution" of
the even states by the odd ones.

From the standard theory of the BCS superconductivity it is well known that
in all expressions the temperature dependence appears as the factor $\tanh
\left( \frac{\xi }{2T}\right) $. In contrast to it, the Eqs.(\ref{1spin}, %
\ref{2-vir}) contain factor $\tanh \left( \frac{\xi }{T}\right) $. This
gives a simple rule to adjust the virial expansion scheme to the case of
vanishing $\Delta _{P}$: 
\begin{equation}
\tanh \frac{E_{i}}{T}\rightarrow \tanh \frac{E_{i}}{2T}  \label{factor2}
\end{equation}%
Thus the temperature $T_{2-3}$ found from the formalism described in
Appendix \ref{Virial expansion in pseudospin subspace} should be simply
divided by 2 to find the corresponding approximation for the transition
temperature in the region of extended single-particle states. In Appendix %
\ref{Virial expansion including single-occupied states} we present the
derivation of the virial susceptibilities for the general case of an
arbitrary paring gap $\Delta _{P}$ (still, neglecting off-diagonal matrix
elements, which mix the "even" and "odd" subspaces of the Hilbert space). In
particular, we show there that as long as the renormalization of energy
levels (i.e. the difference between $E_{+}$ and $\xi _{i}$ in Eq.(\ref{E+-}%
)) can be neglected, one can find (see Eq.(\ref{1-full})) the whole family
of distribution functions $F(\varepsilon ,\Delta _{P})$ that interpolate
between the two limiting cases, $\Delta _{P}\gg T_{c}$ and $\Delta _{P}\ll
T_{c}$.

In the region of extended and weakly localized single-particle states the
renormalization of energy levels is indeed negligible which justifies the
approach made in the derivation of (\ref{1-full}) in this regime. This
renormalization is not small in the region of a strong pseudo-gap but in
this regime one can project the pseudo-spin Hamiltonian Eq.(\ref{HamSpin})
into a truncated Hilbert space that consists of the \textit{even subspace}
only. Thus, both these limiting cases should be well described by a simple
replacement rule: 
\begin{equation}
\tanh \left( \frac{\varepsilon }{T}\right) \rightarrow F(\varepsilon ,\Delta
_{P})=\frac{\sinh \left( \frac{\varepsilon }{T}\right) }{\cosh \left( \frac{%
\varepsilon }{T}\right) +e^{-\Delta _{P}/T}},  \label{interPolate}
\end{equation}%
where $\Delta _{P}=\langle \Delta _{P}\rangle $. Application of this
replacement rule to all the formulae obtained in the pseudo-spin model is
expected to take correctly into account the effect of "dilution" by the odd
sector. With this modification the pseudo-spin Hamiltonian Eq.(\ref{HamSpin}%
) becomes a unified tool to describe (semi-quantitatively or better) the
entire region spanning weakly disordered BCS superconductor to the Anderson
insulator.

\section{Superconducting state very close to mobility edge. \label%
{Superconducting state}}

In this section we derive the main physical properties of the three
dimensional disordered superconductor with the Cooper attraction
characterized by the single-electron states that are critical or very weakly
localized. As explained above, a very important property of these wave
functions is their fractality which is responsible for the unusual
properties of these superconductors. In the regime considered here the
number of paired electrons in localization volume is still large, which
allows us to neglect the level spacing $\delta _{L}$ compared to other
relevant energy scales:\ 
\begin{equation*}
\delta _{L}=E_{0}\left( \frac{E_{c}-E_{F}}{E_{0}}\right) ^{3\nu }\ll T_{c}.
\end{equation*}

The most important quantity that characterizes a superconductor is its
transition temperature and the single particle gap. To compute them we
analyze the modified MFA developed in subsection \ref{Modified mean-field}
and determine the critical temperature $T_{c}$ and the gap function $\Delta
(\xi )$. Then in subsection \ref{Comparison of Tc values} we compare $T_{c}$
obtained by the modified MFA with the results of two other methods based on
the "diagonal" approximation": the spectral analysis of the discrete
modified MF equation with the {\it fluctuating} kernel $\hat{Q}$, and the
virial expansion method. We show that the results for the transition
temperature $T_{c}$ obtained by different methods are in a good mutual
agreement but exhibit a strongly enhancement of $T_{c}$ compared to the
expectations based on Anderson theorem. Finally, in subsections \ref{Pairing
in the modified mean-field}-\ref{Superfluid density and critical current} we
use Ginzburg-Landau functional derived in subsection \ref{Ginzburg-Landau
functional} in order to estimate the fluctuation corrections to these
results, to calculate the distribution function of the local order parameter
near $T_{c}$, the local density of states at low temperatures and the
superfluid density.

\subsection{Pairing in the modified mean-field approximation. \label{Pairing
in the modified mean-field}}

The linearized modified mean-field equation for fractal superconductor was
derived in subsection \ref{Modified mean-field}, see Eq.(\ref{MMFA}). It can
be generalized for an arbitrary $T<T_{c}$ where a non-zero gap function $%
\Delta (\xi )$ develops: 
\begin{equation}
\Delta (\xi )=\frac{\lambda }{2}\int d\zeta \frac{M(\xi -\zeta )\Delta
(\zeta )}{\sqrt{\zeta ^{2}+\Delta ^{2}(\zeta )}}\tanh \frac{\sqrt{\zeta
^{2}+\Delta ^{2}(\zeta )}}{2T}  \label{MMFA2}
\end{equation}%
A straightforward calculation shows that near the transition temperature Eq.(%
\ref{MMFA2}) leads to the same result as the Ginzburg-Landau functional with
coefficients found in section \ref{Ginzburg-Landau functional} above. The
properties of the superconducting phase in the vicinity of the transition
temperature can be found from the solution of the linearized integral
equation, equivalent to (\ref{MMFA}): 
\begin{equation}
\Delta (\xi )=\frac{\lambda }{2}\int d\zeta \frac{M(\xi -\zeta )\Delta
(\zeta )}{\zeta }\tanh \frac{\zeta }{2T}  \label{MMFA_linear}
\end{equation}

\begin{figure}[tbp]
\includegraphics[width=8cm]{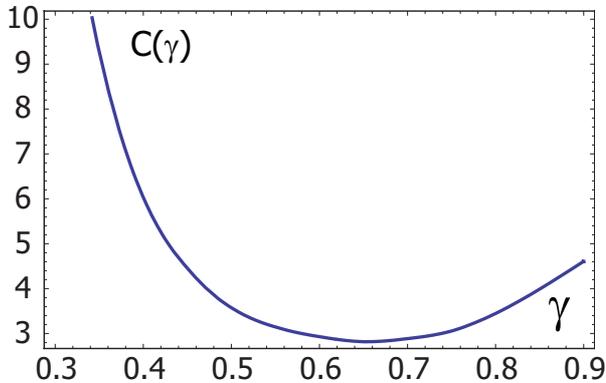}
\caption{Numerical prefactor $C(\protect\gamma )$ appearing in the equation
for the transition temperature (\protect\ref{Tc11}) obtained by numerical
solution of linearized integral equation (\protect\ref{MMFA_linear}).}
\label{Cgamma}
\end{figure}

\begin{figure}[tbp]
\includegraphics[width=6cm]{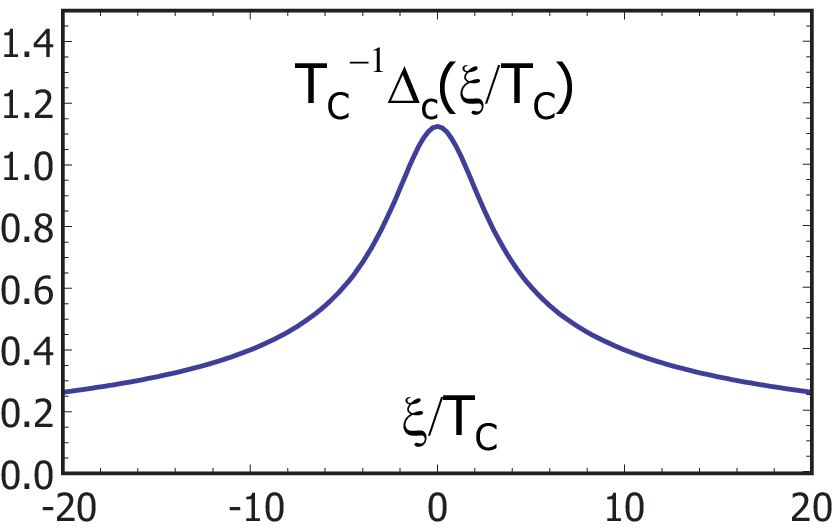} %
\includegraphics[width=6cm]{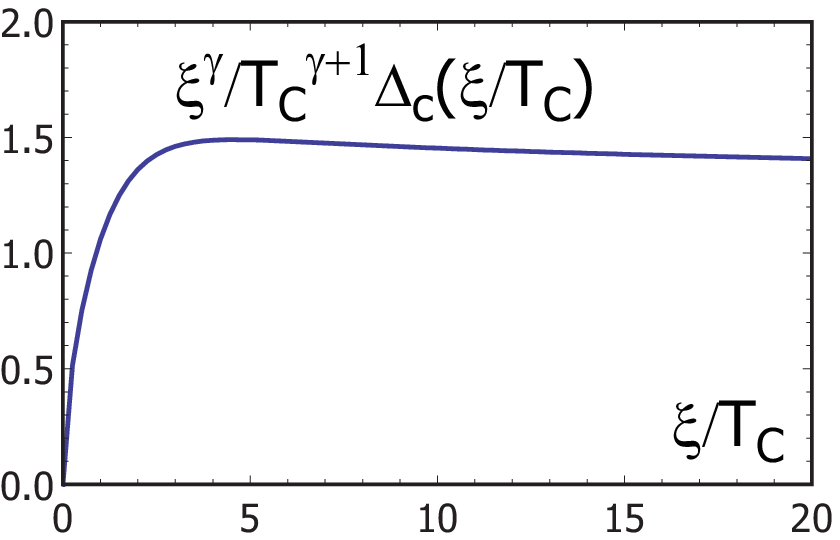}
\caption{(a) Functional dependence of the gap function $\Delta (\protect\xi )
$ for $T=T_{c}$ at $\protect\gamma =0.57$ that corresponds to the largest
eigenvalue of the linearized integral gap equation (\protect\ref{MMFA_linear}%
). (b) Asymptotic behaviour of $\Delta (\protect\xi )$ at large $\protect\xi %
/T$.}
\label{DeltaTc}
\end{figure}

Eqs.(\ref{MMFA2},\ref{MMFA_linear}) with the power-law kernel $M(\omega )$
given by Eq.(\ref{ij}) can be solved numerically. Its solution depends on
the fractal exponent $\gamma $ which controls the power of the kernel. In
particular, the critical temperature is found to be 
\begin{equation}
T_{c}^{0}(\lambda ,\gamma )=E_{0}\lambda ^{1/\gamma }C(\gamma )  \label{Tc11}
\end{equation}%
where the function $C(\gamma )$ is plotted in Fig.\ref{Cgamma}. As was
already mentioned above, the power-law dependence on the interaction
constant $\lambda $, instead of the exponential dependence $e^{-\frac{1}{%
\lambda }}$ in the standard BCS theory, implies a dramatic enhancement of $%
T_{c}$ by disorder if interaction constant $\lambda $ is small. This result
could be important for observation of superconductivity in a system of cold
fermions with weak attraction trapped in a disordered lattice. 
The solution for $\Delta (\xi )$ at $T=T_{c}$ is shown in Fig.\ref{DeltaTc}a
for $\gamma =0.57$ corresponding to the 3D Anderson model. 
At large $\xi \gg T_{c}$ this function decays as $\Delta (\xi )\propto \xi
^{-\gamma }$, as demonstrated in Fig.\ref{DeltaTc}b.

\begin{figure}[tbp]
\includegraphics[width=6cm]{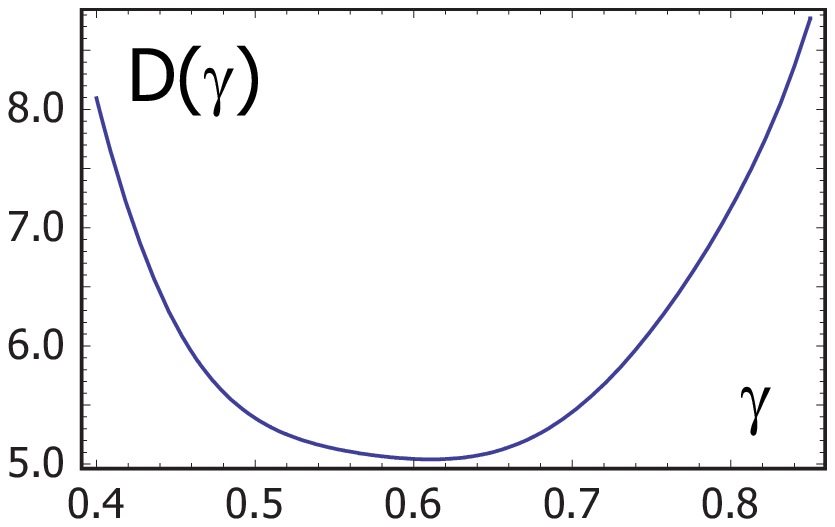}
\includegraphics[width=6cm]{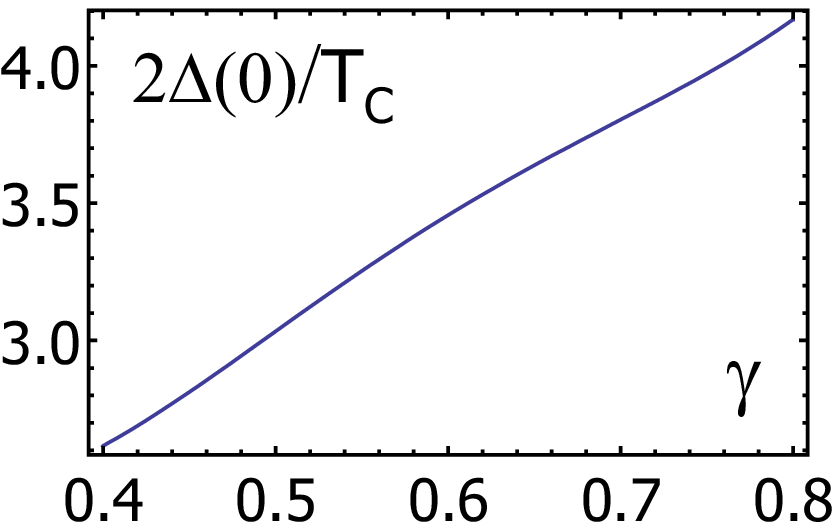}
\caption{(a) Prefactor $D(\protect\gamma )$ in the equation (\protect\ref%
{Delta00}) for the maximum value $\Delta (0)$ of the $T=0$ obtained from the
numerical solution of the non-linear gap equation (\protect\ref{MMFA2}).
(b) Ratio $2\Delta (0)/T_{c}$ that follows from (\ref{Delta00}) with 
$D(\protect\gamma )$ shown here and (\ref{Tc11}) with $C(\protect\gamma )$
shown in Fig.\ref{Cgamma} }
\label{DeltaGamma}
\end{figure}

The maximum value of the function $\Delta (\xi )$ is attained at $\xi=0$
for all temperatures. In the $T\rightarrow 0$ limit we find 
\begin{equation}
\Delta _{0}(0)=E_{0}\lambda ^{1/\gamma }D(\gamma )  \label{Delta00}
\end{equation}%
where the function $D(\gamma )$ is plotted in Fig.\ref{DeltaGamma}a. The
ratio $2D(\gamma )/C(\gamma )$, which characterizes the ratio of
low-temperature spectral gap to the transition temperature, is shown as
function of $\gamma $ in Fig.\ref{DeltaGamma}b. Coincidentally, this ratio for the
3D Anderson model with the fractal exponent $\gamma =0.57$ turns out to be
rather close to the BCS value $3.5$. This implies that unfortunately it is
difficult to distinguish the fractal 3D superconductors from conventional
BCS ones by measuring $2\Delta /T_{c}$ value.

Evident strange feature of $D(\gamma )$ and $2D(\gamma )/C(\gamma )$
behavior is that they do not seem to approach the BCS limit at small $\gamma 
$. The reason is the same as was discussed \ in section \ref{Transition
temperature: coefficient a(T)} around Eq. (\ref{norm2}): in order to get the
correct crossover to the non-fractal limit $\gamma \rightarrow 0$ , one
needs to introduce the upper energy cutoff $\Omega _{D}$. Indeed, the BCS
limit is reached at $\gamma \ll \ln ^{-1}\frac{\Omega _{D}}{T_{c}}\ll 1$
which is never satisfied if the upper energy cutoff is infinite as in the
calculations here.

\subsection{Comparison of $T_{c}$ values obtained in three different
approximations. \label{Comparison of Tc values}}

\begin{figure}[tbp]
\includegraphics[width=8cm]{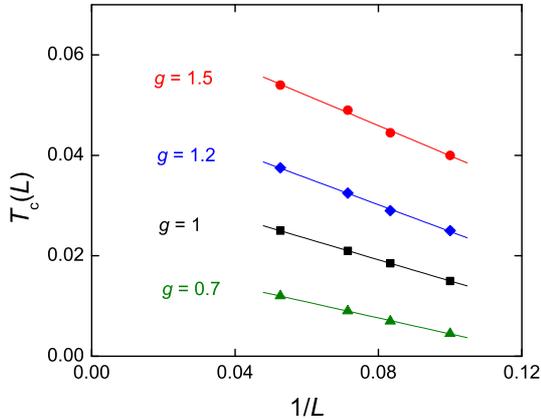}
\caption{Transition temperatures determined via $\protect\rho (k)$ spectrum,
for a number of coupling constants and system sizes between $L=10,12,14$ and 
$19$.}
\label{KvsL}
\end{figure}

General modified MFA equation (\ref{Tc11}), with the parameter $C(\gamma
)\approx 3.1$ evaluated for $\gamma =0.57$, and the estimate (\ref{E00}) for 
$E_{0}$, gives 
\begin{equation}
T_{c}^{0}=(6.5\pm 0.8)\lambda ^{1.77}  \label{Tc12}
\end{equation}%
As we explained above, there are two different types of corrections to this
formula: the corrections caused by the off-diagonal matrix elements in the
original Hamiltonian, and the corrections due to approximations made within
the \textquotedblleft diagonal approximation\textquotedblright , in
particular, the continuum approximation that neglects the fluctuation of
spectrum and the matrix elements which are taken to be a power-law function
of the energy difference. An estimate of the contribution from off-diagonal
matrix elements is given in section \ref{Ginzburg parameters}. In order to
clarify the role of the static fluctuations of the spectrum, we determine
the transition temperature for the same 3D Anderson model using two other
methods. We first determine it from the condition $\lambda _{\max }=1$ (see
section \ref{Modified mean-field}) where $\lambda _{\max }$ is the largest
eigenvalue of the matrix kernel $\hat{Q}$ \ defined by Eq.(\ref{Qmat}) that
contains matrix elements $M_{ij}$ corresponding to the\ specific realization
of disorder\ in finite size samples. This method adapts modified MFA to the
case of strong spatial fluctuations. Second, we apply the virial expansion
method as described in section \ref{Virial expansion}. Because here we
consider the regime $\Delta _{P}\ll T_{c}$, we use the replacement rule Eq.(%
\ref{factor2}) to take account of the dilution of pseudospin states. In both
methods we observe strong finite-size effects as shown in Fig.\ref{KvsL},\ref%
{K1}. 

\begin{figure}[tbp]
\includegraphics[width=8cm]{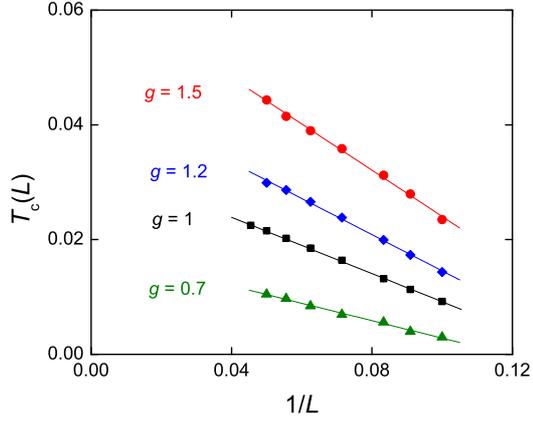}
\caption{Transition temperatures determined via the virial expansion, for a
number of coupling constants and system sizes between $L=10$ and $L=22$.}
\label{K1}
\end{figure}
\begin{figure}[tbp]
\includegraphics[width=8cm]{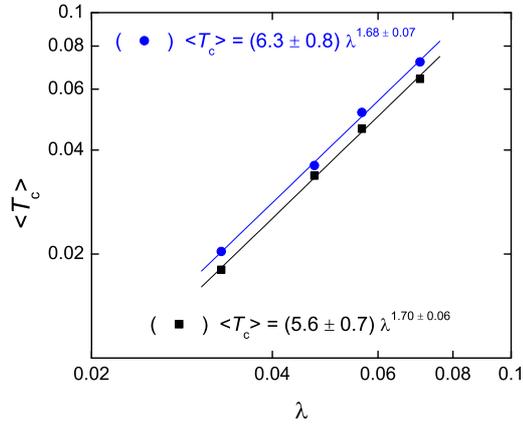}
\caption{(Color online) The results for transition temperature as function
of dimensionless coupling: the $\hat{Q}$-kernel analysis (blue dots) and the
virial expansion (black squares), extrapolated to $L=\infty $; The
Fermi-energy is fixed at the mobility edge.}
\label{VirialTc}
\end{figure}
The extrapolation to infinite system sizes gives the values of $T_{c}$ as
function of the dimensionless coupling constant $\lambda $ shown in Fig.\ref%
{VirialTc}. As one can see, all these methods give the results close to each
other and to the result Eq.(\ref{Tc12}) of the modified MFA at $\gamma =0.57$%
. The conclusion that we draw from this apparent coincidence is that all
approximations involved in these methods are reasonable. Specifically, it
seems that one can neglect the fluctuations of the single-particle DoS and
the matrix elements $M_{ij}$ at the mobility edge; one can also use the
small-$n$ truncated criterion for $T_{c}$ in the virial expansion method.
The main correction to the results of the \textit{analytical} modified MFA
given by Eqs.(\ref{MMFA2}) comes from the off-diagonal matrix elements.
However, even this correction is able to reduce the transition temperature
at most by a factor of the order of unity leaving all the results of
modified MFA \textit{semi-quantitatively correct}.

\subsection{Pairing amplitude in the real space\label{Pairing amplitude}}

The real-space pairing amplitude corresponding to the solution $\Delta _{i}$
of Eq.(\ref{tilde-fin}) can be determined from Eq.(\ref{Deltar}) for $%
T\approx T_{c}$. To demonstrate that $\tilde{\Delta}(\mathbf{r})$
corresponds to strongly spatially inhomogeneous solution, we calculate the
averaged square of the pairing amplitude 
\begin{equation}
\overline{(\tilde{\Delta}(\mathbf{r}))^{2}}\equiv \frac{1}{\mathcal{V}}\int
d^{d}\mathbf{r}\tilde{\Delta}^{2}(\mathbf{r})=\lambda \int_{0}^{\infty }d\xi
\eta (\xi )\Delta _{c}^{2}(\xi )  \label{D2}
\end{equation}%
where we used the definition (\ref{Melements2}) and Eqs.(\ref{tilde-fin},\ref%
{MMFA}) to derive the r.h.s. of the above equation. Then we calculate its
simple average 
\begin{equation}
\overline{\tilde{\Delta}(\mathbf{r})}\equiv \frac{1}{\mathcal{V}}\int d^{d}%
\mathbf{r}\tilde{\Delta}(\mathbf{r})=\lambda \int_{0}^{\infty }d\xi \eta
(\xi )\Delta _{c}(\xi )  \label{D1}
\end{equation}%
>From Eqs.(\ref{D1},\ref{D2}) we conclude that 
\begin{equation}
f=\frac{\left( \overline{\tilde{\Delta}(\mathbf{r})}\right) ^{2}}{\overline{(%
\tilde{\Delta}(\mathbf{r}))^{2}}}=\lambda \mathcal{Q}(\gamma )=\frac{%
\mathcal{Q}(\gamma )}{C^{\gamma }(\gamma )}\left( \frac{T_{c}}{E_{0}}\right)
^{\gamma }\ll 1  \label{D12}
\end{equation}%
where the dimensionless function 
\begin{equation}
\mathcal{Q}(\gamma )=\gamma \left( \int_{0}^{\infty }\frac{d\xi }{\xi }\tanh 
\frac{\xi }{2T_{c}}\frac{\Delta _{c}(\xi )}{T_{c}}\right) ^{2}\,;\quad 
\mathcal{Q}(0.57)=3.1  \label{Q}
\end{equation}%
The small ratio (\ref{D12}) gives the estimate of the space fraction where
pairing correlations are well established. Indeed, consider a toy model
where $\Delta (\mathbf{r})=\Delta _{0}$ in a fraction $f$ of the whole
space, and zero otherwise. Then we find $\overline{\tilde{\Delta}(\mathbf{r})%
}=f\Delta _{0}$ and $\overline{(\tilde{\Delta}(\mathbf{r}))^{2}}=f\Delta
_{0}^{2}$, leading to the ratio (\ref{D12}) equal to $f$. \ In a realistic
case $\Delta (\mathbf{r})$ take a continuum of values, so one can use this
ratio as a proper \textit{definition} of the fraction of space with well
developed superconductive gap.

Note that the regions in space where the superconductive gap is appreciable
constitute a \textit{finite} fraction of the entire space, despite the fact
that fractal support of any \textit{single-particle} wavefunction occupies 
\textit{vanishing} fraction of the entire space. Therefore the global
spatial pattern of superconductivity is not a fractal but rather is
reminiscent of that of the \textit{multi-fractal metal} \cite{CueKra}.

The above estimate was done for a region near $T_{c}$. At temperatures much
below $T_{c}$, (\ref{Deltar}) should be replaced by the similar equation in
which the argument $\xi _{k}$ is replaced by $\sqrt{\xi _{k}^{2}+\Delta
_{k}^{2}}$. As one might expect, one ends up with the same estimates in
which $T_{c}$ gets replaced by $\Delta _{0}(0)$.

Higher moments of the distribution function $P[\tilde{\Delta}(\mathbf{r})]$
can be estimated using the algebra of multi-fractal states discussed in
section \ref{Fractality and correlations}, see Eq.(\ref{multi-ij}).
Straightforward generalization of Eq.(\ref{D12}) gives 
\begin{equation}
\frac{\left( \overline{\tilde{\Delta}(\mathbf{r})}\right) ^{n}}{\overline{(%
\tilde{\Delta}(\mathbf{r}))^{n}}}\propto \left( T_{c}/E_{0}\right)
^{(1-d_{n}/d)(n-1)}  \label{D13}
\end{equation}%
Therefore the moments of $\tilde{\Delta}(\mathbf{r})$ contain information
about all fractal dimensions $d_{n}$. This is because the fractality of
single-particle states manifests itself \textit{locally} at a scale smaller
than $L_{T}\sim (\nu _{0}T_{c})^{-1/d}$ , above this length the order
parameter and other properties of the superconductor become homogeneous,
similar to what happens in a multi-fractal metal \cite{CueKra} which has a
sparse fractal structure within the correlation length. Thus, this state
should be properly named\textit{\ fractal superconductor}.

\subsection{Low-temperature density of states. \label{Low-temperature
density of states}}

Local density of states (DoS)\ in the superconducting state is given by 
\begin{equation}
\nu (\varepsilon ,\mathbf{r})=\frac{1}{2}\sum_{j}\left( 1+\frac{\xi _{j}}{%
\varepsilon }\right) \left[ \delta (\varepsilon -\varepsilon _{j})+\delta
(\varepsilon +\varepsilon _{j})\right] \psi _{j}^{2}(\mathbf{r})
\label{nu00}
\end{equation}%
where $\varepsilon _{j}=\sqrt{\xi _{j}^{2}+\Delta ^{2}(\xi _{j})}$ and $%
\Delta (\xi )$ is the solution of the gap equation (\ref{MMFA2}). We begin
with the calculation of the average DoS: 
\begin{equation}
\nu (\varepsilon )=\nu _{0}\left\vert \frac{d\xi (\varepsilon )}{%
d\varepsilon }\right\vert  \label{nu01}
\end{equation}%
where the function $\xi (\varepsilon )$ is defined as the (positive-valued)
inverse function for $\varepsilon (\xi )=\sqrt{\xi ^{2}+\Delta ^{2}(\xi )}$.

\begin{figure}[tbp]
\includegraphics[width=8cm]{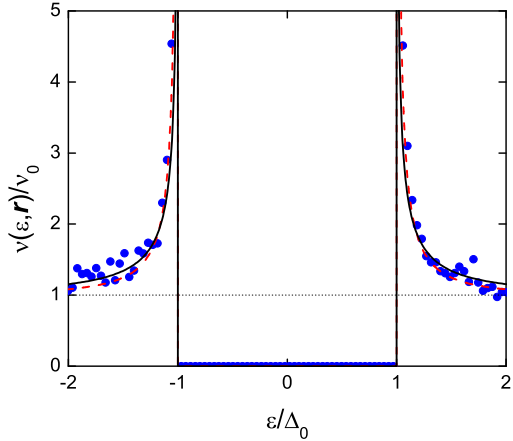}
\caption{Local zero-temperature density of states at the mobility edge of
the 3D Anderson model, averaged over 5x5 square (blue dots); average $T=0$
superconductive DoS at the mobility edge, according to Eq.(\protect\ref{nu01}%
), shown by dashed red line; the BCS density of states with the same gap
value (full black line). }
\label{DoS01}
\end{figure}

We plot in Fig.\ref{DoS01} the local DoS (at $T=0$) obtained numerically
using Eq.(\ref{nu00}), as well as the average DoS obtained from Eq.(\ref%
{nu01}) for $\gamma=0.57$; the usual BCS DoS is shown for comparison. 

Note that the terms $\xi _{j}/\varepsilon $ in the parenthesis in Eq.(\ref%
{nu00}) were irrelevant while calculating the average DoS because
contributions from $\xi =\pm \xi (\varepsilon )$ cancel out. There is no
such cancellation for the higher moments of $\nu (\varepsilon ,\mathbf{r})$.
In particular, these terms might lead to an asymmetry of local DoS: $\nu
(\varepsilon ,\mathbf{r})\neq \nu (-\varepsilon ,\mathbf{r})$. To quantify
this asymmetry we define the antisymmetric part of the DoS: $\nu
_{a}(\varepsilon ,\mathbf{r})=\frac{1}{2}(\nu (\varepsilon ,\mathbf{r})-\nu
(-\varepsilon ,\mathbf{r})$. Using then Eqs.(\ref{nu00},\ref{nu01}), we find
the variance 
\begin{equation}
\overline{\nu _{a}^{2}(\varepsilon ,\mathbf{r})}=\frac{1}{2}\left( \nu
(\varepsilon )\frac{\xi (\varepsilon )}{\varepsilon }\right) ^{2}\left[
M(0)-M(2\xi (\varepsilon ))\right]  \label{nu03}
\end{equation}%
where the function $M(\xi )$ is defined in Eq.(\ref{ij}). Since $M(0)\sim 
\mathcal{V}^{\gamma }\gg M(2\xi )$, the variance (\ref{nu03}) diverges for
an infinite system, and one should consider the distribution function $%
W[\rho (\varepsilon ,\mathbf{r})]$ for the dimensionless variable 
\begin{equation}
\rho (\varepsilon ,\mathbf{r})=\frac{\nu _{a}(\varepsilon ,\mathbf{r})}{\nu
(\varepsilon )}\frac{\sqrt{2}\varepsilon }{\xi (\varepsilon )}  \label{nu04}
\end{equation}%
The distribution function $W[\rho (\varepsilon )]$ coincides with that of
the wavefunction's intensities $\mathcal{P}(\psi ^{2})$, see Eq.(2.33) of
the review~\cite{MirlinNewRep}. It is determined by the "singularity
spectrum" $f(\alpha )$; the shape of this function for the 3D Anderson model
is similar to the one found for the power-law banded matrix model~\cite%
{PRBM,KrMut1997} with the parameter $b\approx 0.4$, see Fig.3 of paper\cite%
{MirlinNewRep}.

It is clear from the definition (\ref{nu04}) that the normalized asymmetric
fluctuations $\nu _{a}(\varepsilon ,\mathbf{r})/\nu (\varepsilon )$ are
small for $\varepsilon \approx \Delta (0)$, where $\xi (\varepsilon
)\rightarrow 0$. However, for a generic value of the energy of the order of $%
\Delta (0)$, the asymmetry in the tunneling spectra is strong; we emphasize
that it occurs due to mesoscopic fluctuations which do not involve any
"regular" mechanism of particle-hole asymmetry.

\subsection{Superfluid density and critical current. \label{Superfluid
density and critical current}}

We define the superfluid density $\rho_s$ via the relation $\mathbf{j} =
-\rho_{s}\mathbf{A}/c$ assuming the transverse gauge with $\mathbf{%
\nabla\cdot A}=0$. To estimate $\rho_s(T)$ near $T_c$, one can use the
expression (\ref{FGL}) for the free energy functional, and make there the
replacement $\nabla \to \nabla - i(2e/\hbar c)\mathbf{A}$. Then taking the
double functional derivative $\delta^2 F_{GL}/\delta \mathbf{A}^2$ one
obtains: 
\begin{equation}
\rho_s(T) = 2\left(\frac{2e}{\hbar}\right)^2\,\nu_0 T_c^2\, C\, \Psi^2(T)
\label{rhos01}
\end{equation}
where $\Psi(T)=\Psi_{\mathrm{MFA}}^2 = \frac{|a(T)|}{b} = 9.5 \left(1-\frac{T%
}{T_c}\right)$ and the coefficient $C \sim L^2_{T}$ is given by Eq.(\ref{CC}%
).

We also estimate the critical current $j_{c}$ in a standard way via the
Ginzburg-Landau functional (\ref{FGL}). By extrapolating the result of
Ginzburg-Landau theory to lower temperatures, one obtains 
\begin{equation}
j_{c}=c\frac{2e}{\hbar }T_{c}(\nu _{0}T_{c})^{2/3}  \label{Jc}
\end{equation}

Extrapolation (\ref{Jc}) of the Ginzburg-Landau result to $T=0$ might be
wrong by a factor of the order of unity, this factor for the fractal
superconductor might be different from the known factor for a conventional
"dirty-limit" superconductor.

The results Eqs.(\ref{rhos01}),(\ref{Jc}) do not contain any
\textquotedblleft fractal\textquotedblright\ specificity and coincide with
estimates of earlier works\cite{MaLee,KapitulnikKotliar1986} that used
scaling arguments. Yet, as we have demonstrated above, the web of
superconductivity is sparse and occupies only a small fraction $f\sim
\lambda \ll 1$ of space. The derivation of Eq.(\ref{CC}) demonstrates that
the fractality-independent result for the coefficient $C$ emerges because of
the compensation of the small parameter $\lambda ^{2}$ by the enhancement
factor $(E_{0}/T_{c})^{2\gamma }$. We interpret this cancellation as the
compensation of the small cross-section area of "superconductive filaments"
by a large current that they support.

\section{Superconductivity with a pseudogap. \label{Superconductivity with a
pseudogap}}

In this section we study the superconductivity formed against the background
of moderately localized single-electron eigenstates. We demonstrate here the
existence of the interesting regime where the local pairing (discussed above
in section \ref{Insulating state} in relation with the hard-gap insulator)
coexists with the long-range superconducting correlations. In this situation
the material demonstrates the "pseudo-gap" phenomenology characterized by
the insulating behavior of $R(T)$ in a significant temperature range above
the superconducting transition. We provide a semi-quantitative description
for the evolution of the pseudo-gap $\Delta _{P}$ and $T_{c}$ while $E_{F}$
moves across the mobility edge $E_{c}$. Surprisingly, we find that $%
T_{c}(E_{F})$ curve is non-monotonic, with a maximum reached at the
Fermi-level on the localized side of the Anderson transition. The most
important outcome of our analysis is the existence of a range of Fermi
energies for which superconductor is characterized by relatively high values
of $T_{c}$ and by a \textit{larger} insulating pseudo-gap $\Delta _{P}\gg
T_{c}$. In this regime the low energy properties such as the formation of
the superconductivity can be described by the truncated pseudo-spin model.
In this model the effective Hamiltonian is of the form (\ref{HamSpin}) while
the Hilbert space contains only the states in which each localized orbital
is either empty or populated by a pair of electrons. The effective
Hamiltonian (\ref{HamSpin}) does not contain $\Delta _{P}$ because the
states that differ in energy by $\Delta _{P}$ are absent from the truncated
Hilbert space. These states, however, appear when single particles are
excited, thus the single particle excitations spectrum contains both
correlations and local pairing effects. Below we discuss the region of
validity of the mean-field treatment of the pseudo-spin Hamiltonian (\ref%
{HamSpin}) and then turn to analysis of several measurable quantities which
may provide one with the \textquotedblleft proof of the
case\textquotedblright\ for realization of the pseudo-gap superconductivity
in real material: the temperature evolution of the local single-particle
density of states $\nu (\varepsilon ,T)$, the Andreev conductance spectra $%
G_{A}(V)$ at $T\ll T_{c}$, and the temperature dependence of the full
spectral weight $K(T)$.

\subsection{Transition temperature and insulating gap as functions of Fermi
energy. \label{Transition temperature}}

We demonstrated in section \ref{Comparison of Tc values} that the transition
temperature found by the virial expansion to the third order provides
results very close to the ones obtained with the modified MFA. These
calculations were performed for the critical case $E_{F}=E_{c}$ where the
function $M(\omega )$ is given by the power law (\ref{ij}). This coincidence
implies that the virial expansion to the third order provide the accurate
result for the transition temperature. Generally, both methods suffer from
the neglect of the off-diagonal matrix elements, these matrix elements mix
the odd and even fermionic sectors which might decrease the $T_{c}$ below
the MFA value by a factor of the order of unity (which would be consistent
with the $\mathrm{Gi}\sim 1$ obtained for the critical case in section \ref%
{Ginzburg parameters}). While there is no general reasons why the
off-diagonal matrix elements can be completely neglected in the metallic and
in the critical region, in the region of well localized single-particle
states the validity of the spin Hamiltonian (\ref{HamSpin}) is ensured by
the large value of the pseudogap $\Delta _{P}$. One thus expects that
pseudospin model provides very good description of the pseudogap state and a
semi-quantitative description of the crossover from the pseudogap state
(where it is exact) to the fractal superconductor (where its results might
differ by a factor of the order of unity). Because the superconductivity
suppression deep in the insulating regime is due to the decrease in the
effective number of neighbors of each pseudospin and this effect is missing
in the modified MFA we shall use the virial expansion applied to the spin
Hamiltonian (\ref{HamSpin}) to obtain the properties in both the crossover
and pseudogap regime.

We begin by applying this method to describe the evolution of $T_{c}$ as
Fermi-level crosses the mobility edge $E_{c}$ and gets into the region of
weakly localized states. In this regime, assuming that single electron
states are characterized by small participation ratio so that corresponding $%
\Delta _{P}<T_{c}$, we can use the simplified version of the virial approach
with the distribution function $\tanh \frac{E_{i}}{T}$ replaced by $\tanh 
\frac{E_{i}}{2T}$, as it was done for critical states at $E_{F}=E_{C}$ in
Sec. V. 
\begin{figure}[tbp]
\includegraphics[width=8cm]{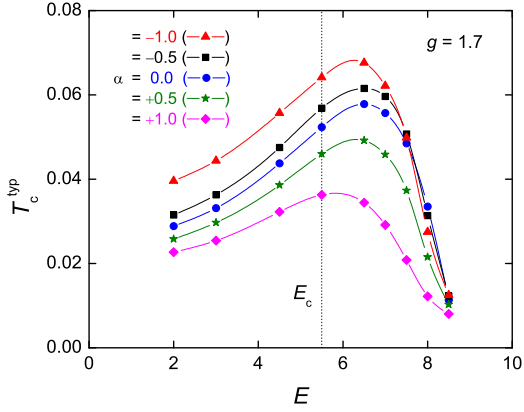}
\caption{(Color online) Transition temperature as a function of the
Fermi-level position, determined by the virial expansion method. Blue dots
correspond to the pseudo-spin Hamiltonian Eq.(\protect\ref{HamSpin}) with $%
g_{\parallel }=0$, whereas other data were obtained after including the
Hartree-type terms proportional to $g_{\parallel }$ of repulsive (red
triangles and black squares) and attractive (green stars and violet
diamonds) sign (see text for details). All calculations were done at a fixed
coupling constant $g_{\perp }=g$.}
\label{TcVersusE}
\end{figure}
The numerical results for the minimal pseudo-spin Hamiltonian Eq.(\ref%
{HamSpin}) with $g_{\parallel }=0$ and coupling matrix $M_{ij}$ generated by
random 3D Anderson model Eq.(\ref{AM}) are shown in Fig.\ref{TcVersusE} by
blue points. They correspond to the fixed coupling constant $g=1.7$ in the
original model and, thus, \textit{dimensionless} coupling constant $\lambda
=g\nu _{0}$ varying as function of $E_{F}$. The energy dependence of density
of states is not negligible around $E_{c}$ as can be seen in Fig.\ref%
{FigDoS1}, as a result $\lambda (E_{F})$ drops considerably in the range of
Fermi-energies covered in Fig.~\ref{TcVersusE}. Nevertheless, the
calculation not only confirms our analytical prediction of the enhancement
of $T_{c}$ near the critical disorder made in section \ref{Modified
mean-field} but also shows that $T_{c}$ continues to increase in a
significant interval of Fermi energies corresponding to the localized
region. This surprising result can be traced back to the behavior of the
correlation function $M(\omega )=\mathcal{V}M_{ij}$ in the domain of weakly
localized states. Indeed, it was found previously (see the discussion at the
end of section VIII B of Ref.~\cite{CueKra}) that the effective decrease of
fractal dimension $d_{2}$, and, thus, the increase of $\gamma =1-d_{2}/d$
occurs while going further into the insulating domain. For small values of $%
\lambda $ it results in the increase of $T_{c}\propto \lambda ^{1/\gamma }$
evident in Fig.\ref{TcVersusE} in the range of $E_{c}<E<6.5$. Note that in
contrast to Fig.\ref{VirialTc} the Fig.\ref{TcVersusE} shows the data for
the fixed value of system size, $L=20$, with no finite-size scaling
adjustment. This is why the values of $T_{c}$ data for the critical states, $%
E=5.5$, differ slightly between these two data sets.

The generic pseudo-spin Hamiltonian Eq.(\ref{HamSpin}) contains also the
Hartree terms 
\begin{equation}
H_{\mathrm{Hartree}}=-g_{\parallel }\sum_{ij}M_{ij}S_{i}^{z}S_{j}^{z}.
\label{HamZZ}
\end{equation}%
Moreover, for the short-range $\delta $-\textit{function} two-body
interaction, the coupling constant $g_{\parallel }$ is equal to the
Cooper-channel coupling constant $g_{\perp }$ and, thus, the terms
containing the spin-spin interaction in pseudo-spin Hamiltonian (\ref%
{HamSpin}) become isotropic. Because the full Hamiltonian (\ref{HamSpin})
remains highly anisotropic due to the presence of the single spin terms, the
effect of the $S_{i}^{+}S_{j}^{-}$ and $S_{i}^{z}S_{j}^{z}$ spin-spin
interactions is dramatically different: the former leads to the
superconducting instability at low temperatures, while the latter term by
itself has only weak effect of spin susceptibility for small $g_{\parallel
}\nu $. \ Because the transition temperature depends strongly on the spin
susceptibility, in the presence of the $S_{i}^{+}S_{j}^{-}$ the Hatree term
can have slightly more significant effect. Below we study the change in the
critical temperatures induced by the $S_{i}^{z}S_{j}^{z}$ interaction for a
generic ratio $\alpha =g_{\parallel }/g_{\perp }$ using the virial expansion
method.

The results are shown in Fig.~\ref{TcVersusE} by green and violet points for 
$\alpha = +0.5$ and $\alpha = +1$, and by red and black points for $\alpha =
- 1$ and $\alpha = - 0.5$, respectively. We conclude that upon the account
for the Hartree terms, all the qualitative features of the original $%
T_c(E_F) $ behavior, including the enhancement of $T_{c}$ by disorder, are
preserved. Therefore below we stick to the simplest version of our model
with $\alpha=0$.

\begin{figure}[tbp]
\includegraphics[width=8cm]{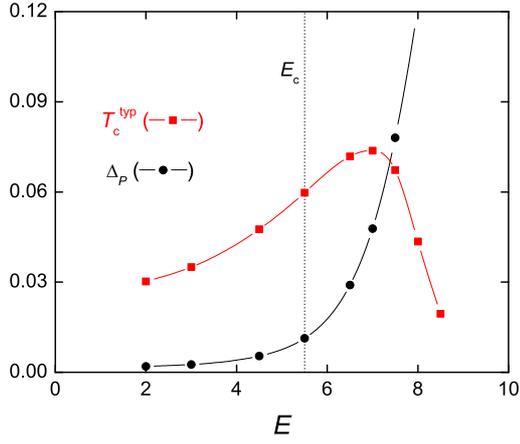}
\caption{(Color online) Virial expansion results for $T_{c}(E_{F})$
\thinspace\ (red squares) and typical pseudogap $\Delta _{P}$ \thinspace\
(black dots) as functions of $E_{F}$. The model with fixed value of the
attraction coupling constant $g=1.7$ was used; pairing susceptibilities were
calculated using equations derived in Appendix B.}
\label{TcDeltaP}
\end{figure}

With a further shift of $E_{F}$ into the localized domain, the two effects
become important. First of them is the growth of the inverse participation
ratio $M_{j}$, and the related development of the parity gap $\Delta _{P}$
\thinspace , see Eq.(\ref{DeltaP}). In some range of $E_{F}$, as we will see
soon, the typical value of $\Delta _{P}$ \thinspace\ becomes larger than $%
T_{c}$, which means that the calculations of the virial coefficients should
be modified, taking into account \textit{exactly} the effect of "dilution"
by the odd states discussed in section \ref{Virial expansion}. The necessary
modification of the formalism is presented in the \ref{Virial expansion
including single-occupied states}. The dilution elimination by the
pseudo-gap leads to the increase of $T_{c}$ values (as compared to the data
for $\Delta _{P}=0$ shown in Fig.\ref{TcVersusE}, where the maximum effect
of dilution was assumed). However,\ as one can see in Fig.~\ref{TcDeltaP} in
the region where superconductivity exists this increase is not more than
30\%. 

The second effect, which takes over sufficiently far in the localized
region, is due to the drop of the effective coordination number $Z_{\mathrm{%
eff}}$ of the spin model defined by Eq.(\ref{HamSpin}). Indeed, the
effective number of states $k$ that are coupled to a given state $j$ can be
roughly estimated by $Z\sim \nu _{0}T_{c}L_{loc}^{3}$ (this estimate misses
the important logarithmic factor that we discuss below around Eq.(\ref{Zeff}%
) ). With decreasing of the localization length $L_{loc}$, the coordination
number $Z_{\mathrm{eff}}$ drops down, and eventually becomes less than
unity, which makes description in terms of average matrix elements $\mathcal{%
V}\overline{M_{ij}}=M(\omega )$ meaningless. When the Fermi energy is
increased further, $T_{c}$ starts to drop down sharply simply because most
individual pseudospins (with a possible exception of rare small spin
clusters) become essentially decoupled from each other. 
\begin{figure}[tbp]
\includegraphics[width=8cm]{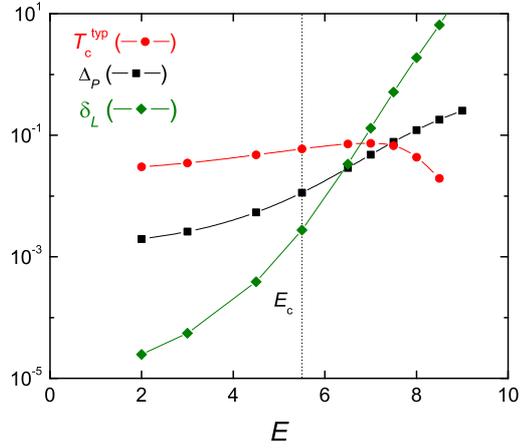}
\caption{(Color online) Virial expansion results for $T_{c}(E_{F})$ (red
dots), the typical pseudo-gap $\Delta _{P}$ (black squares) and the
corresponding level spacing $\protect\delta _{L}$ (green diamonds), as
functions of $E_{F}$ in the semi-logarithmic scale.}
\label{TcDeltaPdeltaL}
\end{figure}

Evolution of both $T_{c}$ and $\Delta _{P}$ with the Fermi-level position
are shown in Fig.\ref{TcDeltaP}, where we show the results obtained by the
virial expansion method, with a finite pseudo-gap $\Delta _{P}$ taken into
account. The most important feature seen in this plot is the existence of
some range of $E_{F}$ where superconductivity with an appreciable $T_{c}$
exists, but the parity gap $\Delta _{P}$ exceeds substantially the
superconductive gap $\Delta \approx 1.7T_{c}$. Below we concentrate
specifically on this range of parameters (approximately, it corresponds to
the range $E_{F}\in (7.5,8.5)$ for the 3D Anderson model studied here),
where the specific features of superconductivity developed on top of the
pseudo-gap are most pronounced. This is precisely the region of the \textit{%
pseudo-gap superconductivity} where the pseudo-spin model Eq.(\ref{HamSpin})
provides a quantitatively correct description of superconductivity.

Note that in this range of parameters the typical level spacing $\delta _{L}$
\thinspace\ inside the localization volume (estimated with the use of Eq.(%
\ref{d-L-IPR})) is much larger than both $T_{c}$ and $\Delta _{P}$, as it is
demonstrated in Fig.\ref{TcDeltaPdeltaL}. The very existence of a nonzero $%
T_{c}$ in this situation is unexpected. \footnote{%
In contrast, superconducting correlations in ultrasmall metall grain are
destroyed at $\delta \sim \Delta $, see discussion in section \ref%
{Ultra-small metallic grain}.} It is related to the fact (noticed in Ref.~%
\cite{CueKra} and discussed in section \ref{Matrix elements}) that the
localized states characterized by the localization length $L_{\mathrm{loc}}$
overlap effectively up to the distance $R({\omega })\gg L_{\mathrm{loc}}$
between them, where $\omega \ll \delta _{L}$ is the relevant energy window $%
(-T_{c},T_{c})$. In the present context we can put $\omega \sim T_{c}$ and
estimate the effective interaction range $R_{0}$ and coordination number $Z_{%
\mathrm{eff}}$ of the spin model (\ref{HamSpin}) using the definition in Eq.(%
\ref{R-X}) in subsection \ref{Matrix elements}, see also Eqs.(21,25) in the
paper~\cite{CueKra}. We obtain 
\begin{equation}
Z_{\mathrm{eff}}=\nu _{0}\,\frac{4\pi }{3}R_{0}^{3}(T_{c})\cdot 2T_{c}
\label{Zeff}
\end{equation}%
where the function $R_{0}(\omega )$ should be determined numerically. Using
data shown in Fig.~\ref{TcDeltaPdeltaL} for the case of $E_{F}=8$ and similar data
for several other values of $E_{F}$ in the pseudogap range, where $T_{c}\ll
\delta _{L}$, we obtain the following values for the effective number of
neighbors in the energy window $(-T_{c},T_{c})$: 
\begin{equation}
\begin{tabular}{|l|l|l|l|l|}
\hline
$E_{F}$  & 7.75 & 8.0 & 8.25 & 8.5\\ \hline
$T_{c}$   & 0.057 & 0.044 & 0.030 & 0.02 \\ \hline
$\delta _{L}$ & 1.0 & 1.9 & 3.5 &  6.5  \\ \hline
$Z_{\mathrm{eff}}$    & 0.08 & 0.09 & 0.09 & 0.07  \\ \hline
\end{tabular}
\label{ZeffTable}
\end{equation}%
Of course, the numerical coefficient in Eq.(\ref{Zeff}) is somewhat
arbitrary; the main conclusion from Eq.(\ref{Zeff}) is that (due to a
presence of the logarithmic enhancement in $R_{0}(\omega )$) the effective
number of neighbors remains nearly constant in a wide range of small $%
T_{c}/\delta _{L}\ll 1$. Together with a weak $T_{c}/\delta _{L}$ dependence
this explains why a significant $T_{c}$ is found even for $\delta
_{L}/T_{c}\sim 300$, see Fig.~\ref{TcDeltaPdeltaL}.

As we explain in more detail below, the possibility to find a pseudogaped
superconductive state with a numerically small effective coordination number
($Z_{\mathrm{eff}}\sim 0.1$) is natural. The qualitative reason is
that our definition of $Z_{\mathrm{eff}}$ implies average over all states.
Therefore, this average includes the states that overlap strongly with each
other and the ones which overlap little with any other states. The former
states form an infinite superconducting cluster while the latter do not contribute to
superconductivity and are largerly irrelevant. In the strongly disordered
regime the superconducting cluster occupies a small part of all states, so
the average is dominated by the states which overlap very little with each
other.

\begin{figure}[tph]
\includegraphics[width=8cm]{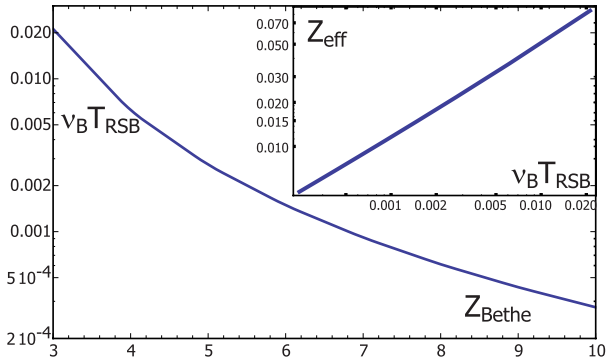}
\caption{Temperature separating high and low temperature regimes of the
simplified pseudospin model (\protect\ref{HamSpin}) on Bethe lattice with $%
Z_{\mathbf{Bethe}}$ neighbors, for which $M_{ij}=1/Z_{\mathbf{Bethe}}$ and
density of states per site $\protect\nu _{B}$. In the high temperature
regime $T>T_{RSB}$ the mean field theory is semi-quantitatively correct, in
the low temperature regime it is qualitatively wrong; the value of $T_{RSB}$
does not depend upon coupling constant $g$ within the Bethe lattice model.
The inset shows the effective number of neighbors in this model defined by $%
Z_{\mathrm{eff}}^{\mathrm{Bethe}}=\protect\nu _{B}T_{c0}Z_{\mathrm{Bethe}}$
at which the mean field solution becomes qualitatively incorrect. }
\label{TRSB}
\end{figure}

Upon a further increase of the Fermi energy into the localized region the
effective number of neighbors slowly decreases leading to a sharp drop in $%
T_{c}^{\mathrm{typ}}(E_{F})$, see Fig.~\ref{TcDeltaP}. Finally, when the
effective number of neighbors becomes less than some critical value $Z_{c}$,
the superconducting instability ceases to occur even at $T=0$. The details
of this quantum transition and the qualitative properties of the phase
formed at high disorder are discussed in \cite{IoffeMezard2009}. This work
solves the simplified model similar to (\ref{HamSpin}) in which spin are
located of a Bethe lattice with coordination number $Z^{\mathrm{Bethe}}$,
i.e. in this model $M_{ij}=1/Z^{\mathrm{Bethe}}$ for $Z^{\mathrm{Bethe}}$
neighbors and $0$ otherwise. \footnote{%
Notice that this definition of the number of neighbors differs a lot from
the (\ref{Zeff}) that counts the neighbors {\it only} in the narrow energy
interval of $T_{c}$.} The result of the solution \cite{IoffeMezard2009}
which is important for the present discussion is that critical fluctuations
become important only in the narrow vicinity of the quantum critical point
at which superconductivity disappears completely. In other words, the
critical fluctuations become important only in the systems in which
transition temperature is already strongly suppressed by disorder. In more
detail, the work \cite{IoffeMezard2009} distinguishes between the high
temperature phase where corrections to mean field are small and a low
temperature phase where mean field solution becomes qualitatively incorrect
due to the strong inhomogeneity of the formed ordered phase. The temperature
($T_{\mathrm{RSB}}$) that separates them corresponds to replica symmetry
breaking (RSB) in the formalism, it turns out to be numerically small even
for modest number of neighbors as illustrated in Fig. \ref{TRSB}. Viewed
differently, it means that the effective number of neighbors in this regime
is very low (see Fig. \ref{TRSB} insert).

As long as the transition temperature obtained in the mean field
approximation is above $T_{\mathrm{RSB}}$ shown in Fig. \ref{TRSB}, the mean
field equations are qualitatively correct. Thus, one can use mean-field
approximation for the Hamiltonian (\ref{HamSpin}) to get semi-quantitative
results for the transition temperature in a large interval of Fermi energies
at which $T_{c}(E_{F})$ decreases significantly from its maximum.

The solution of the Bethe lattice model~\cite{IoffeMezard2009} can be used
to estimate critical value of the effective coordination number $Z_{\mathrm{%
eff}}^{\mathrm{Bethe}}=\nu _{B}T_{c0}$ $Z_{\mathrm{RSB}}^{\mathrm{Bethe}}$
which corresponds to the value $Z_{\mathrm{RSB}}^{\mathrm{Bethe}}$ of the
Bethe-lattice coordination number where $T_{c}=T_{RSB}$ and standard
mean-field solution becomes qualitatively incorrect. The details of the
corresponding calculation will be presented in the extension~\cite%
{FeigelIoffeMezard} of the paper~\cite{IoffeMezard2009}; the result, valid
in the limit $Z_{\mathrm{Bethe}}\gg 1$, is 
\begin{equation}
Z_{\mathrm{eff}}^{\mathrm{Bethe}}=\lambda _{b}e^{1/2\lambda _{b}}\nu
_{B}T_{c0}=\frac{1}{2}\sqrt{\pi }e^{{-\mathrm{\mathbf{C}}}/2}\frac{\sqrt{\nu
_{B}T_{\mathrm{RSB}}}}{|\log (\nu _{B}T_{\mathrm{RSB}})|}\ll 1
\label{ZBethe}
\end{equation}%
where $\lambda _{b}=\nu _{B}g_{B}\ll 1$ is the dimensionless coupling
constant for the Bethe lattice model, with $\nu _{B}T_{c0}=\frac{4}{\pi }e^{%
\mathrm{\mathbf{C}}}e^{-1/\lambda _{b}}$, and $\mathrm{\mathbf{C}}=0.577..$
is the Euler number. Eq.(\ref{ZBethe}) demonstrates again, that typical
number of neighbors with energies within the energy stripe $\sim T_{c}$ can
be very small in the weak-coupling limit, contrary to naive expectations.
Physically, the reason for this is that in this regime a dilute
superconducting cluster is formed by a small number of sites. One expects
that similar phenomena should happen in physical systems. Although our
fractal model differs significantly from the Bethe lattice model due to
power-law dependence of the interaction strength $M(\omega )$, we expect
that qualitative conclusion about $Z_{\mathrm{eff}}^{\mathrm{min}}<1$
survives in the fractal model as well.

The significant number of neighbors that persists up to numerically large
values of $\delta _{L}/T_{c}\sim 300$ implies that Ginzburg parameter
remains of the order of unity in this range. \footnote{%
This does not cotradict our findings in section \ref{Ginzburg parameters}
that Ginzburg parameter becomes $\mathrm{Gi}\sim O(1)$ already for the
fractal superconductor with the Fermi energy at the mobility edge. It only
implies that $\mathrm{Gi}$ grows very slowly and remains $O(1)$ upon a
further increase of the Fermi energy untill the the transition temperature
becomes small. The reason for it is the appearance of a large pseudogap $%
\Delta _{P}\gg T_{c}$ that suppresses the fluctuational processes that
involve off-diagonal terms $M_{ijkl}$.} In this regime the gap below $T_{c}$
can be semi-quantitatively approximated by the solution of the mean field
equation: 
\begin{equation}
\Delta (\xi )=\frac{\lambda }{2}\int d\zeta \frac{M(\xi -\zeta )\Delta
(\zeta )}{\sqrt{\zeta ^{2}+\Delta ^{2}(\zeta )}}\tanh \frac{\sqrt{\zeta
^{2}+\Delta ^{2}(\zeta )}}{T}  \label{MMFA5}
\end{equation}%
where we used the fact $\Delta _{P}\gg T_{c}$ to replace $2T\rightarrow T$
and\ $M(\omega )$ is given, for localized states, by interpolating formula,
Eq.(\ref{interp}).

\begin{figure}[tbp]
\includegraphics[width=8cm]{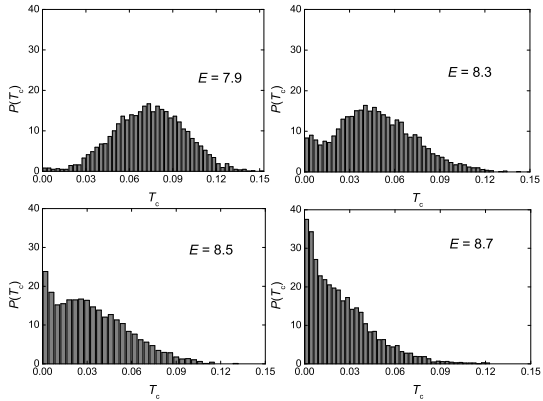}
\caption{Distribution functions $P(T_{c})$ obtained by the virial expansion
for different Fermi energies. All values of $T_{c}$ were obtained in the
pseudo-spin approximation $\Delta _{P}\gg T_{c}$, which leads to some
overestimation of $T_{c}$'s at the Fermi energy $E_{F}=7.9$ where $\Delta
_{P}/T_{c}$ is not very large. The bi-modal character of the distribution
for $8.2<E_{F}<8.7$ indicates on the percolation character of the
superconductive transition.}
\label{TcDistribution}
\end{figure}

One of basic results of weak-coupling BCS theory is the universal relation
between low-temperature gap value and transition temperature, $\Delta
(0)=1.76T_{c}$. As was shown in section \ref{Pairing in the modified
mean-field}, the ratio $\Delta (0)/T_{c}$ stays about the same within MFA
for fractal superconductor with $\gamma $ exponent near $0.6$, corresponding
to 3D Anderson model, see Fig.\ref{DeltaGamma}b. The presence of strong pseudo-gap 
$\Delta _{P}\gg T_{c}$ leads to the doubling of $T_{c}$ for the same value
of $\Delta (0)$; on the other hand, thermal fluctuations beyond MFA always
lead to suppression of $T_{c}$ with respect to $\Delta (0)$. We do not
attempt here an explicit calculation of the $\Delta (0)/T_{c}$ ratio in the
pseudo-gap region, which should take into account both the above effects;
roughly we may expect $\Delta (0)/T_{c}\sim 1.5-2$ in this region.

With a further shift of the Fermi energy deeper into the localized region ($%
E_{F}\geq 8.2$ for our model), a remarkable new feature appears in the
distribution of "transition temperatures" obtained from the truncated
criterion Eq.(\ref{virial3}) via the virial expansion. Namely, it acquires
two maxima as shown in Fig. \ref{TcDistribution}. One of them is located at $%
T_{c}=0$ and corresponds to an insulating phase, while another one
corresponds to a superconducting phase with a nonzero $T_{c}$. The magnitude
of this latter maximum decreases with increasing $E_{F}$, until it
disappears completely at about $E_{F}=8.7$.

Dramatic modification of the character of the distribution of $T_{c}$
implies that the qualitative change in the physics of the superconductive
transition. The appearance of two maxima in the distribution function means
that not-too-large systems available for numerical studies are {\it either}
superconducting or insulating. Larger systems and physical materials become
strongly inhomogeneous with some parts becoming superconducting and some
insulating at low temperatures. This observation is fully consistent with
the result of the analytical theory\cite{IoffeMezard2009} that predicts a
replica symmetry breaking transition upon the decrease of the effective
number of neighbors. This transition signals the absence of a local
self-averaging and a global behavior which is dominated by rare
superconducting paths. Thus, it is likely that for the physical, very large
system the transition in this energy interval occurs by a \textit{percolation%
} scenario.

The inhomogeneity of the superconductor in this regime is very different
from that of the "fractal" superconductivity that we discussed in the
previous section. The latter is \textit{macroscopically homogeneous} because
the strong inhomogeneity corresponding the fractal structure is seen only at
scales smaller than $L_{T}$. In this case, the distribution function of
transition temperatures \textit{for all macroscopic samples} with $L>L_{T}$
appears to be the Gaussian single-peak distribution with the width
decreasing to zero as the system size increases, typical of the
self-averaging quantities. In contrast the bimodal distribution observed in
numerical simulations and in analytic solution\cite{IoffeMezard2009} of the
pseudospin model signals of the macroscopic inhomogeneity and lack of self
averaging.

Finally we note that our treatment above assumed a second-order transition
to the superconducting phase, for instance, we have determined the
transition temperature as a temperature of Cooper instability. We cannot
exclude the first order transition to the superconducting phase although it
seems unlikely, especially in the regime of a large disorder.

\subsection{Tunneling conductance. \label{Tunneling conductance}}

Here we analyze in detail the tunneling conductance into the pseudo-gaped
state, at temperatures above and below the superconductive transition. We
compute the \textit{average} tunneling conductance in these two cases in
sections \ref{Tunneling Normal} and \ref{Tunneling Superconductor}. These
conductances can be probed by large tunneling junctions. Computation of
local tunneling conductance (as measured by scanning tunneling probe) is
more complicated, we are able to make only qualitative predictions for this
quantity in section \ref{Point contact tunneling}.

\subsubsection{Tunneling in a normal state. \label{Tunneling Normal}}

We start by discussing the temperature-dependent differential tunneling
conductance $G(V,T,\mathbf{r})$ in the presence of a pseudo-gap $\Delta _{P}$
\textit{above} the superconducting transition. In this case the total
tunneling current is given by 
\begin{eqnarray}  \label{IV5}
I(V) &=&\frac{G_{0}}{e\nu _{0}}\int \int d\varepsilon d\varepsilon _{1}\nu
_{0}(\varepsilon ,\mathbf{r})\left[ (1-f(\varepsilon ))f(\varepsilon
_{1})\delta (\varepsilon _{1}-\varepsilon +eV-\Delta _{P}^{(j)})\right. 
\notag \\
&&\left. -(1-f(\varepsilon _{1}))f(\varepsilon )\delta (\varepsilon
_{1}-\varepsilon +eV+\Delta _{P}^{(j)})\right]
\end{eqnarray}%
where $G_{0}$ is the bare conductance of the tunnel junction, $\nu
_{0}(\varepsilon ,\mathbf{r})=\sum_{j}\psi _{j}^{2}(\mathbf{r})\delta
(\varepsilon -\tilde{\xi}_{j})$, and $\tilde{\xi}_{j}\equiv \xi _{j}-\Delta
_{P}^{(j)}$. Eq.(\ref{IV5}) was obtained by the following arguments. First,
we consider the tunneling from the probe to the sample, it is given by the
first term in (151). Here the $j$-th level in the sample is empty, the probe
level is full, which gives rise to the usual factor $(1-f(\varepsilon
))f(\varepsilon _{1})$ involving the Fermi distribution function $%
f(\varepsilon )$. In this case the energy conservation requires $\varepsilon
_{1}=\xi _{j}-eV=\varepsilon +\Delta _{P}^{(j)}-eV$. Second, we consider the
tunneling from the sample to the probe: the $j$-th level in the sample is
full, the probe's level is empty, the thermal distribution functions produce
the factor $f(\varepsilon )(1-f(\varepsilon _{1}))$; in this case the energy
of a pair of electrons on the $j$-th level contains the binding energy $%
-2\Delta _{P}^{(j)}$, so that the energy conservation reads: $\varepsilon
_{1}=\xi _{j}-2\Delta _{P}^{(j)}-eV=\varepsilon -\Delta _{P}^{(j)}-eV$.

Note that this calculation assumes the validity of Fermi Golden Rule for the
tunneling rate; it also neglects electron-electron interactions (apart from
the local terms leading to the parity gap $\Delta _{P}$). The simultaneous
validity of both these assumptions is not guaranteed when we consider
localized states with relatively large typical level spacing $\delta _{L}$;
we will discuss this more below in section \ref{Point contact tunneling}.

The differential conductance $G(V)=dI/dV$ corresponding to Eq.(\ref{IV5})
can be represented in the usual form 
\begin{equation}
\frac{G(V,T,\mathbf{r})}{G_{0}}=\nu _{0}^{-1}\int d\varepsilon \nu
(\varepsilon ,\mathbf{r})\left( -\frac{\partial f(\varepsilon -eV)}{\partial
\varepsilon }\right)  \label{G5}
\end{equation}%
where the density of states in the presence of a pseudo-gap (but above $%
T_{c} $) depends explicitly on temperature: {\small 
\begin{eqnarray}
\nu _{n}(\varepsilon ,\mathbf{r}) &=&\sum_{j}\psi _{j}^{2}(\mathbf{r})\left[
\delta (\varepsilon -\tilde{\xi}_{j}-\Delta _{P}^{(j)})(1-f(\varepsilon
-\Delta _{P}^{(j)}))\right.  \label{nu5} \\
&&\left. +\,\,\delta (\varepsilon -\tilde{\xi}_{j}+\Delta
_{P}^{(j)})f(\varepsilon +\Delta _{P}^{(j)})\right]  \notag
\end{eqnarray}%
} Note that Eq.(\ref{nu5}) cannot be represented in terms of the
normal-state DoS, \thinspace \thinspace\ $\nu _{0}(\varepsilon ,\mathbf{r})$%
\thinspace\ , due to the correlations between the eigenfunction intensities $%
\psi _{j}^{2}(\mathbf{r})$ and the local pairing gaps $\Delta _{P}^{(j)}$.
The spatial average of the tunneling DoS can be obtained from Eq.(\ref{nu5})
by integration over $\mathbf{r}$ which replaces of $\mathcal{V}\psi _{j}^{2}(%
\mathbf{r})$ by unity in Eq.(\ref{nu5}).

The next (approximate) step in the simplification of this expression is to
replace the summation over eigenstates $j$ by averaging over distribution of
pairing gaps ${\ P}(\Delta _{P})d\Delta _{P}$, see Fig.~\ref{PDelta}. The
distribution function of the normalized gaps $y=\Delta _{P}/\Delta _{P}^{%
\mathrm{typ}}$ can be fit by analytical expression 
\begin{equation}
P(y)=\frac{A}{y^{2}}\exp \left( -\frac{c}{y}-b_{1}y-b_{2}y^{2}\right) \quad
\label{PDeltafit}
\end{equation}%
where $c=1.1$, $b_{1}=0.16$, $b_{2}=0.03$ and $A=1.8$ for $E=8.0$
(coefficients $b_{1},b_{2}$ are not universal as can be seen from Fig.~\ref%
{FigDoS}).

The result for average density of states then reads: 
\begin{equation}
\nu _{n}(\varepsilon )=\nu _{0}(\varepsilon )\int_{0}^{\infty }\frac{%
e^{-yz}+\cosh \frac{\varepsilon }{T}}{\cosh yz+\cosh \frac{\varepsilon }{T}}%
\cdot {\ P}(y)dy  \label{nu50}
\end{equation}%
where $z=\Delta _{P}^{\mathrm{typ}}/T$ and $\nu _{0}(\varepsilon )$ is the
average (in general, energy-dependent) DoS in the normal state. In the
low-temperature limit Eq.(\ref{nu50}) reduces to Eq.(\ref{DoS}) from section %
\ref{Insulating state}.

Evolution of the average tunneling DoS obtained from Eq.(\ref{nu5}) and its
approximation Eq.(\ref{nu50}) at different temperatures $T<\Delta _{P}^{%
\mathrm{typ}}$ for the 3D Gaussian Anderson model are shown in Fig.\ref%
{TDoS1}. The asymmetry with respect to the sign of $\varepsilon $ is due to
a non-negligible energy-dependence of the bare DoS $\nu _{0}(\varepsilon )$. 
\begin{figure}[tbp]
\includegraphics[width=8cm]{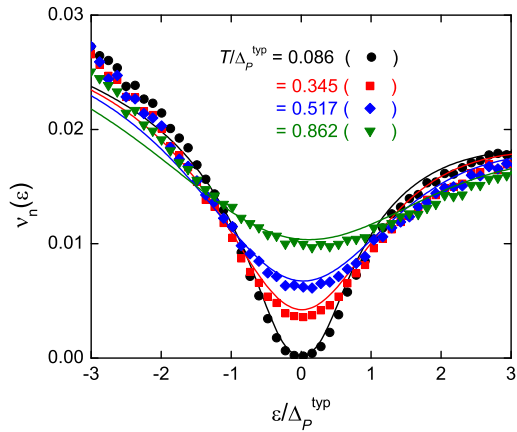}
\caption{(Color online) Global tunneling density of states for the 3D
Gaussian Anderson model at $E_{F}=8$ and the system size $20^{3}$ for
different temperatures normalized to the typical value of the local gap $%
\Delta _{P}^{\mathrm{typ}}$. The data points stand for the all-numerical
evaluation of Eq.(\protect\ref{nu5}) whereas the lines were calculated by
means of Eqs.(\protect\ref{nu50}) and (\protect\ref{PDeltafit}).}
\label{TDoS1}
\end{figure}

\begin{figure}[tbp]
\includegraphics[width=8cm]{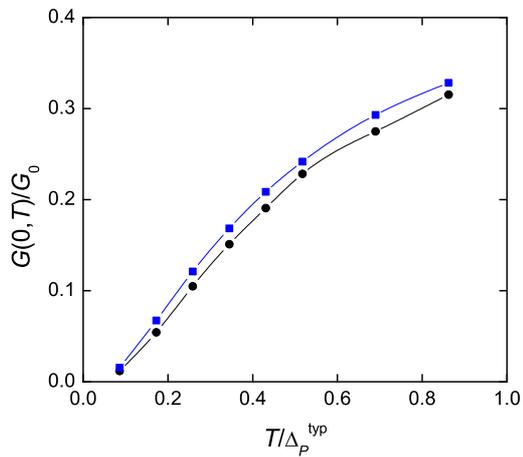}
\caption{(Color online) Zero-bias tunneling conductance as a function of
temperature below the typical pseudo-gap value; results obtained with exact
formula (\protect\ref{nu5}) are shown by black dots, whereas blue squares
correspond to the approximate formula (\protect\ref{nu50}).}
\label{G6}
\end{figure}

The obtained results for the DoS are translated into the measurable
tunneling conductance by Eq.(\ref{G5}). The zero-bias conductances $G(0,T)$
obtained from the "exact" (Eq.(\ref{nu5})) and the approximate (Eq.(\ref%
{nu50})) expressions for the DoS are shown in Fig.~\ref{G6}. It is seen from
Figs.\ref{TDoS1} and \ref{G6} that the approximation (\ref{nu50}) works
reasonably well; below we will use this approximation in the further
analysis of the average tunneling conductance.

\begin{figure}[tbp]
\includegraphics[width=8cm]{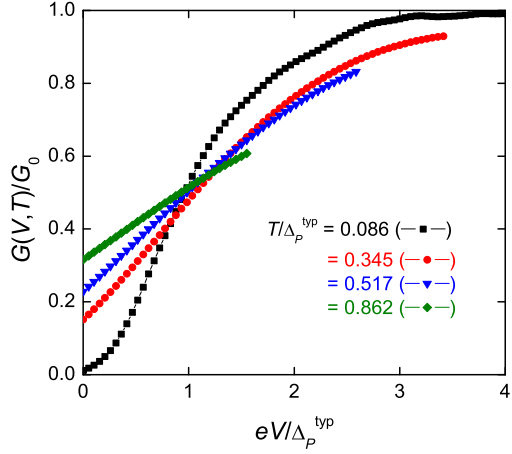}
\caption{(Color online) Ensemble-averaged tunneling conductance, Eq.(\protect
\ref{G5}), for several values of temperature below the typical local gap $%
\Delta _{P}^{\mathrm{typ}}$.}
\label{G7}
\end{figure}

The ensemble-averaged tunneling conductance $G(V,T)$ as a function of
voltage is plotted in Fig.~\ref{G7} for several temperatures. Remarkably,
the curves we obtained demonstrate nearly exact crossing at $eV\approx
\Delta _{P}^{\mathrm{typ}}$. In other terms, the tunneling conductance at
this voltage is nearly $T$-independent. This unexpected feature provides a
simple way for experimental determination of $\Delta _{P}^{\mathrm{typ}}$.
Note that at very low temperatures $T\ll \Delta _{P}^{\mathrm{typ}}$ the
crossing point moves to somewhat lower voltages. Indeed, a simple analysis
of the integral in Eq.(\ref{nu50}) shows that in the $T\rightarrow 0$ limit
the derivative $d\nu (\varepsilon )/dT\propto dP(\varepsilon )/d\varepsilon $
vanishes at $\varepsilon =\varepsilon _{0}$ corresponding to the maximum of
the function $P(\varepsilon /\Delta _{P}^{\mathrm{typ}})$, i.e. $\varepsilon
_{0}\approx 0.5\Delta _{P}^{\mathrm{typ}}$, see Eq.(\ref{PDeltafit}) and
Fig.~\ref{PDelta}.

The averaged tunneling conductance can be measured in the normal state by a
large-area tunnel junction contact. It is important that the temperatures
are not too low with respect to the typical $\Delta _{P}$ because as
temperature drops down, the bulk resistivity grows exponentially, leading to
development of a strong Coulomb gap (not taken into the account above) which
makes tunneling measurement unfeasible. The same problem becomes even more
severe for the STM measurements at low temperatures.

\subsubsection{Tunneling in a superconductor. \label{Tunneling
Superconductor}}

Provided that Fermi Golden Rule remains valid the tunneling current in the
superconductive state is given by the formula similar to Eq.(\ref{IV5}). We
only need to replace in Eq.(\ref{IV5}) the bare normal-state DoS $\nu
_{0}(\varepsilon ,\mathbf{r})$ by its superconductive counterpart given by
Eq.(\ref{nu00}). Then, repeating the derivation of Eq.(\ref{nu5}), we come
to the expression for the tunneling density of states in presence of both
pseudo-gap and the superconducting correlations: 
\begin{eqnarray}
\nu _{sc}(\varepsilon ,\mathbf{r}) &=&\frac{1}{2}\sum_{j}\psi _{j}^{2}(%
\mathbf{r})\left\{ \left( 1+\frac{\tilde{\xi}_{j}}{\varepsilon -\Delta
_{P}^{(j)}}\right) \left[ \delta (\varepsilon -\varepsilon _{j}-\Delta
_{P}^{(j)})+\delta (\varepsilon +\varepsilon _{j}-\Delta _{P}^{(j)})\right]
\right.  \notag \\
&&\times (1-f(\varepsilon -\Delta _{P}^{(j)}))+  \label{nu51} \\
&&\left. \left( 1+\frac{\tilde{\xi}_{j}}{\varepsilon +\Delta _{P}^{(j)}}%
\right) \left[ \delta (\varepsilon -\varepsilon _{j}+\Delta
_{P}^{(j)})+\delta (\varepsilon +\varepsilon _{j}+\Delta _{P}^{(j)})\right]
f(\varepsilon +\Delta _{P}^{(j)})\right\}  \notag
\end{eqnarray}%
where $\varepsilon _{j}=\sqrt{\tilde{\xi}_{j}^{2}+\Delta ^{2}(\tilde{\xi}%
_{j})}$ and the function $\Delta (\xi )$ has to be determined by solving the
gap equation (\ref{MMFA5}). Below we use Eq.(\ref{nu51}) in order to derive
an expression for the \textit{averaged} tunneling conductance; situation
with the \textit{local} tunneling conductance is more complicated and cannot
be described this way, as explained in subsection \ref{Point contact
tunneling}.

The averaged density of states in the superconductive state with a
pseudo-gap, $\nu _{sc}(\varepsilon )$, can be obtained in the same way as it
was done above for the normal state, see Eq.(\ref{nu50}). The odd terms in $%
\tilde{\xi _{j}}$ proportional to $\tilde{\xi}_{j}/(\varepsilon \pm \Delta
_{P}^{j})$ in Eq.(\ref{nu51}) cancel out upon summation over $\tilde{\xi}%
_{j}=\pm |\tilde{\xi}_{j}|$, provided we neglected the energy-dependence of
typical $\Delta _{P}(E)$ inside the energy band of the width $\Delta E\sim
\Delta _{P}$. We get 
\begin{equation}
\nu _{sc}(\varepsilon )=\int_{0}^{\infty }{\ P}(y)\left[ \frac{\nu
_{sc}^{(0)}(\varepsilon -yzT)}{e^{yz-\varepsilon /T}+1}+\frac{\nu
_{sc}^{(0)}(\varepsilon +yzT)}{e^{yz+\varepsilon /T}+1}\right] dy
\label{nu52}
\end{equation}%
where $z=\Delta _{P}^{typ}/T$ and $\nu _{sc}^{(0)}(\varepsilon )$ is the DoS
in the superconductive state without a pseudo-gap, defined by $\nu
_{sc}^{(0)}(\varepsilon )=\nu _{0}(\varepsilon )|d\xi (\varepsilon
)/d\varepsilon |$ (\ref{nu01}), similar to the one shown in Fig.~\ref{DoS01}
for the case of critical superconductor. Note that the function $\xi
(\varepsilon )$ that we need here differs slightly from the one shown in
Fig.~\ref{DoS01} due to the different form of the correlation function $%
M(\omega )$ entering modified MFA equation (\ref{MMFA5}) in the case of
pairing between localized states. However, the effect of this difference
reduces to an overall prefactor, as is seen from Fig.~\ref{Cw2}. We thus do
not expect significant difference between the solution $\Delta (\xi )$ and
the corresponding $\xi (\varepsilon )$ functions calculated within the
modified MFA approximation in the mildly localized region considered here
and for the "critical" case studied in section \ref{Cooper instability}.
Thus we use $d\xi (\varepsilon )/d\varepsilon |$ as it was found in section %
\ref{Low-temperature density of states} but take into account a smooth
energy-dependent DoS pre-factor $\nu _{0}(\varepsilon )$, as it was done
above when computing Eq.(\ref{nu50}) and Fig.~\ref{TDoS1}.

\begin{figure}[tbp]
\includegraphics[width=8cm]{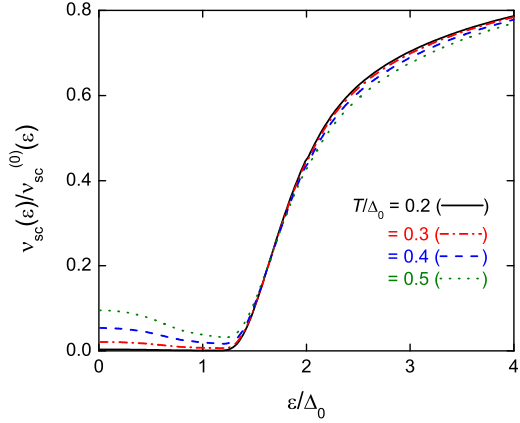}
\caption{(Color online) Ensemble-averaged tunneling DoS, Eq.(\protect\ref%
{nu52}), for several values of temperature much below the value of the
superconductive gap $\Delta _{0}\equiv \Delta _{0}(\protect\xi =0)$. The
latter is chosen to be $\Delta _{0}=0.5\Delta _{P}^{\mathrm{typ}}$, where $%
\Delta _{P}^{\mathrm{typ}}$ is the parity gap.}
\label{nusc}
\end{figure}

\begin{figure}[tbp]
\includegraphics[width=8cm]{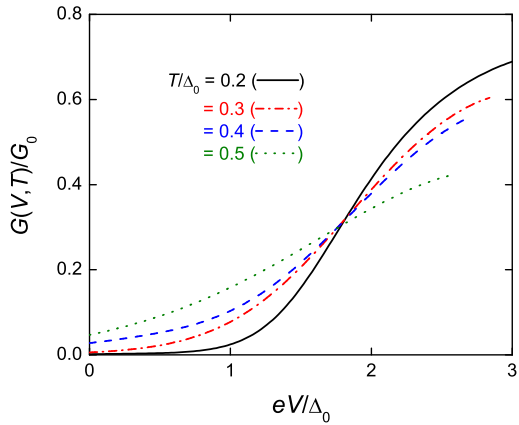}
\caption{(Color online) Average tunneling conductance in the superconductive
state for several values of temperature much below the value of the
superconductive gap $\Delta _{0}=0.5\Delta _{P}^{\mathrm{typ}}$. }
\label{G52}
\end{figure}

\begin{figure}[tbp]
\includegraphics[width=8cm]{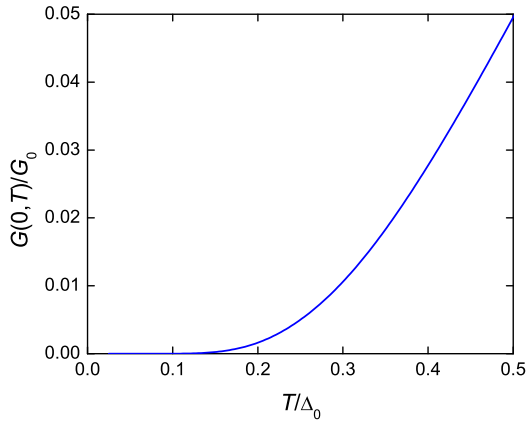}
\caption{Zero-biased conductance as function of temperature for the case of $%
\Delta _{0}=0.5\Delta _{P}^{\mathrm{typ}}$.}
\label{GVTs}
\end{figure}

Evolution of the $\nu _{sc}(\varepsilon )$ shapes as function of temperature
in the low-temperature range $T<0.5T_{c}$ is shown in Fig.\ref{nusc} 
where we use the analytic interpolation formula Eq.(\ref{PDeltafit}) for the
distribution function ${P}(y)$, and the $T=0$ solution for the gap function $%
\Delta (\xi )$. The average tunneling conductance $G(V,T)$ obtained with
this DoS from Eq.(\ref{G5}) is shown in Fig. \ref{G52}. 
The crossing point at this plot, i.e. the $T$-independent conductance at
some specific value of voltage, corresponds to $eV-\Delta _{0}\approx
0.5\Delta _{P}^{\mathrm{typ}}$. In zero energy limit the expression for the
density of states simplifies to 
\begin{equation}
\nu _{sc}(0)=\int_{0}^{\infty }{\ P}(y)\frac{2\nu _{sc}^{(0)}(y)}{e^{y/T}+1}%
dy  \label{nu53}
\end{equation}%
The temperature dependence of the corresponding zero bias tunneling
conductance $G(0,T)$ is shown in Fig.\ref{GVTs}.

The averaged DoS and the tunneling conductance presented in Figs.~\ref{G52}, %
\ref{GVTs} do not show any coherence peaks similar to the one presented in
Fig.~\ref{DoS01} for the critical superconductive state (section \ref%
{Low-temperature density of states}). This is due to the averaging of the
peak position $eV=\Delta _{0}+\Delta _{P}^{(j)}$ over a broad distribution $%
P(\Delta _{P})$ of the local pairing gaps $\Delta _{P}^{(j)}$. Note that the
non-monotonic behavior of the DoS curve in Fig. \ref{nusc} reflects the
(smeared) edges at $\varepsilon =\Delta _{P}^{\mathrm{typ}}\pm \Delta _{0}$
of the $\nu _{\mathrm{sc}}^{(0)}(\varepsilon )$ function shifted by the
paring gap $\Delta _{P}^{\mathrm{typ}}>\Delta _{0}$ and weighted by the
temperature-dependent distribution functions in Eq.(\ref{nu52}).

We emphasize again that the results obtained in this section give the
averaged tunneling conductance which can be measured with a large-area
junctions, such as the ones used in~\cite{Pratap}. They may differ
significantly from the local tunneling conductance measured by a small tip
and discussed in the following subsection.

\subsubsection{Point contact tunneling. \label{Point contact tunneling}}

\bigskip The equation (\ref{IV5}) can be easily generalized to describe
point conductance. One gets using the perturbation theory in the tunneling
amplitude, $t$ (which is equivalent to Fermi Golden Rule) that the most
general case the point conductance is given by equation (\ref{IV5}) where $%
\nu _{0}(\varepsilon ,\mathbf{r})$ is replaced by the exact single particle
Green function at the tunneling position, $\mathrm{Im}G_{R}(\epsilon ,r)$. In
a poor conductor in which single particle states are localized or almost
localized one expects that $\mathrm{Im}G_{R}(\epsilon ,r)$ shows a series of
peaks corresponding to the single particle states located in the vicinity of
the tunneling point. Applying this reasoning to the superconductor-insulator
transition in a situation when $E_{F}$ corresponds to a well developed
paring gap $\Delta _{P}$ and repeating the arguments of the previous
subsection one expects to observe the peaks in the tunneling conductance
with width $T$ and separation $\delta _{L}$.

\bigskip These arguments do not take into account the effect of the
collective modes of the superconductor which smear the peaks in the
tunneling conductance. In fact, the presence of the collective modes make
the observation of the peak structures in the density of states difficult.
Briefly, the experimental observation of tunneling is possible only in the
superconducting state, but in this state the effect of the collective modes
on the single particle density of states is large. We estimate this effect
within a simplified model, namely we use the pseudospin Hamiltonian (\ref%
{HamSpin}) to which we add the possibility of having single particle
excitation at the single site $0$: 
\begin{equation}
H=\xi _{0}n+\Delta _{P}(1-n^{2})-g\left[ c_{\uparrow}^{\dagger }c_{\downarrow}^{\dagger
}\sum_{j}M_{0j}S_{j}^{-}+c_{\downarrow}c_{\uparrow}\sum_{j}M_{0j}S_{j}^{+}\right] +H_{PS}(S)
\label{H-PS-tun}
\end{equation}%
where $n=\sum_{a}c_{a}^{\dagger }c_{a}-1$ where $a=\uparrow ,\downarrow $
are two spin components, and the sum over $j$ runs over the neighbors of the
site $0.$ This model describes the process of electron tunneling in a single
localized state; in a more realistic case the electrons tunnel in few such
states with an amplitude proportional to $\psi _{\alpha }^{2}(r)$ where $r$
is a tunneling point. These processes are incoherent due to randomness of
local binding energies $\Delta _{P}^{(j)}$, so the physical tunneling
conductivity is the sum of the contributions from different eigenstates.

In a good BCS-like superconductor characterized by a large number of
neighbors one can replace pairing field $h_{\perp }=g$ $%
\sum_{j}M_{0j}S_{j}^{\perp }$ by their average values $\left\langle h_{\perp
}\right\rangle =\Delta $. In this case the Green function of the electron at
site $0$ obeys Dyson equation 
\begin{eqnarray}
G_{R}(\epsilon ) &=&\frac{1}{G_{0R}^{-1}(\epsilon )-\Delta
^{2}G_{0A}(-\epsilon )}  \label{Dyson_BCS} \\
G_{0R}(\epsilon ) &=&\frac{1}{\epsilon -\xi _{0}-\Delta _{P}\mathrm{sgn}(\xi
_{0})+i0}. 
\end{eqnarray}%
The resulting single electron Green function has a pole at $\epsilon _{0}=%
\sqrt{(\xi _{0}+\Delta _{P}\mathrm{sgn}(\xi _{0}))^{2}+\Delta ^{2}}$ which
shows the combined effect of the pseudogap and superconducting order
parameter on the electron. As one expects, in the case of a large number of
neighbors the electron density of states has a sharp peak at the pole energy 
$\epsilon _{0}$ which is smeared only by thermal effects by $\delta \epsilon
\sim T$. As the disorder is increased the effective number of neighbors goes
down and fluctuations of the pairing field become relevant. This leads to
two physical effects. The first effect is that tunneling of the electrons
removes the spin from site $0$ which changes the effective spin Hamiltonian.
The new ground state spin wave function is orthogonal to the one realized
before the tunneling process. This leads to a suppression of the tunneling
of the electron with the energy exactly equal to $\epsilon _{0}$. The second
effect is that tunneling might be accompanied by the emission of the
pseudospin excitations. This leads to non-zero tunneling density of states
at energies larger than $\epsilon _{0}$. Both effects are controlled by the
same parameter, the effective number of neighbors which determines the
strength of the quantum fluctuations of the pairing field and both round the
peaks in the observed spectrum. In order to compute local tunneling
conductance one needs the full description of the spin fluctuations in the
regime of the small $Z_{eff}$ which is not presently available.

Below we estimate the second effect, i.e. the magnitude of the level
broadening in the framework of the perturbation theory. In perturbation
theory the emission of the pseudospin excitations is described by the
inclusion of the self energy part in the Dyson equation (\ref{Dyson_BCS}):

\begin{eqnarray}
G_{R}(\epsilon ) &=&\frac{1}{\widetilde{G}_{R}^{-1}(\epsilon )-\Delta ^{2}%
\widetilde{G}_{A}(-\epsilon )},\qquad \widetilde{G}_{R}(\epsilon )=\frac{1}{%
\epsilon -\xi _{0}-\Delta _{P}\mathrm{sgn}(\xi _{0})-\Sigma _{R}(\epsilon )},
\label{Sigma_R} \\
\mathrm{Im}\Sigma _{R}(\epsilon ) &=&g^{2}\sum_{j}M_{0j}^{2}\int_{0}^{\epsilon
}\frac{d\omega }{2\pi }\mathrm{Im}\widetilde{G}_{A}(\omega -\epsilon )\mathrm{Im}%
D_{R}(\omega )
\end{eqnarray}%
where $D(\omega )=$ $\left\langle \left\langle S^{\perp }S^{\perp
}\right\rangle \right\rangle _{\omega }$ is irreducible correlator of the
pairing field. Even in a weakly superconducting phase the pseudospin
excitations are delocalized and thus have a continuos spectrum. Here we
focus on the large energy scale $E\sim \Delta _{P}$ properties of the
spectrum, we do not discuss here the structure of tunneling conductance very
close to the edge, where coherence peaks are expected. At these large
energies, the density of collective modes is featureless, so we estimate the 
$\mathrm{Im}D(\omega )=\nu _{B}\theta (\omega )$.

As it is clear from (\ref{Sigma_R}), the interaction with pseudospins lead
to the decay rate of the single particle excitations, $\Gamma_{el}$ , which
smears the peak structure of $\mathrm{Im}G(\epsilon )$ replacing it with the
threshold behavior. \ In the leading order of perturbation theory we replace 
$\widetilde{G}(\omega -\epsilon )$ in (\ref{Sigma_R}) by $G_{0}(\omega
-\epsilon )$ and get for the level width $\Gamma _{el}=g^{2}\nu
_{B}\sum_{j}M_{0j}^{2}$.

Because the expression for the level width, $\Gamma _{el}$ contains the
square of the interaction constant, $g^{2}$, but only one sum over
neighboring sites, it is small in a good superconductor (where $gM\propto
1/Z $) but may become large in a weak superconductor close to
superconductor-insulator transition. We estimate it in the framework of the
Bethe lattice model for this transition discussed in section \ref{Transition
temperature}. In this model $M_{0j}=1/Z_{\mathrm{Bethe}}$, so we get $\Gamma
_{el}=g^{2}\nu _{B}/Z_{\mathrm{Bethe.}}$ In order to express it in terms of
the physical parameters, we use the equation (\ref{ZBethe}) that relates the
transition temperature to the effective number of neighbors $Z_{\mathrm{%
Bethe.}}$ at the onset of the weak superconductor regime. The onset of this
regime can be defined (and determined experimentally) by the beginning of
the sharp decrease of $T_{c}$ with the increase of disorder. We get: 
\begin{equation}
\Gamma _{el}\sim (g\nu _{B})^{2}\sqrt{T_{c}E_{F}}\gg T_{c}\,.
\label{GammaEl}
\end{equation}%
The apparently large value of the level width in this regime implies that
distinct peaks become absent, being replaced by the threshold. For a
tunneling leading to a single state, $\alpha =0$, the threshold coincides
with the value of energy $\epsilon _{\alpha }\approx \left[ \xi _{\alpha
}^{2}+\Delta ^{2}\right] ^{1/2}+\Delta _{P}^{(\alpha )}$. This discussion
assumed that tunneling happens only in one localized state; in a more
realistic situation when tunneling amplitudes proportional to $\psi _{\alpha
}^{2}(r)$ are significant for a few states, the threshold energy is equal to 
$\Delta $ plus a minimal value of $\min \Delta _{P}^{\alpha }$ available for
a contact tunneling process at a given point.

We conclude that the model Hamiltonian Eq.~(\ref{H-PS-tun}) results in the
threshold dependence of the tunneling conductance. The threshold
distribution (observed when scanning along the surface) is expected to be
similar to $P(\Delta _{P})$ distribution shown in Fig.~\ref{PDelta} whereas
the tunneling conductance $G(V)$ should be similar to the integral of that
distribution shown in Fig.~\ref{FigDoS}. These conclusions are in a
qualitative agreements with the data presented in~\cite{Sacepe2007,Sacepe2}.

The model ~(\ref{H-PS-tun}) neglects the off-diagonal matrix elements of the
form 
\begin{equation*}
\sum_{j}c_{\alpha \uparrow }c_{j\downarrow }\sum_{k}M_{\alpha jkk}\langle
c_{k\uparrow }^{+}c_{k\downarrow }^{+}\rangle
\end{equation*}%
that are present in the full Hamiltonian and might be important for the
quantitative description of the data. Physically these terms lead to the
possibility of non-local process in which electron is converted into a hole
in the adjacent state and emits a collective mode quantum. The presence of a
sum over many states $j$ connected with a pre-selected state $\alpha $ leads
to a partial average over the gap distribution $P(\Delta _{P})$; this
suppresses the relative magnitude of spatial fluctuations of the local
thresholds. As a result, we expect that the threshold voltage is given by 
\begin{equation}
eV_{\mathrm{th}}=\alpha _{\mathrm{th}}\overline{\Delta }_{P}+\Delta
\label{Vth}
\end{equation}%
with $\alpha _{\mathrm{th}}<1$ being numerical factor of order unity, with a
moderate spatial fluctuations.

The arguments leading to an estimate Eq.(\ref{GammaEl}) imply that peak
structure of the density of states might be only observed in a narrow
transient regime that corresponds to a well developed pseudogap combined
with a strong superconductor. It is not clear that this regime can be
realized in physical systems. In particular, it is very likely that
experiment \cite{Sacepe2007,Sacepe2} which reported the gap in the density
of states but no obvious peak structure were performed in a weakly
superconducting state close to the transition into the insulator. As
explained above, in this regime one expects that at a typical point the
spectrum of collective modes is continuos and featureless resulting in the
tunneling conductance characterized by a threshold that corresponds to the
minimal value of $\epsilon _{0}$. However, large inhomogeneity of the
ordered state (discussed in a different context in the section \ref%
{Transition temperature}) implies that at some locations the spectrum of
collective modes might develop large peaks leading to the peak structures in
the tunneling conductance.

Qualitatively, interaction of single electron with pseudospin collective
modes plays the same role as its interaction with other electrons in zero
bias anomaly and Fermi-edge singularity problem~\cite{Nozieres}. The
quantitative theory of these effects and in particular, the predictions for
the shape of the pseudogap require the full theory of the collective modes
appearing in the vicinity of superconductor-insulator transition, this is
beyond the approach developed in this paper.

\subsection{Andreev contact conductance at low temperatures. \label{Andreev
point contact}}

In the regime of a strong pseudo-gap $\Delta _{P}\geq \Delta _{0}$ a
peculiar situation occurs: the energy gap for a single-particle excitation $%
\Delta _{1}=\Delta _{P}+\Delta _{0}$ is \textit{larger} than the
two-particle excitation gap $2\Delta _{0}$. The crucial point here is that
the \textit{local} paring energy $\Delta _{P}$ drops out of the expression
for the two-particle excitation gap which is determined only by the \textit{%
collective} superconductive gap $\Delta _{0}$. Local paring energy becomes
important only when we have to break the pair to produce the single-particle
excitation. This allows to experimentally distinguish between the two types
of gaps: the paring gap $\Delta _{1}$ seen in the \textit{single-electron}
tunneling and the collective gap $\Delta _{0}$ which can be observed in the
experiments where the entire pair is transferred through the tunnel
junction. This type of tunneling experiments is called Andreev-contact
spectroscopy (see \cite{Deutscher} for the review of its application to
similar studies in high-$T_{c}$ cuprates). Since the probability to transfer
a pair is proportional to the square of the tunnel contact transparency $%
\mathcal{T}$, the latter should be not too small for the Andreev conductance
to be observable.

\begin{figure}[tbp]
\includegraphics[width=8cm]{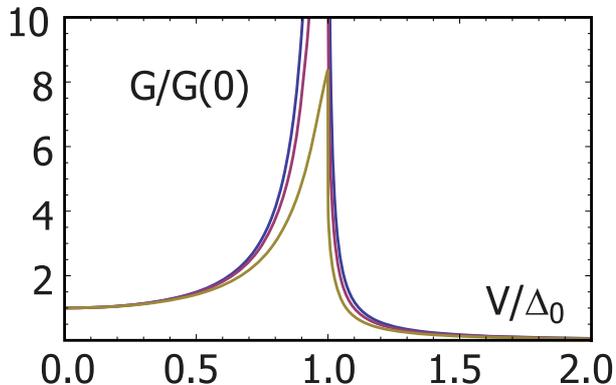}
\caption{(Color online) Andreev point-contact conductance (normalized by its
value at zero bias) is calculated within zero-temperature BTK approximation,
Eq.(\protect\ref{Gns}), for three different values of the contact
transparency, $\mathcal{T} = 0.2$ (blue line) , $0.4$ (red) and $0.6$
(yellow). Voltage $eV$ is normalized by the maximum value $\Delta_0$ of the
gap function $\Delta(\protect\xi)$. It is assumed that single-electron
tunnelling is fully suppressed by large gap $\Delta_P \gg \Delta_0$.}
\label{Andreev}
\end{figure}

Below we provide simplest estimates of Andreev conductance for the
pseudogapped fractal superconductor, based upon Blonder-Tinkham-Klapwijk
(BTK) approach~\cite{BTK}. We consider the case of moderately large
parameter of normal reflection $Z\geq 1$. Charge transport within BTK
approach is characterized by (energy-dependent) probabilities $A$, $B$, $C$
and $D$ of four processes which may occur when normal incident electron
approaches N-S boundary: $A(\varepsilon )$ is the probability of Andreev
reflection, $B(\varepsilon )$ corresponds to normal reflection, $%
C(\varepsilon )$ and $D(\varepsilon )$ correspond to two different processes
of single-electron transmission into superconductor. We are interested here
in the low-temperature limit $T\ll T_{c}\ll \Delta _{P}$ and relatively low
voltages $eV<\Delta _{P}$, thus the processes of electron transmission into
superconductor can be neglected and the full sub-pseudogap current can be
written as 
\begin{equation*}
I_{NS}(V)\propto \int d\varepsilon \,\,2A(\varepsilon )[f(\varepsilon
-eV)-f(\varepsilon )]
\end{equation*}%
This expression is obtained from Eq.(17) of the BTK paper by the
substitution of the integrand $1+A-B\equiv 2A+C+D$ for $2A$, due to
vanishing of normal current. Therefore differential conductance is
proportional to the Andreev reflection probability $A(\varepsilon )$, which
we will write using BTK results, see Table II of Ref.~\cite{BTK}, but in
slightly different notations. To begin with, we introduce normal
transmission coefficient $\mathcal{T}=1/(1+Z^{2})$ and write NS differential
conductance in BTK approximation (at $T=0$) for usual superconductor as 
\begin{eqnarray}
\frac{dI_{NS}}{dV} &=&G_{NS}(V)  \label{Gns} \\
&=&2G_{0}\mathcal{T}\left( \frac{\Delta (eV)}{eV}\right) ^{2}\frac{%
n_{0}^{2}(eV)}{\mathcal{T}^{2}n_{0}^{2}(eV)+(2-\mathcal{T})^{2}+2n_{0}(eV)(2-%
\mathcal{T})\theta (eV-\Delta )}  \notag
\end{eqnarray}%
where $G_{0}=2\mathcal{T}N_{ch}$ is the normal-state conductance of the
contact, $N_{ch}$ is the number of orbital conductance channels (in BTK
approach all channels are characterized by the same transmission coefficient 
$\mathcal{T}$), and $n_{0}(\varepsilon )=\varepsilon /|\varepsilon ^{2}-{%
\Delta }^{2}|^{1/2}.$

In order to generalize Eq.(\ref{Gns}) for the case of fractal
superconductor, we consider subgap region $\varepsilon <\Delta _{0}$ and
region of real excitations $\varepsilon >\Delta _{0}$ separately. In the
subgap region we introduce function $n_{-}(\varepsilon )$ defined by 
\begin{equation}
n_{-}(\varepsilon )=\frac{1}{\pi }\int d\xi \frac{\Delta (\xi )}{%
-\varepsilon ^{2}+\xi ^{2}+\Delta ^{2}(\xi )}  \label{feps}
\end{equation}%
where $\Delta (\xi )$ is the solution to the gap equation (\ref{MMFA5}).
Then NS conductance in the subgap region is: 
\begin{equation}
G_{NS}^{-}(V)=2G_{0}\mathcal{T}\frac{n_{-}^{2}(eV)}{(n_{-}^{2}(eV)-1)%
\mathcal{T}^{2}+(2-\mathcal{T})^{2}}  \label{NSsub}
\end{equation}%
For the region above the gap one should introduce function $\hat{\Delta}%
(\varepsilon )\equiv \Delta (\xi (\varepsilon ))$, and replace in Eq.(\ref%
{Gns}) the function $n_{0}(\varepsilon )$ by the actual normalized density
of states $n_{+}(\varepsilon )=\nu (\varepsilon )/\nu _{0}$, defined by Eq.(%
\ref{nu01}): 
\begin{equation}
G_{NS}^{+}(V)=2G_{0}\mathcal{T}\frac{(\hat{\Delta}(eV)/{eV}%
)^{2}\,\,n_{+}^{2}(eV)}{\mathcal{T}^{2}n_{+}^{2}(eV)+(2-\mathcal{T}%
)^{2}+2n_{+}(eV)(2-\mathcal{T})}  \label{NSover}
\end{equation}%
The comment is in order here: we do not expect Eqs.(\ref{NSsub},\ref{NSover}%
) to be quantitatively exact for the problem considered, due to a number of
oversimplifying assumptions borrowed from the BTK approach (it was developed
for clean superconductors in a contact with clean metal). However,
comparison of BTK expression for $A(\varepsilon )$ with exact calculations
done for dirty NS junctions (see Appendix to the paper~\cite{Bezuglyi} and
references in that paper, and also paper~\cite{Golubov03} ) shows that the
key feature of the result (\ref{NSsub},\ref{NSover}), that is, the peak of
differential conductance at $eV=\Delta _{0}$, is reproduced by these more
adequate calculations; thus we hope it will survive for very strongly
disordered pseudogaped superconductor as well.

In Fig.~\ref{Andreev} we present the results of the computation of the $T=0$
differential Andreev conductivity based on Eqs.(\ref{NSsub},\ref{NSover})
for several values of the contact transparency. Note that we give here the
data for Andreev conductivity normalized to its value at zero voltage $G(0)$%
, which scales itself as $\mathcal{T}^{2}$. Therefore the plots presented in
Fig.\ref{Andreev} illustrate variation of the $G(V)$ peak shape, but do not
show the overall scale of $G(V)$ as function of $\mathcal{T}$. The gap
function $\hat{\Delta}(\varepsilon )$ was computed using the $T=0$ limit of
the solution to the modified MFA equation (\ref{MMFA5}) for the "critical" $%
M(\omega )$, Eq.(\ref{ij}). Voltage is normalized by the maximum value $%
\Delta _{0}$ of the gap function $\Delta (\xi )$.

Note that the plots in Fig.\ref{Andreev} illustrate the limit of a very
large paring gap $\Delta_{P} \gg \Delta_{0}$. In the case of parity gap $%
\Delta_P$ comparable to the superconductive gap $\Delta(0)$, the
point-contact conductance should contain two peaks: at $eV_1 = \Delta(0)$
due to Andreev processes, and at $eV_2 = \Delta_P + \Delta(0)$ due to
single-electron tunnelling. These two voltage values behave differently with
change of temperature and location of the point contact. The voltage $V_1$
is determined by the \textit{collective superconductive gap} $\Delta_{0}$
and, therefore, is position-independent but goes to zero as $T \to T_c$. In
contrast, $V_2$ fluctuates strongly between different contact locations due
to the local paring gap $\Delta_{P}$ but does not vanish as $T$ approaches $%
T_c$ from below. Recent experimental results~\cite{Dubouchet10} seem to
support the above predictions.

\subsection{Spectral weight of high-frequency conductivity and
superconducting density. \label{Spectral weight}}

The studies\cite{Basov} of high $T_{c}$ superconductors has shown that the
unconventional physics of these materials can be probed by the temperature
dependence of the spectral weight $K^{tot}(T)$ of the high-frequency
conductivity defined by 
\begin{equation}
K^{tot}(T)=\frac{2}{\pi }\int_{0}^{\Omega _{max}}\Re \sigma (\omega
,T)d\omega +\rho _{s}(T)\equiv K(T)+\rho _{s}(T).  \label{KT1}
\end{equation}%
Here $\sigma (\omega ,T)$ denotes the regular part of conductivity, whereas
the contribution $\rho _{s}=-c\delta \mathbf{j}/\delta \mathbf{A(\omega =0)}$
of superconductive part (proportional to $\delta (\omega )$) is explicitly
separated in the last term of Eq.(\ref{KT1}); the high-frequency cutoff $%
\Omega _{max}$ is chosen a several times larger than $T_{c}$.

In agreement with the data on conventional superconductors, BCS theory and
its strong coupling generalizations predict that $K^{tot}(T)$ does not
change as temperature decreases below the superconducting transition
temperature $T_{c}$: the appearance of $\rho _{s}(T)$ is compensated by the
decrease of the regular part. In contrast, ~for underdoped cuprates the
regular part $K(T)$ of the c-axis spectral weight changes below $T_{c}$ by
one half of $\rho _{s}(T)$ only, so that $K^{tot}(0)-K^{tot}(T_{c})=\frac{1}{%
2}\rho _{s}(0)$.\cite{Basov}

The observation~\cite{Basov} was explained by Ioffe and Millis~\cite%
{IoffeMillis} as being caused by formation of a pseudo-gap that survives up
to a temperature $T_{PG}$ well above $T_{c}$, and at the same time by the
absence of the inter-layer coherence in the range $T_{c}<T<T_{PG}$ due to
strong quantum and thermal fluctuations. The weight non-conservation occurs
only for ~the c-axis conductivity, whereas the in-plane conductivity behaves
in a usual way in all cuprates.\cite{Basov} The theory~\cite{IoffeMillis}
attributes to the large phase fluctuations between weakly coupled layers and
a tunneling nature of c-axis transport, in contrast with the smooth phase
variations and the continuous electron motion in the planes.

We expect that pseudogapped superconductivity formed near
superconductor-insulator transition shows similar anomaly in the behavior of
the spectral weight $K_{tot}(T)$. In particular, we show that inside the
strong pseudo-gap region $T_{c}\ll \Delta _{P}$ the effect of the $%
K^{tot}(T) $ non-conservation is even stronger: its variation $\Delta
K^{tot}(T)$ is equal (within the mean field approximation) to $\rho _{s}(T)$%
, whereas the variations of the regular part of spectral weight $\Delta K(T)$
are smaller by $\sim 1/Z_{\mathrm{eff}}$ factor. In this derivation we shall
assume that pseudogap sets the largest scale in the problem, i.e. that the
upper frequency cutoff $\Omega _{max}$ satisfies $T_{c}\ll $ $\Omega
_{max}\ll \Delta _{P}$. This assumption allows us to neglect completely
single electron processes (which do not contribute for $\omega <\Delta _{P}$%
) and use pseudospin model for the computation of the conductivity in the
entire frequency range. Furthermore, because, as we shall show below the
conductivity in the pseudospin model decreases fast at frequencies above $%
T_{c}$, we can replace the upper cutoff $\Omega _{max}\rightarrow \infty $
when evaluating the spectral weight in this model.

In order to compute the conductivity in the pseudospin model we need to
include in the spin Hamiltonian (\ref{HamSpin}) the effect of the vector
potential $\mathbf{A}$: 
\begin{equation}
H=\sum_{j}2\xi _{j}S_{j}^{z}-\frac{g}{2}%
\sum_{ij}M_{ij}(S_{i}^{+}S_{j}^{-}e^{i\phi _{ij}}+h.c.)\,,  \label{HamSpinA}
\end{equation}%
where $\phi _{ij}=\frac{2e}{\hbar c}\int_{\mathbf{r}_{i}}^{\mathbf{r}_{j}}%
\mathbf{A}d\mathbf{r}$. Here $i,j$ denote localized single-particle states
(orbitals) with wavefunctions $\psi _{i,j}(\mathbf{r})$. The Hamiltonian in
the form (\ref{HamSpinA}) is applicable provided that the typical distance $%
R_{0}$ between centers of localization of the relevant eigenstates is much
longer than the localization length $L_{loc}$ (the same assumption is used
everywhere in this section, see Eq.(\ref{Zeff})). Then $i,j$ can also be
viewed as sites characterized by the position vectors $\mathbf{r}_{i}$,$%
\mathbf{r}_{j}$ of centers of localization of the corresponding orbitals.
The assumption $R_{0}\gg L_{\mathrm{loc}}$ implies that the theory is not
sensitive to an exact definition of "center of localization" inside the
localization radius. It is this assumption that allows to define the phase $%
\phi _{ij}$ in Eq.(\ref{HamSpinA}). Another assumption implied in (\ref%
{HamSpinA}) is that the magnetic field is weak enough so that its effect
reduces to the introduction of the phase factors $e^{i\phi _{ij}}$ and does
not change the matrix elements $M_{ij}$.

The electrical current across each link $(ij)$ that corresponds to the
Hamiltonian Eq.(\ref{HamSpinA}) is 
\begin{equation}
\hat{\mathcal{I}}_{ij}=\frac{g}{2}M_{ij}\,i(S_{i}^{+}S_{j}^{-}e^{i\phi
_{ij}}-h.c.).  \label{Noetter}
\end{equation}%
The response of the superconductor to the slowly spatially varying
electromagnetic field is fully characterized by $Q(\omega )=-c\delta 
\mathcal{j}/\delta A_{k=0}(\omega )=Q_{0}+\widetilde{Q}(\omega )$ where we
separated the constant part $Q_{0}$ and the frequency dependent one $%
\widetilde{Q}(\omega )$ that satisfies $\widetilde{Q}(\omega \rightarrow
\infty )=0$. At high frequencies the conductivity is purely imaginary with $%
\mathrm{Im}\sigma (\omega )=$ $Q_{0}/\omega $, expressing $\mathrm{Im}\sigma
(\omega )$ through the real part by Kramers-Kronig relations one gets that $%
K^{tot}(T)=Q_{0}$.

The constant term $Q_{0}$ originates from the direct expansion of the
expression for the current (\ref{Noetter}) in $\phi _{ij}$:%
\begin{equation}
K^{tot}(T) = Q_{0}(T) =\frac{g}{2\mathcal{V}}\sum_{ij}\left( \frac{2e}{\hbar 
}x_{ij}\right) ^{2}M_{ij}Q_{ij}^{(0)}  \label{Q_0}
\end{equation}%
where $Q_{ij}^{(0)}=\left\langle S_{i}^{+}S_{j}^{-}+h.c.\right\rangle $.

The equation (\ref{Q_0}) implies that in the general case $Q_{0}$ cannot be
temperature independent because $\left\langle
S_{i}^{+}S_{j}^{-}+h.c.\right\rangle $ is generally temperature dependent.
Furthermore, one expects a strong temperature dependence to appear below $%
T_{c}$ where order parameter induces large spin-spin correlator. In the
leading order in $1/Z_{eff}$ one can replace $\left\langle
S_{i}^{+}S_{j}^{-}\right\rangle \approx \left\langle S_{i}^{+}\right\rangle
\left\langle S_{j}^{-}\right\rangle $ and conclude that the spectral weight
is constant above $T_{c}$ but acquires temperature dependence below $T_{c}$:

\begin{equation}
K^{tot}(T)=\frac{1}{4\mathcal{V}}\sum_{ij}gM_{ij}\left( \frac{2e}{\hbar }%
x_{ij}\right) ^{2}\frac{\Delta _{i}\Delta _{j}\tanh \beta \varepsilon
_{i}\tanh \beta \varepsilon _{j}}{\varepsilon _{i}\varepsilon _{j}}.
\label{K^total}
\end{equation}%
where $\varepsilon _{i}=\sqrt{\xi _{i}^{2}+\Delta _{i}^{2}}$.

We now compute the leading correction in $1/Z_{eff}$ to $\left\langle
S_{i}^{+}S_{j}^{-}\right\rangle $ \ above $T_{c}$ , we find that it does not
result in the spectral weight temperature dependence in this temperature
range. This result does not preclude that temperature dependence appears in
high orders in $1/Z_{eff}$ but it implies that the temperature dependence is
very small in this regime. Note that in this temperature range the
superconducting response is absent, so temperature independence of the full
weight implies temperature independence of its regular part. To compute the
leading contribution to $\left\langle S_{i}^{+}S_{j}^{-}\right\rangle $ it
is sufficient to diagonalize two spin problem with the Hamiltonian

\bigskip 
\begin{equation}
H=\xi _{j}\sigma _{j}^{z}+\xi _{i}\sigma _{i}^{z}-\frac{g}{2}%
M_{ij}(S_{i}^{+}S_{j}^{-}+S_{j}^{+}S_{i}^{-})\,,  \label{Two Spin H}
\end{equation}

which mixes states pairwise. \bigskip We get 
\begin{equation*}
\left\langle S_{i}^{+}S_{j}^{-}+S_{j}^{+}S_{i}^{-}\right\rangle =\frac{g}{2}%
M_{ij}\left[ \frac{\tanh (\beta \xi _{i})-\tanh (\beta \xi _{j})}{4(\xi
_{i}-\xi _{j})}+\frac{\tanh (\beta \xi _{i})+\tanh (\beta \xi _{j})}{4(\xi
_{i}+\xi _{j})}\right]
\end{equation*}

Inserting this expression in (\ref{Q_0}) and averaging over distribution of $%
\xi $ , assuming that it is not correlated with the values of $M_{ij}$ and $%
x_{ij}$we get 
\begin{eqnarray*}
Q_{0} &=&\sum_{ij}\overline{\left( \frac{2e}{\hbar }x_{ij}\right) ^{2}}%
\overline{\;\left( \frac{g}{2}M_{ij}\right) ^{2}}R \\
R &=&\;\overline{\frac{\tanh (\beta \xi _{i})-\tanh (\beta \xi _{j})}{2(\xi
_{i}-\xi _{j})}}
\end{eqnarray*}%
The temperature enters only in the last factor in this product,
differentiating it over the inverse temperature we find%
\begin{equation*}
\frac{dR}{d\beta }\;=\int d\xi _{1}d\xi _{2}\frac{\nu (\xi _{1})\nu (\xi
_{2})}{\xi _{1}-\xi _{2}}\left[ \frac{1}{\cosh ^{2}\beta \xi _{1}}-\frac{1}{%
\cosh ^{2}\beta \xi _{2}}\right]
\end{equation*}%
The integrand decreases exponentially fast for $\xi \gg T$ , neglecting the
contribution from these regions and assuming, as usual, that $\nu (\xi
\lesssim T)=\nu $ is constant at low energies we perform integration and get
the result announced earlier: $dR/d\beta =0$.

We now discuss the regular and superconducting contributions to the spectral
weight below $T_{c}$. The computation above demonstrates that leading terms
in $1/Z_{eff}$ involve only pairs of spins. This remains true below $T_{c}$
as well and allows us to compute easily the regular part of the spectral
weight in this temperature range. We begin by using the Kubo formula to
express it through the spin-spin correlators%
\begin{equation}
K(T)=\frac{1}{\mathcal{V}}\sum_{ij}\left( \frac{2e}{\hbar }x_{ij}\right)
^{2}\left( \frac{g}{2}M_{ij}\right) ^{2}\int_{0}^{\infty }\frac{d\omega }{%
\omega }\mathrm{Re}R_{ij}(\omega )  \label{K_regular}
\end{equation}

where $R_{ij}(\omega )$ is current-current correlator 
\begin{equation}
R_{ij}(\omega )=i\int_{0}^{\infty }dte^{i\omega t}\left\langle \left[ \hat{I}%
_{ij}(t),\hat{I}_{ij}(0)\right] \right\rangle  \label{Rij}
\end{equation}%
evaluated in the absence of the external field: $\hat{I}%
_{ij}(t)=i(S_{i}^{+}S_{j}^{-}-S_{i}^{-}S_{j}^{+})_{t}$. The equation (\ref%
{K_regular}) can be simplified by using the Kramers-Kronig relation for the
correlator $R_{ij}(\omega )$: 
\begin{equation}
K(T)=\frac{1}{\mathcal{V}}\sum_{ij}\left( \frac{ge}{\hbar }%
x_{ij}M_{ij}\right) ^{2}\mathrm{Im}R_{ij}(0)  \label{KT}
\end{equation}

\bigskip\ 
\begin{eqnarray}  \label{K_regular_below}
K(T &<&T_{c})=\frac{1}{8\mathcal{V}}\sum_{ij}\left( \frac{2e}{\hbar }%
x_{ij}\right) ^{2}\left( \frac{g}{2}M_{ij}\right) ^{2}\left[ \left( \frac{%
\xi _{i}}{\varepsilon _{i}}+\frac{\xi _{j}}{\varepsilon _{j}}\right) ^{2}%
\frac{\tanh (\varepsilon _{i}/T)-\tanh (\varepsilon _{j}/T)}{\varepsilon
_{i}-\varepsilon _{j}}\right.  \notag \\
&&\left. +\left( \frac{\xi _{i}}{\varepsilon _{i}}-\frac{\xi _{j}}{%
\varepsilon _{j}}\right) ^{2}\frac{\tanh (\varepsilon _{i}/T)+\tanh
(\varepsilon _{j}/T)}{\varepsilon _{i}+\varepsilon _{j}}\right] .
\end{eqnarray}%
One can immediately see that the regular part (\ref{K_regular_below})
contains an extra factor $\sim \frac{gM_{ij}}{\varepsilon }$ compared to the
total spectral weight $K^{tot}(T)$ in (\ref{K^total}). Thus we conclude that
variation of the regular spectral weight below $T_c$ is smaller than total
spectral weight: 
\begin{equation}
\frac{K(T_c) - K(T)}{K^{tot}(T)}\sim \frac{gM_{ij}}{T_{c}}\sim \frac{1}{Z_{%
\mathrm{eff}}}\, .  \label{comparison}
\end{equation}%
The last estimate in Eq.(\ref{comparison}) may be considered as a practical
definition of $Z_{\mathrm{eff}}$. Indeed, in lattice spin models where each
spin is coupled to $Z_{\mathrm{eff}}$ other spins by a coupling constant $gM$
the transition temperature can be estimated from $gMZ_{\mathrm{eff}}\sim
T_{c}$. As we argued above (see Eqs.(\ref{Zeff},\ref{ZBethe}) and the
corresponding discussion in section \ref{Transition temperature}), the
effective coordination number $Z_{\mathrm{eff}}$ is not too small (and
superconductivity survives) in a wide region of localized single-particle
states where $\delta _{L}/T_{c}$ is larger than 1. Eq.(\ref{comparison})
implies that at $Z_{\mathrm{eff}} \gg 1$ the temperature dependence of the
total spectral weight at $T<T_{c}$ is almost entirely related with the
variation of the superconducting density: 
\begin{equation*}
\delta K^{tot}=K^{tot}(T)-K^{tot}(T_{c})\approx \rho _{s}(T)
\end{equation*}%
which is thus described by Eq.(\ref{Q_0}).

\begin{figure}[tbp]
\includegraphics[width=8cm]{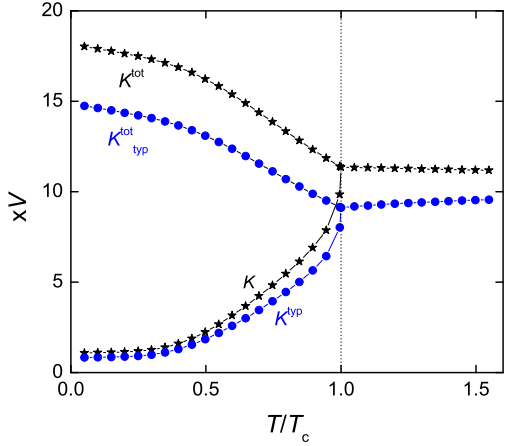}
\caption{(Color online) Temperature dependence of regular spectral weight $%
K(T)$ and full spectral weight $K^{tot}(T)$ computed numerically (with 3D
Gaussian AM of lattice size $L=20$) in the region of pseudogap
superconductivity at $E_F=7.9$. We used Eqs.(\protect\ref{K_regular_below})
and (\protect\ref{K^total}), correspondingly, and employed approximate
expression for the gap function $\Delta(\protect\xi,T)$ in the factorized
form $\Delta(\protect\xi,T) = \Delta_c(\protect\xi)\protect\sqrt{1-T/T_c}$,
where $\Delta_c(\protect\xi)$ is the critical-point solution shown in Fig.~%
\protect\ref{DeltaTc}. Black stars stand for the simple averaging, whereas
blue circles correspond to typical averages. }
\label{FigSpectralWeight}
\end{figure}

We now compute $\rho _{s}(T)$ and $K^{tot}(T)$ as given by (\ref{K^total})
in the mean field approximation developed in this paper. We use modified MFA
equation for the gap function in the form (\ref{MMFA5}) and replace $%
x_{ij}^{2}$ under the sum by its average value $\frac{1}{2}R_{0}^{2}$
determined in Eq.(\ref{Zeff}) in (\ref{K^total}). The result is 
\begin{equation}
\rho _{s}(T)\approx \frac{2\nu _{0}e^{2}R_{0}^{2}}{\hbar ^{2}}%
\int_{0}^{\infty }\frac{d\xi \Delta ^{2}(\xi )}{\sqrt{\xi ^{2}+\Delta
^{2}(\xi )}}\tanh \frac{\sqrt{\xi ^{2}+\Delta ^{2}(\xi )}}{T}  \label{rhos2}
\end{equation}%
which can be further simplified at $T\rightarrow 0$: 
\begin{equation}
\rho _{s}\approx \frac{2e^{2}R_{0}^{2}}{\hbar ^{2}}\int_{\Delta }^{\infty }%
\frac{\nu (\varepsilon )d\varepsilon }{\varepsilon }\hat{\Delta}%
^{2}(\varepsilon )\sim \frac{2\nu _{0}e^{2}R_{0}^{2}}{\hbar ^{2}}\Delta ^{2}
\label{rhos3}
\end{equation}%
This expression allows one to extract the effective interaction range $R_{0} 
$ from the data on the superconducting density in the combination with the
gap $\Delta $ and DoS $\nu (\varepsilon )$ found from the Andreev
spectroscopy measurements (see section \ref{Andreev point contact}).

In the above discussion we assumed that effective number of neighbors $Z_{%
\mathrm{eff}}$ for the pseudospin model is large. It is not always the case
because, as discussed in the end of section~\ref{Transition temperature},
the superconducting state survives in the strongly spatially fluctuating
form in the regime $Z_{\mathrm{eff}}<1$. Qualitative picture of the spectral
weight temperature dependence in the pseudogap range of the Anderson model
formulated in section~\ref{Transition temperature} can be obtained from the
numerical evaluation of the regular part $K(T)$ given by (\ref%
{K_regular_below}) and the total weight $K^{tot}(T)$ given by (\ref{K^total}%
). The results are shown in Fig.\ref{FigSpectralWeight} for $E_{F}=7.9$. We
emphasize that the expressions (\ref{K^total},\ref{K_regular_below}) give
the leading orders in $1/Z_{\mathrm{eff}}$ expansion only, which is the 
\textit{zero order} for Eq.(\ref{K^total}), and the \textit{first order} for
Eq.(\ref{K_regular_below}). As one sees in Fig.\ref{FigSpectralWeight}, both 
$K(T)$ and $K^{tot}(T)$ change by a similar amounts, so in this example $Z_{%
\mathrm{eff}}\sim 1$. We expect that Fig.\ref{FigSpectralWeight} provides a
qualitatively good description of the spectral weight behavior in
pseudogapped superconductors.

\section{Summary of results and unsolved problems. \label{Summary of results}%
}

We presented a generalization of the BCS-like theory of superconductivity
that is appropriate for poor three-dimensional conductors in which Fermi
level is located in the vicinity of the mobility edge. The developed theory
neglects the effects of Coulomb repulsion. The reasons why this
approximation is appropriate to describe many disordered films were
discussed in section \ref{Experimental results}.

The model with BCS-like attraction and no Coulomb repulsion solved in the
bulk of the paper can be also realized by ultra-cold fermionic atomic gases
in optical traps~\cite{atoms,atoms2} with controlled disorder \cite{Aspect}.
The enhancement of $T_{c}$ by disorder predicted theoretically might turn
out to be a useful tool to observe BCS state in these systems because
without such enhancement $T_{c}$ is often too low to be reached
experimentally for the small values of the interaction constant $\lambda $.
Such experiments may be also important for the understanding of
superconductor-insulator transition in general.

The control parameter of the model is position of the Fermi level $E_{F}$
with respect to the mobility edge $E_{c}$. The major new (with respect to
the old works\cite{MaLee,KapitulnikKotliar1986,BulaSad}) ingredient of the
developed theory is the full account of the critical and weakly off-critical
wavefunction's fractality. We identified three qualitatively different
regimes: the hard-gap insulator realized at large disorder, the pseudogap
superconductor that appears when the phase correlations develop between well
localized Cooper pairs and the "fractal" superconductor that appears against
the background of fractal single particle states. Upon a further decrease of
disorder the fractal superconductor smoothly crosses over to a conventional
one.

For the insulating state, our main result is the computation of the
activation energy $T_{I}$ in the Arrhenius temperature dependence of
resistivity, $\ln R(T)\sim T_{I}/T$, at low temperatures. The hard-gap
character of this insulating state is due to the local attractive
electron-electron interaction which leads to formation of localized Cooper
pairs with nonzero binding energies $\Delta _{P}^{(j)}$, specific for each $%
j $-th localized orbital. The probability distribution of these energies is
similar to the distribution of inverse participation ratios for localized
eigenstates of the 3D Anderson model; it drops exponentially fast at low
energies, imitating a hard gap. The estimate for the activation energy $%
T_{I} $ as a function of $E_{F}-E_{c}$ is in a reasonable agreement with
experimental data~\cite{Shahar1992}.

The fractal superconducting state (realized for $E_{F}$ very close to $E_{c}$%
) is characterized by the following features:

\begin{enumerate}
\item {\ In the weak coupling limit, $\lambda \ll 1$, the transition
temperature, }${T}${$_{c}$, becomes a power law function of the interaction
constant, }$\lambda $. This leads to the{\ parametric enhancement of $T_{c}$
with respect to its value deep in the metal state. This conclusion is
different from previous works~\cite{MaLee,KapitulnikKotliar1986,BulaSad}
which assumed the validity of the \textquotedblleft Anderson
theorem\textquotedblright\ in the $E_{F}$ region very close to mobility
edge. The power law exponent in the $T_{c}(\lambda )\propto \lambda
^{3/(3-d_{2})}$ dependence is determined by the fractal dimension $%
d_{2}\approx 1.3$ of the critical eigenfunctions. }

\item {\ Strong local fluctuations of the pairing amplitude $\Delta (\mathbf{%
r})$ coexist with a unique and well-defined critical temperature $T_{c}$
below which a macroscopically coherent state appears.}

\item {\ The thermal Ginzburg parameter $\mathrm{Gi}$ of this transition is
a universal quantity of the order of unity which is independent of the
interaction constant $\lambda $, the same holds for the mesoscopic Ginzburg
number $\mathrm{Gi}_{d}$. This implies that thermal and mesoscopic
fluctuations may change the pre-factor in the power law $T_{c}(\lambda )$ by
a factor $\sim O(1)$ but cannot change the power-law itself.}

\item {\ Local single-particle density of states $\nu (\varepsilon ,\mathbf{r%
})$ fluctuates strongly in real space; these fluctuations lead to a strong
random asymmetry of the tunneling conductance, $G_{T}(V,\mathbf{r})\neq
G_{T}(-V,\mathbf{r})$. }

\item {\ The value of the superfluid response, Eq.(\ref{rhos01}), coincides
with result of ~\cite{MaLee} (Equation (4.1) in this paper), provided that
one inserts in it the correct values of $T_{c}$ and $\Delta _{0}$ given
above. }
\end{enumerate}

The pseudogap superconductor is predicted to occur when the Fermi energy $%
E_{F}$ is deep inside the localized band of single-particle states. Its
major feature is the presence of two distinct energy scales, both
originating from the Cooper attraction between weakly localized electrons:
the collective superconductive gap $\Delta (0)$ and the local binding energy
of a Cooper pair $\Delta _{P}$. The most unusual behavior is expected to
occur in the regime of a strong pseudo-gap $\Delta _{P}\gg (\Delta
(0),T_{c}) $. The very existence of superconducting correlations and a
nonzero $T_{c}$ in this regime is unexpected because it is characterized by
a typical level spacing $\delta _{L}=(\nu _{0}L_{loc}^{3})^{-1}$ which is
larger than $T_{c}$ and $\Delta (0)$: $\delta _{L}/T_{c}\propto (\Delta
_{P}/T_{c})^{3/d_{2}}$ (\ref{hard},\ref{D01}). The appearance of this regime
was not expected in previous studies\cite%
{MaLee,KapitulnikKotliar1986,BulaSad} which concluded that the
superconducting state is stable only up to $\delta _{L}\leq T_{c}$. The
perseverance of superconducting coherence much deeper in the localized
region than was expected previously is the result of the enhancement of the
correlations between the wave-functions intensities \cite{CueKra}, which
occurs due to the Mott's mechanism of resonant mixing of localized states.
The key features of the pseudo-gap superconductor (in addition to the
features of fractal superconductor listed above) are the following: \newline

\begin{enumerate}
\item {\ Insulating behavior of the resistivity in a wide range of
temperatures above $T_{c}$. }

\item {\ Formation of almost hard gap without coherence peaks in the density
of states above $T_{c}$ with the gap value of the gap that fluctuates
significantly from point to point. }

\item {Growth of coherence peaks below a global $T_{c}$. The magnitude of
the coherence peaks (proportional to the local value of the superconducting
order parameter)\ fluctuates from point to point, these fluctuations become
very large close to the superconductor-insulator transition. }

\item {\ Two-peak feature in differential conductance at moderate
transmission probabilities as measured by the Andreev point-contact
spectroscopy below $T_{c}$. The lower peak voltage $V_{1}(T,\mathbf{r})$ is
expected to be $\mathbf{r}$ -independent, but vanishing as temperature
approaches $T_{c}$ from below. In contrast, the higher peak voltage $V_{2}(T,%
\mathbf{r})$ is expected to be almost $T$-independent, but strongly $\mathbf{%
r}$-dependent.}

\item {\ Strong temperature dependence of the total spectral weight $%
K^{tot}(T)$ of the high-frequency conductivity which shows with a loss of
major part of $K^{tot}(T)$ at temperatures $T\leq \Delta _{P}$, and its
re-appearance below $T_{c}$. The regular part $K(T)$ of the spectral weight
is smaller than the superconducting response $\rho _{s}$, see Eq.(\ref%
{comparison}). }
\end{enumerate}

We now briefly discuss possible extensions of the developed theory and open
questions. We begin with purely theoretical questions.

An obvious extension of the present theory would be a consistent account of
Coulomb interaction effects in fractal superconductors near the 3D mobility
edge. Here one should distinguish the effects of short range part of the
Coulomb repulsion that competes with the phonon attraction and the long
range part which might become very important in the insulator resulting in
the formation of the Cooper pair glass and qualitatively new physics.
Decreasing the dielectric constant in a real material would change it from
the fractal superconductor discussed in this paper to the material in which
superconductivity is suppressed by Coulomb interaction. Furthermore,
increasing only the short range part of the Coulomb interaction leads to the
'fermionic' mechanism while increase in the long range part leads to the
Coulomb blockade with completely different properties. The transition
between all these regimes are currently not understood, neither
theoretically nor experimentally. It seems likely that at least one of these
crossovers was observed as the 'region of poor scaling' in 
\cite{Steiner2008}.

The character of the quantum critical point that separates the fractal
superconductivity and the insulator at $T=0$ deserves a further study. As
discussed in section \ref{Superconductivity with a pseudogap} on both sides
of this transition electrons are bound in the localized pairs, and the
transition itself consists in development of phase correlations, well
described in terms of the XY ordering of the Anderson pseudo-spins $\mathbf{S%
}_{j}$. This scenario is different from both "bosonic" and the "fermionic"
mechanisms of superconductor-insulator transition, so we suggest the name
"pseudo-spin" mechanism. The cavity approach to the study of such
transitions was developed recently in \cite{IoffeMezard2009}; the main
qualitative conclusion of this study is self-organized inhomogeneity of the
resulting superconductor in which phase correlations are dominated by the
rare, almost one-dimensional paths. This conclusion is supported by the
results of the virial expansion method shown in Fig. \ref{TcDistribution}
that point out to a percolation-like transition between the pseudo-gap
superconductor and an insulator. Another feature of this solution is a very
rapid decrease in the transition temperature beyond certain value of
disorder which might be seen experimentally as the apparent existence of the
lowest nonzero $T_{c}\approx 0.5\mathrm{K}$ found in amorphous InO$_{x}$
system~\cite{Sacepe2007}.

We now turn to experimental findings that lack (partially or entirely)
theoretical explanation.

As discussed in section \ref{Point contact tunneling} the quantitative
description of point contact tunneling requires the theory of collective
modes in the fractal superconducting state. In the absence of such theory
the spectacular, nearly rectangular shape of the \textit{local} tunneling
conductance $G(V)$ at low temperatures\cite{Sacepe2007} cannot be explained
quantitatively.  

The theory of the hard-gap insulating state presented in section \ref%
{Insulating state} neglects the transport by Cooper pairs and takes into
account only single electron transport. It is therefore limited to the
region relatively far from the transition such as studied in Ref.\cite%
{Shahar1992}. Theory of the incoherent transport by Cooper pairs close to
the transition was discussed in the recent papers \cite%
{IoffeMezard2009,Muller2009}, the main conclusion of these studies is that
very close to the transition the behavior of Cooper pairs is controlled by
the many body mobility edge which becomes zero exactly at the transition.
Away from the transition the theory predicts activated behavior at the
lowest temperatures with the gap that increases fast away with disorder and
becomes infinite a short distance away from the transition signalling the
absence of incoherent transport by Cooper pairs~\cite{IoffeMezard2009}.

Recently the experimental paper~\cite{KowalOvadyahu2} reported the anomalous
size-dependence of the superconductor-insulator transition for extremely
wide range of sizes between 1 and 150 microns. In particular, the workers
observed a slow but significant dependence of the activation energy $%
T_{0}(L) $ on the system size $L$ (the distance between the metal contacts)
on the insulating side of the transition. In one case $T_{0}(L)$ decreased
from 13.5 K to 9.6 K while $L$ was changed from 145 to 12 microns at a fixed
width. Although the theory of the the incoherent transport by Cooper pairs
close to the transition \cite{IoffeMezard2009} predicts a strong size
dependence at mesoscopic scales due to inhomogeneities, the size dependence
at the huge scales observed experimentally are very difficult to explain.
The explanation of this effect seems currently to be beyond the reach of a
theory and presents a real challenge.

A different set of challenges is presented by the superconductor-insulator
transition induced by magnetic field for strongly disordered (almost
insulating) superconducting samples. As one would expect, the critical
field, $H_{c}$ (which corresponds to superconductor-insulator transition at $%
T=0$) is a strong function of disorder, for instance, it varies over two
orders of magnitude in the experiments\cite{Steiner2005}. Surprisingly,
there are indications that phenomenology of the transitions driven by
disorder and by magnetic field look differently~ \cite%
{Shahar1992,Gantmakher1998,Shahar2004,Baturina2007a,Steiner2008,SacepeShahar09}%
. Namely, the critical point in disorder is characterized by the activated
behavior of the resistance, $R\sim \exp (T_{0}/T)$, with a large gap $T_{0}$
even very close to the superconductor-insulator transition. In contrast,
activated behavior, $R\sim \exp (T_{I}(H)/T)$, observed at fields above
critical $H_{c}$ is characterized by a small $T_{I}(H)$ which extrapolates
to zero at $H_{c}$ . In other words, the region in ($H,T$) plane where
transport is characterized by activated behavior with a small gap has a
peculiar wedge like shape as sketched in Figure \ref{PhaseDiagram}).
Generally, one expects that very close to the critical line the transport is
dominated by Cooper pairs; this expectation was experimentally confirmed for
ultrathin Bi films ~\cite{Valles2e}. In this case the experimental phase
diagram implies that the regime of Cooper pair dominated transport is narrow
in disorder but wide in field. Qualitatively this is likely to be due to the
fact that the effect of a small magnetic field is limited to the generation
of local phase differences which have a large effect of the superconducting
state but does not affect the incoherent transport of the pairs. However,
there is currently no consistent theory of this effect. On the experimental
side, it would be important to verify that the transport in this regime is
due to Cooper pairs for the InO and TiN films.

Furthermore, strongly disordered superconductors show puzzling behavior in
very large magnetic fields: their resistivity become temperature independent
and approaches the quantum limit, $h/e^{2}$. The field scale, $H_{P}$, at
which this behavior sets in is not sensitive to the disorder in contrast to
the critical field, $H_{c}.$ The second field scale appears first in the
fractal superconductor: it is the field that destroys local Cooper pairs and
suppresses the hard gap $\Delta _{P}$. In this regime, the parametric
difference between $H_{P}$ and $H_{c}$ is due to the low fractal dimension $%
d_{2}\approx 1.3$ of a single-particle wave-function that suppresses the
orbital effect of magnetic field on the pairing of two electrons localized
on this eigenstate, similarly to a very thin wire. In contrast, long-range
coherence requires the existence of large loops involving many localized
Cooper pairs. The typical size of these loops is larger than the
localization length $L_{loc}$,
leading to relatively small magnetic field scale $H_{c}\leq \Phi
_{0}/L_{loc}^{2}\ll H_{P}$. In the pseudogap superconductor the second field
scale is associated with the suppression of a large gap, $\Delta _{P}$,
responsible for Cooper pair formation whereas the critical field is
associated with the frustration of a weak pseudospin coupling. Thus, one
expects even larger difference between the field scales in this regime.
However, the universal value of the conductance in the regime $B\gg H_{P}$
is not expected theoretically and remains mysterious.

Nernst effect can be potentially a very sensitive probe of the nature of the
superconducting state.\cite{OngNernst} These experiments were performed
recently on InO$_{x}$ samples and show~\cite{Spathis} Nernst signal which
scales as $N(T)\propto T^{-n}$ , with the "Nernst exponent" $n\approx 7.5$
at low magnetic fields. Application of the conventional theory \cite%
{LarkinVarlamovBook} of the superconducting fluctuations to the Nernst
effect in 2D superconductors~\cite{Serbin,Karen} gives Nernst signal that
scales as $1/T\ln T$ at high temperatures. This behavior was indeed observed
in conventional superconductor NbSi~\cite{AubinNernst}. Because Nernst
effect requires a motion of the electrical charge around the plaquette, the
temperature dependence is expected to be different in discrete systems where
motion around the minimal plaquette involve a large number of hops and each
hop implies an extra power of $1/T$ in the high temperature expansion~\cite%
{Podolsky07}. For instance, because minimal plaquette on the hexagonal
lattice contains six sites, one expects that $N(T)\propto T^{-6}$ in this
case. The striking difference between NbSi and InO$_{x}$ behavior observed in~\cite%
{Spathis} on very low $T_{c}$ (very disordered) samples indicates, in our
opinion, the importance of Cooper pair hoping between localized sites. This
behavior should get less pronounced further away from the transition, thus
we expect that slightly less disordered samples will show more
'conventional' exponent in Nernst effect.

A number of papers noted the apparent similarity between phenomenology of 
disordered  films of conventional superconductors discussed in this paper 
and that of high $T_c$ oxides.\cite{Steiner2005,Steiner2004} 
Very briefly, the transport and magnetic measurements show the formation 
of pseudogap at temperatures much higher than superconducting $T_c$ in 
underdoped materials,\cite{TimuskReview} 
while STM measurements display highly inhomogeneous 
order parameter (as measured by coherence peaks) combined with modestly 
homogeneous tunneling gap.\cite{Kapitulnik2001,Davis2001,Davis2002,Yazdani2007}   
Furthemore, a number of indirect evidences points out to the superconducting 
nature of the pseudogap in these materials \cite{Corson1999,Wang2006,Yuli2009}, 
similar to the situation discussed in this paper. 
The crucial difference between s-wave and d-wave pairing characterizing high 
$T_c$ oxides is that, in contrast to s-wave pairing, modest elastic scattering 
with mean free path of the order of superconducting coherence length suppresses 
the d-wave superconductivity. Thus, localization of the wave functions is
incompatible with generic d-wave superconductivity. This conclusion might need 
to be revised for the special case of superconductivity which is due to pairing 
of electrons in a small area of the full Brillouin zone which might be the case
of cuprates \cite{Geshkenbein1997,Galitski2009}. In this case, the electrons 
that are responsible for the pairing belong to two well separated patches on 
the Fermi surface with each patch characterized by a small momentum $p_0$. The
density of states in these patches is large which makes scattering of these 
electrons strong. Furthermore,  because most impurities are located far from the 
conducting copper oxide planes, it is likely that the elastic scattering, though
strong, does not mix different patches and thus does not inhibit d-wave pairing.
In this situation, single electron states inside each patch may become very similar 
to the localized states discussed in this work making the developed theory 
qualitatively correct. This would explain the main phenomenological features
mentioned above: formation of pseudogap far above $T_c$, highly inhomogeneous order
parameter observed in tunneling data, insulating behavior of LaSrCuO superconductors
in high magnetic fields, etc.

Finally, it is not clear why the phenomenology displayed by Josephson
junction networks\cite{FazioZant,Serret2002} was not observed in any
disordered film. In particular, Josephson junction networks close to
superconductor-insulator transition are characterized by the appearance of a
wide region of magnetic fields where resistivity has no temperature
dependence but varies by many orders of magnitude as a function of magnetic
field. This is in a sharp contrast with the films that show more or less
good crossing point of $R(B)$ isotherms. There is very little doubt that
Josephson networks are described by model Hamiltonian (\ref{H_JJ}), so one
has to conclude that it is not the appropriate model for most films. One
possible source of difference discussed above might be that many films are
characterized by a large value of the dielectric constant, while another
reason, relevant for the films with small dielectric constant, might be the
capacitance matrix is not dominated by nearest neighbors in these films.
Nearest neighbor capacitance matrix translates into the Coulomb interaction
that decreases only logarithmically with distance in the Josephson network
as opposed to $1/r$ expected in a film. The physical result might
be the formation of the Cooper pair glass in the Josephson network but not
in the disordered film or vice versa. If true, one expects that Josephson
networks with the Coulomb interaction screened by the ground plate might
display properties more similar to those of the films. Another difference
might be due to the fluxes produced by magnetic field are completely random
in a disordered films but are (almost) identical on the plaquettes of the
Josephson lattice. In this case one expects to see film behavior in
randomized Josephson network.

\bigskip We are grateful to B. L. Altshuler, T. I. Baturina, C. Chapelier,
T. Dubouchet, A. M. Finkelstein, V. F. Gantmakher, A. S. Ioselevich, I. V.
Lerner, A. Millis, A. D. Mirlin, M. Mueller, Z. Ovadyahu, V. V. Ryazanov, B.
Sacepe, V. Schmidt, D. Shahar, M. A. Skvortsov and V. I. Yudson for useful
discussions. This research was supported by Triangle de la physique 2007-36,
grants ANR-06-BLAN-0218, ARO 56446-PH-QC and DARPA HR0011-09-1-0009, by the
grant 07-02-00310 from Russian Foundation for Basic Research, and by the
program "Quantum physics of condensed matter" of Russian Academy of
Sciences. We thank the FEDER and the Spanish DGI for financial support
through Project No. FIS2007-62238. V.E.K acknowledges a hospitality of the
Institute for Nuclear Theory at the University of Washington and the DOE for
partial support during the completion of this work.

\appendix

\section[\hspace{0.7in}Virial expansion in pseudospin subspace]{Virial
expansion in pseudospin subspace}

\label{Virial expansion in pseudospin subspace}

The two-spin susceptibility $\chi _{12}^{(2)}=\chi _{12}^{(2a)}+\chi
_{12}^{(2b)}$ is computed as explained in section \ref{Virial expansion} and
can be easily written analytically even for nonzero Hartree terms. It is a
sum of two terms, one of which does not explicitly depend on $J_{12}^{\perp
} $: 
\begin{eqnarray}
\chi _{12}^{(2,a)} &=&Z_{0}^{-1}\,\left[ \frac{\sinh \left( \frac{E_{+}}{T}+%
\frac{J_{12}^{\parallel }}{2T}\right) }{\left( E_{+}+\frac{J_{12}^{\parallel
}}{2}\right) }\,e^{E_{-}/T}+\frac{\sinh \left( \frac{E_{+}}{T}-\frac{%
J_{12}^{\parallel }}{2T}\right) }{\left( E_{+}-\frac{J_{12}^{\parallel }}{2}%
\right) }\,e^{-E_{-}/T}\right.  \label{reg} \\
&+&\left. \frac{\sinh \left( \frac{E_{-}}{T}+\frac{J_{12}^{\parallel }}{2T}%
\right) }{\left( E_{-}+\frac{J_{12}^{\parallel }}{2}\right) }\,e^{E_{+}/T}+%
\frac{\sinh \left( \frac{E_{-}}{T}-\frac{J_{12}^{\parallel }}{2T}\right) }{%
\left( E_{-}-\frac{J_{12}^{\parallel }}{2}\right) }\,e^{-E_{+}/T}\right] , 
\notag
\end{eqnarray}%
where 
\begin{equation*}
Z_{0}=4\cosh \left( \frac{E_{+}}{T}\right) \,\cosh \left( \frac{E_{-}}{T}%
\right) \,\cosh \left( \frac{J_{12}^{\parallel }}{2T}\right) +\newline
4\sinh \left( \frac{E_{+}}{T}\right) \,\sinh \left( \frac{E_{-}}{T}\right)
\,\sinh \left( \frac{J_{12}^{\parallel }}{2T}\right)
\end{equation*}%
The other term is proportional to $J_{12}^{\perp }$: 
\begin{eqnarray}
\chi _{12}^{(2,b)} &=&\,\frac{Z_{0}^{-1}J_{12}^{\perp }}{(E_{+}-E_{-})}\left[
\frac{\sinh \left( \frac{E_{+}}{T}-\frac{J_{12}^{\parallel }}{2T}\right)
e^{-E_{-}/T}}{\left( E_{+}-\frac{1}{2}J_{12}^{\parallel }\right) }\,-\frac{%
\sinh \left( \frac{E_{+}}{T}+\frac{J_{12}^{\parallel }}{2T}\right)
e^{E_{-}/T}}{\left( E_{+}+\frac{1}{2}J_{12}^{\parallel }\right) }\,\right.
\label{sing-J} \\
&+&\left. \frac{\sinh \left( \frac{E_{-}}{T}+\frac{J_{12}^{\parallel }}{2T}%
\right) }{\left( E_{-}+\frac{1}{2}J_{12}^{\parallel }\right) }\,e^{E_{+}/T}-%
\frac{\sinh \left( \frac{E_{-}}{T}-\frac{J_{12}^{\parallel }}{2T}\right) }{%
\left( E_{-}-\frac{1}{2}J_{12}^{\parallel }\right) }\,e^{-E_{+}/T}\right] 
\notag
\end{eqnarray}

Here we present the details of the formalism for computing the three-spin
susceptibility. It is developed with the goal to use for numerical
computation rather than being a full-analytical approach. The even Hilbert
subspace of the problem of three sites with $2$ or $0$ electron on each site
(equivalent to the spin-$\frac{1}{2}$ problem with three coupled spins)
consists of 8 basic states classified by the total spin. The corresponding
Hamiltonian has a block structure corresponding to the total spin $\frac{3}{2%
}$, $0$, $\frac{1}{2}$, $-\frac{1}{2}$. The first two blocks are just
numbers: 
\begin{equation}
\pm (\xi _{1}+\xi _{2}+\xi _{3})-\frac{1}{2}(J_{12}^{\parallel
}+J_{13}^{\parallel }+J_{23}^{\parallel }).  \label{1X1}
\end{equation}%
The last two blocks are $3\times 3$ matrices: 
\begin{equation}
A=\left( 
\begin{array}{ccc}
\xi _{23}^{1}+\frac{J_{12}^{\parallel }+J_{13}^{\parallel
}-J_{23}^{\parallel }}{2} & -J_{12}^{\perp \ast } & -J_{13}^{\perp \ast } \\ 
-J_{12}^{\perp } & \xi _{13}^{2}+\frac{J_{12}^{\parallel }-J_{13}^{\parallel
}+J_{23}^{\parallel }}{2} & -J_{23}^{\perp \ast } \\ 
-J_{13}^{\perp } & -J_{23}^{\perp } & \xi _{12}^{3}+\frac{-J_{12}^{\parallel
}+J_{13}^{\parallel }+J_{23}^{\parallel }}{2}%
\end{array}%
\right) .  \label{345-J}
\end{equation}%
\begin{equation}
B=\left( 
\begin{array}{ccc}
-\xi _{12}^{3}+\frac{-J_{12}^{\parallel }+J_{13}^{\parallel
}+J_{23}^{\parallel }}{2} & -J_{23}^{\perp \ast } & -J_{13}^{\perp \ast } \\ 
-J_{23}^{\perp } & -\xi _{13}^{2}+\frac{J_{12}^{\parallel
}-J_{13}^{\parallel }+J_{23}^{\parallel }}{2} & -J_{12}^{\perp \ast } \\ 
-J_{13}^{\perp } & -J_{12}^{\perp } & -\xi _{23}^{1}+\frac{J_{12}^{\parallel
}+J_{13}^{\parallel }-J_{23}^{\parallel }}{2}%
\end{array}%
\right) .  \label{678-J}
\end{equation}

where $\xi _{bc}^{a}=\xi _{b}+\xi _{c}-\xi _{a}$ . Such structure suggests
that the eigenvalues of the 3-spin problem are also grouped: 
\begin{equation}
\lambda _{1,2}^{(0)}=\pm (\xi _{i}+\xi _{j}+\xi _{k})-\frac{1}{2}%
(J_{12}^{\parallel }+J_{13}^{\parallel }+J_{23}^{\parallel }),
\label{lambda12}
\end{equation}%
and two groups of eigenvalues $\lambda _{3,4,5}^{(0)}$ and $\lambda
_{6,7,8}^{(0)}$ are found from the solution of the two cubic characteristic
equations $\mathrm{det}(A-\lambda _{3,4,5}^{(0)}\,I)=0$, and $\mathrm{det}%
(B-\lambda _{6,7,8}^{(0)}\,I)=0$.

One can find the solutions to these equations either numerically or from the
Cartan formula for the roots of cubic equations, the former being more
efficient in the numerical calculations.

The 8-eigenvectors $X_1=(1,0,0,0,0,0,0,0)$ and $X_2=(0,0,0,0,0,0,0,1)$ are
simple: they contain unity in the first and eight row respectively, and all
the other elements being zero; the structure of the nontrivial eigenvectors
of the $3\times3 matrices$ is as follows: 
\begin{eqnarray}  \label{EV}
X^T_{\alpha=3,4,5}=\left( 0\, , u_\alpha\, , v_\alpha\, , w_\alpha\, ,0\, ,
0\, ,0\,,0 \right) \\
X^T_{\alpha=6,7,8}= \left( 0\, , 0\, ,0\,,0\, u^{\prime }_\alpha\, ,
v^{\prime }_\alpha\, , w^{\prime }_\alpha\,, 0 \right)  \notag
\end{eqnarray}
where $(u_{\alpha},v_{\alpha},w_{\alpha})$ is the normalized 3-eigenvector
of the matrix $A$ that corresponds to the eigenvalue $\lambda^{(0)}_{%
\alpha=3,4,5}$ and $u^{\prime }_{\alpha},v^{\prime }_{\alpha},w^{\prime
}_{\alpha}$ is the normalized 3-eigenvector of the matrix $B$ that
corresponds to the eigenvalue $\lambda^{(0)}_{\alpha=6,7,8}$.

The perturbation matrix in the same basis takes the form: 
\begin{equation}
V^{(3)}=\left( 
\begin{array}{cccccccc}
0 & \Delta & \Delta & \Delta & 0 & 0 & 0 & 0 \\ 
\Delta ^{\ast } & 0 & 0 & 0 & \Delta & \Delta & 0 & 0 \\ 
\Delta ^{\ast } & 0 & 0 & 0 & \Delta & 0 & \Delta & 0 \\ 
\Delta ^{\ast } & 0 & 0 & 0 & 0 & \Delta & \Delta & 0 \\ 
0 & \Delta ^{\ast } & \Delta ^{\ast } & 0 & 0 & 0 & 0 & \Delta \\ 
0 & \Delta ^{\ast } & 0 & \Delta ^{\ast } & 0 & 0 & 0 & \Delta \\ 
0 & 0 & \Delta ^{\ast } & \Delta ^{\ast } & 0 & 0 & 0 & \Delta \\ 
0 & 0 & 0 & 0 & \Delta ^{\ast } & \Delta ^{\ast } & \Delta ^{\ast } & 0%
\end{array}%
\right)  \label{V-3}
\end{equation}%
With the eigenvalues $\lambda _{\alpha }^{(0)}$ and the eigenvectors $%
X_{\alpha }$ ($\alpha =1...8$) obtained, one can compute $\gamma _{\alpha }$
from Eq.(\ref{gamma}) (the field $\Delta $ cancels out): %\begin{widetext}
\begin{eqnarray}
\gamma _{1} &=&\sum_{\beta =3,4,5}\frac{(u_{\beta }+v_{\beta }+w_{\beta
})^{2}}{\lambda _{1}^{(0)}-\lambda _{\beta }^{(0)}}\qquad \qquad \gamma
_{2}=\sum_{\beta =6,7,8}\frac{(u_{\beta }^{\prime }+v_{\beta }^{\prime
}+w_{\beta }^{\prime })^{2}}{\lambda _{2}^{(0)}-\lambda _{\beta }^{(0)}}
\label{gamma3} \\
\gamma _{\alpha =3,4,5} &=&\sum_{\beta =6,7,8}\frac{u_{\alpha }u_{\beta
}^{\prime }+w_{\alpha }w_{\beta }^{\prime }+u_{\alpha }v_{\beta }^{\prime
}+v_{\alpha }u_{\beta }^{\prime }+v_{\alpha }w_{\beta }^{\prime }+w_{\alpha
}v_{\beta }^{\prime }}{\lambda _{\alpha }^{(0)}-\lambda _{\beta }^{(0)}}+%
\frac{(u_{\alpha }+v_{\alpha }+w_{\alpha })^{2}}{\lambda _{\alpha
}^{(0)}-\lambda _{1}^{(0)}}  \notag \\
\gamma _{\alpha =6,7,8} &=&\sum_{\beta =3,4,5}\frac{u_{\alpha }^{\prime
}u_{\beta }+w_{\alpha }^{\prime }w_{\beta }+u_{\alpha }^{\prime }v_{\beta
}+v_{\alpha }^{\prime }u_{\beta }+v_{\alpha }^{\prime }w_{\beta }+w_{\alpha
}^{\prime }v_{\beta }}{\lambda _{\alpha }^{(0)}-\lambda _{\beta }^{(0)}}+%
\frac{(u_{\alpha }^{\prime }+v_{\alpha }^{\prime }+w_{\alpha }^{\prime })^{2}%
}{\lambda _{\alpha }^{(0)}-\lambda _{2}^{(0)}}.  \notag
\end{eqnarray}%
and then compute $\chi _{ijk}^{(3)}$ from Eq.(\ref{chi-N}). Finally,
subtracting the proper combinations of the two- and one- spin
susceptibilities one obtains the virial coefficient $\chi _{3}(T)$ from Eq.(%
\ref{chi2}).

\section[\hspace{0.7in}Virial expansion including single-occupied states]{%
Virial expansion including single-occupied states}

\label{Virial expansion including single-occupied states}

The Hilbert space we considered so far corresponded to each orbital $%
i,j,k... $ occupied either by two electrons or empty. Now let us consider
the same Hamiltonian, Eq.(\ref{HamSpin}), but extend the Hilbert space to
include possible single-electron occupancies of all orbitals. To avoid
unnecessary complications we omit here the Hartree terms setting $%
g_{\parallel}=0$ in Eq.(\ref{HamSpin}).

\subsection[\hspace{0.7in}One-orbital problem]{One-orbital problem}

The full Hilbert space in this case consists of 
\begin{equation*}
|0\rangle,\;\;\;|2\rangle,\;\;\;|\uparrow\rangle,\;\;\;| \downarrow\rangle.
\end{equation*}
The first two states contain even number of electrons (the "even" states)
and the last two (the "odd" states) contain one electron each. Each "odd"
state has extra energy $\Delta_P$.

It is important that the spin operator acting on the "odd" states $%
|\uparrow \rangle $ or $|\downarrow \rangle $ gives zero: 
\begin{equation}
S^{+/-}|\downarrow \rangle =S^{+/-}|\uparrow \rangle =0.  \label{zero-S}
\end{equation}%
The energies of the four states are (we count all energies with respect to $%
-\varepsilon =-\xi +\Delta _{P}$): 
\begin{equation*}
-\varepsilon ,\;\;\;2\xi -2\Delta _{P}-\varepsilon =+\varepsilon ,\;\;\;\xi
-\varepsilon =\Delta _{P},\;\;\;\;\Delta _{P}.
\end{equation*}%
Eq.(\ref{zero-S}) suggests that the "partial susceptibilities" $\gamma _{%
\mathrm{odd}}$ given by Eq.(\ref{eigen}) are zero for all the odd states.
This is because the Hamiltonian Eq.(\ref{HamSpin}) does not couple the even
and odd sectors of the Hilbert space.

Thus all one has to do to compute the one-site susceptibility is to add $%
2e^{-\Delta_{P}/T}$ to the partition function $Z_{0}$. The result is: 
\begin{equation}  \label{1-full}
\chi^{(1)}=\left(\frac{1}{2\varepsilon}\right)\,F(\varepsilon,\Delta_{P}),\;%
\;\;\; F(\varepsilon,\Delta_{P})= \frac{\sinh\left(\frac{\varepsilon}{T}%
\right)}{\cosh\left(\frac{\varepsilon}{T}\right) +e^{-\Delta_{P}/T}}.
\end{equation}
Note that there is a nice property: 
\begin{equation}  \label{T-2T}
F(\varepsilon,\infty)=\tanh\left( \frac{\varepsilon}{T}\right),\;\;\;\;F(%
\varepsilon,0)=\tanh\left( \frac{\varepsilon}{2T}\right).
\end{equation}

\subsection[\hspace{0.7in}Two-orbital problem]{Two-orbital problem}

The full Hilbert space for the 2-orbital problem consists of the \textit{%
even sector}: 
\begin{equation*}
\psi _{1}=|0,0\rangle ,\;\;\;\;\psi _{2}=|2,2\rangle ,\;\;\;\;\psi
_{3}=|2,0\rangle ,\;\;\;\;\psi _{4}=|0,2\rangle
\end{equation*}%
an the \textit{odd sector}: 
\begin{eqnarray}
\psi _{5,6} &=&|\alpha ,0\rangle ,\;\;\;\;\psi _{7,8}=|0,\alpha \rangle
,\;\;\;\;\psi _{9,10}=|\alpha ,2\rangle ,  \notag \\
\;\;\;\;\psi _{11,12} &=&|2,\alpha \rangle ,\;\;\;\;\psi _{13-16}=|\alpha
,\alpha ^{\prime }\rangle ,  \notag
\end{eqnarray}%
where $\alpha =\uparrow $ or $\downarrow $.

Let us first find the eigenvalues in the odd sector (counted from $%
-(\varepsilon _{1}+\varepsilon _{2})$): 
\begin{eqnarray}
\lambda _{5,6}^{(0)} &=&\xi _{1}-(\varepsilon _{1}+\varepsilon
_{2})=-\varepsilon _{2}+\Delta _{P}^{(1)},  \notag \\
\lambda _{7,8}^{(0)} &=&\xi _{2}-(\varepsilon _{1}+\varepsilon
_{2})=-\varepsilon _{1}+\Delta _{P}^{(2)},  \notag \\
\lambda _{9,10}^{(0)} &=&\xi _{1}+2(\xi _{2}-\Delta _{P}^{(2)})-(\varepsilon
_{1}+\varepsilon _{2})=\varepsilon _{2}+\Delta _{P}^{(1)},  \notag \\
\lambda _{11,12}^{(0)} &=&2(\xi _{1}-\Delta _{P}^{(1)})+\xi
_{2}-(\varepsilon _{1}+\varepsilon _{2})=\varepsilon _{1}+\Delta _{P}^{(2)}.
\notag
\end{eqnarray}

In all the above formulae 
\begin{equation}
\varepsilon _{i}=\xi _{i}-\Delta _{P}^{(i)},  \label{varep}
\end{equation}%
where $\xi _{i}$ is a single-particle energy (eigenvalues of the state $\psi
_{i}(\mathbf{r})$) measured from the Fermi-energy, and $\Delta _{P}^{(i)}$
is a pseudo-gap: 
\begin{equation}
\Delta _{p}^{(i)}=\frac{g}{2}\,\sum_{\mathbf{r}}|\Psi _{i}(\mathbf{r})|^{4}.
\label{D_p}
\end{equation}%
Note that Eq.(\ref{zero-S}) guarantees that the eigenvalues of the odd
sector do not have any $M_{ij}$-dependent renormalization. However, not all
the partial susceptibilities $\gamma _{\mathrm{odd}}$ are equal to zero. The
reason is that the source terms $\Delta \,S^{+}+\Delta ^{\ast }\,S^{-}$
contain only one $S^{+(-)}$ operator whereas the number of sites is two.
This operator may act on the site in an even state and then the result of
action of the source term on a \textit{semi-odd} state $|\mathrm{even},%
\mathrm{odd}\rangle $ will be non-zero. Thus we expect that the structure
constants for the states $\psi _{i}$, ($i=5,6,...12$) are non-zero. A simple
calculation shows that

\begin{eqnarray}  \label{2-site-fin}
\gamma _{5} &=&\gamma _{6}=-\gamma _{9}=-\gamma _{10}=-\frac{1}{2\varepsilon
_{2}},  \notag \\
\gamma _{7} &=&\gamma _{8}=-\gamma _{11}=-\gamma _{12}=-\frac{1}{%
2\varepsilon _{1}}.  \notag
\end{eqnarray}%
Now we are in a position to give an exact result for the 2-orbital
susceptibility:

\begin{equation}
\chi _{12}^{(2)}=\frac{\frac{J_{12}^{\perp }}{2}\,\frac{\mathrm{ssh}(\frac{%
E_{+}}{T},\frac{E_{-}}{T})}{E_{+}E_{-}}\,+\frac{\mathrm{sch}(\frac{E_{+}}{T},%
\frac{E_{-}}{T})}{2E_{+}}\,+\frac{\mathrm{csh}(\frac{E_{+}}{T},\frac{E_{-}}{T})%
}{2E_{-}}\,+\frac{\sinh \left( \frac{\varepsilon _{1}}{T}\right) e^{-\Delta
_{P}^{(2)}/T}}{2\varepsilon _{1}}\,+\frac{\sinh \left( \frac{\varepsilon _{2}%
}{T}\right) \,e^{-\Delta _{P}^{(1)}/T}}{2\varepsilon _{2}}}{\mathrm{cch}\left( 
\frac{E_{+}}{T},\frac{E_{-}}{T}\right) \,+\cosh \left( \frac{\varepsilon _{1}%
}{T}\right) \,e^{-\Delta _{P}^{(2)}/T}+\cosh \left( \frac{\varepsilon _{2}}{T%
}\right) \,e^{-\Delta _{P}^{(1)}/T}+e^{-\Delta _{P}^{(1)}/T}\,e^{-\Delta
_{P}^{(2)}/T}}  \label{2-site-final}
\end{equation}

where we denoted

\begin{equation*}
\mathrm{ssh}(x,y)=\sinh \left( x\right) \sinh \left( y\right) ,\;\mathrm{sch}%
(x,y)=\sinh \left( x\right) \cosh \left( y\right) ,\;\mathrm{cch}(x,y)=\cosh
\left( x\right) \cosh \left( y\right)
\end{equation*}

If one neglects the renormalization of eigenvalues by interaction and sets $%
E_{+}=\varepsilon _{1}$, $E_{-}=\varepsilon _{2},$ the Eq.(\ref{2-site-final}%
) is reduced to a simpler form 
\begin{eqnarray}
\chi _{12}^{(2)} &=&\frac{J_{12}^{\perp }}{2\varepsilon _{1}\varepsilon _{2}}%
\,F(\varepsilon _{1},\Delta _{P}^{(1)})\,F(\varepsilon _{2},\Delta
_{P}^{(2)})+  \notag \\
&&\frac{F(\varepsilon _{1},\Delta _{P}^{(1)})}{2\varepsilon _{1}}+\frac{%
F(\varepsilon _{2},\Delta _{P}^{(2)})}{2\varepsilon _{2}},  \notag
\end{eqnarray}%
where $F(\varepsilon ,\Delta _{P})$ is defined in Eq.(\ref{1-full}). In this
approximation the full susceptibility has the same form as Eq.(\ref{2-vir})
after one replaces $\tanh (\varepsilon /T)$ by the function $F(\varepsilon
,\Delta _{P})$. 
\begin{equation}
\chi ^{(2)}=\sum_{i>j}\frac{J_{ij}^{\perp }}{2\varepsilon _{i}\varepsilon
_{j}}\,F(\varepsilon _{i},\Delta _{P}^{(i)})\,F(\varepsilon _{j},\Delta
_{P}^{(j)}).  \label{chi2full}
\end{equation}%
However, when the renormalization of energy levels is taken into account,
such simple replacement $\tanh (E/T)->F(E,\Delta _{P})$ does not work any
more.

\subsection[\hspace{0.7in}Three-orbital problem]{Three-orbital problem}

In an absolutely similar way one can compute the exact susceptibility of a
three-site problem, considering all 64 states of which 8 are even and 56 odd
states are grouped as follows:\newline
\textit{group I}: 
\begin{equation*}
|\mathrm{two-site \;\;even\;\; state}\rangle \otimes|\alpha\rangle
\end{equation*}
with $\alpha=\uparrow(\downarrow)$ corresponding to the first, second or the
third site. There are $4\times 2\times 3=24$ such states. \newline
\textit{group II}: 
\begin{equation*}
|\mathrm{one-site\;\; even\;\; state}\rangle
\otimes|\alpha\rangle\otimes|\alpha^{\prime }\rangle,
\end{equation*}
with an even state on the first, second, or the third site. There are $%
2\times 4\times 3=24$ such states.\newline
\textit{group III}: 
\begin{equation*}
|\alpha,\alpha^{\prime }\alpha^{\prime \prime }\rangle.
\end{equation*}
There are 8 such states.

The eigenvalues $\lambda _{\alpha }^{(0)}$ can be found exactly like in the
previous subsection; they are  equal to (relative to the vacuum state $%
-(\varepsilon _{1}+\varepsilon _{2}+\varepsilon _{3})$) the eigenvalues of
the corresponding even state plus the sum of $\Delta _{p}^{(i)}$ of the
singly occupied sites. For instance: 
\begin{equation*}
|0,2,\alpha \rangle \rightarrow \lambda ^{(0)}=-\mathrm{sgn}(\varepsilon
_{1}-\varepsilon _{2})\,\sqrt{(\varepsilon _{1}-\varepsilon
_{2})^{2}+|J_{12}|^{2}}\,+\Delta _{p}^{(3)},
\end{equation*}%
\begin{equation*}
|2,\alpha ,\alpha ^{\prime }\rangle \rightarrow \lambda ^{(0)}=\varepsilon
_{1}+\Delta _{p}^{(2)}+\Delta _{p}^{(3)},
\end{equation*}%
\begin{equation*}
|\alpha ,\alpha ^{\prime },\alpha ^{\prime \prime }\rangle \rightarrow
\lambda ^{(0)}=\Delta _{p}^{(1)}+\Delta _{p}^{(2)}+\Delta _{p}^{(3)}.
\end{equation*}%
The three-orbital susceptibility $\chi _{1,2,3}^{(3)}$ can be calculated
using the general Eq.(\ref{chi-N}), in which the partition function $%
Z_{0}=Z_{0}(1,2,3)$ is equal to %\begin{widetext}
\begin{eqnarray}
&&Z_{0}(1,2,3)=Z_{0}^{\mathrm{even}}+8\,\cosh (\bar{E}_{+}^{(1,2)})\,\cosh (%
\bar{E}_{-}^{(1,2)})\,e^{-\bar{\Delta}_{p}^{(3)}}  \notag  \label{PF} \\
&&+8\,\cosh (\bar{E}_{+}^{(1,3)})\,\cosh (\bar{E}_{-}^{(1,3)})\,e^{-\bar{%
\Delta}_{p}^{(2)}}+8\,\cosh (\bar{E}_{+}^{(2,3)})\,\cosh (\bar{E}%
_{-}^{(2,3)})\,e^{-\bar{\Delta}_{p}^{(1)}} \\
&+&8\cosh (\bar{\varepsilon}_{1})\,e^{-\bar{\Delta}_{p}^{(2)}}\,e^{-\bar{%
\Delta}_{p}^{(3)}}+8\cosh (\bar{\varepsilon}_{2})\,e^{-\bar{\Delta}%
_{p}^{(1)}}\,e^{-\bar{\Delta}_{p}^{(3)}}+8\cosh (\bar{\varepsilon}_{3})\,e^{-%
\bar{\Delta}_{p}^{(1)}}\,e^{-\bar{\Delta}_{p}^{(2)}}  \notag \\
&+&8e^{-\bar{\Delta}_{p}^{(1)}}\,e^{-\bar{\Delta}_{p}^{(2)}}\,e^{-\bar{\Delta%
}_{p}^{(3)}}.
\end{eqnarray}

In Eq.(\ref{PF}) each line corresponds to the above group of states, $%
E_{\pm} $ is given by Eq.(\ref{E+-}), and $\bar{E},\bar{\Delta}=E/T,\Delta/T$%
. The partition function of the even states $Z^{\mathrm{even}}_{0}=\sum_{%
\mathrm{even\,\,\,states}}e^{-\lambda_{\alpha}^{(0)}/T}$ has been calculated
above for the 3-spin case.

The numerator of Eq.(\ref{chi-N}) $G(1,2,3)=-\sum_{\alpha}e^{-\lambda_{%
\alpha}^{(0)}/T}\,\gamma_{\alpha}$ is found to be the following:

\begin{eqnarray}
&&G(1,2,3)=G^{\mathrm{even}}(1,2,3)+4\,G_{12}\,e^{-\bar{\Delta}%
_{p}^{(3)}}+4\,G_{13}\,e^{-\bar{\Delta}_{p}^{(2)}}+4\,G_{23}\,e^{-\bar{\Delta%
}_{p}^{(1)}}  \notag  \label{G-num} \\
&+&4\,\frac{\sinh (\bar{\varepsilon}_{1})}{\varepsilon _{1}}\,e^{-\bar{\Delta%
}_{p}^{(2)}-\bar{\Delta}_{p}^{(3)}}\,+4\,\frac{\sinh (\bar{\varepsilon}_{2})%
}{\varepsilon _{2}}\,e^{-\bar{\Delta}_{p}^{(1)}-\bar{\Delta}_{p}^{(3)}}\,+4\,%
\frac{\sinh (\bar{\varepsilon}_{3})}{\varepsilon _{3}}\,e^{-\bar{\Delta}%
_{p}^{(1)}-\bar{\Delta}_{p}^{(2)}}\,,
\end{eqnarray}

where 
\begin{eqnarray}
G_{ij}&=&\frac{J_{ij}^{\perp}}{E_{+}^{(i,j)}E_{-}^{(i,j)}}\, \sinh(\bar{E}%
_{+}^{(i,j)})\,\sinh(\bar{E}_{-}^{(i,j)})+ \\
&+&\frac{\sinh(\bar{E}_{+}^{(i,j)})}{E_{+}^{(i,j)}}\,\cosh(\bar{E}%
_{-}^{(i,j)})+ \frac{\sinh(\bar{E}_{-}^{(i,j)})}{E_{-}^{(i,j)}}\,\cosh(\bar{E%
}_{+}^{(i,j)})  \notag
\end{eqnarray}
is (up to a constant factor) the numerator in Eq.(\ref{chi-N}) for the
2-spin problem. Note also that $\sinh(\bar{\varepsilon}_{i})/\varepsilon_{i}$
is (again up to a constant factor) the numerator of the one-spin problem.
The quantity $G^{\mathrm{even}}(1,2,3)=-\sum_{\mathrm{even\,\,\,states}%
}e^{-\lambda_{\alpha}^{(0)}/T}\,\gamma_{\alpha}$ has been calculated above
for the 3-spin problem.

Using Eqs.(\ref{PF}),(\ref{G-num}) one finds the 3-orbital susceptibility
exactly: 
\begin{equation}  \label{chi-3-site}
\chi^{(3)}_{1,2,3}=\frac{G(1,2,3)}{Z_{0}(1,2,3)}.
\end{equation}


\newpage

\begin{thebibliography}{999}
\bibitem{AndersonLoc} P. W. Anderson, Phys. Rev. \textbf{109}, 492 (1958).


\bibitem{MaLee} M. Ma and P. A. Lee, Phys. Rev. B \textbf{32}, 5658 (1985).

\bibitem{KapitulnikKotliar1986} A. Kapitulnik and G. Kotliar, Phys. Rev.
Lett. \textbf{54}, 473 (1985); G. Kotliar and A. Kapitulnik, Phys. Rev. B 
\textbf{33}, 3146 (1986)

\bibitem{BulaSad} L. N. Bulaevskii and M. V. Sadovskii, Pisma ZhETF \textbf{39}%
, 524 (1984); L. N. Bulaevskii and M. V. Sadovskii, J.Low Temp.Phys. \textbf{59%
}, 89 (1985); M. V. Sadovskii, Phys. Rep, \textbf{282}, 225 (1997).

\bibitem{Ghosal2001} A. Ghosal, M. Randeria and N. Trivedi, Phys. Rev. B 
\textbf{65}, 014501 (2001).

\bibitem{atoms} S. Giorgini, L. P. Pitaevskii and S. Stringari, Rev. Mod.
Phys. \textbf{80}, 1215 (2008).

\bibitem{ReviewGoldman} A. Goldman and N. Markovic, Phys. Today \textbf{51},
39 (1998).

\bibitem{Finkelstein1994} A. M. Finkelstein, Physica B \textbf{197}, 636
(1994)

\bibitem{FazioZant} R. Fazio and H. van der Zant, Phys. Rep. \textbf{355},
235 (2001); H. S. J. van der Zant, W. J. Elion, L. J. Geerlings and J. E. Mooij,
Phys. Rev. B~\textbf{54}, 10081 (1996).

\bibitem{Larkin1999} A. I. Larkin, Ann. Phys. (Leipzig) \textbf{8}, 785
(1999).

\bibitem{Efetov1980} K. B. Efetov, ZhETF \textbf{78}, 2017 (1980)
[Sov.Phys.-JETP \textbf{51}, 1015 (1980)]

\bibitem{LarkinOvchinnikov1983} A. I. Larkin and Yu. N. Ovchinnikov, Phys.
Rev. B \textbf{28}, 6281 (1983)

\bibitem{AES} U. Eckern, G. Schoen and V. Ambegaokar, Phys. Rev.B \textbf{30}%
, 6419 (1984)

\bibitem{FeigKor} M. V. Feigel'man, S. E. Korshunov, and A. B. Pugachev,
Pis'ma v ZhETF \textbf{65}, 541 (1997) [JETP Lett. \textbf{65}, 566 (1997).

\bibitem{Orr1986} B. G. Orr, J. M. Jaeger, A. Goldman and C. G. Kuper, Phys.
Rev. Lett. \textbf{56}, 378 (1986).

\bibitem{MFisher1990a} M. P. A. Fisher, G. Grinstein and S. Girvin, Phys.
Rev. Lett. \textbf{64}, 587 (1990).

\bibitem{MFisher1990b} M. P. A. Fisher, Phys. Rev. Lett. \textbf{65}, 923
(1990).

\bibitem{Zimmerman2008} N. M. Zimmerman \textit{et al}, J. Appl. Phys. 104,
033710 (2008), N. M. Zimmerman and W. H. Huber, Phys. Rev. B \textbf{80}, 195304 (2009).

\bibitem{MullerIoffe} M. Mueller and L. B. Ioffe, arxiv:0711.2668

\bibitem{Gerber1997} A. Gerber, A. Milner, G. Deutscher, M. Karpovsky, and
A. Gladkikh, Phys. Rev. Lett. \textbf{78}, 4277 (1997).

\bibitem{Gantmakher1996} V. F. Gantmakher \textit{et al}, JETP \textbf{82},
951 (1996).

\bibitem{BeloborodovLarkinEfetov} I. S. Beloborodov, K. B. Efetov and A. I.
Larkin Phys. Reb. B \textbf{61}, 9145 (2000).

\bibitem{Beloborodov} I. S. Beloborodov, A. V. Lopatin, V. M. Vinokur, and
K. B. Efetov, Rev. Mod. Phys. \textbf{79}, 469 (2007).

\bibitem{Finkelstein1987} A. M. Finkesltein, Pis'ma ZhETF \textbf{45}, 37
(1987) [Sov. Phys. JETP Lett. \textbf{45}, 46 (1987)].

\bibitem{CoulPert} S. Maekawa and H. Fukuyama J. Phys. Soc. Jap. \textbf{51}%
, 1380 (1982); H. Takagi and Y. Kuroda, Sol. St. Comm. \textbf{41}, 643
(1982).

\bibitem{Oreg1999} Yu. Oreg and A. M. Finkelstein, Phys. Rev. Lett. \textbf{%
83}, 191 (1999).

\bibitem{AndersonMuttalibRamakrishnan1983} P. W. Anderson, K. A. Muttalib
and T. V. Ramakrishnan Phys. Rev. B \textbf{23}, 117 (1983).

\bibitem{Coffey} L. Coffey, K. A. Muttalib and K. Levin, Phys. Rev. Lett. 
\textbf{52}, 783 (1984).

\bibitem{Gershenson} M. E. Gershenson, Yu. B. Khavin, D. Reuter, P.
Schafmeister, and A. D. Wieck, Phys. Rev. Lett. \textbf{85}, 1718 (2000).

\bibitem{Kowal1994} D. Kowal and Z. Ovadyahu, Sol. St. Com. \textbf{90}, 783
(1994).

\bibitem{Skvortsov2005} M. A. Skvortsov and M. V. Feigel'man, Phys. Rev.
Lett. \textbf{95}, 057002 (2005).

\bibitem{SpivakZhou1995} B. Z. Spivak and F. Zhou, Phys. Rev. Lett. \textbf{%
74}, 2800 (1995).

\bibitem{GalitskiiLarkin2001} V. M. Galitskii and A. I. Larkin, Phys. Rev.
Lett. \textbf{87}, 087001 (2001).

\bibitem{FeigelmanLarkinSkvortsov2001} M. V. Feigel'man, A. I. Larkin and M.
A. Skvortsov, Phys. Rev. Lett. \textbf{86}, 1869 (2001); Uspekhi Fiz. Nauk 
\textbf{171}, 76 (2001).

\bibitem{AG1959} A.A. Abrikosov and L.P. Gorkov, Sov. Phys. JETP \textbf{8},
(1958) 1090.

\bibitem{Anderson1959} P.W. Anderson, J.Phys.Chem.Solids \textbf{11}, 26
(1959).

%%%% \bibitem{Meir} Y. Dubi, Y. Meir and Y. Avishai, Nature \textbf{449}, 876-880
%%%% (2007).

\bibitem{Matveev1997} K. A. Matveev and A. I. Larkin, Phys. Rev. Lett. 
\textbf{78}, 3749 (1997).

\bibitem{Shepelyansky2002} B. Srinivasan, G. Benenti and D. L. Shepelyansky, 
Phys. Rev. B \textbf{66}, 172506 (2002); J. Lages and D. L. Shepelyansky
Phys. Rev. B \textbf{64}, 094502 (2001). 


\bibitem{Goldman2002} C. Christiansen, L. M. Hernandez and A. M. Goldman,
Phys. Rev. Lett. \textbf{88}, 037004 (2002); K. H. Sarma, B. Tan, K. A.
Parendo and A. M. Goldman, arXiv:0704.0765

\bibitem{BaturinaReview2007} T. I. Baturina, A. Bilusic, A. Yu. Mironov, V.
M. Vinokur, M. R. Baklanov and C. Strunk, Physics C \textbf{468}, 316 (2007).

\bibitem{Pratap} S. P. Chockalingam, M. Chand, A. Kamlapure, J. Jesudasan,
A. Mishra, V. Tripathi and P. Raychaudhuri, Phys. Rev. B \textbf{79}, 094509
(2009).

\bibitem{Sacepe2007} B. Sacepe, PhD Thesis, CEA Grenoble (2007).

\bibitem{TiN-STM} B. Sacepe, C. Chapelier, T. I. Baturina, V. M. Vinokur, M.
R. Baklanov, and M. Sanquer Phys. Rev. Lett. \textbf{101}, 157006 (2008).


\bibitem{Shahar2004} G. Sambandamurthy, L. W. Engel, A. Johansson, and D.
Shahar, Phys. Rev. Lett. \textbf{92}, 107005 (2004).

\bibitem{Baturina2007a} T. I. Baturina, A. Yu. Mironov, V. M. Vinokur and C.
Strunk, Phys. Rev. Lett. \textbf{99}, 257003 (2007).

\bibitem{Shahar1992} D. Shahar and Z. Ovadyahu, Phys. Rev. B \textbf{46},
10917 (1992).

\bibitem{Zvi-Insulator} Z. Ovadyahu, private communication.

\bibitem{Anisimov2009} V. I. Anisimov, private communication.

\bibitem{Baturina2008} T. I. Baturina, invited talks at Moriond Conference
(2008) and NanoPiter Conference (2008).

\bibitem{Serret2002} E. Serret, Ph.D. Thesis, CNRS, Grenoble (2002).

\bibitem{Wu05} W. Wu and E. Bielejec, cond-mat/0511121.

\bibitem{Shahar2005} G. Sambandamurthy, L. W. Engel, A. Johansson, E. Peled,
and D. Shahar, Phys. Rev. Lett. \textbf{94}, 017003 (2005).

\bibitem{Steiner2005} M. A. Steiner and A. Kapitulnik, Physica C \textbf{422}%
, 16 (2005)


\bibitem{Baturina2004} T. I. Baturina, Islamov D.R., Bentner J., Strunk C.,
Baklanov M.R., Satta A., JETP Letters \textbf{79} 337 (2004).

\bibitem{Baturina2007} T. I. Baturina C. Strunk, M. R. Baklanov and A.
Satta, Phys. Rev. Lett. \textbf{98}, 127003 (2007).

\bibitem{Wu01} E. Bielejec, J. Ruan and W. Wu, Phys. Rev. Lett. \textbf{87},
036801 (2001).


\bibitem{Adams2001} V. Yu. Butko and P. W. Adams, Nature, \textbf{409}, 161
(2001).

\bibitem{Valles2009} H. Q. Nguyen, S. M. Hollen, M. D. Stewart, Jr., 
J. Shainline,  A. Yin, J. M. Xu and J. M. Valles,    Phys. Rev. Lett. \textbf{103}, 
157001 (2009)

\bibitem{Wu012} E. Bielejec, J. Ruan and W. Wu, Phys Rev. B \textbf{63},
100502(R) (2001).

\bibitem{Adams2009} Y.M. Xiong, A.B. Karki, D.P. Young and P. W. Adams,
arXiv:0901.1873.

\bibitem{Gantmakher1998} V. F. Gantmakher, Golubkov M.V., Dolgopolov V.T.,
Tsydynzhapov G.E., Shashkin A.A., JETP Letters \textbf{\ 68}, 363 (1998); V.
F. Gantmakher, Physics-Uspekhi \textbf{41}, 214 (1998).

\bibitem{Steiner2008} M. A. Steiner, N. P. Breznay and A. Kapitulnik Phys.
Rev. B $\mathbf{77}$, 212501 (2008).

\bibitem{Dubi06} Y. Dubi, Y. Meir and Y. Avishai, Phys. Rev. B \textbf{73},
054509 (2006).

\bibitem{Jumps} B. L. Altshuler, V. E. Kravtsov, I. V . Lerner and I. L.
Aleiner, Phys. Rev. Lett. \textbf{102}, 176802 (2009).

\bibitem{SacepeShahar09} M. Ovadia, B. Sacepe and D. Shahar, Phys. Rev.
Lett. \textbf{102}, 176802 (2009).

\bibitem{Sanquer1996} F. Ladieu, M. Sanquer and J. P. Bouchaud, Phys. Rev. B 
\textbf{53}, 973 (1996).

\bibitem{Larkin1965} A. I. Larkin, ZhETF \textbf{48}, 232 (1965)
[Sov.Phys.JETP \textbf{21}, 153 (1965)].

\bibitem{Valles2e} M. D. Stewart Jr, Aijun Yin, J. M. Xu, J. M. Valles Jr, 
\textit{Science}, \textbf{318}, 1273 (2007); arXiv:0712.1076

\bibitem{FIKY2007} M. V. Feigel'man, L. B. Ioffe, V. E. Kravtsov and E. A.
Yuzbashyan, Phys. Rev. Lett. \textbf{98}, 027001 (2007).


\bibitem{Anderson1958} P. W. Anderson, Phys. Rev. \textbf{112}, 1900 (1958).

\bibitem{Chalker1990} J. T. Chalker, Physica A, \textbf{167}, 253 (1990).

\bibitem{KrMut1997} V. E. Kravtsov and K. A. Muttalib, Phys. Rev. Lett., 
\textbf{79} 1913 (1997).

\bibitem{MirlinReview2000} A. D. Mirlin, Phys. Rep., \textbf{326}, 259
(2000).

\bibitem{UltraSmall} Jan von Delft, Annalen der Physik (Leipzig), \textbf{10}%
, 219-276 (2001); cond-mat/0101021.

\bibitem{Kurland00} I. L. Kurland, I. L. Aleiner and B. L. Altshuler, Phys.
Rev. B \textbf{62}, 14886 (2000).

\bibitem{nu} A.MacKinnon. J.Phys.C \textbf{6},2511(1994); T.Ohtsuki,
K.Slevin and T.Kawarabayashi, Ann. Phys. (Leipzig), \textbf{8}, 655 (1999).

\bibitem{CueKra} E. Cuevas and V. E. Kravtsov, Phys. Rev. B \textbf{76},
235119 (2007).

\bibitem{MirlinNewRep} F. Evers and A. D. Mirlin, \textit{Anderson
transitions}, Rev. Mod. Phys. \textbf{80}, 1355 (2008); arXiv:0707.4378.

\bibitem{Roemer} A. Rodriguez, L. J. Vasquez and R. A. Roemer, private
communication

\bibitem{Ioffe1985} L. B. Ioffe, I. R. Sagdeev and V. M. Vinokur, J. Phys.
C, Solid State Phys. \textbf{18}, L641 (1985).

\bibitem{Mirlin2002} A. Mildenberger, F. Evers and A. D. Mirlin, Phys. Rev.
B \textbf{66}, 033109 (2002).

\bibitem{MirlinFyodorov97} A. D. Mirlin and Yan V. Fyodorov, Phys. Rev. B 
\textbf{56}, 13393 (1997).

\bibitem{ZviDan} Z. Ovadyahu and D. Shahar, private communications.

\bibitem{Zvi1986} Z. Ovadyahu, J. Phys. C \textbf{19}, 5187 (1986)


\bibitem{IoffeFeigelman1985} L. B. Ioffe and M. V. Feigel'man, ZhETF \textbf{%
89}, 654 (1985) [Sov.Phys.JETP \textbf{62}, 376 (1985)]

\bibitem{DFI1990} Vik. S. Dotsenko, M. V. Feigel'man, and L. B. Ioffe, Sov.
Sci. Rev. \textbf{15}, edited by I. M. Khalatnikov, Harwood Publishers, 1990.

\bibitem{IoffeLarkin1981} L. B. Ioffe and A. I. Larkin, ZhETF \textbf{81},
707 (1981) [Sov. Phys. JETP \textbf{54}, 378 (1981)].

\bibitem{DeGennesBook} P. De Gennes, "Superconductivity of metals and
alloys", Westview Press (1999).

\bibitem{LarkinVarlamovBook} A. I. Larkin and A. A. Varlamov, in \textit{%
Theory of Fluctuations in Superconductors}, Oxford University Press, New
York, 2002.

\bibitem{Virial} A. I. Larkin and D. E Khmelnitsky, ZhETF \textbf{58}, 1789
(1970).

\bibitem{FeigTsvel1979} A. M. Tsvelik and M. V. Feigelman, ZhETF \textbf{76}%
, 2249 (1979).

\bibitem{PRBM} A. D. Mirlin, Y. V. Fyodorov, F. M. Dittes et al. Phys.Rev.E 
\textbf{54}, 3221 (1996).

\bibitem{IoffeMezard2009} L. B. Ioffe and M. M\'ezard, cond-mat
arXiv:0909.2263 (2009).

\bibitem{FeigelIoffeMezard} M. V. Feigel'man, L. B. Ioffe and M. M\'ezard, to
be published.

\bibitem{Sacepe2} B. Sacepe, T. Dubouchet, C. Chapelier, M. Sanquer, M.
Ovadia, D. Shahar, M. V. Feigel'man and L. B. Ioffe, to be published.

\bibitem{Nozieres} P. Nozieres and C. T. \ De Dominicis, Phys.Rev. \textbf{%
178}, 1097 (1969).

\bibitem{Deutscher} G. Deutscher, Rev. Mod. Phys. \textbf{77}, 109 (2005).

\bibitem{BTK} G. E. Blonder, M. Tinkham and T. M. Klapwijk, Phys. Rev. B 
\textbf{25}, 4515 (1982).

\bibitem{Bezuglyi} E. V. Bezuglyi, E. N. Bratus', V. S. Shumeiko, G. Wendin
and H. Takayanagi, Phys. Rev. B \textbf{62}, 14439 (2000).

\bibitem{Golubov03} Y. Tanaka, A. A. Golubov, and S. Kashiwaya, Phys. Rev. B 
\textbf{68}, 054513 (2003).

\bibitem{Dubouchet10} T. Dubouchet, C. Chapelier, M. Sanquer, B. Sacepe, M.
Ovadia and D. Shahar, to be published.

\bibitem{Basov} D. Basov, T. Timusk, B. Dabrovski and J.D.Jorgenson, Phys.
Rev. B \textbf{50}, 3511 (1994).

\bibitem{IoffeMillis} L. B. Ioffe and A. J. Millis, \textit{Science}, 
\textbf{285}, 1241 (1999).

\bibitem{atoms2} J.K. Chin, D.E. Miller, Y. Liu, C. Stan, W. Setiawan, C.
Sanner, K. Xu, and W. Ketterle, \textit{Nature}, \textbf{443}, 961 (2006).

\bibitem{Aspect} J. Billy, V. Josse, Z. Zuo, A. Bernard, B. Hambrecht, P.
Lugan, D. Clament, L. Sanchez-Palencia, P. Bouyer and A. Aspect, \textit{%
Nature} \textbf{453}, 891 (2008); G. Roati, C. D'Errico, L. Fallani, M.
Fattori, C. Fort, M. Zaccanti, G. Modugno, M. Modugno and M. Inguscio 
\textit{Nature} \textbf{453}, 895 (2008).

\bibitem{Muller2009} M. Mueller, cond-mat arXiv:0909.2260 (2009).

\bibitem{KowalOvadyahu2} D. Kowal and Z. Ovadyahu, Physica C \textbf{468},
322 (2008).

\bibitem{OngNernst} Z. A. Xu, N. P. Ong, Y. Wang, T. Kakeshita and S.
Uchida, Nature, \textbf{\ 406}, 486 (2000); Phys. Rev. Lett. \textbf{88},
257003 (2002).

\bibitem{Spathis} P. Spathis, H. Aubin, A. Pourret and K. Behnia, Europhys.
Lett. \textbf{83}, 57005 (2008); A. Pourret, P. Spathis, H. Aubin and K.
Behnia, arxiv:0902.2732

\bibitem{Serbin} M. N. Serbin, M. A. Skvortsov, A. A. Varlamov and V. M.
Galitskii, Phys. Rev. Lett. \textbf{102}, 067001 (2009).

\bibitem{Karen} K. Michaeli and A. M. Finkelstein, Europhys. Lett. \textbf{86%
}, 27007 (2009); arxiv:0912.4219.

\bibitem{AubinNernst} A. Pourret, H. Aubin, J. Lesueur, C. A.
Marrache-Kikuchi, L. Berge, L. Dumoulin and K. Behnia, Nature Physics, 
\textbf{2}, 683 (2006); Phys. Rev. B \textbf{76}, 214504 (2007).


\bibitem{Podolsky07} D. Podolsky, S. Raghu and A. Vishwanath, Phys. Rev.
Lett. \textbf{99}, 117004 (2007).

\bibitem{Steiner2004} M. A. Steiner, G. Bobinger and A. Kapitulnik, Phys.
Rev. Lett. \textbf{94}, 107008 (2005).


\bibitem{TimuskReview} T. Timusk and B. Statt, Rep. Prog. Phys. \textbf{62}, 61
(1999)

\bibitem{Kapitulnik2001} C. Howald, P. Fournier and A. Kapitulnik, Phys. Rev. B 64, 
100504(R) (2001)

\bibitem{Davis2001} S. H. Pan, J. P. O'Neal1, R. L. Badzey \textit{et al}, Nature  
\textbf{413}, 282 (2001).

\bibitem{Davis2002} K. M. Lang, V. Madhavan, J. E. Hoffman, E. W. Hudson, H.
Eisaki, S. Uchida and J. C. Davis, Nature, \textbf{415}, 412 (2002).

\bibitem{Yazdani2007} K. K. Gomes, A. N. Pasupathy, A. Pushp, S. Ono, Y. Ando and 
A Yazdani, Nature \textbf{447}, 569 (2007).  

\bibitem{Corson1999} J. Corson, R. Mallozzi1, J. Orenstein1, J. N. Eckstein 
and I. Bozovic, Nature \textbf{398}, 221 (1999).

\bibitem{Wang2006} Y. Wang, L. Li and N. P. Ong, Phys. Rev. B \textbf{73},
024510 (2006).

\bibitem{Yuli2009} O. Yuli, I. Asulin, Y. Kalcheim, G. Koren, and O. Millo
Phys. Rev. Lett. \textbf{103}, 197003 (2009). 

\bibitem{Geshkenbein1997} V. B. Geshkenbein, L. B. Ioffe and A. I. Larkin, Phys. Rev. 
B \textbf{55}, 3173 (1997). 

\bibitem{Galitski2009} V. Galitski and S. Sachdev, Phys. Rev. Lett. \textbf{79}, 
134512 (2009)


\end{thebibliography}
\end{document}